\documentclass[paper]{JHEP3}
\usepackage{epsfig,subfigure}
\usepackage{graphicx}
\usepackage{amssymb}
\usepackage{amsmath}
\usepackage{booktabs,multirow,tabularx}
\usepackage{slashed}
\usepackage{cite}
\usepackage{caption}
\usepackage{float}
\usepackage{placeins}
\usepackage{rotating}
\usepackage{lscape}

\def\db#1{\bar D_{#1}}

\newcommand{\bqa}{\begin{eqnarray}}
\newcommand{\eqa}{\end{eqnarray}}

\def\beq{\begin{equation}}
\def\eeq{\end{equation}}
\def\beqn{\begin{eqnarray}}
\def\eeqn{\end{eqnarray}}
\def\abs#1{\left|#1\right|}

\def\remove#1#2{#1\hspace{-#2truecm}\backslash}

\newcommand\sss{\scriptscriptstyle}
\newcommand\mydot{\!\cdot\!}
\newcommand\ep{\epsilon}
\newcommand\half{\frac{1}{2}}

\newcommand\as{\alpha_{\sss S}}
\newcommand\gs{g_{\sss S}}

\newcommand\gW{g_{\sss W}}
\newcommand\aem{\alpha}
\newcommand\MadFKS{{\sc\small MadFKS}}
\newcommand\CutTools{{\sc\small CutTools}}
\newcommand\MadLoop{{\sc\small MadLoop}}
\newcommand\MadGraph{{\sc\small MadGraph}}
\newcommand\madgraph{{\sc\small MadGraph}}
\newcommand\MadGraphq{{\sc\small MadGraph4}}
\newcommand\MadGraphf{{\sc\small MadGraph5}}
\newcommand\madwidth{{\sc\small MadWidth}}
\newcommand\madweight{{\sc\small MadWeight}}
\newcommand\syscalc{{\sc\small SysCalc}}
\newcommand\asperge{{\sc\small AsperGe}}

\newcommand\delphes{{\sc\small Delphes}}
\newcommand\HELAS{{\sc\small HELAS}}
\newcommand\helas{\HELAS}
\newcommand\mcatnlo{{\sc\small MC@NLO}}
\newcommand\amcatnlo{{\sc\small aMC@NLO}}
\newcommand\aNLO{{\sc\small MadGraph5\_aMC@NLO}}

\newcommand\FJ{{\sc\small FastJet}}
\newcommand\HW{{\sc\small Herwig}}
\newcommand\HWs{{\sc\small HERWIG6}}
\newcommand\HWpp{{\sc\small Herwig++}}
\newcommand\PY{{\sc\small Pythia}}
\newcommand\PYs{{\sc\small Pythia6}}
\newcommand\PYe{{\sc\small Pythia8}}
\newcommand\Sherpa{{\sc\small SHERPA}}
\newcommand\Alpgen{{\sc\small Alpgen}}
\newcommand\Madspin{{\sc\small MadSpin}}
\newcommand\MadSpin{{\sc\small MadSpin}}
\newcommand\ML{\MadLoop}
\newcommand\MLq{{\sc\small MadLoop4}}
\newcommand\MLf{{\sc\small MadLoop5}}
\newcommand\ALOHA{{\sc\small ALOHA}}

\newcommand\UFO{{\sc\small UFO}}
\newcommand{\FeynRules}{{\sc \small FeynRules}}
\newcommand{\feynrules}{\FeynRules}
\newcommand\OL{{\sc\small OpenLoops}}
\newcommand\PJF{{\sc\small PJFry++}}
\newcommand\IREGI{{\sc\small IREGI}}
\newcommand\prompt{{\tt MG5\_aMC>}}
\newcommand\Lcut{L-cut}

\newcommand\qloop{\ell}
\newcommand\bqloop{\bar{\ell}}
\newcommand\tqloop{\tilde{\ell}}
\newcommand\ident{{\cal I}}
\newcommand\amp{{\cal A}}
\newcommand\ampnt{\amp^{(n,0)}}
\newcommand\ampnl{\amp^{(n,1)}}

\newcommand\ampnX{\amp^{(n,{\sss\rm X})}}

\newcommand\muF{\mu_{\sss F}}
\newcommand\muR{\mu_{\sss R}}
\newcommand\QES{Q_{\sss\rm ES}}
\newcommand\muFz{\mu_{{\sss F}0}}
\newcommand\muRz{\mu_{{\sss R}0}}
\newcommand\QESz{Q_{{\sss\rm ES}0}}
\newcommand\ffF{f_{\sss F}}
\newcommand\ffFd{\ffF^\downarrow}
\newcommand\ffFc{\ffF^c}
\newcommand\ffFu{\ffF^\uparrow}
\newcommand\ffR{f_{\sss R}}
\newcommand\ffRd{\ffR^\downarrow}
\newcommand\ffRc{\ffR^c}
\newcommand\ffRu{\ffR^\uparrow}
\newcommand\ffQES{f_{\sss\rm ES}}
\newcommand\bu{\bar{u}}
\newcommand\bd{\bar{d}}
\newcommand\bb{\bar{b}}
\newcommand\bt{\bar{t}}
\newcommand\bq{\bar{q}}

\newcommand\diag{{\cal C}}
\newcommand\idiag{C}
\newcommand\unr{{\rm\sss U}}

\newcommand\pt{p_{\sss T}}
\newcommand\kt{k_{\sss T}}
\newcommand\ptcut{p_{\sss T}^{\rm (cut)}}
\newcommand\Ht{H_{\sss T}}
\newcommand\NC{N_{\sss c}}

\newcommand\clH{{\mathbb H}}
\newcommand\clS{{\mathbb S}}
\newcommand\lum{{\cal L}}
\newcommand\luma{\lum^{(\alpha)}}
\newcommand\lumMC{\lum^{\sss\rm (MC)}}
\newcommand\GenMC{{\cal F}_{\mbox{\tiny MC}}}

\newcommand\GenNLO{{\cal F}_{\mbox{\tiny MC@NLO}}}

\newcommand\GenFxFx{{\cal F}_{\mbox{\tiny FxFx}}}

\newcommand\showxi{\xi^{(l)}}
\newcommand\showz{z^{(l)}}
\newcommand\showxiij{\showxi_{ij}}
\newcommand\showzij{\showz_{ij}}
\newcommand\stepfMC{\Theta^{\sss\rm (MC)}}
\newcommand\RED{{\rm Red}}
\newcommand\OPP{{\rm OPP}}

\newcommand\nonRt{{{\rm non}-R_2}}

\newcommand\tree{{\cal T}}
\newcommand\tV{\widetilde{V}}

\newcommand\Wa{W^{(\alpha)}}
\newcommand\conf{{\cal K}}

\newcommand\confnpo{\conf_{n+1}}

\newcommand\confnpoE{\conf_{n+1}^{(E)}}
\newcommand\confnpoC{\conf_{n+1}^{(C)}}
\newcommand\confnpoS{\conf_{n+1}^{(S)}}
\newcommand\confnpoSC{\conf_{n+1}^{(SC)}}

\newcommand\confHn{\conf_n^{(\clH)}}
\newcommand\confSn{\conf_n^{(\clS)}}

\newcommand\confHip{\conf_{p+i}^{(\clH)}}
\newcommand\confSip{\conf_{p+i}^{(\clS)}}
\newcommand\confHNp{\conf_{p+N}^{(\clH)}}
\newcommand\confSNp{\conf_{p+N}^{(\clS)}}
\newcommand\sigmaLO{\sigma^{\sss\rm (LO)}}
\newcommand\sigmaNLO{\sigma^{\sss\rm (NLO)}}
\newcommand\sigmaNLOa{\sigma^{{\sss (\rm NLO},\alpha{\sss )}}}
\newcommand\sigmaNLOE{\sigma^{\sss {(\rm NLO},E)}}
\newcommand\sigmaNLOS{\sigma^{\sss {(\rm NLO},S)}}
\newcommand\sigmaH{\sigma^{\sss (\clH)}}
\newcommand\sigmaS{\sigma^{\sss (\clS)}}

\newcommand\bsigmaHn{\bar{\sigma}_n^{\sss (\clH)}}
\newcommand\bsigmaSn{\bar{\sigma}_n^{\sss (\clS)}}
\newcommand\bsigmaHip{\bar{\sigma}_{p+i}^{\sss (\clH)}}
\newcommand\bsigmaSip{\bar{\sigma}_{p+i}^{\sss (\clS)}}
\newcommand\bsigmaHNp{\bar{\sigma}_{p+N}^{\sss (\clH)}}
\newcommand\bsigmaSNp{\bar{\sigma}_{p+N}^{\sss (\clS)}}
\newcommand\hsigmaHip{\hat{\sigma}_{p+i}^{\sss (\clH)}}
\newcommand\hsigmaSip{\hat{\sigma}_{p+i}^{\sss (\clS)}}
\newcommand\sigmaMC{\sigma^{\sss\rm (MC)}}

\newcommand\sigmaMCz{\sigma^{{\sss (\rm MC},0{\sss )}}}
\newcommand\sigmaMCD{\sigma^{{\sss (\rm MC},{\sss D)}}}
\newcommand\hsigmaMC{\hat{\sigma}^{\sss\rm (MC)}}
\newcommand{\meas}{\chi}
\newcommand\Gfun{{\cal G}}

\newcommand\xii{\xi_i}
\newcommand\yij{y_{ij}}

\newcommand\Sfun{{\cal S}}
\newcommand\Sfunij{\Sfun_{ij}}
\newcommand\stepf{\Theta}
\newcommand\phsp{d\phi}
\newcommand\phspn{\phsp_{n}}
\newcommand\phspnpo{\phsp_{n+1}}

\newcommand\Phsp{\Phi}

\newcommand\Phspnpo{\Phsp^{(n+1)}}
\newcommand\isubrmv{\remove{i}{0.125}}

\newcommand\FKSpairs{{\cal P}_{\sss\rm FKS}}
\newcommand\FKSpairsred{\overline{{\cal P}}_{\sss\rm FKS}}
\newcommand\proc{r}
\newcommand\procB{r_{\sss B}}
\newcommand\procR{r_{\sss R}}
\newcommand\allproc{{\cal R}}
\newcommand\allprocnpo{\allproc_{n+1}}
\newcommand\allprocn{\allproc_{n}}
\newcommand\symm{\varsigma}

\newcommand\symmnpoij{\symm_{ij}^{(n+1)}}
\newcommand\ampsq{{\cal M}}

\newcommand\ampsqnt{\ampsq^{(n,0)}}
\newcommand\ampsqnpot{\ampsq^{(n+1,0)}}

\newcommand\muSimn{\mu_{i,{\min}}^{(\clS)}}
\newcommand\muSimx{\mu_{i,{\max}}^{(\clS)}}
\newcommand\muHimn{\mu_{i,{\min}}^{(\clH)}}
\newcommand\muHimx{\mu_{i,{\max}}^{(\clH)}}

\newcommand{\perc}{\%}

\newcommand{\QME}{Q_\text{cut}^\text{ME}}
\newcommand{\Qmatch}{Q_\text{match}}
\newcommand{\Qhardest}{Q_\text{hardest}^\text{PS}}
\newcommand{\QMElow}{Q_\text{softest}^\text{ME}}

\preprint{
 CERN-PH-TH/2014-064\\
 CP3-14-18, LPN14-066\\
 MCNET-14-09, ZU-TH 14/14
 }
\title{The automated computation of tree-level and next-to-leading
order differential cross sections, and their matching to parton shower 
simulations
}
\author{J.~Alwall$^a$, R.~Frederix$^b$, S.~Frixione$^b$,
V.~Hirschi$^c$, F.~Maltoni$^d$, O.~Mattelaer$^d$, 
H.-S.~Shao$^e$, T.~Stelzer$^f$, P.~Torrielli$^g$,
M.~Zaro$^{hi}$\\
 $^a$ Department of Physics, National Taiwan
 University, Taipei 10617, Taiwan\\
 $^b$ PH Department, TH Unit, CERN, CH-1211 Geneva 23, Switzerland\\
 $^c$ SLAC National Accelerator Laboratory, 2575 Sand Hill Road,\\
 $\phantom{^e}$ Menlo Park, CA 94025-7090 USA\\
 $^d$ CP3, Universit\'e Catholique de Louvain, B-1348 Louvain-la-Neuve, 
 Belgium\\
 $^e$ Department of Physics and State Key Laboratory of Nuclear Physics 
 and Technology,\\
 $\phantom{^e}$ Peking University, Beijing 100871, China\\
 $^f$ University of Illinois, Urbana, IL 61801-3080, USA\\
 $^g$ Physik-Institut, Universit\"at Z\"urich, Winterthurerstrasse 190, 
8057 Zurich, Switzerland\\
 $^h$ Sorbonne Universit\'es, UPMC Univ. Paris 06, UMR 7589, LPTHE,\\
 $\phantom{^h}$ F-75005, Paris, France\\
 $^i$ CNRS, UMR 7589, LPTHE, F-75005, Paris, France\\
}

\abstract{
We discuss the theoretical bases that underpin the automation of the
computations of tree-level and next-to-leading order cross sections,
of their matching to parton shower simulations, and of the merging
of matched samples that differ by light-parton multiplicities.
We present a computer program, \aNLO, capable of handling all these
computations -- parton-level fixed order, shower-matched, merged -- 
in a unified framework whose defining features are flexibility,
high level of parallelisation, and human intervention limited to 
input physics quantities. We demonstrate the potential of the program
by presenting selected phenomenological applications relevant to the LHC
and to a 1-TeV $e^+e^-$ collider.
While next-to-leading order results are restricted to QCD corrections
to SM processes in the first public version, we show that from the user
viewpoint no changes have to be expected in the case of corrections due
to any given renormalisable Lagrangian, and that the implementation
of these are well under way.
}

\begin{document}

\section{Introduction\label{sec:intro}}
Quantum Chromo Dynamics is more than forty years old~\cite{Gross:1973id,
Politzer:1973fx}, and perturbative calculations of observables beyond the 
leading order are almost as old, as is clearly documented in several 
pioneering works (see e.g.~refs.~\cite{Appelquist:1973uz,Sterman:1977wj,
Buras:1977qg,Bardeen:1978yd,Altarelli:1978id,Celmaster:1980ji,Ellis:1980wv}), 
where the implications of asymptotic freedom had been quickly 
understood. The primary motivation for such early works was
a theoretical one, stemming from the distinctive features of QCD
(in particular, the involved infrared structure, and the fact that
its asymptotic states are not physical), which imply the need of 
several concepts (such as infrared safety, hadron-parton duality,
and the factorisation of universal long-distance effects) that come
to rescue, and supplement, perturbation theory. On the other hand,
the phenomenological necessity of taking higher-order effects into 
account was also acknowledged quite rapidly, in view of the structure
of jet events in $e^+e^-$ collisions and of the extraction of
$\as$ from data.

Despite this very early start, the task of computing observables
beyond the Born level in QCD has remained, until very recently,
a highly non-trivial affair: the complexity of the problem, due
to both the calculations of the (tree and loop) matrix elements
and the need of cancelling the infrared singularities arising from
them, has generated a very significant theoretical activity by
a numerous community. More often than not, different cases
(observables and/or processes) have been tackled in different
manners, with the introduction of ad-hoc solutions. This situation
has been satisfactory for a long while, given that beyond-Born 
results are necessary only when precision is key (and, to a lesser extent,
when large $K$ factors are relevant), and when many hard and well-separated
jets are crucial for the definition of a signature; these conditions have
characterized just a handful of cases in the past, especially in hadronic
collisions (e.g., the production of single vector bosons, jet pairs,
or heavy quark pairs). 

The advent of the LHC has radically changed the picture since, in
a still relatively short running time, it has essentially turned
hadronic physics into a high-precision domain, and one where events
turning up in large-$\pt$ tails are in fact not so rare, in spite of
being characterised by small probabilities. Furthermore, the absence
so far of clear signals of physics beyond the Standard Model implies an
increased dependence of discovery strategies upon theoretical predictions
for known phenomena. These two facts show that presently the phenomenological
motivations are extremely strong for higher-order and multi-leg computations 
of all observables of relevance to LHC analyses.

While a general solution is not known for the problem of computing exactly
the perturbative expansion for any observable up to an arbitrarily large 
order in $\as$, if one restricts oneself to the case of the first order 
beyond the Born one (next-to-leading order, NLO henceforth), then such
a solution does actually exist; in other words, there is no need for
ad-hoc strategies, regardless of the complexity of the process under 
study. This remarkable fact results from two equally important theoretical
achievements. Namely, a universal formalism for the cancellation of 
infrared singularities~\cite{Frixione:1995ms,Catani:1996vz,Frixione:1997np,
Kosower:1997zr,Campbell:1998nn}, and a technique for the algorithmic 
evaluation of renormalised one-loop amplitudes~\cite{Bern:1994zx,
delAguila:2004nf,Bern:2005cq,Ossola:2006us,Anastasiou:2006jv,
Anastasiou:2006gt,Ellis:2007br,Ellis:2008ir,Giele:2008ve,
Cascioli:2011va,Mastrolia:2012bu}, both of which 
must work in a process- and observable independent manner. At the NLO 
(as opposed to the NNLO and beyond) there is the further advantage 
that fixed-order computations can be matched to parton-shower event 
generators (with either the MC@NLO~\cite{Frixione:2002ik} or the
POWHEG~\cite{Nason:2004rx} method -- see also refs.~\cite{Dobbs:2001gb,
Chen:2001nf,Kurihara:2002ne,Nagy:2005aa,Bauer:2006mk,Nagy:2007ty,
Giele:2007di,Bauer:2008qh,Hoeche:2011fd,Hamilton:2013fea}
for early, less-developed, or newer related approaches), 
thus enlarging immensely the scope of the former, 
and increasing significantly the predictive power of the latter.

It is important to stress that while so far we have explicitly
considered the case of QCD corrections, the basic theoretical
ideas at the core of the subtraction of infrared singularities,
of the computation of one-loop matrix elements, and of the matching
to parton showers will require no, or minimal, changes in the context
of other renormalisable theories, QCD being essentially a worst-case
scenario. This is evidently true for tree-level multi-leg computations,
as is proven by the flexibility and generality of tools such as
\MadGraphf~\cite{Alwall:2011uj}, that is able to deal with basically 
any user-defined Lagrangian.

In summary, there are both the phenomenological motivations and the
theoretical understanding for setting up a general framework for the
computation of (any number of) arbitrary observables in an arbitrary
process at the tree level or at the NLO, with or without the 
matching to parton  showers. We believe that the most effective way of 
achieving this goal is that of automating the whole procedure, whose 
technological challenges can be tackled with high-level computer 
languages capable of dealing with abstract concepts, and which 
are readily available.

The aim of this paper is that of showing that the programme sketched
above has been realised, in the form of a fully automated and public 
computer code, dubbed \aNLO. As the name suggests, such a code merges
in a unique framework all the features of 
\MadGraphf\ and of \amcatnlo, and thus supersedes both of them (and must
be used in their place). It also includes several new capabilities
that were not available in these codes, most notably those relevant to
the merging of event samples with different light-parton multiplicities.
We point out that \aNLO\ contains {\em all} ingredients (the very few
external dependencies that are needed are included in the package)
that are necessary to perform an NLO, possibly plus shower 
(with the MC@NLO formalism), computation:
it thus is the first public (since Dec.~16$^{{\rm th}}$, 
2013) code, and so far also the only one, with these characteristics.
Particular attention has been devoted to the fact that calculations
must be doable by someone who is not familiar with Quantum Field Theories, 
and specifically with QCD. We also show, in general as well as with
explicit examples, how the construction of our framework lends itself 
naturally to its extension to NLO corrections in theories other than QCD, in
keeping with the fact that such a flexibility is one of the standard features
of the tree-level computations which were so far performed by \MadGraph, and
that has been inherited by \aNLO.

It is perhaps superfluous to point out that approaches to automated
computations constitute a field of research which has a long history,
but which has had an exponential growth in the past few years, out
of the necessities and possibilities outlined above. The number of
codes which have been developed, either restricted to leading order 
(LO henceforth) predictions~\cite{Stelzer:1994ta,Caravaglios:1995cd,
Yuasa:1999rg,Kanaki:2000ey,Moretti:2001zz,Krauss:2001iv,Mangano:2002ea,
Fujimoto:2002sj,Boos:2004kh,Tsuno:2006cu,Cafarella:2007pc,Kilian:2007gr,
Alwall:2007st,Gleisberg:2008fv,Alwall:2011uj,Belyaev:2012qa}, 
or including NLO capabilities~\cite{Hahn:1998yk,Hahn:2000kx,Gleisberg:2007md,
Berger:2008sj,Frederix:2008hu,Giele:2008bc,Czakon:2009ss,Frederix:2009yq,
Hasegawa:2009tx,Hoche:2010pf,Alioli:2010xd,Mastrolia:2010nb,Frederix:2010cj,
Becker:2010ng,Hirschi:2011pa,Bevilacqua:2011xh,Becker:2011vg,Cullen:2011ac,
Binoth:2011xi,Agrawal:2011tm,Bern:2011ep,Actis:2012qn,
Badger:2012pg,GoncalvesNetto:2012yt,Badger:2013vpa,Bern:2013pya,
vanDeurzen:2013saa,Cullen:2014yla,Peraro:2014cba} is truly staggering. 
The level of automation and the physics scope of such codes, not to mention 
other, perhaps less crucial, characteristics, is extremely diverse, and we 
shall make no attempt to review this matter here.

We have organized this paper as follows. In sect.~\ref{sec:basis}, 
we review the theoretical bases of our work, and discuss
new features relevant to future developments. In sect.~\ref{sec:howto}
we explain how computations are performed. Section~\ref{sec:res}
presents some illustrative results, relevant to a variety of situations:
total cross sections at the LHC and future $e^+e^-$ colliders,
differential distributions in $pp$ collisions, and benchmark
one-loop pointwise predictions, in the Standard Model and beyond.
We finally conclude in sect.~\ref{sec:conc}. Some technicalities are 
reported in appendices~\ref{sec:depen} to~\ref{sec:tpc}.

\section{Theoretical bases and recent progress\label{sec:basis}}
At the core of \aNLO\ lies the capability of computing tree-level
and one-loop amplitudes for arbitrary processes. Such computations
are then used to predict physical observables with different perturbative
accuracies and final-state descriptions. Since there are quite
a few possibilities, we list them explicitly here, roughly in order 
of increasing complexity, and we give them short names that will
render their identification unambiguous in what follows.
\begin{enumerate}
\item 
fLO: this is a tree- and parton-level computation, where the exponents
of the coupling constants are the smallest for which a scattering
amplitude is non zero. No shower is involved, and observables 
are reconstructed by means of the very particles that 
appear in the matrix elements. 
\item 
fNLO: the same as fLO, except for the fact that the perturbative
accuracy is the NLO one. As such, the computation will involve
both tree-level and one-loop matrix elements.
\item 
LO+PS: uses the matrix elements of an fLO computation, but matches them
to parton showers. Therefore, the observables will have to be reconstructed
by using the particles that emerge from the Monte Carlo simulation.
\item 
NLO+PS: same as LO+PS, except for the fact that the underlying computation
is an NLO rather than an LO one. In this paper, the matching of the NLO matrix 
elements with parton showers is done according to the MC@NLO formalism.
\item 
MLM-merged: combines several LO+PS samples, which differ by final-state 
multiplicities (at the matrix-element level). In our framework, two 
different approaches, called $\kt$-jet and shower-$\kt$ schemes, 
may be employed.
\item 
FxFx-merged: combines several NLO+PS samples, which differ
by final-state multiplicities.
\end{enumerate}
We would like to stress the fact that having all of these different
simulation possibilities embedded in a single, process-independent framework 
allows one to investigate multiple scenarios while being guaranteed of 
their mutual consistency (including that of the physical parameters such 
as coupling and masses), and while keeping the technicalities to a minimum 
(since the majority of them are common to all types of simulations). 
For example, one may want to study the impact of perturbative corrections 
with (NLO+PS vs LO+PS) or without (fNLO vs fLO) the inclusion of a parton 
shower. Or to assess the effects of the showers at the LO (LO+PS vs fLO) 
and at the NLO (NLO+PS vs fNLO). Or to check how the inclusion of
different-multiplicity matrix elements can improve the predictions
based on a fixed-multiplicity underlying computation, at the LO
(MLM-merged vs LO+PS) and at the NLO (FxFx-merged vs NLO+PS). 

In the remainder of this section we shall review the theoretical
ideas that constitute the bases of the computations listed above
in items 1--6. Since such a background is immense, we shall sketch the
main characteristics in the briefest possible manner, and rather
discuss recent advancements that have not yet been published.

\subsection{Methodology of computation\label{sec:method}}
The central idea of \aNLO\ is the same as that of the \MadGraph\ family.
Namely, that the {\em structure} of a cross section, regardless
of the theory under consideration and of the perturbative order,
is essentially independent of the process, and as such it can be
written in a computer code once and for all. For example, phase
spaces can be defined in full generality, leaving only the 
particle masses and their number as free 
parameters (see e.g.~ref.~\cite{Byckling:1971vca}). Analogously, in order to
write the infrared subtractions that render an NLO cross section
finite, one just needs to cover a handful of cases, which can
be done in a universal manner. Conversely, matrix 
elements are obviously theory- and process-dependent, but
can be computed starting from a very limited number of formal 
instructions, such as Feynman rules or recursion relations.
Thus, \aNLO\ is constructed as a meta-code, that is a (Python) code that
writes a (Python, C++, Fortran) code, the latter being the one specific to 
the desired process. In order to do so, it needs two ingredients:
\begin{itemize}
\item a theory model;
\item a set of process-independent building blocks.
\end{itemize}
A theory model is equivalent to the Lagrangian of the theory plus its
parameters, such as couplings and masses. Currently, the method
of choice for constructing the model given a Lagrangian is 
that of deriving its Feynman rules, that \aNLO\ will 
eventually use to assemble the matrix 
elements. At the LO, such a procedure is fully automated in 
\FeynRules~\cite{Christensen:2008py,Christensen:2009jx,Christensen:2010wz,
Duhr:2011se,Alloul:2013bka,Alloul:2013fw}. NLO cross sections
pose some extra difficulties, because Feynman rules are not sufficient
for a complete calculation -- one needs at least UV counterterms,
possibly plus other rules necessary to carry out the reduction of
one-loop amplitudes (we shall generically denote the latter by $R_2$,
adopting the notation of the Ossola-Papadopoulos-Pittau
method~\cite{Ossola:2006us}). These NLO-specific
terms are presently not computed by \FeynRules\footnote{We expect they
will in the next public version~\cite{Degrande:nloct}, since development 
versions exist that are already capable of doing so -- see 
e.g.~sect.~\ref{sec:next}.} and have to be supplied by 
hand\footnote{Note that these are a finite and typically small
number of process-independent quantities.},
as was done for QCD corrections to SM processes. Therefore, while the 
details are unimportant here, one has to bear in mind that there are ``LO''
and ``NLO'' models to be employed in \aNLO\ -- the former being those
that could also be adopted by \MadGraphf, and the latter being the
only ones that permit the user to exploit the NLO capabilities 
of \aNLO.

Given a process and a model, \aNLO\ will build the process-specific
code (which it will then proceed to integrate, unweight, and so forth)
by performing two different operations. {\em a)}~The writing of the
matrix elements, by computing Feynman diagrams in order to define
the corresponding helicity amplitudes, using the rules specified 
by the model. {\em b)}~Minimal editing of the process-independent
building blocks. In the examples given before, this corresponds to
writing definite values for particles masses and the number of particles,
and to select the relevant subtraction terms, which is simply
done by assigning appropriate values to particle identities. The building
blocks modified in this manner will call the functions constructed
in {\em a)}. Needless to say, these operations are performed automatically,
and the user will not play any role in them.

We conclude this section by emphasising a point which should already
be clear from the previous discussion to the reader familiar with
recent \MadGraph\ developments. Namely that, in keeping with the
strategy introduced in \MadGraphf~\cite{Alwall:2011uj}, we do {\em not}
include among the process-independent building blocks the routines
associated with elementary quantities (such as vertices and
currents), whose roles used to be played by the \HELAS\ 
routines~\cite{Murayama:1992gi} in previous \MadGraph\ 
versions~\cite{Stelzer:1994ta,Alwall:2007st}. Indeed, the analogues
of those routines are now automatically and dynamically created 
by the module \ALOHA~\cite{deAquino:2011ub} (which is embedded in
\aNLO), which does so by gathering directly the relevant
information from the model, when this is written in the Universal
\FeynRules\ Output (\UFO~\cite{Degrande:2011ua}) format. See 
sect.~\ref{sec:treegen} for more details on this matter.

\subsection{General features for SM and BSM physics\label{sec:gen}}
Since the release of \MadGraphf\ a significant
effort, whose results are now included in \aNLO, went into extending the 
flexibility of the code at both the input and the output levels. While
the latter is mostly a technical development (see appendix~\ref{sec:MG5out}), 
which allows one to
use different parts of the code as standalone libraries and to
write them in various computer languages, the former extends 
the physics scope of \aNLO\ w.r.t.~that of \MadGraphf\
in several ways, and in particular for what concerns the capability
of handling quantities (e.g., form factors, particles with spin
larger than one, and so forth) that are relevant to BSM theories.
Such an extension, to be discussed in the remainder of this section
and partly in sect.~\ref{sec:treegen}, goes in parallel with the 
analogous enhanced capabilities of \FeynRules, and focuses on models
and on their use. Thus, it is an overarching theme of relevance to
both LO and NLO simulations, in view of the future complete automation
of the latter in theories other than the SM. Some of the topics to which
significant work has been lately devoted in \aNLO, and which 
deserve highlighting, are the following:
\begin{enumerate}
\item Complex mass scheme (sect.~\ref{sec:CMS}).
\item Support of various features, of special relevance
to BSM physics  (sect.~\ref{sec:BSM}).
\item Improvements to the \FeynRules/\UFO/\ALOHA\ chain 
(sect.~\ref{sec:treegen}).
\item Output formats and standalone libraries (appendix~\ref{sec:MG5out}).
\item Feynman gauge in the SM (sect.~\ref{sec:OPP}).
\item Improvements to the front-end user interface (the \aNLO\ shell --
appendix~\ref{sec:depen}).
\item Hierarchy of couplings: models that feature more than one coupling
constant order them in terms of their strengths, so that for processes with 
several coupling combinations at the cross section level only the assumed
numerically-leading contributions will be simulated (unless the user
asks otherwise) -- sect.~\ref{sec:NLO} and appendix~\ref{sec:verbose}.
\end{enumerate}

\noindent
We would finally like to emphasise that in the case of LO computations, 
be they fLO, LO+PS, or merged, one can always obtain from the short-distance
cross sections a set of {\em physical} unweighted events. The same is not
true at the NLO: fNLO cross sections cannot be unweighted, and unweighted
MC@NLO events are not physical if not showered. This difference
explains why at the LO we often adopt the strategy of performing 
computations with small, self-contained modules whose inputs are Les Houches 
event (LHE henceforth) files~\cite{Alwall:2006yp,Butterworth:2010ym}, 
while at the NLO this is typically not worth the effort --
compare e.g.~appendices~\ref{sec:errors} and~\ref{sec:LOerrors},
where the computation of scale and PDF uncertainties at the NLO
and LO, respectively, is considered. Further examples of modules 
relevant to LO simulations are given in sect.~\ref{sec:rwgt}.

\subsubsection{Complex mass scheme\label{sec:CMS}}
In a significant number of cases, the presence of unstable particles 
in perturbative calculations can be dealt with by using the Narrow 
Width Approximation (NWA)\footnote{See sect.~\ref{sec:madspin} for
a general discussion of the NWA and of other related approaches.}.
However, when one is interested in studying either those kinematical 
regions that correspond to such unstable particles being very off-shell,
or the production of broad resonances, or very involved final states,
it is often necessary to go beyond the NWA. In such cases, one needs 
to perform a complete 
calculation, in order to take fully into account off-shell effects, spin
correlations, and interference with non-resonant backgrounds in the
presence of possibly large gauge cancellations. Apart from technical
difficulties, the inclusion of all resonant and non-resonant diagrams
does not provide one with a straightforward solution, since the (necessary)
inclusion of the widths of the unstable particles  -- which amounts to a
resummation of a specific subset of terms that appear at all orders in
perturbation theory -- leads, if done naively, to a violation of gauge
invariance. While this problem can be evaded in several ways at the
LO (see e.g.~refs.~\cite{Argyres:1995ym,Beenakker:1996kn,Passarino:1999zh,
Beenakker:1999hi,Beenakker:2003va,Beneke:2003xh}),
the inclusion of NLO effects complicates things further.
Currently, the most pragmatic and widely-used solution is the so-called 
complex mass scheme~\cite{Denner:1999gp,Denner:2005fg}, that
basically amounts to an analytic continuation in the complex plane of the 
parameters that enter the SM Lagrangian, and which are related to the 
masses of the unstable particles. Such a scheme can be shown to maintain 
gauge invariance and unitarity at least at the NLO, which is precisely 
our goal here.

In \aNLO\ it is possible to employ the complex mass scheme in the context
of both LO and NLO simulations, by instructing the code to use
a model that includes the analytical continuation mentioned above
(see the {\bf Setup} part of appendix~\ref{sec:verbose} for an explicit
example). For example, at the LO this operation simply amounts 
to upgrading the model that describes SM physics in the following 
way~\cite{Denner:1999gp}:
\begin{itemize}
\item The masses $m_k$ of the unstable particles are
replaced by $\sqrt{m_k^2-im_k\Gamma_k}\,$.
\item The EW scheme is chosen where $\alpha(m_Z)$, $m_Z$, and $m_W$ 
(the former a real number, the latter two complex numbers defined
as in the previous item) are input parameters.
\item All other parameters (e.g., $G_F$ and $\theta_W$) assume complex
values. In particular, Yukawa couplings are defined by using the complex
masses introduced in the first item.
\end{itemize}
At the NLO, the necessity of performing UV renormalisation introduces
additional complications. At present, the prescriptions of 
ref.~\cite{Denner:2005fg} have been explicitly included and validated.
As was already mentioned in sect.~\ref{sec:method}, this operation will 
not be necessary in the future, when it will be replaced by an automatic 
procedure performed by \FeynRules.

\subsubsection{BSM-specific capabilities\label{sec:BSM}}
One of the main motivations to have very precise SM predictions, and therefore
to include higher order corrections, is that of achieving a better experimental
sensitivity in the context of New Physics (NP) searches. At the same time, 
it is necessary to have as flexible, versatile, and accurate simulations 
as is possible not only for the plethora of NP models proposed so far, but
for those to be proposed in the future as well. These capabilities have been 
one of the most useful aspects of the \MadGraphf\ suite; they are 
still available in \aNLO\ and, in fact, have been further extended.

As was already mentioned, the required flexibility is a direct consequence
of using the \UFO\ models generated by dedicated 
packages such as \feynrules\ or 
{\sc\small SARAH}~\cite{Staub:2012pb}, and of making \aNLO\ compatible
with them. Here, we limit ourselves to listing the several extensions 
recently made to the \UFO\ format and the \aNLO\ code, which have a direct 
bearing on BSM simulations.

\begin{itemize}
\item The possibility for the user to define the analytic expression for 
the propagator of a given particle in the model~\cite{Christensen:2013aua}.
\item The implementation of the analytical formulae for two-body decay
widths~\cite{Alwall:2014bza}, which allows one to save computing time
when the knowledge of the widths of all unstable particles in the
model is needed (see sect.~\ref{sec:madwidth}).
\item The possibility to define form factors (i.e., coupling constants 
depending on the kinematic of the vertex) directly in the \UFO\ model.
Note that, since form factors cannot be derived from a Lagrangian,
they cannot be dealt with by programs like \feynrules, and have therefore
to be coded by hand in the models. In order to be completely generic,
the possibility is also given to go beyond the format of the current
\UFO\ syntax, and to code the form factors directly in Fortran\footnote{Which 
obviously implies that this option is not available should other type of 
outputs be chosen (see appendix~\ref{sec:MG5out}). Further details
are given at:\\
{\tt https://cp3.irmp.ucl.ac.be/projects/madgraph/wiki/FormFactors}.}.
\item Models and processes are supported that feature massive and massless 
particles of spin $3/2$~\cite{Christensen:2013aua}. This implies that
all spins are supported in the set $\{0,1/2,1,3/2,2\}$.
\item The support of multi-fermion interactions, including the case of 
identical particles in the final state, and of \UFO\ models that 
feature vertices with more than one fermion flow. Multi-fermion 
interactions with fermion-flow violation, such as in the presence of 
Majorana particles, are not supported. Such interactions, however, can 
be implemented by the user by splitting the interaction in multiple pieces
connected via heavy scalar particles, a procedure that allows one to define
unambiguously the fermion flow associated with each vertex.
\end{itemize}
While not improved with respect to what was done in \MadGraphf, we remind 
the reader that the module responsible for handling the colour algebra 
is capable of treating particles whose SU$_{\rm c}(3)$ representation
and interactions are non-trivial, such as the sextet and
$\epsilon^{ijk}$-type vertices respectively.

\subsection{LO computations\label{sec:LOcomp}}
The general techniques and strategies used in \aNLO\ to integrate
a tree-level partonic cross section, and to obtain a set of 
unweighted events from it, have been inherited from \MadGraphf;
the most recent developments associated with them have been presented
in ref.~\cite{Alwall:2011uj}, and will not be repeated here. After
the release of \MadGraphf, a few optimisations have been introduced
in \aNLO, in order to make it more efficient and flexible than
its predecessor. Here, we limit ourselves to listing the two which
have the largest overall impact.
\begin{enumerate}
\item The phase-space integration of decay-chain topologies has been
  rewritten, in order to speed up the computations and to deal with extremely
  long decay chains (which can now easily extend up to sixteen particles).
  In addition, the code has also been optimised to better take into account 
  invariant-mass cuts, and to better handle the case where interference 
  effects are large.
\item It is now possible to integrate matrix elements which are not
  positive-definite\footnote{We stress that this statement is non-trivial just
    because it applies to LO computations. In the context of NLO simulations
    this is the standard situation, and \amcatnlo\ has obviously been always
    capable of handling it.}. This is useful e.g.~when one wants to study a
  process whose amplitude can be written as a sum of two terms, which one
  loosely identifies with a ``signal'' and a ``background''. Rather than
  integrating $\abs{S+B}^2$, one may consider
  \mbox{$\abs{S}^2+2\Re(SB^\star)$} and $\abs{B}^2$ separately, which is a
  more flexible approach (e.g.~by giving one the possibility of scanning a
  parameter space, in a way that affects $S$ while leaving $B$ invariant),
  that also helps reduce the computation time by a significant amount. One
  example of this situation is the case when $\abs{B}^2$ is numerically
  dominant, and thus corresponds to a very large sample of events which can
  however be generated only once, while smaller samples of events, that
  correspond to \mbox{$\abs{S}^2+2\Re(SB^\star)$} for different parameter
  choices, can be generated as many times as necessary.  Another example is
  that of an effective theory where a process exists ($S$) that is also
  present in the SM ($B$). In such a case, it is typically
  \mbox{$\abs{B}^2+2\Re(SB^\star)$} which is kept at the lowest order in
  $1/\Lambda$ (with $\Lambda$ being the cutoff scale). 
  Finally, this functionality is needed in order to study the
  large-$\NC$ expansion in multi-parton amplitudes, where beyond the leading
  $1/\NC$ terms the positive definiteness of the integrand is not guaranteed.
\end{enumerate}
In the following sections, we shall discuss various topics relevant
to the calculation of LO-accurate physical quantities. 
Sect.~\ref{sec:treegen} briefly reviews the techniques employed
in the generation of tree-level amplitudes, emphasising the role of
recent \UFO/\ALOHA\ developments. Sect.~\ref{sec:madwidth} presents
the module that computes the total widths of all unstable particles
featured in a given model. Sect.~\ref{sec:rwgt} describes reweighting
techniques. Finally, in sect.~\ref{sec:CKKW} we review the situation
of MLM-merged computations.

\subsubsection{Generation of tree-level matrix elements\label{sec:treegen}}
The computation of amplitudes at the tree level in \aNLO\ has a scope
which is in fact broader than tree-level physics simulations, since
{\em all} matrix elements used in both LO and NLO computations are
effectively constructed by using tree-level techniques. While this is
obvious for all amplitudes which are not one-loop ones, in the case of
the latter it is a consequence of the L-cutting procedure, which was
presented in detail in ref.~\cite{Hirschi:2011pa} and which, roughly
speaking, stems from the observation that any one-loop diagram can be turned
into a tree-level one by cutting one of the propagators that enter the
loop. Furthermore, as was also explained in ref.~\cite{Hirschi:2011pa}
and will be discussed in sect.~\ref{sec:OPP}, all of the companion operations
of one-loop matrix element computations (namely, UV renormalisation and
$R_2$ counterterms) can also be achieved through tree-level-like 
calculations, which are thus very central to the whole \aNLO\ framework.

The construction of tree-level amplitudes in \aNLO\ is based on three 
key elements: Feynman diagrams, helicity amplitudes, and colour
decomposition. Helicity amplitudes~\cite{Berends:1981rb,DeCausmaecker:1981bg,
Kleiss:1985yh,Gastmans:1990xh,Xu:1986xb,Gunion:1985vca,Hagiwara:1985yu}
provide a convenient and effective way to evaluate matrix elements for 
any process in terms of complex numbers, which is quicker and less involved
than one based on the contraction of Lorentz indices. As the name implies, 
helicity amplitudes are computed with the polarizations of the external
particles fixed.  Then, by employing colour
decompositions~\cite{Mangano:1990by,DelDuca:1999rs,Maltoni:2002mq}, they can 
be organised into gauge-invariant subsets (often called dual amplitudes),
each corresponding to an element of a colour basis. In this way, the
complexity of the calculation grows linearly with the number of diagrams
instead of quadratically; furthermore, the colour matrix that appears in the 
squared amplitude can be easily computed automatically (to any order in 
$1/\NC$) once and for all, and then stored in memory. If the number of 
QCD partons entering the scattering process is not too large (say, up
to five or six), this procedure is manageable notwidthstanding the fact 
that the number of Feynman diagrams might grow factorially. Otherwise,
other techniques that go beyond the Feynman-diagram expansion have to be 
employed~\cite{Mangano:2002ea,Duhr:2006iq,Cafarella:2007pc,Gleisberg:2008fv}. 
The algorithm used in \aNLO\ for the determination of the Feynman diagrams 
has been described in detail in ref.~\cite{Alwall:2011uj}. There, it has
been shown that it is possible to efficiently ``factorise" diagrams, such 
that if a particular substructure shows up in several of them, it only needs 
to be calculated once, thus significantly increasing the speed of the 
calculation. In addition, a not-yet-public version of the algorithm
can determine directly dual amplitudes by generating only the relevant
Feynman diagrams, thereby reducing the possible factorial growth to 
less than an exponential one.

The diagram-generation algorithm of \aNLO\ is completely general, though it
needs as an input the Feynman rules corresponding to the Lagrangian of a 
given theory. The information on such Feynman rules is typically
provided by \FeynRules, in a dedicated format (\UFO). 
We remind the reader that \feynrules\ is a {\sc\small Mathematica}-based 
package that, given a theory in the form of a list of 
fields, parameters and a Lagrangian, returns the associated Feynman rules
in a form suitable for matrix element generators.  It now supports 
renormalisable as well as non-renormalisable theories, two-component 
fermions, spin-$3/2$  and spin-$2$ fields, superspace notation and
calculations, automatic mass diagonalization and the \UFO\ interface.  
In turn, a \UFO\ model is a standalone Python 
module, that features self-contained definitions for all classes which 
represent particles, parameters, and so forth.
With the information from the \UFO, the dedicated routines that will
actually perform the computation of the elementary blocks
that enter helicity amplitudes are built by \ALOHA. 
Amplitudes are then constructed by initializing a set of external
wavefunctions, given their helicities and momenta. The wavefunctions 
are next combined, according to the interactions present in the Lagrangian,
to form currents attached to the internal lines. Once all of the currents are 
determined, they are combined to calculate the complex number that corresponds
to the amplitude for the diagram under consideration. Amplitudes associated
with different diagrams are then added (as complex numbers), and squared 
by making use of the colour matrix calculated previously, so as to give
the final result. We point out that versions of \MadGraph\ earlier than
\MadGraphf\ used the \helas~\cite{Murayama:1992gi,Hagiwara:2008jb} library
instead of \ALOHA. By adopting the latter, a significant number of 
limitations inherent to the former could be lifted. A few examples 
follow here. \ALOHA\ is not forced to deal with 
pre-coded Lorentz structures; although its current
implementation of the Lorentz algebra assumes four space-time dimensions,
this could be trivially generalised to any even integer, as the algebra is 
symbolic and its actual representation enters only at the final stage 
of the output writing; its flexibility has allowed the implementation
of the complex mass scheme (see sect.~\ref{sec:CMS}), and of generic
UV and $R_2$ computations (see sect.~\ref{sec:OPP}); it includes 
features needed to run on GPU's, and the analogues of the {\sc Heget}
\cite{Hagiwara:2009aq, Hagiwara:2009cy} libraries can be automatically
generated for any BSM model; finally, it caters to 
matrix-element generators other than \madgraph, such as those now used 
in \HWpp~\cite{Bahr:2008pv,Bellm:2013lba} and \PYe~\cite{Sjostrand:2007gs}. 
Since its release in 2011~\cite{deAquino:2011ub},
several important improvements have been made in \ALOHA. 
On top of the support to the complex mass scheme and to the Feynman gauge,
and of specific features relevant to one-loop computations (which are 
discussed in sects.~\ref{sec:CMS} and~\ref{sec:OPP}), 
we would like to mention here that there 
has been a conspicuous gain in speed, both at the routine-generation phase 
as well as in the actual evaluation, thanks to the extensive use of caching.
In addition, the user is now allowed to define the form of a propagator 
(which is not provided by \feynrules), while previously this 
was determined by the particle spin: non-trivial forms, such as those 
relevant to spin-2 particles in ADD models~\cite{ArkaniHamed:1998rs}, 
or to unparticles, can now be used.

\subsubsection{Width calculator\label{sec:madwidth}}

Among the basic ingredients for Monte Carlo simulations of new-physics models
are the masses and widths of unstable particles; which particle is stable
and which unstable may depend on the particular parameter
benchmark chosen. Masses are typically obtained by going to the mass 
eigenbasis and, if necessary, by evolving boundary conditions from a large
scale down to the EW one. Very general codes exist that perform these 
operations starting from a \FeynRules\ model, such as 
\asperge~\cite{Alloul:2013fw}. The determination of the corresponding 
widths, on the other hand, requires the explicit calculation
of all possible decay channels into lighter (SM or BSM) states.  
The higher the number of the latter, the more daunting it is to
accomplish this task by hand. Furthermore, depending on the mass hierarchy and
interactions among the particles, the computation of two-body decay rates
could be insufficient, as higher-multiplicity decays might be the dominant
modes for some of the particles. The decay channels that are
kinematically allowed are highly dependent on the mass spectrum of the model,
so that the decay rates need to be re-evaluated for every choice of the input
parameters. The program \madwidth~\cite{Alwall:2014bza} has been
introduced in order to address the above issues. In particular, 
\madwidth\ is able to compute partial widths for $N$-body 
decays, with arbitrary values of $N$, at the tree-level and 
by working in the narrow-width approximation\footnote{Even if those two 
assumptions are quite generic, there are particles for which they 
do not give sufficiently-accurate results, such as the Standard Model 
Higgs, which has significant loop-induced decay modes.}. The core of 
\madwidth\ is based on new routines for diagram generation that have 
been specifically designed to remove certain classes of diagrams:
\begin{itemize}
\item {\bf Diagrams with on-shell intermediate particles}. If the kinematics
  of an $A \to n$-particles decay allows an internal particle $B$, that
  appears in an $s$-channel, to be on shell, the corresponding diagram can be
  seen as a cascade of two decays, $A \to B + (n-k)$-particles followed by $B
  \to k$-particles. It is thus already taken into account in the calculation
  of lower-multiplicity decay channels, and the diagram is discarded.
\item {\bf Radiative diagrams.}  Roughly speaking, if one or more zero-mass
  particles are radiated by another particle, the diagram is considered to be
  a radiative correction to a lower-multiplicity decay -- the interested reader
  can find the precise definition in ref.~\cite{Alwall:2014bza}. Such a diagram
  is therefore discarded, because it should be considered only in the context
  of a higher-order calculation. Furthermore, all diagrams with the same
  coupling-constant combination and the same external states are also
  discarded, so that gauge invariance is preserved.
\end{itemize}
\madwidth\ begins by generating all two-body decay diagrams, and 
then iteratively adds extra final state particles with the condition
that any diagram belonging to either of the classes above is forbidden.
This iterative procedure stops when all $N=4$ modes have been considered,
or estimated to be numerically irrelevant\footnote{Both of these
conditions can be controlled by the user.}.
All diagrams thus generated are integrated numerically.
\madwidth\ uses several methods to reduce significantly the overall
computation time. Firstly, it features two fast (and conservative)
estimators, one for guessing the impact of adding one extra final-state 
particle {\em before} the actual diagram-generation phase, and another one
for evaluating the importance of a single integration channel. Both of these
estimators are used to neglect parts of the computation which are numerically
irrelevant. Secondly, if the model is compatible with the recent \UFO\
extension of ref.~\cite{Alwall:2014bza}, and thus includes the 
analytical formulae for two-body decays, then the code automatically uses
those formulae and avoids the corresponding numerical integrations.

We conclude this section by remarking that, although essential for
performing BSM cross-section computations, \madwidth\ should be
seen as a {\em complement} for the existing tools that generate models.
This is because the information it provides one with must be available 
before the integration of the matrix elements is carried out but, at
the same time, really cannot be included in a model itself, since
it depends on the chosen benchmark scenario. For more details on the 
use of this model-complementing feature in \aNLO, and of its mass-matrix 
diagonalisation analogue to which we have alluded above,
see the {\bf Setup} part of appendix~\ref{sec:verbose}.

\subsubsection{Event reweighting\label{sec:rwgt}}

The generation of large samples of events for experimental analyses can be a
very time-consuming operation, especially if it involves a full simulation
of the detector response. It is therefore convenient, whenever possible, 
to apply corrections, or to study systematics of theoretical or modelling 
nature, by using reweighting techniques. The reweighting of either 
one-dimensional distributions or that performed on a event-by-event basis are
equally-common practices in experimental physics. Although the basic idea
(that a non-null function can be used to map any other function
defined in the same domain)
behind these two procedures is the same, one must bear in mind that they
are not identical, and in particular that the former can never be proven
to be formally correct, since correlations (with other, non-reweighted 
variables) may be lost or modified, while the latter is correct in general,
at least in the limit of a large number of events. For this reason,
we consider only event-by-event reweighting approaches in what follows.

Thanks to its flexibility and the possibility of accessing a large amount
of information in a direct and straightforward manner (model and running
parameters, matrix elements, PDFs, and so forth) \aNLO\ provides one with 
an ideal framework for the implementation of such approaches. The 
strategy is rather simple: one starts with a set of hard events,
such as those contained in an LHE file, and rescales their weights:
\beq
w_{\rm new} =   r \, w_{\rm old}\, 
\label{LOrwgt}
\eeq
without modifying their kinematics. The rescaling factor $r$ is not
a constant, and may change on an event-by-event basis. This implies
that, even when the original sample of events is unweighted (i.e., the
$w_{\rm old}$'s are all equal), the reweighted one will be in general 
weighted (i.e., the $w_{\rm new}$'s may be different), and therefore 
degraded from a statistical point of view. If, however, the spread 
of the new weights is not too large (i.e., the $r$'s are close to each 
other, and feature a small number of outliers), the reweigthing is 
typically advantageous with respect to generating a new independent 
sample from scratch.

While eq.~(\ref{LOrwgt}) is completely general, its practical implementation
depends on the kind of problems one wants to solve. We shall consider 
three of them in this section, two of which constitute a direct application
of the basic formula~(\ref{LOrwgt}), and a third one which is more
sophisticated. Since these procedures address different types of physics,
they are conveniently associated with different modules in \aNLO, but
they are all fully embedded into our framework and easily accessible
through it, as we shall briefly explain in what follows and in 
appendices~\ref{sec:LOerrors} and~\ref{sec:appLO}.

The simplest example is that of the evaluation of the uncertainties through
to the variations of the renormalisation and factorisation scales, and
of the PDFs. In such a case $r$ is easily determined by using the identities
$i$ and $j$ of the initial-state partons, and the power ($b$) of $\as$ in
the Born matrix elements\footnote{\label{ft:one}There may be cases where 
such matrix elements do not factorise a single $\as$ factor -- see
e.g.~sect.~\ref{sec:NLO}. We limit ourselves to discussing the
simplest, and more common, situation here.}:
\beq
r =  \frac{ f_i^{\rm new}(x_1,\muF^{\rm new}) 
f_j^{\rm new}(x_2,\muF^{\rm new}) \as^b(\muR^{\rm new})}
{f_i^{\rm old}(x_1,\muF^{\rm old}) 
f_j^{\rm old}(x_2,\muF^{\rm old})\, \as^b(\muR^{\rm old})} \,.
\label{rff}
\eeq
It should be stressed that, although scale and PDF systematics can
also be computed with reweighting techniques at the NLO 
(see~ref.~\cite{Frederix:2011ss}), in general they cannot be
written in the very simple form of eq.~(\ref{LOrwgt}), which is
that of an overall rescaling. For a direct comparison of the NLO
and LO techniques employed in this context by \aNLO, and for fuller
details about them, see appendices~\ref{sec:errors} and~\ref{sec:LOerrors} 
respectively.

Another rather simple example is that of the case where one is interested
in studying the implications of changing the modelling of a process,
with the sole constraint that its initial and final states be
the same. Such a situation can for example occur when the numerical
values of the coupling constants are modified, or when the contributions
of classes of diagrams are included or eliminated (e.g., Higgs exchange
in EW vector boson scattering). The common feature of all examples of
this kind is that they are associated with changes to matrix elements,
all computed with the same kinematic configuration. Therefore,
one can simply write:
\beq
r  = |\amp_{\rm new}|^2 /|\amp_{\rm old}|^2\,,
\label{rwgtME}
\eeq
which for obvious reasons is dubbed matrix-element reweighting.
Despite its simplicity, this method has very many different applications, 
from parameter scanning in new-physics models (where one starts with
a single sample of events, that corresponds to a given benchmark point,
and generates as many new ones as the number of parameter configurations
of interest), to more advanced approaches, such as the inclusion of 
exact loop effects ($\amp_{\rm new}$) in processes that can be also, 
and more simply, described by effective theories ($\amp_{\rm old}$) --
see refs.~\cite{Alwall:2011cy,Frederix:2014hta} for recent results
of the latter type that make use of \MadGraphf\ and \aNLO. Some further 
comments on matrix-element reweighting and its practical usage in \aNLO\ 
are given in appendix~\ref{sec:appLO}.

We finally turn to discussing the matrix-element method~\cite{Kondo:1988yd,
Dalitz:1991wa,Kondo:2006ar}, which can be seen as a reweighting one because 
the weights determined at the parton level through matrix-element computations 
are possibly modified by a convolution to take into account a variety of 
blurring effects (such as those due to
a detector). On the other hand, from the practical point of view it
turns out to be more convenient, rather than talking about reweighting
factors, to introduce a likelihood $P({\bf q}|\alpha)$ for the observation 
of a given kinematic configuration ({\bf q}) given a set of theoretical 
assumptions ($\alpha$). By doing so, one can just re-express 
eq.~(\ref{rwgtME}) in a different language:
\beq
P({\bf q}|\alpha)=\frac{{\cal V}}{\hat{\sigma}_\alpha}
\abs{\amp_\alpha({\bf q})}^2\,,
\label{rwgtLI}
\eeq
with ${\cal V}$ a suitable volume factor, and $\hat{\sigma}_\alpha$ the 
total rate associated with the given assumptions $\alpha$.
The advantage of eq.~(\ref{rwgtLI}) is that it is suitable to 
handle cases which are much more complicated than the purely-theoretical
exercise of eq.~(\ref{rwgtME}), which has led to its introduction here.
For example, in the first approximation one may think of ${\bf q}$ as
the actual kinematic configuration measured by an experiment,
whose accuracy is such that it can be directly used as an argument
of the matrix elements, as is done in eq.~(\ref{rwgtLI}). Note that
this is a rather strong assumption, that in practice identifies
hadron-level with parton-level quantities, and assumes that the
knowledge of the final state is complete (such as that which one can
ideally obtain in Drell-Yan production, \mbox{$pp\to Z\to \ell^+\ell^-$}).
It is clear that there are many ways in which this simple approximation
(which is used, for example, in refs.~\cite{Gao:2010qx,Avery:2012um,
Campbell:2012cz,Gainer:2013rxa,Plehn:2013paa})
can break down: the effects of radiation, of the imperfect knowledge of
the detector, of the impossibility of a strict identification of parton-
with hadron-level quantities, the uncertainties that plague the latter,
the fact that all the relevant four-momenta cannot in general be
measured, are but a few examples. It is therefore necessary to generalise
eq.~(\ref{rwgtLI}), which one can do as follows:
\beq
P({\bf q}|\alpha)=\frac{1}{\sigma_\alpha}
\int dx_1 dx_2 d\phi({\bf p})f^{(1)}(x_1)f^{(2)}(x_2)
\abs{\amp_\alpha({\bf p})}^2 W({\bf q},{\bf p})\,.
\label{rwgtLI2}
\eeq
In eq.~(\ref{rwgtLI2}), we have denoted by {\bf p} the parton kinematic
configuration. All effects that may turn {\bf p} into the detector-level
quantity {\bf q} (whose dimensionality therefore need not coincide with
that of {\bf p}) are parametrised by $W({\bf q},{\bf p})$, called
transfer function. As for any hadron-collision measurable quantity,
eq.~(\ref{rwgtLI2}) features the convolution with the PDFs.
The likelihood introduced in this way can be used in the context
e.g.~of an hypothesis test in order to determine which among various
choices of $\alpha$ is the most probable.

Although this method is both conceptually simple and very attractive, 
the numerical evaluation of $P({\bf q}|\alpha)$ is difficult
because the transfer function $W({\bf q},{\bf p})$ behaves in a way 
which cannot be probed efficiently by phase-space parametrisations that
work well for just the matrix elements. In order to address this issue,
a dedicated program, dubbed \madweight~\cite{Artoisenet:2010cn}, has
been introduced that includes an optimised phase-space treatment
specifically designed for eq.~(\ref{rwgtLI2}). The new version of the 
code~\cite{Artoisenet:2014_MW} embedded in \aNLO\ features several 
improvements w.r.t.~that of ref.~\cite{Artoisenet:2010cn}. It includes
the method for the approximate description of higher-order effects due
to initial-state radiation, as proposed in ref.~\cite{Alwall:2010cq}. 
Furthermore, several optimizations have been achieved that render
the computations much faster (sometimes by orders of magnitude); 
this allows one to use this approach
also in the case of very involved final states, such as those relevant
to Higgs production in association with a $t\bt$ 
pair~\cite{Artoisenet:2013vfa}. Further details on \madweight\
and its use within \aNLO\ can be found in appendix~\ref{sec:appLO}.

\subsubsection{Tree-level merging\label{sec:CKKW}}
The goal of merging is that of combining samples associated with 
different parton multiplicities in a consistent manner, that avoids
double counting after showering, thus allowing one to effectively
define a single fully-inclusive sample.  The tree-level merging
algorithms implemented in \aNLO\ are a hybrid version of those available 
in \Alpgen~\cite{Mangano:2002ea} and \Sherpa~\cite{Gleisberg:2003xi};
they work for both SM and BSM hard processes, but are fully automated
only when the shower phase is performed with either
\PYs~\cite{Sjostrand:2006za} or \PYe~\cite{Sjostrand:2007gs} 
(however, there are no reasons in principle which prevents these schemes
from working with \HWs~\cite{Corcella:2000bw,Corcella:2002jc} or 
\HWpp~\cite{Bahr:2008pv,Bellm:2013lba}). 
They are based on the use of a
$\kt$-measure~\cite{Catani:1993hr} to define hardness and to separate
processes of different multiplicities, and do not perform any
analytic-Sudakov reweighting of events; rather, this operation is
effectively achieved by rejecting showered events under certain
conditions (see later), which implies a direct use of the well-tuned
showering and hadronization mechanisms of the parton shower Monte Carlos.

There are two merging schemes that can be used in conjunction with
\PYs\ and \PYe; in the case of the latter, one is also given the 
possibility of considering CKKW-L 
approaches~\cite{Lonnblad:2001iq,Lonnblad:2011xx,Lonnblad:2012ng}
(after having generated the samples relevant to various parton 
multiplicities with \aNLO); in what follows, we shall limit ourselves to
briefly describe the former two methods. 
Firstly, one has the $\kt$-jet MLM scheme~\cite{Alwall:2007fs}, where
final-state partons at the matrix-element level are clustered 
according to a $\kt$ jet algorithm to find the ``equivalent parton shower 
history'' of the event. In our implementation the Feynman diagram information 
from \aNLO\ is used to retain only those clusterings that correspond to 
actual Feynman diagrams. In order to mimic the behaviour of the parton shower, 
the $\kt$ value for each clustering vertex associated with a QCD branching 
is used as the renormalisation scale for $\as$ in that vertex. All
factorisation scales squared, and the renormalisation scale squared for 
the hard process (the process with the zero-extra-parton multiplicity), are
constructed by clustering back to the irreducible $2\to 2$ system and by 
using the transverse mass in the resulting frame: 
\mbox{$\mu^2=p_{{\sss T}}^2+m^2$}. The smallest $\kt$
value found in the jet-reconstruction procedure is restricted to be larger
than some minimum cutoff scale, which we denote by $\QME$; if this condition
is not satisfied, the event is rejected.
The hard events are then showered by \PY: at the end of the 
perturbative-shower phase, final-state partons are clustered into jets, 
using the very same $\kt$ jet algorithm as before, with the jets required
to have a transverse momentum larger than a given scale $\Qmatch$, with
$\Qmatch>\QME$. The resulting jets are compared to the partons at the
hard subprocess level (i.e., those that result from the matrix-element
computations): a jet $j$ is said to be matched to a parton $p$ if the
distance between the two, defined according to ref.~\cite{Catani:1993hr},
is smaller than the minimal jet hardness:  $\kt(j,p)<\Qmatch$.
The event is then rejected unless each jet is matched to a parton, 
except in the case of the largest-multiplicity sample, where extra 
jets are allowed if softer than the $\kt$ of the softest matrix-element 
parton in the event, $\QMElow$. Secondly, and with the aim to give 
one a non-parametric way to study merging systematics, one has the
shower-$\kt$ scheme, which can be used only with \PY's $\pt$-ordered shower.
In this case, events are generated by \aNLO\ as described above and
then showered, but information is also retained on the
hardest (which is also the first, in view of the $\pt$-ordered nature
of \PY\ here) emission in the shower, $\Qhardest$; furthermore, one
sets $\Qmatch=\QME$, which cannot be done in 
the context of the $\kt$-jet MLM scheme. For all samples but 
the largest-multiplicity one, events are rejected if $\Qhardest>\Qmatch$, 
while in the case of the largest-multiplicity sample events are rejected
when $\Qhardest>\QMElow$. This merging scheme is simpler than the $\kt$-jet 
MLM one, but it rather effectively mimics it. Furthermore, it probes the
Sudakov form factors used in the shower in a more direct manner. Finally, the
treatment of the largest-multiplicity sample is fairly close to that used in
the CKKW-inspired merging schemes. In both the $\kt$-jet MLM and 
shower-$\kt$ methods, merging systematics are associated with variations
of $\Qmatch$; in the former case, changes to $\Qmatch$ must be done by
keeping \mbox{$\Qmatch-\QME$} a constant. For applications of the two
schemes described here, see 
e.g.~refs.~\cite{Alwall:2007fs,Alwall:2008qv,Alwall:2011cy,
deAquino:2012ru,Artoisenet:2013puc}.

\subsection{NLO computations\label{sec:NLO}}
When discussing the problem of perturbative corrections,
one should bear in mind that one usually considers an expansion in 
terms of a single quantity (which is a coupling constant for fixed-order
computations). However, this is just a particular case of the more 
general scenario in which that expansion is carried out simultaneously
in two or more couplings, all of which are thus treated as ``small"
parameters -- we shall refer to such a scenario as mixed-coupling
expansion. Despite the fact that there is typically a clear numerical
hierarchy among these couplings, a mixed-coupling situation is far
from being academic; in fact, as we shall show in the following,
there are cases when one is obliged to work with it. In order to 
study a generic mixed-coupling expansion without being too abstract, 
let us consider an observable $\Sigma$ which receives contributions
from processes that stem from both QCD and QED interactions. The
specific nature of the interactions is in fact not particularly
relevant (for example, QED here may be a keyword that also understands 
the pure-EW contributions); what matters, for the sake of the present 
discussion, is that $\Sigma$ may depend on more than one coupling constant.
We assume that the regular function
\beq
\Sigma(\as,\aem)
\eeq
admits a Taylor representation:
\beq
\Sigma(\as,\aem)=\sum_{n=0}^\infty\sum_{m=0}^\infty
\frac{\as^n\aem^m}{n!m!}
\left[\frac{\partial^{n+m}\Sigma}
{\partial^n\as\partial^m\aem}\right]_{(\as,\aem)=(0,0)}\,,
\label{taylor}
\eeq
which is by definition the perturbative expansion of $\Sigma$.
The first few terms of the sums in eq.~(\ref{taylor}) 
will be equal to zero, with the number of such vanishing terms increasing
with the complexity of the process under consideration -- this is because
$n$ and $m$ are directly related to the number of vertices that enter
a given diagram. In general, it is clear that for a given pair $(n,m)$ 
which gives a non-vanishing contribution to eq.~(\ref{taylor}), there
may exist another pair $(n^\prime,m^\prime)$, with $n\ne n^\prime$,
$m\ne m^\prime$, and $n+m=n^\prime+m^\prime$ whose contribution to
eq.~(\ref{taylor}) is also non zero. It appears therefore convenient
to rewrite eq.~(\ref{taylor}) with a change of variables:
\beq
k=n+m\,,\;\;\;\;\;\;\;\;
l=n-m\,,
\eeq
whence:
\beq
\Sigma(\as,\aem)=\sum_{k=0}^\infty\sum_{l=-k}^k 
\frac{{\cal P}_{k,l}\,\as^{(k+l)/2}\aem^{(k-l)/2}}{((k+l)/2)!((k-l)/2)!}
\left[\frac{\partial^{k}\Sigma}{\partial^{(k+l)/2}\as
\partial^{(k-l)/2}\aem}\right]_{(\as,\aem)=(0,0)}\,,
\label{taylor2}
\eeq
where
\beq
{\cal P}_{k,l}=\delta\Big({\rm mod}(k,2),{\rm mod}(l,2)\Big)\,,
\label{Pkldef}
\eeq
which enforces the sum over $l$ to run only over those integers
whose parity is the same as that of $k$ (therefore, there are $k+1$ terms 
in each sum over $l$ for a given $k$). Equation~(\ref{taylor2}) implies
that we need to call {\em Born} the sum (over $l$) of all the contributions 
with the smallest $k\equiv k_0$ which are non-vanishing. Hence, the 
{\em NLO corrections} will correspond to the sum over $l$ of all terms 
with $k=k_0+1$. This notation
is compatible with the usual one used in the context of the perturbation 
theory of a single coupling: the QCD- or QED-only cases are 
recovered by considering $l=k$ or $l=-k$ respectively. 
In a completely general case, for any given $k$ there will exist
two integers $l_m(k)$ and $l_M(k)$ which satisfy the following 
conditions:
\beq
-k\le l_m(k)\le l_M(k)\le k\,,
\eeq
and such that all contributions to eq.~(\ref{taylor2}) with
\beq
l<l_m(k)\;\;\;\;\;\;{\rm or}\;\;\;\;\;\;
l>l_M(k)
\eeq
vanish, and $l=l_m(k)$, $l=l_M(k)$ are both non-vanishing. 
This implies that in the range
\beq
l_m(k)\le l\le l_M(k)\,,
\label{lrange}
\eeq
there will be at least one (two if $l_m(k)\ne l_M(k)$)
non-null contribution(s) to eq.~(\ref{taylor2})
(the typical situation being actually that where all the terms 
in eq.~(\ref{lrange}) are non-vanishing). Given eq.~(\ref{lrange}), 
one can re-write eq.~(\ref{taylor2}) in the following way:
\beq
\Sigma(\as,\aem)=\sum_{k=k_0}^\infty\as^{c_s(k)}\aem^{c(k)}
\sum_{q=0}^{\Delta(k)}\Sigma_{k,q}\,
\as^{\Delta(k)-q}\aem^q\,,
\label{taylor3}
\eeq
where
\beqn
c_s(k)&=&\half\left(k+l_m(k)\right)\,,
\\
c(k)&=&\half\left(k-l_M(k)\right)\,,
\\
\Delta(k)&=&\half\left(l_M(k)-l_m(k)\right)\,.
\label{Deltadef}
\eeqn
The coefficients $\Sigma_{k,l}$ of eq.~(\ref{taylor3}) can be
expressed in terms of the quantities that appear in eq.~(\ref{taylor2}),
but this is unimportant here. A typical situation is where:
\beqn
l_M(k+1)&=&l_M(k)+1\,,
\label{nextlM}
\\
l_m(k+1)&=&l_m(k)-1\,,
\label{nextlm}
\eeqn
so that:
\beqn
c_s(k+1)&=&c_s(k)\,,
\label{cskpo}
\\
c(k+1)&=&c(k)\,,
\label{ckpo}
\\
\Delta(k+1)&=&\Delta(k)+1\,,
\label{Delkpo}
\eeqn
whence:
\beq
\Sigma(\as,\aem)=\as^{c_s(k_0)}\aem^{c(k_0)}\sum_{p=0}^\infty
\sum_{q=0}^{\Delta(k_0)+p}\Sigma_{k_0+p,q}\,
\as^{\Delta(k_0)+p-q}\aem^q\,,
\label{taylor4}
\eeq
where the Born and NLO contributions correspond to $p=0$ and $p=1$
respectively. Note that eq.~(\ref{taylor4}) is the most general form
of the observable $\Sigma(\as,\aem)$ if one allows the possibility of having 
$\Sigma_{k_0+p,0}=0$ or $\Sigma_{k_0+p,\Delta(k_0)+p}=0$ (or both) for 
$p>k_0$, since this renders eqs.~(\ref{nextlM}) and~(\ref{nextlm}) 
always true. Equation~(\ref{taylor4}) has the advantage of a
straightforward interpretation of the role of NLO corrections.

\begin{figure}[h]
  \begin{center}
    \epsfig{figure=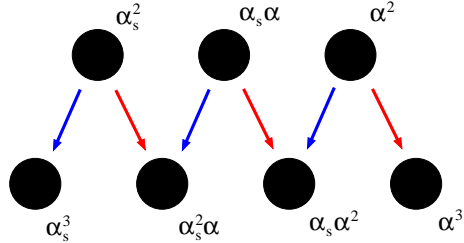,width=0.4\textwidth}
\caption{\label{fig:corr}
QCD (blue, right-to-left arrows) corrections and
QED (red, left-to-right arrows) corrections
to dijet production. See the text for details.
}
  \end{center}
\end{figure}

\begin{figure}[h]
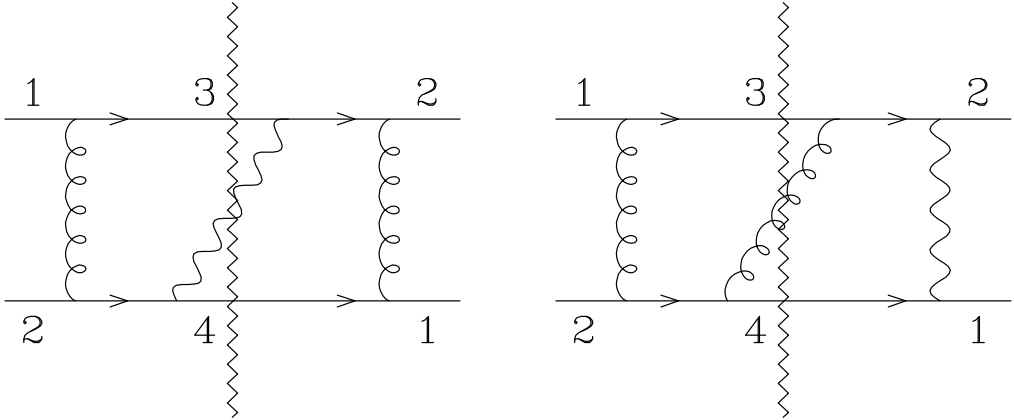

  \begin{center}
    \epsfig{figure=uuuuggp.eps,width=0.4\textwidth}
$\phantom{aaaaa}$
    \epsfig{figure=uuuugpg.eps,width=0.4\textwidth}
\caption{\label{fig:diag1} 
Real-emission contributions to dijet production at the NLO
and ${\cal O}(\as^2\aem)$. The saw-shaped lines represent
Cutkosky cuts.
}
  \end{center}
\end{figure}

\begin{figure}[h]
  \begin{center}
    \epsfig{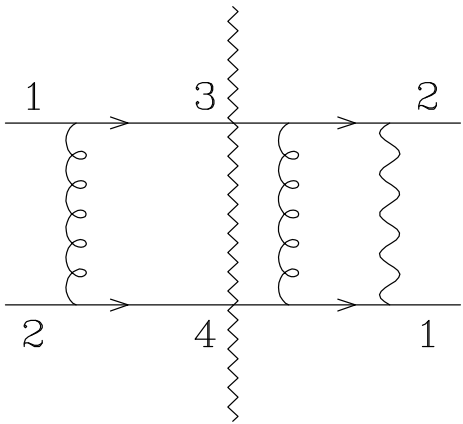}
\caption{\label{fig:diag2} 
Virtual contribution to dijet production at the NLO
and ${\cal O}(\as^2\aem)$. The saw-shaped line represents
a Cutkosky cut.
}
  \end{center}
\end{figure}

An example may help make the points above more explicit.
Consider the contribution to dijet production due to the partonic
process $uu\to uu$; the corresponding lowest-order $t$- and 
$u$-channel Feynman diagrams feature the exchange of either a gluon or 
a photon (or a $Z$, but we stick to the pure-$U(1)$ theory here). 
The Born matrix elements will therefore be the sum of terms 
that factorise the following coupling combinations:
\beq
\as^2\,,\;\;\;\;\;\;\;\;
\as\aem\,,\;\;\;\;\;\;\;\;
\aem^2\,,
\label{cpl1}
\eeq
which implies $k_0=2$, $\Delta(2)=2$,  and $c_s(2)=c(2)=0$. Therefore,
according to eq.~(\ref{taylor4}), the NLO contribution $p=1$ will 
feature the following coupling combinations:
\beq
\as^3\,,\;\;\;\;\;\;\;\;
\as^2\aem\,,\;\;\;\;\;\;\;\;
\as\aem^2\,,\;\;\;\;\;\;\;\;
\aem^3\,.
\label{cpl2}
\eeq
From the procedural point of view, it is convenient to identify QCD
and QED corrections according to the relationship between one coupling
combination in eq.~(\ref{cpl1}) and one in eq.~(\ref{cpl2}),
as follows:
\beqn
&&\as^n\aem^m\;\;\;\stackrel{\rm QCD}{\longrightarrow}
\;\;\;\as^{n+1}\aem^m\,,
\label{QCDcorr}
\\
&&\as^n\aem^m\;\;\;\stackrel{\rm QED}{\longrightarrow}
\;\;\;\as^n\aem^{m+1}\,,
\label{QEDcorr}
\eeqn
which has an immediate graphic interpretation, depicted in
fig.~\ref{fig:corr}.
Such an interpretation has a Feynman-diagram counterpart in
the case of real-emission contributions, which is made explicit
once one considers cut-diagrams, like those presented in 
fig.~\ref{fig:diag1}.
Loosely speaking, one can indeed identify the diagram on the left 
of that figure as representing QED (since the photon is cut) real-emission
corrections to the $\as^2$ Born contribution. On the other hand, the diagram 
on the right represents QCD (since the gluon is cut) real-emission 
corrections to the $\as\aem$ Born contribution. This immediately shows
that, in spite of being useful in a technical sense, QCD and QED 
corrections are not physically meaningful if taken separately:
in general, one must consider them both in order to arrive at a
sensible, NLO-corrected result. This corresponds to the fact that
a given coupling combination in the bottom row of fig.~\ref{fig:corr} 
can be reached by means of two different arrows when starting from
the top row (i.e., the Born level).
Therefore, fig.~\ref{fig:corr} also immediately shows that
when one considers {\em only} the Born term associated with the highest 
power of $\as$ ($\aem$), then QCD-only (QED-only) corrections are 
sensible (because only a right-to-left or left-to-right arrow
is relevant, respectively): they coincide with the NLO corrections 
as defined above (see the paragraph after eq.~(\ref{Pkldef})). 
It also should be clear that the above arguments have a general validity, 
whatever the values of $c_s(k_0)$, $c(k_0)$, and $\Delta(k_0)$ in
eq.~(\ref{taylor4}) -- the former two quantities never play a role
in the analogues of fig.~\ref{fig:corr}, while by increasing $\Delta(k_0)$
one simply inserts more blobs (i.e., coupling combinations) in both of the
rows of fig.~\ref{fig:corr}. Finally, note that reading eqs.~(\ref{QCDcorr})
and~(\ref{QEDcorr}) in terms of diagrams, as has been done for those
of fig.~\ref{fig:diag1}, becomes much harder when one
considers virtual contributions. For example, the one whose 
${\cal O}(\as^2\aem)$ cut-diagram is shown in fig.~\ref{fig:diag2} 
(and its analogues) can indeed be equally well interpreted as a QED 
loop correction to a QCD$\times$QCD ${\cal O}(\as^2)$ Born cut-diagram, 
or as a QCD loop correction to a QCD$\times$QED ${\cal O}(\as\aem)$ 
Born cut-diagram. 

\aNLO\ has been constructed by having eq.~(\ref{taylor4}) in mind;
although the majority of the relevant features are not yet available 
in the public version of the code, all of them have already 
been thoroughly tested in the module responsible for computing 
one-loop matrix elements (see sects.~\ref{sec:OPP} and~\ref{sec:next}), 
which is by far the hardest from this point of view, and the checks
on the real-emission part are also at quite an advanced stage. The basic 
idea is that of giving the user the choice of which coupling combinations
to retain either at the Born or at the NLO level; this corresponds to
choosing a set of blobs in the upper or lower row of fig.~\ref{fig:corr},
respectively. \aNLO\ will then automatically also consider the blobs in
the row not involved in the selection by the user, in order to
construct a physically-meaningful cross section, compatible with both
the user's choices, and the constraints due to a mixed-coupling expansion
(the arrows in fig.~\ref{fig:corr}). It should be stressed that, although
the results for the coefficients $\Sigma_{k_0+p,q}$ can be handled
separately by \aNLO, such coefficients are not (all) independent
from each other from a computational viewpoint, because a single Feynman
diagram (an amplitude-level quantity) may contribute to several
$\Sigma_{k_0+p,q}$'s (the latter being amplitude-squared quantities).
For this reason, as far as the CPU load is concerned the choice of
which coupling combinations to consider can be equivalently made
at the amplitude level. Indeed, this is the only option presently available
in the public version of \aNLO; more detailed explanations are given
in appendix~\ref{sec:verbose}.

\subsubsection{NLO cross sections and FKS subtraction: MadFKS\label{sec:FKS}}

In this section, we briefly review the FKS
subtraction~\cite{Frixione:1995ms,Frixione:1997np} procedure, and
emphasise the novelties of its implementation in \aNLO\ w.r.t.~its
previous automation in \MadFKS~\cite{Frederix:2009yq}, the dedicated
module included in \amcatnlo.

We shall denote by $n$ the number of final-state
particles relevant to the Born contributions to a given cross section.
The set of all the partonic $2\to n$ subprocesses that correspond to
these contributions will be denoted by $\allprocn$; each of these subprocesses
can be represented by the ordered list of the identities of its $2+n$ 
partons, thus:
\beq
\proc=\left(\ident_1,\ldots\,\ident_{n+2}\right)\in\allprocn\,.
\label{Bproc}
\eeq
The first operation performed by \aNLO\ is that of constructing 
$\allprocn$, given the process and the theory model.
For example, if one is interested in the hadroproduction of a $W^+Z$
pair in association with a light jet
\beq
pp\;\longrightarrow\;W^+Zj
\label{pptoWZj}
\eeq
as described by the SM, \aNLO\ will obtain:
\beq
\allproc_3=\left\{\left(u,\bd,W^+,Z,g\right),\ldots
\left(u,g,W^+,Z,d\right),\ldots\right\}.
\eeq
Since the processes in $\allprocn$ are tree-level, \aNLO\ will
construct them very efficiently using the dedicated algorithms 
(see sect.~\ref{sec:treegen}).
Beyond the Born level, an NLO cross section receives contributions
from the one-loop and real-emission matrix elements. As is well known,
the set of the former subprocesses coincides\footnote{This is because we
are considering here only those cases where one-loop matrix elements
are obtained by multiplying the one-loop amplitudes times the Born ones.
Loop-induced processes, in which the LO contribution is a one-loop
amplitude squared, are not to be treated as part of an NLO computation.}
with that of the Born, $\allprocn$. Real-emission processes are by nature
tree-level, and can therefore be obtained by using the very same algorithms 
as those employed to generate the Born contributions. This is achieved
by making the code generate all tree-level processes that have the same 
final-state as the Born's, plus one light jet -- using the example of
eq.~(\ref{pptoWZj}), these would correspond to:
\beq
pp\;\longrightarrow\;W^+Zjj\,.
\label{pptoWZjj}
\eeq
Such was the strategy adopted in the original \MadFKS\ 
implementation~\cite{Frederix:2009yq}. There is however an alternative 
procedure, which we have implemented in \aNLO\ because it is
more efficient than the previous one in a variety of ways. Namely,
for any given $r_0\in\allprocn$, one considers all possible $a\to bc$
branchings for each non-identical $a\in r_0$ with $a$ strongly-interacting 
(i.e., $g\to gg$, $g\to q\bq$, and $q\to qg$, but also $Q\to Qg$, with 
$Q$ a quark with non-zero mass). For each of these branchings, a new
list is obtained by removing $a$ from $r_0$, and by inserting the pair $(b,c)$
in its place. By looping over $r_0$ one thus constructs the 
set\footnote{The exceedingly rare cases of non-singular real-emission
contributions can be obtained by crossing; one example is $q\bq\to Hg$,
which is the crossed process of $qg\to Hq$.} of real-emission processes 
$\allprocnpo$. 
As a by-product, one also naturally obtains, for each $r\in\allprocnpo$,
the pairs of particles which are associated with a soft and/or a collinear
singularity of the corresponding matrix element (which we denote
by $\ampsqnpot(\proc)$); by definition~\cite{Frederix:2009yq}, these pairs 
form the set of FKS pairs, $\FKSpairs(\proc)$, which is central in
the FKS subtraction procedure. We point out, finally, that regardless
of the type of strategy adopted to construct $\allprocnpo$ and $\FKSpairs$,
it is immediate to apply it in \aNLO\ to theories other than QCD, 
such as QED.

After having obtained $\FKSpairs(\proc)$, \aNLO\ constructs the 
$\Sfun$ functions that are used by FKS in order to achieve what is
effectively a dynamic partition of the phase space: in each sector
of such a partition, the structure of the singularities of the matrix
elements is basically the simplest possible, amounting (at most) to one
soft and one collinear divergence. The properties of the $\Sfun$ functions 
are:
\beqn
&&\phantom{aaaaaaaa}\,\Sfunij(\proc) \longrightarrow 1
\phantom{aaaa}i,j~~{\rm collinear}\,,
\label{Sfuncoll}
\\
&&\sum_{\stackrel{j}{(i,j)\in\FKSpairs(\proc)}}\!\!\!\!\Sfunij(\proc) 
\longrightarrow 1
\phantom{aaaa}i~~{\rm soft}\,,
\label{Sfunsoft}
\\
&&\phantom{aaaaaaaa}\,\Sfunij(\proc) \longrightarrow 0
\phantom{aaaa}{\rm all~other~IR~limits}\,,
\label{Sfunnonsing}
\\
&&\sum_{(i,j)\in\FKSpairs(\proc)}\!\!\!\!\Sfunij(\proc) = 1\,.
\label{Sfununit}
\eeqn
One exploits eq.~(\ref{Sfununit}) by rewriting the real matrix elements
as follows:
\beq
\ampsqnpot(\proc)=\sum_{(i,j)\in\FKSpairs(\proc)}
\Sfunij(\proc)\ampsqnpot(\proc)
\;\;\equiv \sum_{(i,j)\in\FKSpairs(\proc)}\ampsqnpot_{ij}\,.
\label{Realident}
\eeq
Thanks to eqs.~(\ref{Sfuncoll})-(\ref{Sfunnonsing}), $\ampsqnpot_{ij}$
has the very simple singularity structure mentioned above. Furthermore,
the terms in the sum on the r.h.s.~of eq.~(\ref{Realident}) are independent
of each other, and \aNLO\ is thus able to handle them in parallel.

The FKS method exploits the fact that phase-space sectors associated 
with different $\Sfunij$ functions are independent of each other by choosing
different phase-space parametrisations in each of them. There is ample
freedom in such a choice, bar for two integration variables: the
rescaled energy of parton $i$ (denoted by $\xii$), and the cosine of the
angle between partons $i$ and $j$ (denoted by $\yij$), both defined in 
the incoming-parton c.m.~frame. The idea is that these quantities
are in one-to-one correspondence with the soft ($\xii\to 0$) and 
collinear ($\yij\to 1$) singularities respectively, which renders it
particularly simple to write the subtracted cross section.
The $(n+1)$-body phase space is then written as follows:
\beqn
\phspnpo=\Phspnpo_{ij}\left(\confnpo(\meas_{n+1}^{(ij)})\right) 
d\meas_{n+1}^{(ij)}\,,
\label{phspnpoij}
\eeqn
where $\meas_{n+1}^{(ij)}$ collectively denote the $3n-1$ independent
integration variables, with
\beq
\left\{\xii,\yij\right\}\subset\meas_{n+1}^{(ij)}\,,
\label{xyinset}
\eeq
and where
\beq
\confnpo=\left\{k_3,k_4,\cdots k_{n+3}\right\}
\eeq
is the set of final-state momenta. Given eq.~(\ref{xyinset}),
\aNLO\ chooses the other $3n-3$ integration variables and thus determines
$\Phspnpo_{ij}$ and the functional dependence $\confnpo(\meas_{n+1}^{(ij)})$
according to the form of the integrand, gathered from the underlying Feynman
diagrams. In general, this implies splitting the computation in several
integration channels, which are independent of each other and can be
dealt with in parallel. Such multi-channel technique is irrelevant
here, and will be understood in the following. More details can be found
in refs.~\cite{Maltoni:2002qb,Kleiss:1994qy} and~\cite{Frederix:2009yq}.
Implicit in eq.~(\ref{phspnpoij}) are the maps that allow one to construct
soft and collinear kinematic configurations starting from a non-special
configuration (i.e., one where no parton is soft, and no two partons
are collinear). This we shall denote as follows. Given:
\beq
\confnpoE=\left\{k_3,k_4,\cdots k_{n+3}\right\}\;\;\;\;\;\;
{\rm non~special}
\eeq
\aNLO\ constructs its soft, collinear, and soft-collinear limits with:
\beqn
\confnpoS&=&\confnpoE(\xii=0)\,,\;\;\;\;
\\
\confnpoC&=&\confnpoE(\yij=1)\,,\;\;\;\;
\\
\confnpoSC&=&\confnpoE(\xii=0,\yij=1)\,.
\eeqn
Furthermore, all the phase-space parametrisations employed
by \aNLO\ are such that\footnote{Equation~(\ref{cntequal}) holds for
all particles except the FKS-pair partons; for the latter, it is the
sum of their four-momenta that is the same in the three configurations. 
This is sufficient owing to the underlying infrared-safety conditions.}:
\beq
\confnpoS=\confnpoC=\confnpoSC\,,
\label{cntequal}
\eeq
which is beneficial from the point of view of the numerical stability 
of observables computed at the NLO, and is necessary in view of matching 
with the parton shower according to the MC@NLO formalism. As is usual
in the context of NLO computations, we call the non-special and the IR-limit
configurations (and, by extension, the corresponding cross-section
contributions) the event and the soft, collinear, and soft-collinear
counterevents respectively.

Given a real-emission process $\procR\in\allprocnpo$ and an $\Sfunij$ 
contribution, the FKS-subtracted cross section consists of four terms:
\beqn
d\sigmaNLO_{ij}(\procR)&\longleftrightarrow&\Big\{d\sigmaNLOa_{ij}(\procR)
\Big\}_{\alpha=E,S,C,SC}
\label{xsecNLO}
\\
d\sigmaNLOa_{ij}(\procR)&=&\luma\!\left(\procR;\meas_{Bj}^{(ij)}\right) 
\Wa_{ij}(\procR)\,\,d\meas_{Bj}^{(ij)}\,d\meas_{n+1}^{(ij)}\,,
\label{fact3}
\eeqn
where $d\meas_{Bj}^{(ij)}$ is the integration measure over Bjorken $x$'s,
$\luma$ is the corresponding parton-luminosity factor\footnote{Whose
dependence on $\alpha$ is a consequence of eq.~(\ref{cntequal}) when
$j=1,2$ -- see ref.~\cite{Frixione:2002ik} for more details.},
and the short-distance weights $\Wa_{ij}$ are reported in
refs.~\cite{Frederix:2009yq,Frederix:2011ss}. In ref.~\cite{Frederix:2009yq},
in particular, an extended proof is given that all contributions to
an NLO cross section which are {\em not} naturally $(n+1)$-body ones 
(such as the Born, virtual, and initial-state collinear remainders)
can be cast in a form formally identical to that of the soft or
collinear counterterms, and can thus be dealt with simultaneously 
with the latter. The fully differential cross section that emerges
from eqs.~(\ref{xsecNLO}) and~(\ref{fact3}) is:
\beqn
\frac{d\sigmaNLO_{ij}(\procR)}{d\conf}&=&
\delta\left(\conf-\confnpoE\right)
\frac{d\sigmaNLOE_{ij}(\procR)}{d\meas_{Bj}^{(ij)}d\meas_{n+1}^{(ij)}}\,
d\meas_{Bj}^{(ij)}d\meas_{n+1}^{(ij)}
\nonumber\\*&+&
\delta\left(\conf-\confnpoS\right)
\sum_{\alpha=S,C,SC} 
\frac{d\sigmaNLOa_{ij}(\procR)}{d\meas_{Bj}^{(ij)}d\meas_{n+1}^{(ij)}}\,
d\meas_{Bj}^{(ij)}d\meas_{n+1}^{(ij)}\,,
\label{NLOdiff}
\eeqn
where we have understood the complete integration over the measures
on the r.h.s.. \aNLO\ scans the phase space by generating randomly
\mbox{$\{\meas_{Bj}^{(ij)},\meas_{n+1}^{(ij)}\}$}. For each of
these points, an event kinematic configuration $\confnpoE$ and
its weight, and a counterevent kinematic configuration $\confnpoS$ and
its weight are given in output; with these, any number of (IR-safe)
observables can be constructed. As can be seen from eq.~(\ref{NLOdiff}),
the weight associated with the single counterevent kinematics is the sum
of the soft, collinear, soft-collinear, Born, virtual, and initial-state
collinear remainders contributions, which reduces the probability of
mis-binning and thus increases the numerical stability of the result.

In order for the results of eq.~(\ref{NLOdiff}) to be physical, they
must still be summed over all processes $\procR\in\allprocnpo$ and
all FKS pairs \mbox{$(i,j)\in\FKSpairs(\procR)$}. As far as the latter
sum is concerned, it is easy to exploit the symmetries due to identical
final-state particles, and thus to arrive at the set~\cite{Frederix:2009yq}:
\beq
\FKSpairsred\subseteq\FKSpairs\,,
\label{PFKSreddef}
\eeq
whose elements give non-identical contributions to the sum over
FKS pairs. Therefore:
\beq
\sum_{\procR\in\allprocnpo}\sum_{(i,j)\in\FKSpairs(\procR)}
\frac{d\sigmaNLO_{ij}(\procR)}{d\conf}=
\sum_{\procR\in\allprocnpo}\sum_{(i,j)\in\FKSpairsred(\procR)}
\symmnpoij\frac{d\sigmaNLO_{ij}(\procR)}{d\conf}\,,
\label{sumsymm}
\eeq
with $\symmnpoij$ a suitable symmetry factor. The sum on
the r.h.s.~of eq.~(\ref{sumsymm}) is obviously more convenient
to perform than that on the l.h.s.; it is customary to include the 
symmetry factor in the short-distance weights 
(see e.g.~ref.~\cite{Frederix:2011ss}). The number of elements
in $\FKSpairsred(\procR)$ can indeed be much smaller than that 
in $\FKSpairs(\procR)$. For example, when $\procR$ is a purely gluonic
process, we have $\#(\FKSpairsred(\procR))=3$ (i.e., independent
of $n$), while $\#(\FKSpairs(\procR))=(n+1)(n+2)$. While the former figure
typically increases when quarks are considered, it remains true that,
for asymptotically large $n$'s, $\#(\FKSpairsred)$ is a constant
while $\#(\FKSpairs)$ scales as $n^2$.

The sum on the r.h.s.~of eq.~(\ref{sumsymm}) is what has been
originally implemented in \MadFKS. It 
emphasises the role of the real-emission processes, which implies
that for quantities which are naturally Born-like (such as the
Born matrix elements themselves) one needs to devise a way to 
map unambiguously $\allprocn$ onto $\allprocnpo$. The interested
reader can find the definition of such a map in sect.~6.2 of
ref.~\cite{Frederix:2009yq}, which we summarise here using the
Born cross section $d\sigmaLO$ as an example:
\beq
d\sigmaLO(\procR)\,=
\sum_{(i,j)\in\FKSpairsred(\procR)}
\delta_{g\ident_i}\Sfunij(\xii=0)d\sigmaLO(\procR^{\isubrmv})\,,
\label{BfromR}
\eeq
where
\beq
\procR=
\left(\ident_1,\ldots\ident_i,\ldots\ident_j,
\ldots\ident_{n+3}\right)
\;\;\;\;\Longrightarrow\;\;\;\;
\procR^{\isubrmv}=
\left(\ident_1,\ldots\remove{\ident}{0.2}_i,\ldots\ident_j,
\ldots\ident_{n+3}\right)\,.
\label{sfproc}
\eeq
In other words, through the r.h.s.~of eq.~(\ref{BfromR}) one is able
to define a Born-level quantity as a function of a real-emission process.
In \aNLO\ we have followed the opposite strategy, namely that of
defining real-emission level quantities as functions of a Born process.
The proof that this can indeed be done is given in appendix~E of
ref.~\cite{Frederix:2009yq}; here, we limit ourselves to summarising
it as follows, using the event contribution $d\sigmaNLOE_{ij}$ 
to the NLO cross section as an example. One has the identity:
\beq
\sum_{\procR\in\allprocnpo}\sum_{(i,j)\in\FKSpairsred(\procR)}
\symmnpoij\,d\sigmaNLOE_{ij}(\procR)=
\sum_{\procB\in\allprocn} d\sigmaNLOE(\procB)\,.
\eeq
Here we have defined:
\beq
d\sigmaNLOE(\procB)=
\sum_{\procR\in\allprocnpo}\sum_{(i,j)\in\FKSpairsred(\procR)}
\delta\left(\procB,\procR^{j\oplus i,\isubrmv}\right)
\symmnpoij(\procR)\, 
d\sigmaNLOE_{ij}(\procR)\,,
\label{RfromB}
\eeq
where the generalised Kronecker symbol $\delta(\ldots)$ is equal to one if 
its arguments are equal, and to zero otherwise, and
\beq
\procR^{j\oplus i,\isubrmv}=
\left(\ident_1,\ldots\remove{\ident}{0.2}_i,
\ldots\ident_{j\oplus i},\ldots\ident_{n+3}\right)\,.
\label{clproc}
\eeq
Although the first sum in eq.~(\ref{RfromB}) might seem to involve
a very large number of terms, \aNLO\ knows immediately which terms
will give a non-zero contribution, thanks to the procedure used to
construct $\allprocnpo$ which was outlined at the beginning of this
section. On top of organising the sums over processes and FKS pairs in
a different way w.r.t.~the first version of \MadFKS, 
\aNLO\ also performs some of these
(and, specifically, those in eq.~(\ref{RfromB})) using MC techniques
(whereas all sums are performed explicitly in ref.~\cite{Frederix:2009yq}). 
In particular, for any given $\procB\in\allprocn$, one real-emission
process and one FKS pair are chosen randomly among those which contribute
to eq.~(\ref{RfromB}); these choices are subject to importance sampling,
and are thus adaptive. In summary, while the procedure adopted
originally in \MadFKS\ takes a viewpoint from the real-emission level,
that adopted in \aNLO\ emphasises the role of Born processes. 
The two are fully equivalent, but the latter is more efficient
in the cases of complicated processes, and it offers some further
advantages in the context of matching with parton showers.

\subsubsection{One-loop matrix elements: MadLoop\label{sec:OPP}}
Both \aNLO\ and its predecessor \amcatnlo\ are capable of computing
the virtual contribution to an NLO cross section in a completely 
independent manner (while still allowing one to interface to a 
third-party one-loop provider if so desired), through a module 
dubbed \ML~\cite{Hirschi:2011pa}. However, there are very significant 
differences between the \ML\ embedded in \amcatnlo\ (i.e., the one documented
in ref.~\cite{Hirschi:2011pa}), and the \ML\ currently available
in \aNLO; hence, in order to avoid confusion between the two, we
shall call the former \MLq, and the latter \MLf. The aim of this
section is that of reviewing the techniques for automated one-loop 
numerical computations, and of giving the first public documentation of \MLf.

As the above naming scheme suggests, core functionalities relevant
to the handling of tree-level amplitudes were inherited from 
\MadGraphq\ in \MLq, while \MLf\ uses \MadGraphf. 
This was a necessary improvement
in view of the possibility of computing virtual corrections
in arbitrary renormalisable models (i.e., other than the SM).
More in general, one can identify the following three items as strategic 
capabilities, that were lacking in \MLq, and that are now available in 
\ML5\footnote{Some of them are not yet public, but are fully tested.}:
\begin{itemize}
\item[{\em A.}] The adoption of the procedures introduced with 
\MadGraphf, and in particular the \UFO/\ALOHA\ chain for constructing
amplitudes starting from a given model.
\item[{\em B.}] The possibility of switching between two reduction
methods for one-loop integrals, namely the Ossola-Papadopoulos-Pittau
(OPP~\cite{Ossola:2006us}) and the Tensor Integral Reduction
(TIR~\cite{Passarino:1978jh,Davydychev:1991va}) procedures.
\item[{\em C.}] The organization of the calculation in a way consistent
with the mixed-coupling expansion described in sect.~\ref{sec:NLO},
and in particular with eq.~(\ref{taylor4}).
\end{itemize}
It should be clear that these capabilities induce an extremely 
significant broadening of the scope of \ML\ (in particular, extending 
it beyond the SM). As a mere by-product,
they have also completely lifted the limitations affecting \MLq,
which were described in sect.~4 of ref.~\cite{Hirschi:2011pa}.
It is instructive to see explicitly how this has happened.
Item {\em A.} is responsible for lifting
\MLq\ limitation \#1 (\MLq\ cannot generate a process whose Born
contains a four-gluon vertex, because the corresponding $R_2$ routines
necessary in the OPP reduction were never validated, owing to the 
technically-awkward procedure for handling them in \MadGraphq; this
step is now completely bypassed thanks to the \UFO/\ALOHA\ chain).
Limitation \#2 (\MLq\ cannot compute some loops that feature
massive vector bosons, which is actually a limitation of 
\CutTools~\cite{Ossola:2007ax}, in turn due to the use of the unitary 
gauge) is now simply absent because of the possibility of adopting 
the Feynman gauge, thanks again to item {\em A}. Limitation \#4 
(\MLq\ cannot handle finite-width effects in loops) is removed thanks 
to the implementation of the complex mass scheme, a consequence of
item {\em A}. Finally, item {\em C.} lifts \MLq\ limitation \#3 
(\MLq\ cannot generate a process if different contributions to the 
Born amplitudes do not factorise the same powers of all the relevant 
coupling constants).

The advances presented in items {\em A.}--{\em C.} above are underpinned
by many technical differences and improvements w.r.t.~\MLq. Here, we
limit ourselves to listing the most significant ones: 
\begin{itemize}
\item \MLf\ is written in Python, the same language which was adopted
for \MadGraphf\ (\MLq\ was written in C++).
\item The UV-renormalisation procedure has been improved and rendered
fully general (that of \MLq\ had several hard-coded simplifying solutions,
motivated by QCD).
\item An extensive use is made of the optimisations proposed in
ref.~\cite{Cascioli:2011va} (\OL).
\item The self-diagnostic numerical-stability tests, and the procedures
for fixing numerically-unstable loop-integral reductions, have been 
redesigned.
\end{itemize}
More details on these points will be given in what follows.
Before turning to that, we shall discuss the basic principles
used by \MLf\ for the automation of the computation of one-loop integrals.

\vskip 0.4truecm
\noindent
{\bf Generalities}

\noindent
Given a $2\to n$ partonic process $r$ (see eq.~(\ref{Bproc})),
\ML\ computes the quantity:
\beqn
V(\proc)=\mathop{\sum_{\rm colour}}_{\rm spin}
2\Re\left\{\ampnl(\proc){\ampnt(\proc)}^{\star}\right\},
\label{Vdef}
\eeqn
with $\ampnt$ and $\ampnl$ being the relevant tree-level and
UV-renormalised one-loop amplitudes respectively; the averages
over initial-state colour and spin degrees of freedom are understood.
The result for $V(\proc)$ is given as a set of three numbers,
corresponding to the residues of the double and single IR poles,
and the finite part, all in the 't~Hooft-Veltman 
scheme~\cite{'tHooft:1972fi}. In the case
of a mixed-coupling expansion, each of these three numbers is replaced
by a set of coefficients, in the form given by eq.~(\ref{taylor4}).
There may be processes for which $\ampnt$ is identically equal to zero,
and $\ampnl$ is finite; for these processes (called loop-induced),
\ML\ computes the quantity:
\beqn
V_{\rm LI}(\proc)=\mathop{\sum_{\rm colour}}_{\rm spin}
\abs{\ampnl(\proc)}^2.
\label{VLIdef}
\eeqn
Only eq.~(\ref{Vdef}) is relevant to NLO computations proper,
and we shall mostly deal with it in what follows. In the current version
of \aNLO, the loop-induced $V_{\rm LI}(\proc)$ cannot be automatically
integrated (for want of an automated procedure for multi-channel 
integration), and hence in this case the code output by \MLf\ must be 
interfaced in an ad-hoc way to any MC integrator (including \aNLO\ -- see 
e.g.~ref.~\cite{Frederix:2014hta} for a recent application). 

The basic quantity which \ML\ needs to compute and eventually 
renormalise in order to obtain $\ampnl$ that enters eq.~(\ref{Vdef}) 
is the one-loop UV-unrenormalised amplitude:
\beqn
\ampnl_\unr=\sum_{\rm diagrams} \diag\,,
\label{Csum}
\eeqn
where $\diag$ denotes the contribution of a single Feynman diagram
after loop integration, whose possible colour, helicity, and Lorentz 
indices need not be specified here, and are understood; they will be
re-instated later. A standard technique for the evaluation of $\diag$ 
is a so-called reduction procedure, pioneered by Passarino and 
Veltman~\cite{Passarino:1978jh}, which can be written as follows:
\beq
\diag=\RED\left[\diag\right]=\sum_i c_i(\diag) 
{\cal J}_i^{\sss(\RED)}+R(\diag)\,.
\label{REDdef}
\eeq
The quantities ${\cal J}_i^{\sss(\RED)}$ are one-loop integrals, independent
of $\diag$. The essence of any reduction procedure is that it is
an {\em algebraic} operation that determines the coefficients $c_i$ 
and $R$ (which are functions of external momenta and of masses, and some 
of which may be equal to zero); the intricacies of loop integration are 
dealt once and for all in the computations of the ${\cal J}_i^{\sss(\RED)}$'s
(which are much simpler than any $\diag$). As the leftmost equality
in eq.~(\ref{REDdef}) indicates, from the operator point of view
$\RED[\,]$ is the identity; its meaning is that of replacing $\diag$
with the linear combination in the rightmost member of eq.~(\ref{REDdef}).
As the notation $\{{\cal J}_i^{\sss(\RED)}\}$ suggests, different reduction 
procedures can possibly make use of different sets of one-loop integrals.

Equation~(\ref{REDdef}) is basically what one would do if one were to
compute $\diag$ in a non-automated manner. In automated approaches,
however, additional problems arise, for example due to the necessity of 
relying on numerical methods, which have obvious difficulties in dealing with
the non-integer dimensions needed in the context of dimensional 
regularisation, and with the analytical information on the integrand
of $\diag$, which is extensively used in non-automated reductions.
In order to discuss how these issues can be solved,
let us write $\diag$ in the following form:
\beqn
\diag=\int d^d \bqloop\,\bar{\idiag}(\bqloop)\,,
\;\;\;\;\;\;\;\;
\bar{\idiag}(\bqloop)=\frac{\bar{N}(\bqloop)}
{\prod_{i=0}^{m-1}\db{i}}\,,
\label{Cbardef}
\eeqn
where $d=4-2\ep$, and we have assumed the diagram to have $m$ 
propagators in the loop and have defined:
\beqn
\db{i} = (\bqloop + p_i)^2 - m_i^2\,,
\;\;\;\;\;\;\;\;
0\le i\le m-1\,,
\label{Dbardef}
\eeqn
with $m_i$ the mass of the particle relevant to the $i^{th}$ loop
propagator, and $p_i$ some linear combination of external momenta.
For any four-dimensional quantity $x$, its $(4-2\ep)$-dimensional
counterpart is denoted by $\bar{x}$, and its $(-2\ep)$-dimensional
one by $\tilde{x}$. The fact that $p_i$, rather than $\bar{p}_i$,
enters eq.~(\ref{Dbardef}) is a consequence of the use of the
't~Hooft-Veltman scheme. The loop momentum is decomposed as follows:
\beqn
\bqloop=\qloop+\tqloop
\;\;\;\;\;\;{\rm with}\;\;\;\;\;\;
\qloop\mydot\tqloop=0\,,
\label{lbardec}
\eeqn
with similar decompositions holding for the Dirac matrices 
$\bar{\gamma}^\mu$ and metric tensor $\bar{g}^{\mu\nu}$.
One can thus define~\cite{Ossola:2008xq} the purely four-dimensional 
part of the numerator that appears in eq.~(\ref{Cbardef}):
\beq
N(\qloop)=\lim_{\ep\to 0}\bar{N}(\bqloop=\qloop;\,
\bar{\gamma}^\mu=\gamma^\mu,\,\bar{g}^{\mu\nu}=g^{\mu\nu})\,,
\label{N4def}
\eeq
from whence one can obtain its $(-2\ep)$-dimensional counterpart:
\beqn
\widetilde{N}(\qloop,\tqloop)=\bar{N}(\bqloop)-N(\qloop)\,.
\label{N2epdef}
\eeqn
The quantity defined in eq.~(\ref{N4def}), not involving non-integer
dimensions, can be treated by a computer with ordinary techniques.
By using eq.~(\ref{N2epdef}) in eq.~(\ref{Cbardef}) one obtains:
\beq
\diag=\diag_{\nonRt}+R_2\,,
\label{Cdef}
\eeq
where
\beqn
\diag_{\nonRt}&=&\int d^d \bqloop\,
\frac{N(\qloop)}{\prod_{i=0}^{m-1}\db{i}}\,,
\label{ccpR1def}
\\
R_2&=&\int d^d \bqloop\,
\frac{\widetilde{N}(\qloop,\tqloop)}{\prod_{i=0}^{m-1}\db{i}}\,.
\label{R2def}
\eeqn
Both integrals in eq.~(\ref{ccpR1def}) and eq.~(\ref{R2def}) still
depend on $(4-2\ep)$-dimensional quantities, but they do so in a way
that allows one to further manipulate and cast them in a form
suitable for a fully numerical treatment. In particular, one can
show~\cite{Ossola:2008xq} that the computation of $R_2$ is equivalent
to that of a tree-level amplitude,
constructed with a universal set of theory-dependent rules
(see ref.~\cite{Draggiotis:2009yb},  
refs.~\cite{Garzelli:2009is,Garzelli:2010qm,Shao:2011tg}, and
refs.~\cite{Pittau:2011qp,Shao:2012ja,Page:2013xla} for the 
QCD, QED+EW, and some BSM cases respectively), analogous to the Feynman 
ones and that can be derived once and for all (for each model)
by just considering the
one-particle-irreducible amplitudes with up to four external 
legs~\cite{Binoth:2006hk}. On the other hand, eq.~(\ref{ccpR1def})
is still potentially divergent in four dimensions. The details
of how this is dealt with may vary, but the common characteristic is that
all of them are entirely defined by a reduction procedure. In
other words, we shall use the following identity:
\beq
\diag=\RED\left[\diag_{\nonRt}\right]+R_2\,,
\label{Cdec}
\eeq
which is a consequence of eqs.~(\ref{REDdef}) and~(\ref{Cdef}). The 
general idea is that all things that are inherently $(4-2\ep)$-dimensional
in \mbox{$\RED[\diag_{\nonRt}]$} can be parametrized in terms of 
the one-loop integrals ${\cal J}_i^{\sss(\RED)}$, so that any
piece of computation that would require an analytical knowledge of
the integrand and an analytical treatment of the $(-2\ep)$-dimensional
terms is indeed treated analytically, but in a universal manner
through ${\cal J}_i^{\sss(\RED)}$.

We emphasise that, although the decomposition of eq.~(\ref{Cdef})
is inspired by the OPP reduction method, it is universal, in the
sense that the operator $\RED[\,]$ in eq.~(\ref{Cdec}) does not
need to be OPP-inspired, and that the definition of $R_2$ has nothing
to do with the OPP procedure as such, but rather with the interplay of 
$(4-2\ep)$-dimensional quantities and their four-dimensional
counterparts. There are of course several alternative approaches, 
but the majority of them do not lend themselves to the {\em numerical} 
computation of the rational part $R(\diag_{\nonRt})+R_2$. 
The two methods which have been used for complicated numerical 
simulations are bootstrap~\cite{Bern:2005cq} and $D$-dimensional 
unitarity~\cite{Giele:2008ve,Anastasiou:2006jv,Anastasiou:2006gt}.
However, they also involve
rather non-trivial issues, such as the presence of spurious singularities
(for bootstrap), or the necessity of performing additional computations
in 6 and 8 dimensions (for $D$-dimensional unitarity). The latter problem
can be bypassed by means of a mass shift~\cite{Badger:2008cm}, which
however might imply additional complications in the case of axial
couplings in massive theories. In summary, while it is true that
there are advantages and disadvantages in each of these approaches,
we point out that $R_2$ must not really be seen as an extra issue
in the context of a complete calculation, simply because one has to
carry out UV renormalisation anyhow, which is similar to $R_2$ 
but more involved.

\vskip 0.4truecm
\noindent
{\bf Integral reduction in MadLoop}

\noindent
We now turn to discussing the way in which the previous formulae
are handled by \ML. In order to do so, we shall re-instate in the
notation the dependence of the amplitudes on the relevant quantities;
in particular, we work with scalar sub-amplitudes that have definite 
helicities (i.e., all Lorentz indices are contracted away, and all 
Dirac matrices are sandwiched between spinors), and that factorise 
a single colour factor. The latter condition implies that, in general,
our sub-amplitudes are not in one-to-one correspondence with Feynman
diagrams (typically when these feature at least a four-gluon vertex),
but that they can be written as follows:
\beq
\ampnt_h=\sum_b\lambda^{(0)}_b {\cal B}_{h,b}\,,
\label{Borndec}
\eeq
for the Born amplitude. Here, $h$ denotes a given helicity configuration, 
and $b$ runs over all possible single-colour factors. The quantity
$\lambda^{(0)}_b$ is one such colour factor, that collects all the
colour indices, which are understood. Hence, ${\cal B}_{h,b}$ is a
scalar quantity which does {\em not} contain any colour index.
In the case of one-loop diagrams, we shall use a similar notation,
thus replacing the quantities that appear in eqs.~(\ref{ccpR1def})
and~(\ref{R2def}) with:
\beqn
N(\qloop)&\longrightarrow&\lambda^{(1)}_l {\cal N}_{h,l}(\qloop)\,,
\label{Ndec}
\\
\db{i}&\longrightarrow&\db{i,l}\,,
\label{dbdec}
\\
R_2&\longrightarrow&R_{2,h,l}\,.
\label{R2dec}
\eeqn
The quantity $\lambda^{(1)}_l$ in eq.~(\ref{Ndec}) has the same meaning
as $\lambda^{(0)}_b$ in eq.~(\ref{Borndec}), but is relevant to one-loop
amplitudes rather than tree-level ones, hence the different notation.
Since the index $l$ unambiguously identifies a single-colour-structure
subamplitude, and the latter has a non-trivial kinematic dependence,
different $l$'s may correspond to different one-loop Feynman diagrams, 
and thus the necessity of inserting a dependence on $l$ on the r.h.s.~of
eq.~(\ref{dbdec}). Finally, we did not factor out the colour structure
in eq.~(\ref{R2dec}), since this will not be relevant in the following.
We remark that decompositions such as those on the r.h.s.'s of 
eqs.~(\ref{Borndec}) and~(\ref{Ndec}) are easily handled by
\aNLO, which has inherited the treatment of the colour algebra
from \MadGraphf\ (see sect.~2.3 of ref.~\cite{Alwall:2011uj}). It should
be stressed that such a treatment is symbolic and that, although
introduced to deal with tree-level quantities, is perfectly capable
of computing one-loop ones such as $\lambda^{(1)}_l$. Furthermore,
the colour algebra is internally decomposed in terms of colour flows
(for which several different representations are available); 
although this information is presently
not used in the integral-reduction procedure, it will be trivial to
exploit it in the future, should this need arise.
By using eqs.~(\ref{Borndec})--(\ref{R2dec}) one obtains:
\beq
\ampnl_\unr {\ampnt}^{\star}=\sum_{\rm colour}\sum_h
\left(\sum_l \lambda^{(1)}_l\int d^d \bqloop\,
\frac{ {\cal N}_{h,l}(\qloop)}{\prod_{i=0}^{m_l-1}\db{i,l}}+
\sum_l R_{2,h,l}\right)
\left(\sum_b \lambda^{(0)}_b {\cal B}_{h,b}\right)^\star.
\label{VBtemp}
\eeq
As was already mentioned, the $R_2$ term gives rise to what is
effectively a tree-level computation; therefore,
we shall drop it in what follows, and just deal with:
\beqn
\left.\ampnl_\unr {\ampnt}^{\star}\right|_{\nonRt}&=&
\sum_h\sum_l\sum_b \int d^d \bqloop\,
\frac{ {\cal N}_{h,l}(\qloop)}{\prod_{i=0}^{m_l-1}\db{i,l}}\,\,
\Lambda_{lb}\,{\cal B}_{h,b}^\star\,,
\label{nonR2}
\\
\Lambda_{lb}&=&\sum_{\rm colour}\lambda^{(1)}_l {\lambda^{(0)}_b}^\star\,.
\eeqn
The integral-reduction procedure of \MLq~\cite{Hirschi:2011pa} can be read 
directly from eq.~(\ref{nonR2}), and is as follows:
\beq
\left.\ampnl_\unr {\ampnt}^{\star}\right|_{\nonRt}=
\sum_h\sum_l\sum_b \RED\left[\int d^d \bqloop\,
\frac{ {\cal N}_{h,l}(\qloop)}{\prod_{i=0}^{m_l-1}\db{i,l}}\right]
\Lambda_{lb}\,{\cal B}_{h,b}^\star\,,
\label{redML4}
\eeq
where $\RED\equiv\OPP$, since in \MLq\ only OPP reduction has been
considered. Equation~(\ref{redML4}) is not particularly satisfactory
from an efficiency point of view, for two reasons, both of which have
to do with the fact that the integral-reduction operation is quite 
time-consuming. Firstly, the $\RED[\,]$ operator is called $\#h\times\#l$
number of times (we recall that, by construction, $\#l$ is equal to or
larger than the number of loop diagrams). Secondly, each of these calls
involves the recomputation of ${\cal N}_{h,l}(\qloop)$ a large number
of times (determined by the OPP procedure), but which involve only
changing the numerical value of $\qloop$, without affecting any other
quantity entering it. The first issue is solved by reducing the number
of integral-reduction operations, while the second by rendering more
efficient the computation of ${\cal N}_{h,l}(\qloop)$. The strategies
adopted in \MLf\ are the following. One begins by observing that the
operator $\RED[\,]$ acts on the ``space" of one-loop integrals. Therefore,
Born amplitudes must be seen as c-numbers as far as this operator is
concerned. One can thus exploit the fact that $\RED[\,]$ is linear.
Furthermore, what really drives integral reduction is the structure
of the denominators (numerators are just numbers computed with suitable
values of $\qloop$, specific to the given $\RED[\,]$ operator).
Hence, one can organize loop integrals in sets of {\em topologies}, the 
latter being defined as subsets of integrals with the same denominator
combinations:
\beq
\prod_{i=0}^{m_l-1}\db{i,l}=\prod_{i=0}^{m_p-1}\db{i,p}\,,\;\;\;\;\;\;
\forall\,l,p\in t\,,\;\;\;\;\;\;t\in~{\rm topologies}\,.
\label{topdef}
\eeq
By exploiting eq.~(\ref{topdef}), one can rewrite eq.~(\ref{redML4})
as follows:
\beq
\left.\ampnl_\unr {\ampnt}^{\star}\right|_{\nonRt}=
\sum_t\RED\left[\int d^d \bqloop\,
\frac{\sum_h\sum_{l\in t}\sum_b {\cal N}_{h,l}(\qloop)\,
\Lambda_{lb}\,{\cal B}_{h,b}^\star}
{\prod_{i=0}^{m_{l_t}-1}\db{i,{l_t}}}
\right]\,,
\label{redML5}
\eeq
for any $l_t\in t$. Eq.~(\ref{redML5}) is optimal from the viewpoint
of reducing the number of calls to $\RED[\,]$, and thus addresses the
first of the issues mentioned before. As far as the second of those
issues is concerned, \ML5\ makes a systematic use of the fact that
any numerator ${\cal N}_{h,l}(\qloop)$ admits the following 
representation:
\beq
{\cal N}_{h,l}(\qloop)=\sum_{r=0}^{r_{\max}}
C^{(r)}_{\mu_1\ldots\mu_r;h,l}\,\qloop^{\mu_1}\ldots\qloop^{\mu_r}\,,
\label{Nhlexp}
\eeq
where the coefficients $C^{(r)}$ are independent of the loop momentum;
when $r=0$, we understand that no Lorentz indices and no loop momenta
appear on the r.h.s.~of eq.~(\ref{Nhlexp}). The quantity $r_{\max}$ is 
the largest rank in ${\cal N}_{h,l}$, and in the Feynman gauge in
renormalisable theories is always 
lower than or equal to the number of loop propagators. However,
since different ${\cal N}_{h,l}$ functions appear in the inner sums
in eq.~(\ref{redML5}), it is actually convenient (just for the sake
of using the present simplified notation) to regard it as the
largest rank in the whole one-loop amplitude; for a given ${\cal N}_{h,l}$,
this simply implies that some of the $C^{(r)}$ coefficients will be
equal to zero. Equation~(\ref{Nhlexp}) can be exploited in two ways.
One starts by determining the $C^{(r)}$'s once and for all. Then,
in the context of the OPP reduction, the numerical computation of
${\cal N}_{h,l}(\qloop)$ becomes much faster, as suggested originally
in \OL\ (see ref.~\cite{Cascioli:2011va}), because for each new
value of $\qloop$ generated within the OPP reduction one simply needs
to perform the sums and multiplications explicit in eq.~(\ref{Nhlexp}),
without recomputing the $C^{(r)}$'s. Furthermore, when 
eq.~(\ref{Nhlexp}) is symbolically (as opposed to numerically) replaced
in eq.~(\ref{redML5}), the structures of tensor integrals naturally
emerge. Thus, eq.~(\ref{Nhlexp}) paves the way to performing
a Tensor Integral Reduction (TIR) as well. Therefore, regardless of 
whether an OPP or TIR procedure will be applied, the inputs to the 
$\RED[\,]$ operator in \ML5\ are the sets:
\beq
\left\{\int d^d \bqloop\,\frac{\qloop^{\mu_1}\ldots\qloop^{\mu_r}}
{\prod_{i=0}^{m_{l_t}-1}\db{i,{l_t}}}\,,\;
\sum_h\sum_{l\in t}\sum_b
C^{(r)}_{\mu_1\ldots\mu_r;h,l}\,\Lambda_{lb}\,{\cal B}_{h,b}^\star
\right\}_{r=0}^{r_{\max}}.
\label{ML5inputs}
\eeq
Then, when using OPP the Lorentz indices of the two members of these 
sets are contracted in order to give OPP the scalar functions it needs.
On the other hand, when using TIR the first members in eq.~(\ref{ML5inputs})
are all that is needed for this type of reduction to work. 
It is clear that the practical success of the 
decomposition in eq.~(\ref{Nhlexp}) relies on the capability of a
fast and efficient computation of the coefficients $C^{(r)}$.
\ML5\ has a fully independent implementation of the recursion-construction
procedure presented in ref.~\cite{Cascioli:2011va} (thanks to 
a dedicated treatment by \ALOHA), and its own
internal system of caching and retrieving the $C^{(r)}$'s.
At variance with what is done in ref.~\cite{Cascioli:2011va}, \MLf\ 
does not assume $r_{\max}$ to be less than or equal to the number of loop 
propagators, i.e.~it admits the possibility that the contribution of 
any given vertex to the rank of ${\cal N}_{h,l}(\qloop)$ be larger 
than one. This is useful, for example, in the context of the computation
of QCD corrections to processes stemming from a Higgs EFT Lagrangian.
See appendix~\ref{sec:pol} for more details.

We conclude this part with a few diverse observations. Firstly, given
the advantages of an efficient caching and recycling of the coefficients
$C^{(r)}$, in the case of a mixed-coupling expansion \MLf\ starts from
determining all of the coefficients that will contribute to $\Sigma_{k_0+1,0}$
(we recall that we always associate the largest power of the dominant
coupling constant, as defined by the hierarchy of the model, with the
terms $q=0$: see sect.~\ref{sec:NLO}). Some of the $C^{(r)}$'s thus
computed will also contribute to $\Sigma_{k_0+1,1}$, for which \MLf\ will 
only calculate those $C^{(r)}$'s not yet found, and just recycle the others. 
This procedure is then iterated till necessary, in the sense that it
can be stopped when reaching the largest $q$'s among those selected
by the user; it is maximally efficient when the smallest user-selected
$q$ is equal to zero, which is justified from a physics viewpoint
given the coupling hierarchy of the model. 
Secondly, in general the kinematic configurations which are 
potentially unstable in TIR are rather different than with OPP;
therefore, the possibility of using TIR as an alternative to OPP
before turning to quadruple-precision calculations (see below)
is very beneficial for reducing the overall computing time when 
numerically-unstable situations are encountered. Finally, since 
TIR essentially performs integral reduction at the level of amplitudes,
rather than at that of amplitude squared as OPP, it allows one to
use efficiently the decomposition of eq.~(\ref{Nhlexp}) and its
caching-and-recycling system also in the case of loop-induced
matrix elements, eq.~(\ref{VLIdef}).

\vskip 0.4truecm
\noindent
{\bf UV renormalisation and $R_2$ contribution}

\noindent
In order to obtain $V(\proc)$ as defined in eq.~(\ref{Vdef}),
\ML\ must sum to the result of the integral-reduction procedure, 
eq.~(\ref{redML5}), the $R_2$ (see eq.~(\ref{VBtemp})) and the
UV-renormalisation contributions. Both of these can be cast in the
form of a tree-level-like amplitude $\ampnX$ times the Born
amplitude, whose contribution to eq.~(\ref{Vdef}) will therefore be:
\beqn
\mathop{\sum_{\rm colour}}_{\rm spin}
2\Re\left\{\ampnX(\proc){\ampnt(\proc)}^{\star}\right\},
\;\;\;\;\;\;\;\;
{\rm X}=R_2\,,\;{\rm UV}\,.
\label{VR2UV}
\eeqn
In an automated approach, the computation of $\ampnX$ may be performed
in the same manner as that of $\ampnt$, provided that the usual Feynman
rules are supplemented by new UV and $R_2$ rules, and by imposing
that $\ampnX$ contain one and only one UV- or $R_2$-type 
vertex\footnote{Note that in this context a ``vertex" can have
two external legs.}. This was indeed the procedure adopted by
\MLq\ for the $R_2$ computation. As far as UV renormalisation was
concerned, the fact that \MLq\ was limited to considering QCD
corrections to SM processes allowed significant simplifications,
and eq.~(\ref{VR2UV}) was effectively computed in a simpler way,
by taking the Born amplitude squared multiplied by suitable UV
factors. Mass insertions cannot be accounted for in this way; however,
their structure being identical to that of the $R_2$ two-point vertex,
the two could always be treated together (see ref.~\cite{Hirschi:2011pa} 
for more details).

The above solution is not tenable when considering an arbitrary
renormalisable theory, and therefore \MLf\ must explicitly compute
eq.~(\ref{VR2UV}) for both the UV and $R_2$ contributions.
This is not the only significant difference between \MLq\ and \MLf.
According to the general philosophy of the current \aNLO\ approach,
the UV and $R_2$ rules are part of the NLO theory model\footnote{In fact,
it is their very presence that tells NLO and LO models apart. See
also sect.~\ref{sec:method}.} chosen by the user, and that \ML\ adopts 
when performing a computation: they are not (as opposed to what happened 
with \MLq) hard-coded UV or $R_2$ computer routines corresponding to
$n$-point counterterms, but a set of instructions in \UFO, that \ALOHA\ 
will dynamically translate into the latter routines. 

After including the UV and $R_2$ rules into a \UFO\ model (an operation
that, we remind the reader, has to be performed once and for all per 
theory and per type of corrections, and which
\FeynRules\ will soon be able to perform automatically), it should
be clear that the computation by \aNLO\ of $\ampnX$ becomes 
identical to that of a regular tree-level amplitude for which only
Feynman rules are relevant; thus, we shall call this a tree-matching
construction. In order to increase the flexibility of \MLf, in
particular for what concerns the exclusion of the contributions of
certain loop integrals, and to allow developers more freedom when
debugging, an alternative but fully equivalent procedure has 
been implemented, which we shall call loop-matching construction. 
The tree-matching and loop-matching constructions (an explanation
of which will be given below)
can be understood by considering the physics contents of the generic
UV and $R_2$ counterterm $G$, which we shall 
denote as follows:
\beq
G=\left\{\left\{\ident_1^{(e)},\ldots\ident_m^{(e)}\right\};
\left\{\ident_1^{(l_k)},\ldots\ident_{n_k}^{(l_k)}\right\}_{k=1}^{k_{loop}};
{\cal W},L,\lambda,c,X\right\}\,.
\label{cntdef}
\eeq
The quantity $X$ determines the kind of counterterm one is working
with -- UV wave-function, or internal $n$-point UV and $R_2$ functions;
it also allows the builder of the model to specify whether $G$ will
be used in the context of a tree-matching or loop-matching construction.
In other words, and in order to stress this point again: \MLf\ can
handle both types of construction, which are simply seen as attributes
of the model used for the computation. This flexibility is important
for example because models constructed ``by hand" are set up
in a different way w.r.t.~that adopted by \FeynRules.
The quantity $c$ is symbolically (i.e., not numerically) set equal
to a coupling constant, with that implying that $G$ is a counterterm
relevant to NLO corrections in the theory whose perturbative expansion
is governed by that coupling. Note that in a model there may be several
subsets of counterterms, each associated with a different type of correction;
in the example of the mixed QCD-QED case discussed at the beginning of
sect.~\ref{sec:NLO}, when $c=\as$ one has counterterms for QCD
corrections (eq.~(\ref{QCDcorr})), and when $c=\aem$ one has counterterms 
for QED corrections (eq.~(\ref{QEDcorr})). The quantities $L$ and
$\lambda$ in eq.~(\ref{cntdef}) denote the Lorentz and colour structures
of $G$ respectively. 
The set \mbox{$\{\ident_1^{(e)},\ldots\ident_m^{(e)}\}$} is the
list of ``external" particle identities associated with $G$. For example,
if $G$ is the contribution to $\as$ renormalisation due 
to the $gu\bu$ vertex, then this set is equal to 
\mbox{$\{g,u,\bu\}$}; if $G$ corresponds to the mass insertion for
the top quark, then one has\footnote{A mass insertion is treated
in a similar manner as vertices; therefore, all external particles must 
be outgoing, whence the $\bt$.} \mbox{$\{t,\bt\}$}. The counterterm $G$
in general receives contributions from $k_{loop}\ge 1$ different types
of loop diagrams, and \mbox{$\{\ident_1^{(l_k)},\ldots\ident_{n_k}^{(l_k)}\}$}
is called the $k^{th}$ loop topology, i.e.~the set of the identities 
of the particles circulating in the
$k^{th}$ type of loop\footnote{The term ``set" implies that two or more 
identical particles contribute {\em one} particle identity to this set, 
as opposed to the case of external particles, which are all explicitly 
present in the
{\em list} \mbox{$\{\ident_1^{(e)},\ldots\ident_m^{(e)}\}$}.
Note that, when circulating in a loop, a quark and its antiquark
can be identified. Furthermore, it should be clear that the present
``topology" has nothing to do with that introduced in eq.~(\ref{topdef}).}.
When counting these loops, one needs to take into account the physical
meaning of $G$. By using again the example of the $gu\bu$ vertex, one may
be tempted to conclude that $k_{loop}=1$, which corresponds to the triangle
corrections to such a vertex. This would be incorrect: in fact, since $G$
is a contribution to $\as$ renormalisation, it must include wave-function
renormalisation factors; therefore, in the $k_{loop}$ types of loops one
must include the bubble diagrams relevant to the $g$, $u$, and $\bu$
external legs. Thus, in this example one will need to consider both
\mbox{$\{g,u\}$} (for triangles and the $u$ self-energy) and $\{g\}$, 
$\{q\}$, $\{b\}$, $\{t\}$ (for the gluon self-energy; $q$ is a massless
quark, and $b$ and $t$ are heavy quarks). It should be clear that this
complication (there are more loop topologies that contribute to $G$
than the list of its external particles would suggest) is basically due
to wave-function renormalisation, which physically corresponds to the
fact that renormalised couplings are defined in terms of renormalised
Green functions. When this is not the case, and notably for mass insertions,
and for $R_2$ contributions (because the latter are directly defined
in terms of specific loop integrals -- see eq.~(\ref{R2def})), 
loop topologies are indeed in one-to-one 
correspondence with those naively deduced from the list of external 
particles. Finally, in eq.~(\ref{cntdef}) ${\cal W}$ represents the
actual value of the counterterm $G$. By following the usual textbook
procedure which makes use of renormalised coupling constants, and by
ignoring the Lorentz and colour structures, one may e.g.~have:
\beq
{\cal W}\propto 1-Z_{\rm coupling}^{-1}=
1-Z_{\rm vertex}^{-1}\prod_i Z_{{\rm wf},i}\,,\;\;\;\;\;\;\;\;
{\cal W}=0\,,
\label{GZs1}
\eeq
for coupling and internal 2-point (bubble) renormalisation 
respectively, and
\beq
{\cal W}\propto \delta m\,,\;\;\;\;\;\;\;\;
{\cal W}\propto W_{R_2,{\rm vertex}}\,,
\label{GZs2}
\eeq
for UV mass insertions and an $R_2$ correction, respectively. 
From the discussion presented before, it follows that the 
$Z$ and $\delta m$ terms in eqs.~(\ref{GZs1}) and~(\ref{GZs2}) will 
be related to the sets
\mbox{$\{\ident_1^{(l_k)},\ldots\ident_{n_k}^{(l_k)}\}$}.
On the other hand, \MLf\ leaves the model builder the possibility
of implementing coupling-constant renormalisation by directly working
at the level of vertex renormalisation. This can be simply done for
example by adopting:
\beq
{\cal W}\propto 1-Z_{\rm vertex}^{-1}\,,\;\;\;\;\;\;\;\;
{\cal W}\propto 1-Z_{{\rm wf},i}\,,
\label{GZs3}
\eeq
instead of the settings of eq.~(\ref{GZs1}). One may (slightly
improperly, since the physics contents are exactly the same) refer
to the procedures induced by eq.~(\ref{GZs1}) and eq.~(\ref{GZs3}) 
as coupling-constant and vertex renormalisation respectively. The latter 
is the current method of choice for the extension of \FeynRules\ to NLO.
Note that, when working with vertex renormalisation, wave-function
factors for external legs need not be included (as opposed to the usual
case of coupling-constant renormalisation), which can be easily specified
in the model definition. Again, no assumption is made in \MLf\ on the
treatment of external legs, and full flexibility is maintained by
reading the relevant information from the model.

Now we suppose that an NLO model is given, and thus eq.~(\ref{cntdef}) 
is fully specified for all counterterms relevant to all types of corrections
dealt with by the model. We shall now discuss how \MLf\ exploits such
information. First of all, one may want to exclude some loops
from the calculation (this can be motivated by physics requirements,
for example when leaving out heavy-flavour contributions, or done for
debugging purposes); we call this operation {\em loop-content filtering}
(not to be confused with diagram filtering, which discards over-counting L-cut 
diagrams at generation time -- see sect.~3.2.1 of ref.~\cite{Hirschi:2011pa}).
This filtering is basically trivial in the generation of diagrams: one
simply does not include the undesired particles in the list of 
L-cut particles. The presence of 
\mbox{$\{\ident_1^{(l_k)},\ldots\ident_{n_k}^{(l_k)}\}$} 
in eq.~(\ref{cntdef}) allows one to do the same when computing
the UV and $R_2$ contributions: if this set contains the particle(s)
to be discarded, the corresponding contributions are not included
in eqs.~(\ref{GZs1})--(\ref{GZs3}). Apart from loop-content filtering, 
eq.~(\ref{cntdef}) can be exploited in the context of the tree-matching 
construction in the same way as all other elementary building blocks
derived from ordinary Feynman rules, by using the information on
external particles, and Lorentz and colour structures. As far as the
loop-matching construction is concerned, one starts from a given loop
integral $\diag$, and determines what we call its associated tree topology:
\beq
\diag\;\longrightarrow\;\Gamma^{(\diag)}=\left(
\left\{\tree_1,\ldots \tree_{T(\diag)}\right\},
\left\{\ident_1^{(l_\diag)},\ldots\ident_{n}^{(l_\diag)}\right\}\right)\,.
\label{treetop}
\eeq
The notation understands that there are $T(\diag)$ trees $\tree_\alpha$ 
attached to the loop\footnote{A four-gluon vertex, with two gluons belonging
to the loop, gives rise to two trees, both of which are attached to 
the loop at the same point.} (see fig.~2 of 
ref.~\cite{Hirschi:2011pa} for a graphical example of such trees);
\mbox{$\{\ident_1^{(l_\diag)},\ldots\ident_{n}^{(l_\diag)}\}$}
is the set of particles that flow in the loop. Since different loops
can have the same associated tree topology, \MLf\ first collects
all of the different tree topologies relevant to the computation being
performed. Next, for each of these, all counterterm vertices $G$
(that have survived loop-content filtering) are found that fulfill 
the following equation:
\beq
\left\{\ident_1^{(e)},\ldots\ident_m^{(e)}\right\}=
\left\{{\cal R}\left(\tree_1\right),\ldots 
{\cal R}\left(\tree_{T(\diag)}\right)\right\}\,,
\label{treevsG}
\eeq
where by ${\cal R}(\tree_\alpha)$ we have denoted the root of
the $\alpha^{th}$ tree $\tree_\alpha$ (the root being obviously the
single particle that stems from the loop, and branches into the tree);
note that eq.~(\ref{treevsG}) implies $m=T(\diag)$.
Among the counterterms thus found, one further considers the UV mass
insertion and $R_2$ ones, and discards those that do not fullfil 
the equation:
\beq
\left\{\ident_1^{(l_k)},\ldots\ident_{n_k}^{(l_k)}\right\}=
\left\{\ident_1^{(l_\diag)},\ldots\ident_{n}^{(l_\diag)}\right\}\,.
\label{loopvsG}
\eeq
Equation~(\ref{loopvsG}) guarantees a rather strict correlation 
between a generated diagram and its UV and $R_2$ counterterms, which 
is quite useful for example when establishing the correctness of a model.
It should be pointed out, however, that such a correlation can
never be turned into a one-to-one map, because 
of coupling-constant or vertex renormalisation, in which case
eq.~(\ref{loopvsG}) cannot be imposed (since vertex corrections
and wave-function renormalisation are always strictly related,
and this is true also when carrying out a vertex-renormalisation
procedure). 
Finally, for each counterterm $G$ selected as explained above,
\MLf\ builds the corresponding tree amplitude by attaching to $G$ 
the off-shell currents relevant to the tree structures
\mbox{$\{\tree_1,\ldots \tree_{T(\diag)}\}$}.
When performing coupling-constant renormalisation, \MLf\
also constructs additional tree-amplitude counterterms by multiplying
each Born amplitude by the suitable combination of external 
wave-function factors.

We conclude this part by re-iterating its main message: \MLf\ has
a fully-flexible structure that allows it to handle on equal footing
several different strategies, such as tree-matching vs loop-matching
construction, or coupling vs vertex renormalisation, thanks to its
capability of obeying the relevant instructions encoded in the
model. This includes the prescription for the renormalisation scheme,
which now must be simply seen as one of the model characteristics.

\vskip 0.4truecm
\noindent
{\bf Checks, stability, and recovery of numerically-unstable 
integral reductions}

\noindent
\MLq\ featured many self-consistency checks, that served to establish
the correctness of any one-loop matrix element generated by the code;
they are described extensively in sect.~3.3 of ref.~\cite{Hirschi:2011pa},
and all of them are inherited by \MLf. On top of those, in \MLf\ we
have included two new checks: firstly, we verify that the 
matrix elements are Lorentz scalars, by recomputing them using a kinematic 
configuration obtained by boosting and rotating the original one;
and secondly, in the case of QCD corrections we test whether the matrix
elements computed in the unitary gauge are identical to those computed
in the Feynman gauge\footnote{Incidentally, these tests can also be
performed in the context of tree-level matrix-element computations.}. 
The experience with \MLq\ and \MLf\ has shown 
that there is a vast amount of redundancy in all of these checks; therefore,
in \MLf\ we have decided to perform only one of them before proceeding
to integrate the matrix elements. We have chosen the most complete one,
namely that on the residues of the infrared poles (where the numerical
values of such residues as returned by \ML\ are compared to those known
analytically from the subtraction of real-emission singularities), which
has the virtue of being an indirect test on the UV-renormalisation
procedure as well. The other checks can still be performed if need be,
by simply executing a single command from the \aNLO\ interactive
shell\footnote{Apart from that on the dependence on the mass of 
an heavy quark, which is still available but too process-specific to 
be worth automating.}.

Integral-reduction procedures are fairly involved, and some kinematic
configurations may give rise to numerically-unstable results. Any automated
approach must therefore have solid self-diagnostic and recovery strategies:
those of \MLq\ have been presented in sect.~3.4 of ref.~\cite{Hirschi:2011pa}.
In view of the extended scope of \MLf\ w.r.t.~that of \MLq, both of these 
strategies have been completely redesigned, for the technical reasons
which we shall now discuss.
Firstly, \MLq\ based the instability diagnostics on the results of
tests performed within \CutTools, on a loop-by-loop basis. 
Therefore, the capability of \MLf\ to exploit both the OPP and TIR
methods has forced us to set up a \CutTools-independent diagnostic tool.
Furthermore, a loop-by-loop method is in any case not ideal, because it
is never trivial to determine the threshold that decides when a computation
is flagged as unstable (e.g., a single loop integral may be unstable, but
give a totally negligible contribution to the full amplitude). This
problem is exacerbated when increasing the final-state multiplicity,
and indeed suggests to use an ``inclusive" (i.e., at the level of the
amplitude, rather than of the individual loop integral) type of test
in \MLf, which can handle more complicated processes than \MLq.
Secondly, the recovery strategy used by \MLq\ (which involved a
small deformation of the kinematics) is on the one hand related
to the loop-by-loop diagnostic (because when one considers two or more
integrals simultaneously, the deformation of the kinematics that renders
stable an unstable integral can turn a stable integral into an unstable
one), and on the other hand not particularly
satisfactory when increasing the process multiplicity (because it
becomes more difficult to obtain a stable result out of the deformed
kinematic configuration).

The diagnostic procedure of \MLf\ works as follows. Let us write
the result of eq.~(\ref{Vdef}), obtained by \ML\ after integral
reduction and UV renormalisation, as follows:
\beqn
V(\proc)=\frac{(4\pi)^\ep}{\Gamma(1-\ep)}
\left(\frac{\muF^2}{Q^2}\right)^\ep
\left(\frac{c_{-2}}{\ep^2}+\frac{c_{-1}}{\ep}+c_0\right)\,.
\label{Vexpr}
\eeqn
For any given kinematic configuration, the coefficients $c_j$
of eq.~(\ref{Vexpr}) are evaluated \mbox{$1+n_{test}$} times
(in a way specified below), which we denote as follows:
\beq
c_j\;\longrightarrow\; c_j^{(i)}\,,\;\;\;\;\;\;\;
i=0,\ldots n_{test}\,,\;\;\;\;\;\;{\rm for}~j=-2,-1,0\,.
\label{cji}
\eeq
These coefficients are used to define the following quantities:
\beqn
\bar{c}_j&=&\half\left(\max\left\{\abs{c_j^{(i)}}\right\}_{i=0}^{n_{test}}+
\min\left\{\abs{c_j^{(i)}}\right\}_{i=0}^{n_{test}}\right)\,,
\label{cjbar}
\\
\Delta c_j&=&\max\left\{\abs{c_j^{(i)}}\right\}_{i=0}^{n_{test}}-
\min\left\{\abs{c_j^{(i)}}\right\}_{i=0}^{n_{test}}\,,
\label{Delcj}
\eeqn
which in turn enter the definition of the relative accuracy
of the \ML\ evaluation:
\beq
\chi=\frac{\sum_{j=-2}^0 \Delta c_j}{\sum_{j=-2}^0 \bar{c}_j}\,.
\label{chidef}
\eeq
A computation is deemed unstable, and the corresponding kinematic
configuration called an Unstable Phase-Space point (UPS), when:
\beq
\chi\,>\,\varepsilon\,,
\label{chicond}
\eeq
with $\varepsilon$ a quantity which can be defined by the user, but whose 
default value is $10^{-3}$. The $c_j^{(i=0)}$ results in 
eqs.~(\ref{cji})--(\ref{Delcj}) are those obtained by applying the OPP 
reduction with the given kinematic configuration. The $c_j^{(i>0)}$ 
are obtained in two different ways, by performing again the integral
reduction either: {\em a)} by using a kinematic configuration obtained 
by rotating the original one (hence, by following the same 
procedure as is used in one of the self-consistency checks previously 
discussed); or {\em b)} by using a different ordering of the loop
propagators $\db{i}$ as input to OPP (this changes the inner workings
of the reduction procedure, and is thus numerically different from, 
although physically completely equivalent to, what one does with the
original ordering). These two re-computation procedures are called
{\em Lorentz test} and {\em Direction test} respectively. By default,
\MLf\ sets $n_{test}=2$, and performs one Lorentz test and one Direction
test. Both $n_{test}$ and the type of tests performed can be controlled
by the user. Note that any Direction test re-uses the coefficients $C^{(r)}$ 
of eq.~(\ref{Nhlexp}) computed in the context of the first evaluation
($c_j^{(i=0)}$), and is thus less time-consuming than Lorentz tests, despite
the fact that both require the integral reduction to be performed from 
scratch. In the case of a mixed-coupling expansion, each of the $c_j^{(i)}$ 
is expanded as is done in eq.~(\ref{taylor4}), so that it will correspond 
to a set $\{c_{j,q}^{(i)}\}$. By fixing $q$ (which is associated with 
a given combination of coupling constants, see sect.~\ref{sec:NLO}), 
one defines $\chi_q$ as in eqs.~(\ref{cjbar})--(\ref{chidef});
a kinematic configuration is a UPS if, following eq.~(\ref{chicond}),
$\chi_q>\varepsilon$ for any $q$. 

When a UPS is found, \MLf\ has two main methods for recovery,
which are attempted in turn. It starts by changing the 
integral-reduction procedure, from OPP to TIR. The results of
TIR depend on the specific TIR library \MLf\ is linked to.
In principle, any library might be used; in practice, so far we 
have considered \IREGI~\cite{ShaoIREGI} and
\PJF~\cite{Fleischer:2012et,YundinPhd:2012}. 
A given TIR library has a maximal number of
propagators it can handle (presently, up to hexagons for \IREGI\ and up to 
pentagons for \PJF); in the case one particular loop integral exceeds
that number, OPP is used again for it and only for it\footnote{This
being a single integral, it should be clear that this procedure
will not necessarily result again in being classified as a UPS --
the original UPS was due to {\em all} integrals being reduced with OPP.}.
More than one TIR library can be linked to \MLf\ at the same time.
After having found a UPS with OPP, \MLf\ switches to the first of
such TIR libraries, and repeats the diagnostic tests mentioned above.
If the result is again classified as a UPS, the next TIR library
is used, and so forth. If none of the available TIR libraries 
is able to give a numerically-stable reduction, \MLf\ resorts
to the second method of recovery, namely the OPP integral reduction
with all relevant quantities (${\cal N}_{h,l}$ and the internal
\CutTools\ algebra) computed in quadruple precision. This is usually
extremely effective, but has the disadvantage of being extremely slow.
In the case when the recovery in quadruple precision fails as well, \MLf\
gives up, sets $c_0=0$, and proceeds to the next kinematic configuration;
the user is warned when this happens.
We emphasise that the order in which the various integral-reduction
procedures are used in the context of UPS recovery (OPP, TIR library \#1 
to TIR library \#$n$) can be controlled through an input card. So in the 
present public version of \aNLO, where TIR reduction is not yet included, 
only OPP and quadruple-precision calculations are employed.

\subsubsection{Integration of one-loop contributions\label{sec:int}}

\noindent
The way in which the virtual contributions are integrated by \aNLO\
in either an fNLO or an NLO+PS computation is quite different w.r.t.~what
was done in \amcatnlo. In the latter, one-loop
matrix elements were integrated separately from the other contributions,
and eventually combined with them at the level of either distributions
(in the case of fNLO), or unweighted events (in the case of NLO+PS);
on the other hand, in \aNLO\ all contributions are integrated simultaneously.
The original strategy of \amcatnlo\ had been adopted because it allowed 
one to control, 
in a very direct manner, the number of phase-space points for which the 
virtual corrections were computed, and by doing so to reduce such a number,
without this implying a degradation of the overall accuracy of the 
physical results\footnote{The reduction of the number of evaluations of
the one-loop matrix elements was (and still is) highly desirable because
virtual contributions are typically the numerical bottleneck in our
NLO computations (owing to the efficiency of the FKS subtraction,
which leads to a relatively fast convergence of the real-emission 
contributions).}. The fact that the accuracy of the final result does 
not change significantly despite the reduction mentioned above stems
from the following two observations. 
{\em a)} $n$-body phase-space integrals are significantly simpler 
than $(n+1)$-body ones, and therefore require to be sampled a smaller
number of times than the latter.
{\em b)} Virtual corrections are usually smaller than the Born, which
implies that a smaller number of phase-space points has to be used
to integrate the former than the latter, in order to obtain the
same absolute precision for the two resulting integrals.
The possibility of exploiting observation {\em a)} in a flexible
manner was the main reason why in \amcatnlo\ the virtual contributions
were integrated separately. In fact, without a separate integration,
$n$-body matrix elements were previously evaluated the same number
of times as $(n+1)$-body ones (see eq.~(\ref{BfromR}) for an explicit
example, relevant to the Born). This could of course be bypassed in
several ways, none of which however is simpler than a separate treatment,
and better suited to an adaptive multi-channel integration.
As explained in sect.~\ref{sec:FKS}, \aNLO\ combines $n$- and $(n+1)$-body
contributions in the opposite way w.r.t.~that of \amcatnlo, taking
an $n$-body viewpoint. This is what allows \aNLO\ to naturally use 
observation {\em a)} while integrating one-loop matrix elements
together with all other contributions.

The simultaneous versus separate integration is only one of the differences
between the current treatment in \aNLO\ and what was done previously.
While the former has several advantages over the latter\footnote{On top of 
item {\em a)} discussed above: all other things being equal, two
or more contributions integrated together lead to a better accuracy
than when integrated separately, in the case of cancellation among
them, as it often happens with NLO cross sections; also, a smaller
number of integration channels implies a reduction of negative-weighted
NLO+PS events.}, if applied straightforwardly it implies that the same number
of evaluations are performed for the one-loop as for the Born matrix
elements, which is not ideal in view of observation {\em b)}. In order
to amend this situation, and thus to increase the speed of \aNLO\ without
a loss of accuracy, several solutions have been devised. They are
based on the properties of the following quantity\footnote{With some abuse
of notation, $V$ here denotes only the finite part of the one-loop
contribution, i.e.~the coefficient $c_0$ of eq.~(\ref{Vexpr}) up to 
overall factors, which are irrelevant for the present discussion.}:
\beq
\frac{V_h}{\abs{\ampnt_h}^2}\equiv
\frac{2\Re\big\{\ampnl_h{\ampnt_h}^{\star}\big\}}{\abs{\ampnt_h}^2}\,,
\label{VoBh}
\eeq
where $\ampnt_h$ has been introduced in eq.~(\ref{Borndec}), and $V_h$ 
can be obtained e.g.~from eq.~(\ref{VBtemp}) by not performing the sum 
over $h$ there (and similarly for the corresponding UV counterterms).
The ratio in eq.~(\ref{VoBh})
is a slowly-varying function over the phase space (behaving essentially 
as logarithms or dilogarithms), and is to a good extent independent
of the helicity configuration $h$ (note that the same helicity 
configuration $h$ is used in the numerator and in the denominator):
in other words, one-loop and Born matrix elements have very similar 
dependencies on helicity configurations and, in particular, there are 
no helicity configurations for which $\ampnt_h$ is null while $V_h$ 
is not. This also implies that the ratio of eq.~(\ref{VoBh}) 
is numerically of the same order as its helicity-summed counterpart:
\beq
\frac{V}{\abs{\ampnt}^2}\equiv
\frac{\sum_h V_h}{\sum_{h^\prime}\abs{\ampnt_{h^\prime}}^2}\,.
\label{VoB}
\eeq
In \aNLO\ we exploit the behaviour w.r.t.~to $h$ of eq.~(\ref{VoBh})
by performing the sum over helicities implicit in $V$ by
means of MC methods: for each phase-space point, a single helicity 
configuration is chosen, according to the relative weights of 
$\abs{\ampnt_h}^2$. What has been said above guarantees the 
efficiency and the fast convergence of this procedure, as well as a
reduction of the time spent in computing the one-loop contribution
(see eq.~(\ref{ML5inputs}) -- such a reduction is due to the fact that
the numerator is simpler and therefore less time-consuming: the time
spent carrying out the integral reduction is not affected).

Let us finally see how the fact that the quantity in eq.~(\ref{VoB})
is a slowly-varying function of the kinematics
helps reduce further the CPU load necessary
to compute the integral of $V$. In order to shorten the notation
introduced in sect.~\ref{sec:FKS}, we symbolically write the integral
of the NLO cross section as follows:
\beq
\int \phspn \left(E_V+V\right)\,,
\label{NVpVint}
\eeq
where $E_V$ denotes all contributions other than $V$ (the integration over
the extra degrees of freedom relevant to the real matrix elements plays
no role here, and is understood). Integrals such as that of 
eq.~(\ref{NVpVint}) are performed by adaptive methods, which entail
successive estimates (called iterations) of quantities relevant to
the integrals. For the generic integral:
\beq
\int \phspn F
\label{Fint}
\eeq
we shall denote by
\beq
I_k(F)\,,\;\;\;\;\;\;\;\;
\sigma_k(F)\,,
\eeq
the results of the $k^{th}$ iteration for the mean (i.e., the integral
itself) and the standard deviation. One expects that:
\beq
\lim_{k\to\infty}I_k(F)=\int \phspn F\,,\;\;\;\;\;\;\;\;
\lim_{k\to\infty}\sigma_k(F)=0\,.
\eeq
It will be convenient for what follows to have an explicit expression
for the mean:
\beq
I_k(F)=\frac{1}{p_k}\sum_{i=1}^{p_k}\Phi_n\left(\phi_n^{(k,i)}\right)
F\left(\phi_n^{(k,i)}\right)\,.
\label{Ikexpl}
\eeq
Here, we have denoted by $\phi_n^{(k,i)}$ the $i^{th}$ phase-space point,
generated at random during the course of the $k^{th}$ iteration; a total
of $p_k$ points are considered. The quantity $\Phi_n$ collects all 
normalisation and jacobian factors. We point out that, when applying
eqs.~(\ref{Fint})--(\ref{Ikexpl}) to the case of interest,
eq.~(\ref{NVpVint}), one is able to obtain not only the integral
of the sum \mbox{$E_V+V$}, but also those of $E_V$ and $V$ individually,
by keeping track of $I_k(E_V)$ and $I_k(V)$ respectively (despite the
fact that the two terms are still integrated simultaneously). This
is useful in view of the following manipulation: we introduce an
approximant of $V$, that we denote by $\tV_k$, which we use in 
the identity:
\beq
\int \phspn V = \int \phspn\left[\tV_k+\left(V-\tV_k\right)\right]\,.
\label{VtViden}
\eeq
As the notation suggests, the approximant $\tV_k$ is a function of
the adaptive-integration iteration, where it is used according
to the following formula:
\beq
I_k(V)=\frac{1}{p_k}\sum_{i=1}^{p_k}\Phi\left(\phi_n^{(k,i)}\right)
\tV_k\left(\phi_n^{(k,i)}\right)+
\frac{1}{p_kf_k}\sum_{i=1}^{p_kf_k}\Phi\left(\phi_n^{(k,i)}\right)
\left[V\left(\phi_n^{(k,i)}\right)-\tV_k\left(\phi_n^{(k,i)}\right)\right]\,.
\label{IkVmVt}
\eeq
A number $0<f_k\le 1$ has been introduced in eq.~(\ref{IkVmVt}), which
implies that the difference \mbox{$V-\tV_k$} is computed only in a fraction
$f_k$ of the total number of point thrown\footnote{Although eq.~(\ref{IkVmVt})
literally implies that these are the first $p_kf_k$ points, in the actual 
computation
they are chosen randomly in the whole set of the $p_k$ points, so that biases
are avoided.}. For an explicit evaluation of eq.~(\ref{IkVmVt}), we need
to define what enters it. We have:
\beq
\tV_k=c_k\abs{\ampnt}^2\,,
\label{tVkdef}
\eeq
with $c_k$ a quantity to be determined iteration-by-iteration, similarly
to what happens for $f_k$. The initial conditions are:
\beq
f_1=1\,,\;\;\;\;\;\;\;\;
c_1=0\,,
\label{incond}
\eeq
and for $k>1$ we define:
\beqn
c_k&=&\frac{{\rm grid}\left\{I_{k-1}(V)\right\}}
{{\rm grid}\left\{I_{k-1}\left(\abs{\ampnt}^2\right)\right\}}\,,
\label{ckdef}
\\
f_k&=&f_{k-1}\max\left\{
\min\left\{\frac{2\sigma_{k-1}(V-\tV_{k-1})}{\sigma_{k-1}(E_V+V)},2\right\},
\frac{1}{4}\right\}\,.
\label{fkdef}
\eeqn
The value of $f_k$ obtained from eq.~(\ref{fkdef}) is further constrained
to be in the range
\beq
0.005\le f_k\le 1\,.
\label{fkrange}
\eeq
A few explanations are in order. Firstly, $c_k$ and $f_k$ are dynamically
constructed, using the information that the numerical integrator
(we use a modified version of 
{\sc\small MINT}~\cite{Nason:2007vt}) has gathered during
the previous iteration. One such piece of information is a grid,
which among other things stores the averages of the function that
is being integrated in non-overlapping phase-space regions which cover
the whole phase space. Therefore, $c_k$ as defined in eq.~(\ref{ckdef})
is a piecewise-constant function. Because of the properties of
eq.~(\ref{VoB}), we expect it to be close to an overall constant,
and $\tV_k$ defined in eq.~(\ref{tVkdef}) to be a good approximant
of $V$. Secondly, if indeed $\tV_k$ is an increasingly (with $k$) good
approximant of $V$, we expect the quantity \mbox{$\sigma_{k-1}(V-\tV_{k-1})$}
that appears in eq.~(\ref{fkdef}) to decrease faster than the estimated
error on the integral of \mbox{$E_V+V$}, thus inducing the values of
$f_k$ to decrease. On the other hand, eq.~(\ref{fkdef}) prevents the
series of $f_k$'s to be fluctuating: w.r.t.~the preceding value $f_{k-1}$,
$f_k$ can be at most a factor of 2 larger, or a factor $1/4$ smaller --
these values are simply sensible, but can of course be changed, as
the absolute minimum for $f_k$ given in eq.~(\ref{fkrange}).

The rationale behind eqs.~(\ref{VtViden})--(\ref{fkrange}) should
now be clear, and it has to do with the fact that one can compute 
$\tV_k$ much faster than $V$. One starts in the first iteration 
by always computing $V$; while doing so, \aNLO\ gathers the information
that will allow it to construct the approximant $\tV_2$ to be used
in the next iteration. While this procedure is iterated, the 
relative\footnote{\aNLO, following \MadGraph, starts with a relatively
small number of points $p_1\simeq 80N_{dim}$, and doubles it at each 
iteration.} number of times $V$ ($\tV_k$) is computed is decreased 
(increased). The procedure is exact, being based on the {\em local}
identity~(\ref{VtViden}).
Furthermore, the code is protected against any pathological behaviour:
if, for example, $\tV_k$ does not turn out to be a good approximant
of $V$, one will have $f_k\simeq 1$ for all $k$'s, so that $\tV_k$ will
not play any role (see eq.~(\ref{IkVmVt})). In practice, this situation
has not been encountered so far.

\subsubsection{Matching to showers: MC@NLO\label{sec:mcatnlo}}

In this section, we review the MC@NLO formalism~\cite{Frixione:2002ik}
and its implementation in \aNLO, making extensive use of the results
given in sect.~\ref{sec:FKS}. We start by considering the formulation
where the short-distance cross sections are defined for a given
real-emission process $\procR\in\allprocnpo$, which also allows one
to symplify the notation, since the dependence on $\procR$ can thus be
easily understood. We shall eventually arrive at expressions which
lend themselves to the same manipulations as those carried out 
at the end of sect.~\ref{sec:FKS}, which \aNLO\ exploits in order
to deal with MC@NLO cross sections defined at given Born processes,
precisely as for their NLO counterparts.

In essence, MC@NLO defines two short-distance cross sections, associated
with real-emission-type kinematics (i.e., $(n+1)$-body) and Born-type 
kinematics (i.e., $n$-body), dubbed $\clH$- and $\clS$-event contributions
respectively. Their forms are written as follows:
\beqn
d\sigmaH&=&\sum_{(i,j)\in\FKSpairs}d\sigmaNLOE_{ij}-d\sigmaMC\,,
\label{xsecH}
\\
d\sigmaS&=&d\sigmaMC+\sum_{(i,j)\in\FKSpairs}
\sum_{\alpha=S,C,SC} d\sigmaNLOa_{ij}\,,
\label{xsecS}
\eeqn
where $d\sigmaNLOE_{ij}$ and $d\sigmaNLOa_{ij}$ are {\em exactly} the same 
quantities (eqs.~(\ref{xsecNLO}) and~(\ref{fact3})) that appear
in the NLO cross section (eq.~(\ref{NLOdiff})). The only new 
(w.r.t.~the NLO) ingredient in MC@NLO is thus $d\sigmaMC$, which
is the cross section one obtains from the parton shower Monte Carlo
(PSMC) one interfaces to by truncating the perturbative expansion at 
${\cal O}(\as^{b+1})$ (the Born matrix elements being of ${\cal O}(\as^b)$),
in the case of resolved emission (eq.~(\ref{xsecH})) and of no
resolved emission (eq.~(\ref{xsecS})) -- indeed, as is implicit in
the notation these two cases result in the same cross section, up
to a sign. The crucial point is that, since the leading IR behaviour
of any PSMC must be the same as that resulting from an exact 
matrix-element computation in QCD, eqs.~(\ref{xsecH}) and~(\ref{xsecS})
are locally finite\footnote{In fact, this locality property may be spoiled
by certain approximations inherent in the PSMC. It is not difficult to
restore it~\cite{Frixione:2002ik}, as we shall briefly discuss later.}.
This is the reason why the $d\sigmaMC$ terms are called the MC counterterms,
and the MC@NLO cross section, contrary to the NLO one, can be unweighted.

The definition of the MC counterterms immediately implies that their
actual expressions depend on the specific PSMC one interfaces to. These
expressions have therefore to be worked out case-by-case, which has been
done for the following PSMCs, whose matching with NLO calculations has 
been fully validated in \aNLO: \PYe~\cite{Sjostrand:2007gs}, 
\HWpp~\cite{Bahr:2008pv,Bellm:2013lba}, 
\HWs~\cite{Corcella:2000bw,Corcella:2002jc}, and \PYs~\cite{Sjostrand:2006za} 
(in the case of $\pt$-ordered \PYs, only processes with no 
strongly-interacting particles in the final state are supported).
The details of the construction of the MC counterterms for some of
these PSMCs are given in refs.~\cite{Frixione:2002ik,Frixione:2003ei,
Frixione:2005vw,Torrielli:2010aw,Frixione:2010ra}. 
On the other hand, the general structure of the MC counterterms
is actually PSMC-independent, and it is easy to convince oneself
that they can always be written in the following way:
\beq
d\sigmaMC=\sum_{(i,j)\in\FKSpairs}d\sigmaMC_{ij}\,,
\label{MCcntij}
\eeq
since FKS pairs are in one-to-one correspondence with IR singularities,
which in turn are at the core of the shower mechanism. It is important
to bear in mind that this implies that eq.~(\ref{MCcntij}) is therefore
valid not only for those PSMCs based on a $1\to 2$ branching picture
(such as those just mentioned), but more generally for any PSMC consistent 
with QCD (in particular, those that adopt a dipole 
picture~\cite{Lonnblad:1992tz,Nagy:2006kb,Giele:2007di,Dinsdale:2007mf,
Schumann:2007mg,Winter:2007ye,Platzer:2011bc,Ritzmann:2012ca}).
The functional form of the terms $d\sigmaMC_{ij}$ is the same for
all of the PSMCs considered here\footnote{Different classes of PSMCs may be 
conceived, for example dipole-shower-based or by going beyond the 
leading-$\NC$ approximation (see e.g.~refs.~\cite{Friberg:1996xc,
Giele:2011cb,Platzer:2012np,Nagy:2012bt}), which could induce a different 
form. However, the idea of MC counterterms in general, and of
eq.~(\ref{MCcntij}) in particular, would still be valid.}, and we
shall briefly describe its construction in what follows. One starts
with the PSMC cross section that results from a single 
branching\footnote{In order to simplify the notation, we understand
the universal, azimuthal-dependent part of the branching (see 
e.g.~appendix~B of ref.~\cite{Frixione:1995ms}).}:
\beqn
d\sigmaMCz_{ij}&=&\sum_c\sum_{l\in c}d\sigmaMCz_{ij,cl}\,,
\label{MCcnt0}
\\
d\sigmaMCz_{ij,cl}&=&\lumMC\left(x_{1,2}^{(l)}\right)
\frac{\delta_{i\oplus j\in l}}{N_{i\oplus j}}\frac{\as}{2\pi}
\frac{P_{\ident_j\ident_{i\oplus j}}(\showzij)}{\showxiij}\ampsqnt_c 
\stepfMC d\showxiij d\showzij \frac{d\varphi}{2\pi}
\phspn\,.
\label{MCcnt1}
\eeqn
As the notation suggests, although $d\sigmaMCz_{ij}$ does not necessarily
coincide with $d\sigmaMC_{ij}$ (the possible differences between the two
will be explained below), it does fully include its physics contents,
which we now turn to describing.

The sums in eq.~(\ref{MCcnt0}) run over all possible planar colour
configurations ($c$), and the individual colour lines belonging to 
them ($l$). In \aNLO\ we represent a colour configuration as a list,
\mbox{$c=\{l_1,\ldots l_m\}$}, where the individual colour line
is represented as an ordered pair, \mbox{$l_k=(s(k),e(k))$}, whose
meaning is that of a connection between particle $\ident_{s(k)}$
(the starting point of the line) and particle $\ident_{e(k)}$
(the end point of the line). This implies that, for any given $c$,
a quark or an antiquark will belong to a single colour line (through
its colour or anticolour respectively), while a gluon will belong
to two colour lines (one for colour and one for anticolour). This is 
the reason for the factor $N_{i\oplus j}$ in eq.~(\ref{MCcnt1}),
which is equal to 1(2) if $i\oplus j$ is a quark or an antiquark
(a gluon). \aNLO\ constructs the colour configurations during
an initialisation phase, by gathering the relevant information
from the underlying matrix elements. $\showxiij$ and $\showzij$
are the PSMC shower variables; as the notation indicates, in general 
their forms depend on the branching particle $i\oplus j$ (in particular,
on whether it is in the final or initial state), and on the colour 
line (which determines the colour partner of $i\oplus j$). The
actual shower variables are very PSMC-dependent, and they are coded 
in \aNLO\ for all the PSMCs one may match with. The colour connections
in general also determine the choice of Bjorken $x$'s made by the
PSMC (see e.g.~ref.~\cite{Frixione:2002ik}), which is the reason for
the dependence on $l$ in the argument of the luminosity factor $\lumMC$
in eq.~(\ref{MCcnt1}). $P_{ba}(z)$ is the Altarelli-Parisi one-loop
kernel~\cite{Altarelli:1977zs}, for parton $b$ emerging from the branching 
of parton $a$ with momentum fraction $z$. $\ampsqnt_c$ is the Born matrix
element multiplied by a factor determined by the colour configuration $c$,
according to the prescription of ref.~\cite{Odagiri:1998ep}. 
Finally, $\stepfMC$ symbolically denotes all kind of kinematics 
constraints, such as generation-level cuts (an $n$-body Born must 
have $n$ well-separated partons), possible dead-zone conditions, 
and so forth.

Implicit in eq.~(\ref{MCcnt1}) is the choice of a shower scale,
which roughly speaking sets an upper bound for the hardness of
each branching. Since PSMCs are based on a small-scale approximation,
it is clear that the larger the shower scale, the worse the description
of physics by any PSMC. While in the context of standalone-PSMC simulations
it may be necessary to consider shower scales that stretch that approximation
(simply to fill phase-space regions otherwise inaccessible), such an
attitude is not justified when PSMC are matched with NLO computations,
since the latter provide a much better description of hard-emission regions.
Note that in MC@NLO these undesirable large-shower-scale effects are
indeed removed completely at ${\cal O}(\as^{b+1})$ by the MC counterterms
(see eqs.~(\ref{xsecH}) and~(\ref{xsecS})). However, at ${\cal O}(\as^{b+2})$ 
and beyond the PSMC may still radiate in the hard regions, potentially
giving effects which are simply not sensible from the physics viewpoint.
Furthermore, even at ${\cal O}(\as^{b+1})$ it does not make much sense
to allow the PSMC to produce radiation only to eventually remove it.
Fortunately, it is possible to give
the PSMC an external mass scale in input; during the course of the shower,
the PSMC will generate branchings after choosing the smallest between 
this external scale and its internally-generated shower scale.
In \aNLO, we exploit this possibility in the following 
way\footnote{This technique has been used sparingly in \mcatnlo\ v3.3
and higher.}. Firstly, we introduce a function of a mass scale $\mu$:
\beqn
D(\mu)=\left\{
\begin{array}{ll}
1               &\phantom{aaaa} \mu\le\mu_1\,,\\
{\rm monotonic} &\phantom{aaaa} \mu_1<\mu\le\mu_2\,,\\
0               &\phantom{aaaa} \mu > \mu_2\,,\\
\end{array}
\right.
\label{Ddef}
\eeqn
with $\mu_1\le\mu_2$ two given mass scales. While we typically regard $D$ 
as a smooth function, it is perfectly fine to consider its sharp version:
\beq
D(\mu)=\stepf\left(\mu_Q-\mu\right)\,,\;\;\;\;\;\;\;\;\mu_Q=\mu_1=\mu_2\,,
\label{Dsharp}
\eeq
which is a particular case of eq.~(\ref{Ddef}). Secondly, on an 
event-by-event basis we determine a mass scale by using:
\beq
\mu_r=D^{-1}(r)\,,
\eeq
with $r$ a flat random number (note that with eq.~(\ref{Dsharp}) one
obtains $\mu_r\equiv\mu_Q$). Thirdly, we give $\mu_r$ in input to
the PSMC, where it acts as an upper bound to the internally-generated
shower scales as explained before. The physical meaning of $\mu_r$
(be it a relative transverse momentum, a virtuality, or whatever else)
depends on the specific PSMC chosen, but is irrelevant here and need
not be specified. The crucial thing is the following: by means of this
procedure, we are effectively {\em changing} the shower w.r.t.~what the
PSMC would do if left alone. This change must therefore correspond to
a change in the MC counterterms, because of the very definition of the
latter. This amounts to:
\beq
d\sigmaMCD_{ij}=D\left(\mu(\conf_{n+1})\right)d\sigmaMCz_{ij}\,.
\label{MCcntD}
\eeq
Note that the argument of $D$ in eq.~(\ref{MCcntD}) is computed
by using the underlying kinematic configuration (after having taken
into account its PSMC-specific form: $\pt$, $\sqrt{Q^2}$, and so forth),
and must not be generated randomly. Equation~(\ref{MCcntD}) can
always be used in place of eq.~(\ref{MCcnt1}), the latter being
a particular case of the former, which one can formally obtain
by setting $\mu_1=\mu_2=\infty$.

As was discussed before, the function $D$ controls perturbative
effects higher than NLO;
hence, its variations can be used to assess the NLO+PS matching systematics
(which is, by definition, the size of terms beyond the formal accuracy
of the computation) of the MC@NLO method. Although this is expected
to be small\footnote{The main reason being that MC@NLO short-distance
cross sections have no contributions of ${\cal O}(\as^{b+2})$ or higher;
terms of these orders in the physical cross sections can only be generated
through MC radiation.}, its actual size is observable-, process- 
and (especially)
PSMC-dependent\footnote{On top of being, obviously, matching-method
dependent. We stress that the results of $D$ variations in \aNLO\ can
not be used, even as a mere indication, of the matching systematics
that affects other matching methods, such as POWHEG.}, and it is therefore
convenient to be able to study it in a straightforward way. This is
the case in \aNLO, where one can control the values of the scales
$\mu_i$ of eq.~(\ref{Ddef}) through the external parameters $f_i$,
with $\mu_i=f_i\sqrt{\hat{s}_0}$, and $\hat{s}_0$ the Born-level 
partonic c.m. energy squared.

Equation~(\ref{MCcntD}) would give the desired MC counterterms if
the corresponding PSMC behaved as expected in the IR regions, namely
if it gave exactly the same result as a QCD matrix-element computation in 
{\em both} the collinear and the soft limits, whence the local cancellations
in eqs.~(\ref{xsecH}) and~(\ref{xsecS}). Unfortunately, this is not
the case in the soft limit, at least for \HW\ and \PY, where this 
deficiency is basically a consequence of the necessity of having 
a Markovian shower. What is true, however, is that the amount
of soft radiation predicted by the PSMCs is correct; in other words,
only its angular pattern is not consistent with the one required by QCD.
Fortunately, such an undesirable feature of certain PSMCs will not
have dramatic consequences on physical observables,
because of the infrared-safety of the latter (the interested reader can 
find a fuller discussion of this issue in sect.~A.5 of 
ref.~\cite{Frixione:2002ik}). The technical problem of the local
finiteness of the MC@NLO short-distance cross sections can be
solved by the following definition of the MC counterterms:
\beq
d\sigmaMC_{ij}=\left(1-\Gfun\right)d\sigmaMCD_{ij}
+\Gfun\, d\sigmaNLOS_{ij}|_{\rm real}\,.
\label{MCcntG}
\eeq
Here, $\Gfun$ is a smooth function defined so that $\Gfun\to 1$ in
the soft limit, and $\Gfun=0$ outside of the soft region; 
$d\sigmaNLOS_{ij}|_{\rm real}$ is the soft part of the NLO
cross section, eq.~(\ref{fact3}), where 
only the real-emission matrix element contribution is 
kept\footnote{An analogous solution is adopted when the azimuthal
part of the PSMC branching kernel does not agree with that
predicted by the matrix elements.}.

A few comments concerning eq.~(\ref{MCcntG}) are in order.
Given that what the PSMC is supposed to do is
$d\sigmaMCD_{ij}$, while what MC@NLO assumes the PSMC does
is $d\sigmaMC$ of eq.~(\ref{MCcntG}), there is a mismatch of
${\cal O}(\as^{b+1})$ between the two. This mismatch, however,
is utterly irrelevant for several reasons. Firstly, because
of the properties of $\Gfun$, it is confined to the soft regions,
where effects of all orders in $\as$ are equally important.
Secondly, in practice in the soft region the PSMC does not even
correspond to $d\sigmaMCD_{ij}$ if not in a fully inclusive sense,
since the PSMC is unable to handle emissions below the IR cutoffs, which 
are of the order of the typical hadron mass (and this for a very fundamental
reason: QCD does not have infinite resolution power). Thirdly, because
of the previous point {\em all} NLO+PS matching schemes are liable to
have ${\cal O}(\as^{b+1})$ effects in small-scale regions which are
not in formal agreement with fixed-order results at the NLO, even
if the second term on the r.h.s.~of eq.~(\ref{MCcntG}) were not 
present (see appendix~B.3 of ref.~\cite{Frixione:2002ik} for a discussion
specific to the MC@NLO formalism). Ultimately, then, the differences
driven by the $\Gfun$ function are power suppressed (see 
e.g.~ref.~\cite{Nason:2012pr}); eq.~(\ref{MCcntG})
is nothing but a formal trick to render the ${\cal O}(\as^{b+1})$ 
MC@NLO cross sections non-divergent (which is important in view of
their numerical integration) in a region where not even the perturbative 
predictions of PSMCs are sensible, let alone those of fixed-order
computations.

Equation~(\ref{MCcntG}) also gives us the opportunity of commenting
briefly on the alternative implementation of the MC@NLO method presented
in ref.~\cite{Hoeche:2011fd}. There, the function  $\Gfun$ does not appear, 
for the simple reason that the two short-distance
cross sections on the r.h.s.~of eq.~(\ref{MCcntG}) coincide in
the soft limit and, as explained above, this is a sufficient condition 
for not having to introduce $\Gfun$. In turn, this situation occurs
because the shower used in ref.~\cite{Hoeche:2011fd}, and in subsequent
papers, in the context of NLO-matched simulations
is {\em constructed} to have the same soft behaviour of the matrix 
elements (see, in particular, eq.~(2.5) of ref.~\cite{Hoeche:2014lxa}).
Once this is done, the form of the MC counterterms is uniquely
determined, lest one has a mismatch of ${\cal O}(\as^{b+1})$ (everywwhere 
in the phase space). One may opt (as is done in ref.~\cite{Hoeche:2011fd})
to see this as a choice made at the level of short-distance
cross sections (the soft behaviour of the MC counterterms), that 
forces one to modify the shower in order to preserve the perturbative
accuracy. While the point of view of ref.~\cite{Frixione:2002ik} and 
of this paper is the opposite one (namely that it is the chosen PSMC 
which determines the MC counterterms), the fundamental idea is just the 
same, and therefore the MC@NLO subtractions of ref.~\cite{Hoeche:2011fd} 
do not differ in any significant way from those that had been 
adopted in \mcatnlo\ and are now used in \aNLO.
What has been changed in the approach of ref.~\cite{Hoeche:2011fd} 
(w.r.t.~the previous default) is the {\em shower} adopted in conjunction
with the NLO matching performed there (which happens not to be the same 
as that used in simulations which are not matched to NLO results). 
We point out that should a version of the \PY\ or \HW\ showers become available 
with a matrix-element-type soft behaviour, the relevant MC counterterms would
be constructed without a $\Gfun$ function. Furthermore, we stress
that the NLO accuracy of the MC@NLO method  (including, in particular, the 
whole $1/\NC$ expansion) is maintained, and that the ${\cal O}(\as^{b+1})$ 
results are thus in agreement with those of the corresponding matrix elements 
(up to power-suppressed effects, as explained above), regardless of the 
behaviour of the PSMC and thus of the presence of a $\Gfun$ function.
On the other hand, if some aspect of the PSMC is deficient (for example the
treatment of subleading-colour contributions), the MC@NLO method itself cannot
provide an improvement in the MC-dominated kinematic regions -- any issue of
this kind must be addressed at the level of the PSMC itself.

Equation~(\ref{MCcntG}) is the final form to be used in eq.~(\ref{MCcntij});
when the latter is in turn replaced in eqs.~(\ref{xsecH}) and~(\ref{xsecS}),
one immediately realises that it is convenient to define the $\clH$- and
$\clS$-event contributions at fixed FKS pair:
\beqn
d\sigmaH_{ij}&=&d\sigmaNLOE_{ij}-d\sigmaMC_{ij}\,,
\label{xsecHij}
\\
d\sigmaS_{ij}&=&d\sigmaMC_{ij}+
\sum_{\alpha=S,C,SC} d\sigmaNLOa_{ij}\,.
\label{xsecSij}
\eeqn
It is easy to see that these quantities are locally finite 
(by construction of the shower variables $\showxiij$ and $\showzij$),
precisely as their summed counterparts of eqs.~(\ref{xsecH}) 
and~(\ref{xsecS}). Eqs.~(\ref{xsecHij}) and~(\ref{xsecSij}) can 
now be used to define the MC@NLO generating functional.
In order to do that we also introduce, for consistency with the standard
notation and in view of the discussion to be given in sect.~\ref{sec:FxFx},
the $\clH$- and $\clS$-event kinematic configurations:
\beq
\confHn\equiv\confnpoE\,,\;\;\;\;\;\;\;\;
\confSn\equiv\confnpoS\,,
\eeq
so that the generating functional finally reads as follows:
\beq
\GenNLO=\sum_{\procR\in\allprocnpo}
\sum_{(i,j)\in\FKSpairs}\left\{\GenMC\left(\confHn\right)
\frac{d\sigmaH_{ij}(\procR)}{d\meas_{Bj}^{(ij)}d\meas_{n+1}^{(ij)}} 
+\GenMC\left(\confSn\right)
\frac{d\sigmaS_{ij}(\procR)}{d\meas_{Bj}^{(ij)}d\meas_{n+1}^{(ij)}}
\right\}\,,
\label{genMCatNLO}
\eeq
where $\GenMC$ is the generating functional of the PSMC one interfaces
to, and its argument indicates the parton configuration to be adopted
as the starting condition for the shower. In eq.~(\ref{genMCatNLO})
we have reinstated the formal dependence of the short-distance
cross sections on the real-emission process $\procR$. In this way,
the complete similarity between the MC@NLO and NLO cross sections
in terms of sums over partonic processes and FKS pairs (i.e., between
the r.h.s.~of eq.~(\ref{genMCatNLO}) and the l.h.s.~of eq.~(\ref{sumsymm}))
is evident. Thus, as was anticipated at the beginning of this section,
all the manipulations performed at the end of sect.~\ref{sec:FKS}
apply to the MC@NLO case as well. In particular, the formulation
where the MC@NLO short-distance cross sections are defined by fixing
the Born-level process is the one adopted in \aNLO. Among other things,
this renders it particularly easy to obtain $\clS$ events by integrating 
over $\xii$ and $\yij$ before unweighting (which is expected to reduce the
number of negative-weight events, as advocated in ref.~\cite{Frixione:2002ik} 
in the context of the MC@NLO formalism; the same idea, dubbed ``folding",
has been independently proposed and implemented in {\small\sc POWHEG}).

Before concluding this section, we present two variants of what was 
discussed so far, that we shall want to consider for future \aNLO\ 
developments. The first one concerns the definition of the MC 
counterterms. The advantage of using eqs.~(\ref{xsecHij}) 
and~(\ref{xsecSij}) is a one-to-one correspondence between
the shower variables $(\showzij,\showxiij)$ and the FKS integration
variables $(\xii,\yij)$. Apart from a transparent way of identifying
the IR singular structure (which in turn is related to $d\sigmaH_{ij}$
and $d\sigmaS_{ij}$ being locally finite at fixed $(i,j)$), this implies
that, when the integration measure over the MC variables is expressed
in terms of $d\meas_{n+1}^{(ij)}$, as it must according to 
eq.~(\ref{genMCatNLO}), one can factorise the jacobian
\beq
\frac{\partial\left(\showzij,\showxiij\right)}
{\partial\left(\xii,\yij\right)}\,;
\label{jac}
\eeq
in the current version of \aNLO, this jacobian is computed analytically.
The definitions of eqs.~(\ref{xsecHij}) and~(\ref{xsecSij}) might however
have a numerical drawback, due to the presence of the factor
$\Sfunij$ only in the terms $d\sigmaNLOa_{ij}$. Its absence in the
MC counterterms could induce differences in the damping of singularities
not due to the FKS pair $(i,j)$, that in turn could result in an 
unnecessary large fraction of events with negative weights.
This situation can be amended as follows: by using eq.~(\ref{Sfununit}) 
one obtains the identity:
\beq
d\sigmaMC=\sum_{(i,j)\in\FKSpairs}d\sigmaMC_{ij}=
\sum_{(i,j)\in\FKSpairs}\sum_{(k,p)\in\FKSpairs}
\Sfun_{kp}\,d\sigmaMC_{ij}
\label{MCcntSfun}
\eeq
which implies
\beqn
d\sigmaMC&=&\sum_{(i,j)\in\FKSpairs}d\hsigmaMC_{ij}\,,
\\
d\hsigmaMC_{ij}&=&\Sfunij\sum_{(k,p)\in\FKSpairs}
d\sigmaMC_{kp}\,.
\eeqn
By using $d\hsigmaMC_{ij}$ in place of $d\sigmaMC_{ij}$ in 
eqs.~(\ref{xsecHij}) and~(\ref{xsecSij}), one can factor out
a term $\Sfunij$. The disadvantage
of course is that the pair $(\showz_{kp},\showxi_{kp})$ is not
in one-to-one correspondence with $(\xii,\yij)$ any longer, which
implies that the relevant jacobians will be much more involved than
that in eq.~(\ref{jac}). However, this is clearly only a technical
problem, which can be overcome by giving up the requirement that
jacobians be computed analytically. With modern routines for the
numerical evaluation of derivatives this appears definitely feasible,
and it would also pave the way for a leaner interface to new PSMCs.

We now turn to the second variant, which concerns the MC@NLO formulation
proper. A good feature of the MC@NLO cross section is that it gives
a clear separation of matrix element and Monte Carlo effects. 
A drawback is that one is forced to ignore the fact that in the
MC-dominated region (i.e., at small scales) real-emission matrix elements 
do not give a good description of the underlying physics, which implies 
that the contribution of $\clH$ events there is important only in terms of
total rate, but not in terms of shapes, which indeed are dominated by showered
$\clS$ events. From the viewpoint of final (physical) results this is
irrelevant, but it entails a loss of efficiency, since typically $\clH$ 
events have negative weights in the MC-dominated region.
A possible way to address this problem is the following.
Consider a function $\Delta$ with the properties:
\beqn
\Delta&=&1+{\cal O}(\as)\,,
\label{Dprop1}
\\
\Delta&\longrightarrow&0\;\;\;\;\;\;{\rm IR~limits}\,.
\label{Dprop2}
\eeqn
It is immediate to see that the following definitions:
\beqn
d\sigmaH_{ij}&=&\left(d\sigmaNLOE_{ij}-d\sigmaMC_{ij}\right)\Delta\,,
\label{xsecHD}
\\
d\sigmaS_{ij}&=&d\sigmaMC_{ij}\Delta+
\sum_{\alpha=S,C,SC} d\sigmaNLOa_{ij}+
d\sigmaNLOE_{ij}\left(1-\Delta\right)\,.
\label{xsecSD}
\eeqn
result in a generating MC@NLO functional with the {\em same} formal
accuracy as that of eq.~(\ref{genMCatNLO}). It is clear that, while
this conclusion holds regardless of the form of $\Delta$, provided that
the conditions in eqs.~(\ref{Dprop1}) and~(\ref{Dprop2}) are satisfied,
from the physical viewpoint one would identify $\Delta$ with a suitable
combination of Sudakovs, whose explicit forms would ideally be extracted
from the same PSMC one interfaces to. Although at present there is no
straightforward way to obtain (numerically) the Sudakovs from a PSMC,
there is no reason of principle which prevents the implementation of such
a possibility in future versions of modern PSMCs. In the meanwhile,
\aNLO\ will have the means to test eqs.~(\ref{xsecHD})
and~(\ref{xsecSD}), by using the analytical expression for the
Sudakovs which are currently used in the context of the FxFx
NLO merging method (see sect.~\ref{sec:FxFx}).

\subsubsection{Merging samples at the NLO: FxFx\label{sec:FxFx}}

In this section, we present a brief review of the FxFx 
procedure~\cite{Frederix:2012ps}, whose aim is that of improving,
by systematically including PSMC matching at the NLO, the multi-leg
tree-level merging techniques established in the course of the
past decade, such as CKKW, CKKW-L, and MLM (see refs.~\cite{Catani:2001cc,
Lonnblad:2001iq,Krauss:2002up,Mrenna:2003if,Lavesson:2005xu,Alwall:2007fs,
Hoeche:2009rj,
Hamilton:2009ne,Lonnblad:2011xx,Lonnblad:2012ng}). The key word here is 
merging, which identifies the following problem. If one obtains hard events
from the processes:
\beq
\ident_1+\ident_2\;\longrightarrow\;S+i~{\rm partons}\,,
\label{partproc}
\eeq
where $S$ is a set of $p$ particles which does not contain any QCD 
massless partons, how does one match them to PSMCs when different
$i$ values (i.e., different final-state hard-process multiplicities)
are {\em simultaneously} considered? The core of the problem is 
the avoidance of double counting; note that this is on top of, and more
complicated than, the double-counting problem that one faces when
fixing $i$ in eq.~(\ref{partproc}), and which is a matching (not a
merging) issue. While the formulation of the problem of merging is
independent of the perturbative order (i.e., of the accuracy to which
the cross sections of the processes in eq.~(\ref{partproc}) are computed),
its solution is not, being strictly connected with the adopted matching 
strategy. NLO (and beyond) mergings are thus inherently more complicated than
LO ones, because LO-matching is basically trivial; they have attracted
a significant amount of attention lately~\cite{Lavesson:2008ah,
Hamilton:2010wh,Hoche:2010kg,Giele:2011cb,Alioli:2011nr,Hoeche:2012yf,
Frederix:2012ps,Platzer:2012bs,Alioli:2012fc,Lonnblad:2012ix,Hamilton:2012rf,
Alioli:2013hqa}. The quickest way
to realise this is that of considering the tree-level matrix element
associated with the process in eq.~(\ref{partproc}); this will give
the Born contribution to that process, but at the same time it will 
also be needed at the NLO as real-emission contribution to a process 
whose Born features a \mbox{$S+(i-1)~{\rm partons}$} final state.
Such a double role of a given matrix element is specific to an NLO
merging, and is absent at the LO.

This example of the tree-level matrix-element double role suggests
a way to tackle the NLO-merging problem. In particular, in view of the
fact that, in the context of a given calculation, the perturbative accuracy 
of the prediction for an observable is larger the more inclusive the 
observable, one wants to use as much as is possible an $i$-parton
tree-level matrix element as a Born, rather than as a real-emission,
contribution. The role of hard emissions will thus be mainly played
by the Born's associated with processes with larger multiplicities,
while for any given multiplicity the real-emission contributions will 
mostly provide the correct (NLO) normalisation. Because the 
MC@NLO formalism is designed
to perturb in a minimal way both the underlying matrix-element description
and the PSMC one uses for showering events, the above scheme essentially
corresponds to limiting the hardness of $\clH$-event emissions.
Technically, this can be achieved by simply exploiting the $D$ function
introduced in eq.~(\ref{Ddef}), and by applying analogous conditions
at the matrix element level. In order to do so, it is convenient 
(and particularly sensible physics-wise) to re-interpret the parameters
$\mu_i$ of eq.~(\ref{Ddef}) in the following way:
\beqn
\mu\le\mu_1&&\phantom{aaaaaa}{\rm soft~(MC-dominated)},
\nonumber\\
\mu_1 < \mu\le\mu_2&&\phantom{aaaaaa}{\rm intermediate},
\nonumber\\
\mu > \mu_2&&\phantom{aaaaaa}{\rm hard~(ME-dominated)}.
\nonumber
\eeqn
Note that in the case of a sharp $D$ function, eq.~(\ref{Dsharp}),
the intermediate region collapses to a zero-measure set.
Once one has defined hard and soft mass scales, one needs to define
a way to measure the hardness; because of the fact that NLO corrections
will be computed, it is mandatory that such a measure be IR-safe. The
easiest way to achieve this is that of employing quantities that
arise from a jet-reconstruction algorithm. We denote by $d_j$
the scale (with canonical dimension equal to one, i.e.~mass) 
at which a given $S$+partons configuration passes from being reconstructed 
as a $j$-jet one to being reconstructed as a $(j-1)$-jet one, according 
to a $\kt$ jet-finding algorithm~\cite{Catani:1993hr} (in other words, 
there are $j$ jets of hardness $d_j-\varepsilon$, and $(j-1)$ jets of 
hardness $d_j+\varepsilon$, with $\varepsilon$ arbitrarily small). 
In general, for $n$ final-state partons one will have
\beq
d_n\le d_{n-1}\le\ldots\le d_2\le d_1\,.
\label{dord}
\eeq
It will also turn out to be convenient to define
\beq
d_j=\sqrt{\hat{s}}\,,\;\;\;\;\;\;\;\;j\le 0\,,
\label{dmax}
\eeq
with $\sqrt{\hat{s}}$ the parton c.m.~energy, i.e.~the largest energy
scale available event-by-event. Since the function $D$ determines
rather directly the way in which the various partonic processes
of eq.~(\ref{partproc}) are combined, the results of variations of 
the parameters that enter into it can be associated with
the systematics of the merging procedure (and not of the matching one,
as is the case for the unmerged cross sections discussed in 
sect.~\ref{sec:mcatnlo}). This is particularly straightfoward, and
totally analogous to what is typically done at the LO, when one
chooses a sharp $D$ function, in which case $\mu_Q$ has to be seen
as the merging scale.

We now limit ourselves to reporting the final forms of the short-distance
cross sections necessary to implement the FxFx merging scheme; the 
interested reader can find more details in ref.~\cite{Frederix:2012ps}.
We denote by $N$ the largest light-parton multiplicity at the Born
level that we consider (therefore, $N$ is the largest value that $i$
can possibly assume in eq.~(\ref{partproc})). The cross sections given
below are the analogues of those in eqs.~(\ref{xsecHij}) and~(\ref{xsecSij}),
i.e., at fixed real-emission process and FKS pair; in the present section,
we understand these quantities, lest we clutter the notation with 
unnecessary details. On the other hand, each $\clH$- or $\clS$-event
contribution or short-distance cross section will carry an index, equal 
to the number of final-state particles in the corresponding hard process 
(at the Born level), this information being
crucial in all merging procedures. We have:
\beqn
d\bsigmaSip&=&\Big[\sum_{\alpha=S,C,SC} d\sigmaNLOa_{p+i}
+d\sigmaMCz_{p+i} D(d_{i+1}(\confHip))\Big]
\label{HiSfin}
\\*&&\times
\left(1-D(d_i(\confSip))\right)\,
\stepf\left(d_{i-1}(\confSip)-\mu_2\right)\,,
\nonumber
\\
d\bsigmaHip&=&\Big[d\sigmaNLOE_{p+i}\left(1-D(d_i(\confHip))\right)
\stepf\left(d_{i-1}(\confHip)-\mu_2\right)
\label{HiHfin}
\\*&&-
d\sigmaMCz_{p+i} \left(1-D(d_i(\confSip))\right)
\stepf\left(d_{i-1}(\confSip)-\mu_2\right)\Big]\,D(d_{i+1}(\confHip))\,,
\nonumber
\eeqn
\beqn
d\bsigmaSNp&=&\Big[\sum_{\alpha=S,C,SC} d\sigmaNLOa_{p+N}
+d\sigmaMCz_{p+N} \Big]
\label{HNSfin}
\\*&&\times
\left(1-D(d_N(\confSNp))\right)\stepf\left(d_{N-1}(\confSNp)-\mu_2\right)\,,
\nonumber
\\
d\bsigmaHNp&=&d\sigmaNLOE_{p+N}\left(1-D(d_N(\confHNp))\right)
\stepf\left(d_{N-1}(\confHNp)-\mu_2\right)
\label{HNHfin}
\\*&-&
d\sigmaMCz_{p+N}\left(1-D(d_N(\confSNp))\right)
\stepf\left(d_{N-1}(\confSNp)-\mu_2\right)\,,
\nonumber
\eeqn
where in eqs.~(\ref{HiSfin}) and~(\ref{HiHfin}) one has $0\le i<N$,
and in writing the MC counterterms we have understood the $\Gfun$
dependence as given in eq.~(\ref{MCcntG}), which is irrelevant
for the sake of the present discussion. The MC@NLO cross section
defined by eqs.~(\ref{HiSfin}) and~(\ref{HiHfin}) (eqs.~(\ref{HNSfin}) 
and~(\ref{HNHfin})), and its analogue that we shall introduce later,
is called the $i$-parton ($N$-parton) sample. With the cross sections 
above, one defines the FxFx generating functional:
\beqn
\GenFxFx&=&\sum_{n=p}^{p+N}\GenFxFx^{(n)}\,,
\label{genFxFx}
\\
\GenFxFx^{(n)}&=&\GenMC\left(\confHn\right)
\frac{d\bsigmaHn}{d\meas_{Bj}d\meas_{n+1}} 
+\GenMC\left(\confSn\right)
\frac{d\bsigmaSn}{d\meas_{Bj}d\meas_{n+1}}\,.
\label{genFxFxn}
\eeqn
Note that eq.~(\ref{genFxFxn}) and eq.~(\ref{genMCatNLO}) are 
formally identical (as was said above, the sums on the r.h.s.~of
eq.~(\ref{genMCatNLO}) are understood here), the difference being
in the form of the short-distance cross sections. Equation~(\ref{genFxFx})
implies that the FxFx generating functional is the incoherent sum
of MC@NLO generating functionals, each of which incorporates 
FxFx-specific type of cuts but that are otherwise fully analogous
to their non-merged counterparts. This renders it straightforward
to implement the FxFx merging prescription into \aNLO. 

As was discussed in ref.~\cite{Frederix:2012ps}, the formulation
of FxFx according to eq.~(\ref{genFxFxn}) can be supplemented
by a Sudakov suppression, in keeping with what is done at the LO 
in the CKKW(-L) and MLM methods. The modifications of the short-distance
cross sections are in fact rather minimal, and amount to what follows:
\beqn
d\hsigmaSip&=&\Big[d\bsigmaSip+
d\sigma_{p+i}^{(\Delta)}\Big]\,\Delta_i\!\left(\muSimn,\muSimx\right)\,,
\label{SudiS}
\\
d\hsigmaHip&=&d\bsigmaHip\,\,
\Delta_i\!\left(\muHimn,\muHimx\right)\,,
\label{SudiH}
\eeqn
with $d\bsigmaSip$ and $d\bsigmaHip$ defined in eq.~(\ref{HiSfin})
(for $i<N$) or eq.~(\ref{HNSfin}) (for $i=N$), and in eq.~(\ref{HiHfin})
(for $i<N$) or eq.~(\ref{HNHfin}) (for $i=N$) respectively.
$\Delta_i$ is a suitable combination of Sudakov form factors, the
construction of which follows closely the CKKW prescription. Further
details, including the definition of the hard scales that enter
these formulae can be found in ref.~\cite{Frederix:2012ps}. Here,
we limit ourselves to stressing the following fact: while for the
processes studied in ref.~\cite{Frederix:2012ps} the flavour structure
of $\Delta_i$ had been worked out by hand, it has now been fully automated
in \aNLO. The term $d\sigma_{p+i}^{(\Delta)}$ is necessary in order not
to spoil the formal NLO accuracy of the formalism:
\beq
d\sigma_{p+i}^{(\Delta)}=-d\sigmaNLOS_{p+i}|_{\rm Born}
\left(1-D(d_i(\confSip))\right)\,
\stepf\left(d_{i-1}(\confSip)-\mu_2\right)
\Delta_i^{(1)}\!\left(\muSimn,\muSimx\right)\,,
\label{Delsig}
\eeq
where $d\sigmaNLOS_{p+i}|_{\rm Born}$ is the soft part of the NLO
cross section, eqs.~(\ref{xsecNLO}) and~(\ref{fact3}), where 
only the Born matrix element contribution is kept. By $\Delta_i^{(1)}$ 
we have denoted the ${\cal O}(\as)$ term in the perturbative expansion 
of $\Delta_i$. It should be clear that $\Delta_i$ satisfies 
eqs.~(\ref{Dprop1}) and~(\ref{Dprop2}); therefore, a by-product 
of the Sudakov-improved FxFx merging procedure (which we regard
as our ``best" FxFx scheme, and is thus the default in \aNLO)
is the possibility of testing the alternative (non-merged) MC@NLO 
formulation presented in eqs.~(\ref{xsecHD}) and~(\ref{xsecSD}).

We conclude this section by discussing two arguments of general
relevance to NLO-merging techniques, and which may have significant
bearings on their phenomenological predictions: unitarity,
and merging systematics. In the context of merging, imposing
a unitarity condition means that the fully-inclusive merged 
cross section (i.e., the sum over all $i$'s of the $i$-parton-sample 
predictions) is equal to the total rate of the {\em un}merged 
$0$-parton sample\footnote{We neglect here possible complications due
to heavy-flavour thresholds, and to contributions to $i$-parton samples
that cannot be shower-generated starting from a lower multiplicity.}.
In FxFx unitarity is not imposed, for the reasons we shall now explain.
One of the advantages of unitarity is
the fact that the merging scale $\mu_Q$ can be chosen\footnote{In 
principle; in practice, there may be efficiency issues, which are
however not of concern here.} in an arbitrarily-large range, whereas
in non-unitary approaches this is not possible, and a sensible choice
(which takes into account the hardness of the process, and the fact
that $\mu_Q$ must itself be hard) is always arbitrary to a certain extent.
Such a benefit of unitarity has however a less-pleasant side. Namely,
in the general context of resummed computations matched with 
fixed orders, constraints on total rates contribute to significant 
modifications of matrix-elements predictions, in shape and absolute
value, also in regions where one would expect large-logarithms effects 
to be suppressed (the Higgs $\pt$ spectrum is a spectacular example 
of this phenomenon -- see e.g.~figs.~1 and~2 of ref.~\cite{Bozzi:2005wk}). 
This is of course acceptable, and actually constitutes a defining 
prediction of a matched formalism, if all contributions to the latter are 
perturbatively consistent: an example is that of an MC@NLO unmerged
cross section, where the total rate is of ${\cal O}(\as^{b+1})$, which
is the same perturbative order to which both $\clS$ and $\clH$ events
contribute. When unitarised-merging is considered, 
the total rate is still imposed
to be that of ${\cal O}(\as^{b+1})$; however, $\clS$- and $\clH$-event
$i$-parton samples (and their analogues within any merging formalism)
are of ${\cal O}(\as^{b+i+1})$; this mismatch of perturbative orders
present for any $i\ge 1$ might result in a bias (uniquely due to the
total-rate constraint) at the level of shapes, the stronger the 
larger $i$. Note that one could impose the unitarity constraint
using an NNLO total-rate prediction (${\cal O}(\as^{b+2})$), were
that available. This would marginally alleviate the problem above, 
but not solve it: firstly, it merely shifts it to $i\ge 2$, and secondly
one still uses NLO-matched (and not NNLO-matched) $i$-parton samples
which thus require, at the very least, to be re-normalised.
The counter argument is that, although a bias might indeed be present,
it appears at perturbative orders which are in any case beyond
accuracy. This is correct, but does not have any implications on the
presence of large logarithmic terms, that could enhance such contributions.
More importantly, it is also an argument that can be used in the context
of a non-unitary approach, where it would apply chiefly to total rates
(and, by construction, in non-unitary approaches there is no bias due
to total-rate constraints). In fact, the dependence on $\mu_Q$ of
inclusive results can be used effectively in non-unitary mergings
as a way to arrive at a sensible $\mu_Q$ range, to be employed to
assess the merging systematics of differential distributions.
Examples of this, and of the fact that FxFx exhibits a rather 
small $\mu_Q$ dependence, will be given in sect.~\ref{sec:diff}.
In conclusion, the arguments above can be argued in different ways;
it should be clear, however, that regardless of whether a merging
approach imposes or not unitarity, in those phase-space regions where
matrix elements and PSMCs will be vastly different the $\mu_Q$
dependence is bound to be large. In order to be able to study such
effects as locally as is possible, we prefer not to use
unitarity arguments in FxFx.

For what concerns the study of merging-scale systematics, we emphasise
that the discussion presented above by no means justifies the practice 
of not being conservative with the choice of $\mu_Q$ in non-unitary
approaches. There is a particularly common misconception, relevant
when the physics one wants to study features a threshold for jet
hardness (that we shall denote by $\ptcut$ henceforth): such a
misconception entails choosing $\mu_Q<\ptcut$. The (implicit) 
argument for this choice is that a tagged jet is by definition
a hard quantity, and by setting $\mu_Q>\ptcut$ one might spoil the
underlying NLO accuracy\footnote{Or the tree-level matrix element
accuracy in the case of LO mergings, where this argument is also
applied.} for the corresponding jet cross section. Several observations
are in order here. Firstly, the use of $\ptcut$ as a criterion to
determine $\mu_Q$ is a contradiction in terms: by definition, a merging
prescription is what allows one to use samples of hard events without
knowing {\em a priori} for which observables they will be employed,
in particular which minimal jet hardness will be imposed. The fact that
the merging is not perfect (i.e., it does not have zero systematics)
just implies that both $\mu_Q\ll\ptcut$ and $\mu_Q\gg\ptcut$ are not
particularly sensible, and nothing else. Secondly, a criterion based
solely on $\ptcut$ misunderstands the meaning of hardness, which is
not absolute, but relative. For example, a jet with $\ptcut=40$~GeV
can rightly be defined to be hard in inclusive $W$ production; the same jet 
is less hard in Higgs production, basically soft in $t\bar{t}$ production,
and certainly soft in the production of a 1-TeV $Z^\prime$ resonance.
Indeed, it should be obvious that it is always a ratio of mass scales
(one of which is of the order of the intrinsic hardness of the
production process, essentially defined by the masses of the
final-state particles, and the other of the order of $\ptcut$),
and never the absolute value of one such scale, that matters:
the argument of a logarithm is a dimensionless quantity.
Thirdly, and related to the previous item. When choosing $\mu_Q>\ptcut$
one might indeed spoil some underlying NLO description, and for a very
good reason: such matrix-element-driven prediction may simply be irrelevant 
in the case of a strong scale hierarchy where $\ptcut$ is much smaller
than the intrinsic hardness of the process, because there one
believes the correct type of prediction for jet observables to be
rather a PSMC-dominated one. Even if the presence of Sudakov
suppression factors in the merged matrix elements may allow one to
employ a merging scale which is smaller than what naive expectations
would suggest, still by choosing $\mu_Q<\ptcut$ one runs the risk of
spoiling the underlying PSMC description of the lowest-multiplicity
sample. In conclusion, given that a merging-scale choice necessarily
represents a non-perfect compromise between a matrix-element- and a
PSMC-dominated prediction, for processes where a scale hierarchy is
not overwhelmingly clear (i.e., when $\ptcut$ is a non-negligible
fraction of the hardness of the production process), one must not decide
beforehand whether it is either a matrix-element or a PSMC description
which is suited best: they are in principle both valid alternatives,
and a sufficiently large range of $\mu_Q$ in the surrounding of $\ptcut$
must be probed, lest one underestimates the merging-scale systematics.

A final technical remark: given that the work of ref.~\cite{Frederix:2012ps}
has used \HWs\ as PSMC, and that the FxFx formalism naturally matches
a $\pt$-ordered shower, its use with \PYe\ and \HWpp\ does not pose
any conceptual problems. In fact, a fully automatic FxFx interface with
\PYe\ has now been achieved, but is not part of the current public \aNLO\
release (the related, specific routine inside the \PYe\ code has become
available starting from v8.185). For this reason, the sample
FxFx-merged results which will be presented in sect.~\ref{sec:diff}
will make use of \HWs. The FxFx-\PYe\ interface also paves the way 
for a fully analogous procedure, that will be carried out with \HWpp.
We also remind the reader that the FxFx method has so far been formulated
only for processes that do not feature light jets at the Born level of the
lowest-multiplicity sample, and that the merging of $b$-quark production
processes has not been explicitly studied. Although we believe that 
these cases can be treated with only minor modifications (if any)
w.r.t.~the present implementation, we postpone their discussion to 
a future work.

\subsection{Spin correlations: MadSpin\label{sec:madspin}}
In both SM physics and BSM searches the role of unstable particles,
that are not directly observable but (some of) whose decay products 
may be seen in detectors, is very prominent. Let us consider the
production of $p$ unstable particles $u_k$ ($k=1,\ldots p$; for
example, $u_1=Z$, $u_2=t$, $u_3=\bt$, and so forth), each of which
decays into $n_k$ particles $d_{1,k},\ldots d_{n_k,k}$, in association
with $l$ stable particles $s_1,\ldots s_l$:
\beqn
x+y&\longrightarrow&u_1(\to d_{1,1}+\ldots d_{n_1,1}+X_1)+\ldots
u_p(\to d_{1,p}+\ldots d_{n_p,p}+X_p)+
\nonumber\\*&&
s_1+\ldots s_l+X_0\,.
\label{unsproc}
\eeqn
It is convenient to regard eq.~(\ref{unsproc}) as a parton-level
quantity, so that $x$, $y$, and all the particles in the sets $X_k$
are gluons or light quarks; the contents of $X_k$ depend on the 
perturbative order considered\footnote{And, in general, on the
type of corrections. In order to be definite, we consider here
the case of QCD.}, and need not be specified here.
Equation~(\ref{unsproc}) does not properly define a process,
but has the following intuitive meaning: it corresponds
to the contributions to the process
\beqn
x+y &\longrightarrow& d_{1,1}+\ldots d_{n_1,1}+\ldots
d_{1,p}+\ldots d_{n_p,p}+
s_1+\ldots s_l+X\,,
\label{procdec}
\\
X&=&\bigcup_{k=0}^p X_k\,,
\eeqn
whose Feynman diagrams feature an $s$-channel propagator for each
of the $p$ unstable particles $u_k$, with one end of the propagator
attached to a subdiagram that contains at least the decay products
$d_{1,k},\ldots d_{n_k,k}$ (such a subdiagram is a tree at the
leading order). These diagrams may be called $p$-resonant diagrams
and, by extension, any diagram that features $n$ such 
propagator-plus-subdiagram structures will be called $n$-resonant
(so that, by including the case $n=0$, all diagrams contributing
to eq.~(\ref{procdec}) can be classified in this way). This implies
that it would be natural to associate with eq.~(\ref{unsproc}) the
matrix elements obtained by considering only the $p$-resonant diagrams
in their computations. Unfortunately, this is not straightforward, since 
in general it violates gauge invariance. So the only possibility 
is that of an operative meaning: thus, eq.~(\ref{unsproc}) 
stands for the matrix elements relevant to the process of 
eq.~(\ref{procdec}), subject to selection cuts whose purpose is 
that of forcing the $p$-resonant diagrams to be numerically dominant.
While this approach is the cleanest possible from a theoretical viewpoint,
it has an obvious problem of efficiency: the non-$p$-resonant contributions
to eq.~(\ref{procdec}) might swamp the $p$-resonant ones; furthermore,
the matrix elements of eq.~(\ref{procdec}) are usually very involved.
The difficulties mentioned above can be overcome by observing that
a computation that uses only $p$-resonant diagrams is formally correct
in the limit where all the widths of unstable particles vanish,
$\Gamma_{u_k}\to 0$, $\forall k$ (narrow-width approximation, which
can be systematically improved in the context of a pole 
expansion~\cite{Stuart:1991xk,Aeppli:1993rs,Beneke:2003xh}).
An immediate consequence of the narrow-width approximation is that the
amplitudes associated with parton emissions (i.e., beyond leading order) from 
the decay products $d_{i_k,k}$ do not interfere with those associated with 
emissions either from $d_{i^\prime_{k^\prime},k^\prime}$, 
for any $k^\prime\ne k$,
or from any particle which is not a decay product. Hence, higher-order 
corrections factorise, and can thus be sensibly considered separately
for production and decay. This results in a significant simplification 
of the calculation, since in the narrow-width approximation 
one can therefore write:
\beqn
x+y&\longrightarrow& u_1+\ldots u_p+s_1+\ldots s_l+X_0\,,
\label{procundec}
\\
u_k&\longrightarrow& d_{1,k}+\ldots d_{n_k,k}+X_k
\;\;\;\;\;\;\;\;\;\;\;\;\;\;\;\;\;\;
k=1,\ldots p\,.
\label{decays}
\eeqn
As this notation suggests, the particles $u_k$ in eqs.~(\ref{procundec}) 
and~(\ref{decays}) are regarded as final- and initial-state objects 
respectively, rather than as intermediate ones as in eq.~(\ref{procdec}).  
The calculation of amplitudes in the narrow-width approximation can be done 
by employing well-established spin-density-matrix techniques, which allow 
one to account for all spin correlations (we remind the reader that
the process of eq.~(\ref{procdec}) is said to have decay spin correlations
if its matrix elements depend non-trivially on the invariants
$d_{i,k}\mydot d_{j,k}$ for some $k$; production spin correlations
are present in the case of non-trivial dependences on
$d_{i_k,k}\mydot d_{i^\prime_{k^\prime},k^\prime}$, 
for any $i_k$ and $i^\prime_{k^\prime}$
with $k^\prime\ne k$, or on $d_{i_k,k}\mydot s_q$, $d_{i_k,k}\mydot x$, 
$d_{i_k,k}\mydot y$ and $d_{i_k,k}\mydot X_0$ for some $i_k$ and $k$).

When spin correlations effects are small or can be neglected, a further 
simplification can be made, where one replaces the {\em squared} 
amplitudes associated with the $p$-resonant diagrams relevant to 
eq.~(\ref{procdec}) with those relevant to the 
production (eq.~(\ref{procundec})) and decays (eq.~(\ref{decays}))
separately\footnote{In other words, the narrow-width approximation allows 
one to get rid of the non-$p$-resonant diagrams, whereas the simplification
mentioned here gives a prescription for the actual computation of the 
$p$-resonant diagrams where one does not use spin-density matrices.}. 
Despite being a priori rather crude, such a simplification is very
widely used in the context of PSMCs, since the narrow-width approximation 
is more difficult to automate already at LO, and more so beyond LO. Besides, 
unweighted-event generation is significantly more efficient for the process 
of eq.~(\ref{procundec}) than it is for that of eq.~(\ref{procdec}).
The procedure is therefore that of generating (unweighted) events 
that correspond to eq.~(\ref{procundec}), and then let the PSMC decay
the unstable particles according to eq.~(\ref{decays}) (with $X_k=\emptyset$, 
i.e., at the leading order) during the shower phase. Note that the PSMC
must know how to handle these decays, which sometimes involve non-trivial
matrix elements (e.g., in top decays, or for $H^0\to 2\ell 2\nu$).
If this is not the case, decay spin correlations are also incorrectly
predicted.

In order to retain the advantages of the separation of production from 
decays at the level of squared amplitudes,
such as efficiency and ease of automation, without losing the capability
of predicting spin correlations, \aNLO\ uses the method introduced
in ref.~\cite{Frixione:2007zp}\footnote{For an alternative method, based
on the knowledge of the polarization states of the unstable particles and
on spin-density matrices, and which is adopted in \HWpp, see 
ref.~\cite{Richardson:2001df}.}, and studied there for top 
and $W$ decays in the SM, which has been fully automated and extended 
to generic models in ref.~\cite{Artoisenet:2012st}; the corresponding 
module in the code has been dubbed \Madspin\footnote{When working
at the LO, production spin correlations can be recovered not only
by using \Madspin, but also by adopting the so-called decay-chain
syntax -- see appendix~\ref{sec:verbose} for more details.}. The method 
is based on the following identity:
\beq
\lim_{\{\Gamma_{u_k}\}\to 0}
\frac{\ampsq\left(x+y\to d_{1,1}+\ldots d_{n_p,p}+
s_1+\ldots s_l+X\right)}
{\ampsq\left(x+y\to u_1+\ldots u_p+s_1+\ldots s_l+X\right)}
\prod_{k=1}^p\Delta_{u_k}^{-1}
\;\le\;
{\cal U}\left(\{u_k\}_{k=1}^p\right)\,,
\label{FLMW}
\eeq
where ${\cal U}$ is a universal factor,
and the matrix elements at the numerator and denominator on the l.h.s.~are
the tree-level ones relevant to the processes of eqs.~(\ref{procdec}) 
(with $p$-resonant diagrams only) and~(\ref{procundec}) (with $X_0=X$)
respectively, and:
\beq
\Delta_{u_k}^{-1}=(p_{u_k}^2-m_{u_k}^2)^2+(m_{u_k}\Gamma_{u_k})^2\,.
\label{delmo}
\eeq
Note that the factors $\Delta_{u_k}^{-1}$ in eq.~(\ref{FLMW}) cancel
exactly the denominators of the unstable-particle propagators, so that
the limit is indeed a finite quantity. The fact that the set $X$ of 
radiated partons is the same at the numerator and denominator in
eq.~(\ref{FLMW}) implies that the two matrix elements are computed
at the same relative order w.r.t.~the leading one; this allows one
to correctly take into account the effects of hard radiation at
the production level. Finally, the factor ${\cal U}$ 
depends only on the identities of the unstable particles
(and possibly on the decay kinematics), but is independent of the
production process. It is computed by considering the decays of
eq.~(\ref{decays}) at the LO (i.e., with $X_k=\emptyset$), in a
fully numerical manner by \Madspin.

Equation~(\ref{FLMW}) is used within a standard hit-and-miss procedure,
that determines the kinematics of the decay products $d_{i_k,k}$ given
that of the unstable particles $u_k$. In practice, unweighted events
are first obtained for the process in eq.~(\ref{procundec}); then,
for each of these the phase-space of the decay products is sampled,
and through hit-and-miss unweighted events for the process in 
eq.~(\ref{procdec}) are obtained. It should be clear that the latter
events thus correctly incorporate the information on both production
and decay spin correlations. Furthermore, there is evidence that, at
the NLO, eq.~(\ref{FLMW}) gives a general better description of the 
exact result for eq.~(\ref{procdec}) than that of an NLO narrow-width 
prediction, in spite of the LO-only treatment of the decays in the
former. This has
to do with the fact that, in the narrow-width approximation, configurations
where the virtuality of an unstable particle (as reconstructed from
final-state objects) is larger than the particle mass are suppressed
both by $\as$ (being due to hard NLO corrections) and by a kinematical 
factor -- for a recent discussion relevant to top physics, see sect.~3 of 
ref.~\cite{Papanastasiou:2013dta}. Furthermore, the NLO origin of these
configurations implies that, in the case of the production of more than 
one unstable particle, such off-shell effects can be possibly included 
only for one particle at a time in the NWA.
On the other hand, eq.~(\ref{FLMW}) trivially allows one to include
off-shell effects without any kinematics suppression, already at
the leading order, and for all unstable particles simultaneously. 
In particular, one starts by generating the virtualities of $u_k$
according to a Breit-Wigner distribution (rather than using the pole 
masses), and employs those in the generation of the momenta of the decay
particles that enter the hit-and-miss procedure mentioned above.
In so doing, one may introduce back into the problem gauge-violating
terms formally of ${\cal O}(\Gamma_{u_k}/m_{u_k})$, i.e., within the
accuracy of the method. So the only potential issue might result
from the numerical coefficients in front of such terms not being
of ${\cal O}(1)$. However, one can use eq.~(\ref{FLMW}) to check that
this is not the case; indeed, the bound of that equation is a fairly
good one, even for configurations where the virtuality of some unstable
particle is more than ten widths away from the pole mass.

We have so far tacitly assumed to deal with the most common and
numerically-relevant case in the SM, namely that of QCD corrections
to the production of weakly-decaying unstable particles. When one
starts considering EW NLO effects, one may face a self-consistency
problem, due e.g.~to the fact that the loop propagators relevant to 
EW vector bosons would have to feature non-zero widths (through the
complex mass scheme prescription), while the same particles would
have to be treated as on-shell (owing to the $\Gamma_{u_k}\to 0$
limits) when appearing in the final state. Hence, in the context
of a mixed-coupling expansion, or in general when width effects
are induced by the same type of interactions which are responsible
for higher-order corrections, the method discussed in this section
has to be employed by carefully considering the characteristics of
the production process.

The \Madspin\ module included in \aNLO\ features a number of upgrades 
w.r.t.~that presented in ref.~\cite{Artoisenet:2012st}. The most
important of these is that the phase-space generation is now
handled by the Fortran code (rather than by a Python routine). The
determination of the maximum weight has been fully restructured
as well; on top of that, it has been noticed that using more events, 
but less phase-space points per event, is more efficient (the defaults 
have changed from $20$ and $10^4$ to $75$ and $400$, respectively).
These two major improvements have resulted in a dramatic decrease of the
running time (by a factor of about 10). Furthermore, the former one
has paved the way for lifting the present limitation that restricts
the code to dealing only with (a succession of) $1\to 2$ decays.
From the phenomenology viewpoint, two important novelties are the
following. Firstly, \Madspin\ now always writes the information
on the polarization of final-state particles, gathered
from the matrix elements, onto the LHE files. This implies,
in particular, that PSMCs, if equipped with a suitable
module, can handle $\tau$ decays including exactly all decay spin
correlations, and also the production ones due to the diagonal terms
of the spin-density matrix. Secondly, one can now use a \UFO\ model in 
the determination of the decays which can be different from the one adopted 
in the computation of the short-distance undecayed matrix elements.

\section{How to perform a computation\label{sec:howto}}
The theoretical background discussed in sect.~\ref{sec:basis}
can be ignored if one is only interested in 
obtaining phenomenological predictions. This is the attitude that
will be taken in this section, whose aim is that of giving the
briefest documentation for the basic running of the code, and thus
to show its extreme simplicity and level of automation.

The \aNLO\ package is self-contained (third-party codes are included
in the tarball -- see appendix~\ref{sec:tpc} for their complete list).
The local directory where its tarball is unpacked is called main directory; 
there is a miniminal (and optional) 
setup to be done, as described in appendix~\ref{sec:depen}.
In essence, the computation of a cross section by \aNLO\ consists of 
three steps: {\em generation}, {\em output}, and {\em running}.
All three steps are performed by executing on-line commands
in the \aNLO\ shell\footnote{As in \MadGraphf, scripting commands
is of course possible.}, which can be accessed by executing from
a terminal shell in the main directory the following command:

\noindent
{\tt ~./bin/mg5\_aMC}

\noindent
The prompt now reads:

\noindent
~\prompt\

\noindent
which signals the fact that one is inside the \aNLO\ shell.
The three steps mentioned above correspond to the following commands:

\noindent
~\prompt\ {\tt ~generate} {\em process}

\noindent
~\prompt\ {\tt ~output} 

\noindent
~\prompt\ {\tt ~launch} 

\noindent
respectively; here, {\em process} denotes the specific process one
is interested into generating (see appendix~\ref{sec:verbose} for full
details on the syntax).

When generating a process, one must decide whether he/she is interested
in including or not including NLO effects. In fact, although by definition
an NLO cross section does include an LO part (the Born contribution),
if NLO effects are not an issue it does not make much sense to
include their contributions only to discard them at run time, 
especially in view of their being much more involved than their
Born counterparts. Furthermore, the majority of physics models
(see sect.~\ref{sec:method}) have not yet been extended to include
NLO corrections, and in these cases the whole procedure could not
even be conceived.

For these reasons, we talk about LO-type and NLO-type generations,
and we note that these two types of generation give access to 
different running options. We shall discuss their different merits
in the following sects.~\ref{sec:LOgen} and~\ref{sec:NLOgen} respectively. 
Before going into that, we
point out that an LO-type generation produces the
same short-distance cross section code (and hence the same type of
physics) as that one would have obtained by running \MadGraphf.
All possibilities that were available in \MadGraphf\ are still
available in \aNLO; at the same time, it should be clear that,
even in the context of an LO-type generation, \aNLO\ has a much
wider scope than \MadGraphf.

\subsection{LO-type generation and running\label{sec:LOgen}}

\vskip 0.4truecm
\noindent
$\blacklozenge$ {\em Generation}

\noindent
In the generation phase, \aNLO\ constructs the cross section
relevant to the process given as input by the user, and thus performs
the operations sketched in sect.~\ref{sec:method}. For example, if
one is interested in $t\bt W^+$ production in $pp$ collisions
at the LO, one will need to execute the following command:

\noindent
~\prompt\ {\tt ~generate p p > t t\~{} w+} 

\noindent
When generating a process, \aNLO\ assumes the model to be the SM 
(with massive $b$ quarks); a different model can be adopted, 
by ``importing'' it before the generation (see appendix~\ref{sec:verbose}).
For example, the generation of a pair of top quarks in association
with a pair of neutralinos (the latter denoted by {\tt n1}) can be
achieved as follows:

\noindent
~\prompt\ {\tt ~import model mssm}

\noindent
~\prompt\ {\tt ~generate p p > t t\~{} n1 n1}

\vskip 0.4truecm
\noindent
$\blacklozenge$ {\em Output}

\noindent
When the process generation is complete, the relevant information is still
in the computer memory, and needs to be written on disk in order
to proceed. This is done by executing the command:

\noindent
~\prompt\ {\tt ~output MYPROC}

\noindent
where {\tt MYPROC} is a name chosen by the user\footnote{\label{ft:ft}
Such a name may be omitted; in this case, \aNLO\ will choose one. On the 
other hand, there are a few names which are reserved, since they are
interpreted as options of the {\tt output} command. See
appendix~\ref{sec:MG5out} for more details.}
that \aNLO\ will in turn assign to the directory under whose tree the cross
section of the process just generated is written. We call such
a directory the current-process directory, which is where all
subsequent operations will be performed. For more details on
this and for an overview of the structure of the current-process 
directory, see appendix~\ref{sec:depen}.

\vskip 0.4truecm
\noindent
$\blacklozenge$ {\em Running}

\noindent
The running stage allows one to accomplish a variety of tasks,
the most important of which are the production of unweighted events,
and the plotting of user-defined physical observables. Regardless
of the final product(s) of the run, \aNLO\ will start by integrating
the cross section generated and written in the two previous steps.
In order to run \aNLO, one executes the command:

\noindent
~\prompt\ {\tt ~launch}

\noindent
What is prompted afterwards opens an interactive talk-to (which, again,
can be scripted) that allows the user to choose among various options.
This looks as follows:

\vskip 0.4truecm
\noindent
\begin{minipage}{1.1\textwidth}
{\tt
The following switches determine which programs are run:\\
 1 Run the pythia shower/hadronization:
 \hspace*{2.8truecm}pythia=OFF\\
 2 Run PGS as detector simulator:
 \hspace*{4.0truecm}pgs=OFF\\
 3 Run Delphes as detector simulator:
 \hspace*{3.2truecm}delphes=OFF\\
 4 Decay particles with the MadSpin module:
 \hspace*{2.0truecm}madspin=OFF\\
 5 Add weight to events based on coupling parameters:
 \hspace*{0.0truecm}reweight=OFF\\
  Either type the switch number (1 to 5) to change its default setting,\\
  or set any switch explicitly (e.g. type 'madspin=ON' at the prompt)\\
  Type '0', 'auto', 'done' or just press enter when you are done.\\
 {[0, 4, 5, auto, done, madspin=ON, madspin=OFF, madspin, reweight=ON, ... ]}
}
\end{minipage}

\vskip 0.4truecm
\noindent
We would like to emphasise that the structure of the above prompt
will evolve in the near future, and be made more similar to its NLO
counterpart (see sect.~\ref{sec:NLOgen}); however, the general idea
that underpins its use will remain the same, so that what follows
has to be regarded as an exemplification of the general features
of the \aNLO\ talk-to phase.

By entering {\tt 1}, {\tt 2}, {\tt 3}, {\tt 4}, or {\tt 5} 
at the prompt one toggles between the two
values {\tt ON} and {\tt OFF} of the corresponding feature. For example, 
by entering {\tt 1} one is prompted again with the display above, except for 
the fact that {\tt pythia=OFF} has now become {\tt pythia=ON}. By entering
{\tt 1} again, one gets back to {\tt pythia=OFF}. By entering {\tt 0}, 
or {\tt done},
or by simply hitting return, the talk-to phase ends, and \aNLO\ starts
the actual run. The defaults shown above imply that \aNLO\ will simply
limit itself to integrating the cross section, and to producing the
required number of unweighted events. On the other hand, by turning the
various switches above to {\tt ON}, one enables the following
features. 

{\tt pythia=ON}: with such a setting \aNLO\ will steer the showering of 
the hard events previously generated, by employing \PYs. However, it is
crucial to bear in mind that the same hard events can be showered 
with PSMCs other than \PYs, but in this case \aNLO\ is not capable of 
steering the shower\footnote{This is not the case for NLO simulations -- 
see later.} (the steering of \PYe\ will soon become available). 
The user may have an independent installation of \PYs, but 
also install the code using the \aNLO\ shell, by executing the command 
{\tt install pythia-pgs}. 

{\tt pgs=ON}: in this case, \aNLO\ will also steer the run of the
Pretty Good Simulator (PGS)~\cite{PGS} after that of \PYs\ (i.e., first
all events are showered and hadronised, and next they are passed through the 
basic detector simulation as implemented by PGS). For this reason, when
{\tt pgs=ON} \aNLO\ automatically sets {\tt pythia=ON}. Note, also,
that when the \aNLO\ shell is used to install \PY\ with the
{\tt install pythia-pgs} command, PGS is installed too.

{\tt delphes=ON}: this allows one to steer the run of 
\delphes~3~\cite{deFavereau:2013fsa} for a fast detector simulation.  
\delphes\ can also be installed through the \aNLO\ shell with the
command {\tt install Delphes}\footnote{Or {\tt install Delphes2},
if one wanted to use the older \delphes~2~\cite{Ovyn:2009tx} version.}.
As for switch number 2, when {\tt delphes=ON} \aNLO\ sets automatically 
{\tt pythia=ON}.

{\tt madspin=ON}: by doing so, one includes production spin correlations
by means of \Madspin\ (see sect.~\ref{sec:madspin}). Note that the 
decay-chain syntax (see appendix~\ref{sec:verbose}) is actually faster 
and features a better approximation of the exact cross section than
\Madspin, which is thus more conveniently used in the context of
NLO simulations.

{\tt reweight=ON}: instructs \aNLO\ to store in the LHE file information
to be used later within the matrix-element reweighting procedure, for 
example relevant to assessing the impact of different theoretical 
assumptions (see sect.~\ref{sec:rwgt} and appendix~\ref{sec:appLO}).

When switches {\tt 1}--{\tt 5} 
are set as desired by the user and {\tt 0}, or {\tt done},
or {\tt <return>} are entered, \aNLO\ proceeds with the run, whose first
stage is that of giving the user the possibility of modifying the various
inputs relevant to the options selected above. Such inputs, and the 
{\tt *\_card.dat} files where they are stored, have a self-explanatory
meaning, and we will not discuss them in detail here.

\subsection{NLO-type generation and running\label{sec:NLOgen}}

\vskip 0.4truecm
\noindent
$\blacklozenge$ {\em Generation}

\noindent
The generation phase of an NLO-type generation has the same conceptual
meaning of that relevant to the LO case, described at the beginning
of sect.~\ref{sec:LOgen}. The syntax is also {\em very} similar:
adopting again the example of $t\bt W^+$ production in $pp$ collisions,
one will need to execute the following command in order to include 
QCD NLO effects:

\noindent
~\prompt\ {\tt ~generate p p > t t\~{} w+ [QCD]} 

\noindent
As one can see, the only difference w.r.t.~the case of the 
LO-generation is in the presence of the keyword {\tt [QCD]} here.
Fuller details on the syntax for NLO-type generation are given
in appendix~\ref{sec:verbose}.

\vskip 0.4truecm
\noindent
$\blacklozenge$ {\em Output}

\noindent
There is no conceptual or technical difference w.r.t.~the case of
an LO-type generation in the output phase. Specifically, the command
one will need to execute is the same as that described in 
sect.~\ref{sec:LOgen}:

\noindent
~\prompt\ {\tt ~output MYPROC}

\noindent
The same comments concerning the choice of {\tt MYPROC} as made
for LO simulations apply here (see footnote~\ref{ft:ft}).

\vskip 0.4truecm
\noindent
$\blacklozenge$ {\em Running}

\noindent
Also in the case of an NLO-type generation, the running phase begins
by integrating the cross section generated and written in the 
two previous steps. There is an important difference
w.r.t.~the LO case that must be stressed here. Namely, at the LO
the short-distance cross sections relevant to LO+PS and to fLO computations
are identical. This is not the case at the NLO: the MC@NLO cross sections
used in NLO+PS computations are different from the fixed-order ones 
used in fNLO computations (the former contain the MC counterterms, 
the latter do not; the former can be unweighted, the latter cannot -- 
see sect.~\ref{sec:mcatnlo} for more details). However, when \aNLO\ 
generates and outputs a process, it writes {\em both} cross sections,
and in so doing allows the user to choose at runtime which type of 
computation to perform. In other words, after having generated a process
and used it to obtain (say) fNLO results, there is no need to re-generate it
to obtain NLO+PS results: it is sufficient to run \aNLO\ again.

\noindent
In order to run \aNLO, one executes the same command as that 
relevant to the LO-type generation:

\noindent
~\prompt\ {\tt ~launch}

\noindent
However, what is prompted afterwards opens an interactive talk-to 
which is different from the one of the LO-type generation.
In particular, one now obtains what follows:

\vskip 0.4truecm
\noindent
\begin{minipage}{1.1\textwidth}
{\tt
The following switches determine which operations are executed:\\
 1 Perturbative order of the calculation:
 \hspace*{4.25truecm}order=NLO\\
 2 Fixed order (no event generation and no MC@[N]LO matching):
 \hspace*{0.0truecm}fixed\_order=OFF\\
 3 Shower the generated events:
 \hspace*{6.3truecm}shower=ON\\
 4 Decay particles with the MadSpin module:
 \hspace*{3.9truecm}madspin=OFF\\
  Either type the switch number (1 to 4) to change its default setting,\\
  or set any switch explicitly (e.g. type 'order=LO' at the prompt)\\
  Type '0', 'auto', 'done' or just press enter when you are done.\\
  {[0, 1, 2, 3, 4, auto, done, order=LO, order=NLO, ... ]}
}
\end{minipage}

\vskip 0.4truecm
\noindent
By entering {\tt 1}, {\tt 2}, {\tt 3}, or {\tt 4} 
at the prompt one toggles between the two
values of the corresponding feature (which are {\tt NLO} and {\tt LO}
for 1, and {\tt ON} or {\tt OFF} for 2--4). For example, by entering 
{\tt 2} one is prompted again what is displayed above, except for the fact 
that {\tt fixed\_order=OFF} has now become {\tt fixed\_order=ON}. By entering
{\tt 2} again, one gets back to {\tt fixed\_order=ON}. By entering {\tt 0}, 
or {\tt done},
or by simply hitting return, the talk-to phase ends, and \aNLO\ starts
the actual runs.

It is the combinations of the values of the switches 1 and 2 that 
control which kind of computation the program will perform. More
explicitly, we have:

\noindent
\hspace*{1.0truecm}
({\tt order=NLO},{\tt fixed\_order=OFF})
$\phantom{aaa}\longrightarrow\phantom{aaa}$
NLO+PS and FxFx-merged\footnote{We remind the reader that $i$-parton
samples, the contributions to an FxFx cross section of a given multiplicity,
are nothing but unmerged NLO+PS samples with some extra damping factors,
which are included by \aNLO\ through a parameter ({\tt ickkw}) in an 
input card.}

\noindent
\hspace*{1.0truecm}
({\tt order=NLO},{\tt fixed\_order=ON})
$\phantom{aaaa}\longrightarrow\phantom{aaa}$
fNLO

\noindent
\hspace*{1.0truecm}
({\tt order=LO},{\tt fixed\_order=OFF})
$\phantom{aaaa}\longrightarrow\phantom{aaa}$
LO+PS

\noindent
\hspace*{1.0truecm}
({\tt order=LO},{\tt fixed\_order=ON})
$\phantom{aaaaa}\longrightarrow\phantom{aaa}$
fLO

\noindent
One need not be suprised by the fact that LO+PS and fLO results
can be obtained following an NLO-type generation, since all the LO 
information is obviously there, being part of the NLO cross section.
Rather, one may wonder why these options are not disabled, since they
lead to the same physics as a run that follows an LO-type generation.
A discussion on this point will be given in sect.~\ref{sec:LO}; here,
we limit ourselves to giving the shortest answer, which is the following:
the access to both LO and NLO results within the same generation
procedure (i.e., in the same current-directory process) guarantees
an extremely easy and swift comparison between them.

Switch 3 controls whether one employs \aNLO\ to steer
the showering of hard-subprocess events previously generated
({\tt shower=ON}) or not ({\tt shower=OFF}). We shall give a more
extended discussion about this point in sect.~\ref{sec:shower}.
Here, we would like to stress one crucial point:
if the hard events are relevant to an NLO+PS run, they do not have
any physical meaning unless they are showered. Hence, one is free to
use or not use \aNLO\ for showering them, but shower them he/she must.
Note that this is not the case for LO+PS hard events, although of course
shower or not shower them leads to different types of physics (in particular,
observables constructed with unshowered LO+PS hard events are the same
as those resulting from an fLO computation).

Finally, switch 4 allows one to decide whether to include production
spin correlations by means of \MadSpin\ (see sect.~\ref{sec:madspin}).
Since the method works starting from unweighted events, \MadSpin\ is
disabled when a fixed-order run is selected (in other words, the
inputs {\tt fixed\_order=ON} and {\tt madspin=ON} are incompatible,
and \aNLO\ will automatically prevent the user from making such a 
choice; this is not necessary when {\tt order=LO}, but it is done
anyway in order to simplify things).

As in the case of an LO-type generation, after setting switches 1--4
to the desired values and entering 0, or {\tt done}, or {\tt <return>},
\aNLO\ proceeds with the run by giving the user the possibility of modifying 
the various input cards relevant to the options selected.

\subsubsection{(N)LO+PS results\label{sec:shower}}
The aim of this section is that of giving some details on
the (N)LO+PS runs that follow an NLO-type generation.
We recall that (N)LO+PS results are obtained by setting 
{\tt fixed\_order=OFF}, and the perturbative order is assigned
according to the value of the switch {\tt order} (see sect.~\ref{sec:NLOgen}).

\noindent
There are two types of objects\footnote{There are actually several other
auxiliary files produced by \aNLO, whose role is however not important
here. See appendix~\ref{sec:depen} for more details.} that may be obtained 
with an (N)LO+PS run:
\begin{enumerate}
\item One or two files of unweighted events at the hard-subprocess level.
\item One file that collects results at the end of the shower, be
them in the form of histograms, or n-tuples, or events;
we call it {\tt MC.output}.
\end{enumerate}
The first of the files in 1 is the result of the integration of the 
short-distance partonic cross section performed by \aNLO. The second
file is present only in the case when \MadSpin\ is used ({\tt madspin=ON}),
and results from feeding the former file to \MadSpin. These two files
will be found under the current-process directory tree:

\noindent
{\tt ~MYPROC/Events/run\_nn/events.lhe.gz}

\noindent
{\tt ~MYPROC/Events/run\_nn\_decayed\_mm/events.lhe.gz}

\noindent
Although of course \MadSpin\ may not be used (or it may simply be not
relevant, if the generated process does not feature particles that decay), 
the above structure is the most general, and hence we shall discuss it
so as to encompass all cases. We shall call the two files above
the undecayed and decayed (hard-subprocess) event files, respectively;
both files are fully compliant with the
recent Les Houches Accord v3.0~\cite{LHEv30}.

The undecayed event file will contain
events whose particle content is the same as that given as input
during the generation step (up to one final-state parton at the NLO),
i.e., using the example of sect.~\ref{sec:NLOgen}:
\beq
x+y\;\longrightarrow t+\bt+W^+ (+z)\,,
\eeq
where $x$, $y$, and $z$ are quarks or gluons\footnote{The parton $z$ 
is present in the case of $\clH$ events, and absent in $\clS$ 
events~\cite{Frixione:2002ik}.}. The integration of the cross section
that results in the actual events is performed once the values of the
relevant parameters (such as particles masses, collider energy, PDFs)
are given in input by the user. The two primary input cards are:

\noindent
{\tt ~MYPROC/Cards/run\_card.dat}

\noindent
{\tt ~MYPROC/Cards/param\_card.dat}

\noindent
One can obtain an undecayed event file with some parameter settings,
then change these settings, integrate the cross section again, and
obtain a second event file; the procedure can be iterated as many times
as one likes. Each run is identified by an integer number, chosen
by \aNLO, which unambiguously names the directory where the
event files will be stored. So, in the example given above, we shall
have {\tt nn=01} for the first run, {\tt nn=02} for the second run,
and so forth.

Each undecayed event file can be fed to \MadSpin\ in order to obtain
a decayed event file. In order to do so, the user is expected to
give in input to \MadSpin\ the actual decay(s) he/she is interested
into, which can be done by means of the input card:

\noindent
{\tt ~MYPROC/Cards/madspin\_card.dat}

\noindent
Using again the generation example given before, and supposing
that one wants to study the decays:
\beq
t\;\to\;b\,e^+\nu_e\,,
\;\;\;\;\;\;
\bt\;\to\;\bar{b}\,e^-\bar{\nu}_e\,,
\;\;\;\;\;\;
W^+\;\to\;\mu^+\nu_\mu\,,
\label{ttWlepdec}
\eeq
then the decayed event file will contain events of the following kind:
\beq
x+y\;\longrightarrow b+e^+ +\nu_e +\bar{b}+e^- +\bar{\nu}_e + 
\mu^+ +\nu_\mu\, (+z)\,.
\eeq
It is possible to run \MadSpin\ multiple times, by feeding it with the
same undecayed event file and by changing the type of decays considered
(e.g., semileptonic top decays after the di-leptonic ones of 
eq.~(\ref{ttWlepdec})). For each \MadSpin\ run, \aNLO\ will store
the decayed event file in a different directory -- this is the reason
for the integer number {\tt mm} in the example above, which is automatically
assigned to each new run: so {\tt mm=1} will identify the first \MadSpin\
run, {\tt mm=2} the second, and so forth.

In summary, directories that contain undecayed event files are identified 
by an integer {\tt nn}, which roughly speaking corresponds to a given 
choice of settings in {\tt run\_card.dat} and {\tt param\_card.dat}.
On the other hand, directories that contain decayed event files are 
identified by a pair of integers $(${\tt nn},{\tt mm}$)$, which correspond 
to a given choice of settings in {\tt run\_card.dat}, {\tt param\_card.dat},
and {\tt madspin\_card.dat}.

Both undecayed and decayed hard-subprocess event files are non
physical (at the NLO), and {\em must} be showered in order to obtain 
physical results. Such a showering can be done without using 
\aNLO, since it is nothing but the nowadays typical PSMC run starting from 
an external (to the PSMC) LH-compliant event file. On the other hand, 
\aNLO\ can steer PSMC runs: this is convenient for a varienty of reasons, 
among which the most important are probably those of ensuring a full 
consistency among the parameters used when integrating the cross section 
and those adopted by the PSMC, and the correct setting of a few control 
switches in the PSMC itself\footnote{See {\tt http://amcatnlo.cern.ch} 
$\longrightarrow$ {\tt Help and FAQs} $\longrightarrow$ 
{\tt Special settings for the parton shower}.}. As was explained in 
sect.~\ref{sec:NLOgen}, \aNLO\ will steer the shower when {\tt shower=ON}:
in this case, several of the features of the PSMC run can be controlled
through a \aNLO\ input card:

\noindent
{\tt ~MYPROC/Cards/shower\_card.dat}

\noindent
The parameters not explicitly included in this card must be controlled
directly in the relevant PSMC, in the same way as one would follow
in a PSMC standalone run.

The steering of the PSMC by \aNLO\ also guarantees that the PSMC
adopted is consistent with that assumed during the integration of
the short-distance cross section. Indeed, it is important to keep
in mind that NLO+PS events obtained with the MC@NLO formalism are
PSMC-dependent, and for this reason one of the input parameters
in {\tt run\_card.dat} is the name of the PSMC which will be eventually
used to shower the hard-subprocess events. This kind of consistency
is the user's responsibility in the case of a PSMC not steered 
by \aNLO. Finally, we point out that the most recent versions of the
source codes of the Fortran77 PSMCs (\HWs\ and \PYs) are included in 
the \aNLO\ tarball\footnote{These codes are essentially frozen, so we
expect no or very minor changes in the future. Should the need arise
to use versions different w.r.t.~those included here, copy them to
{\tt MYPROC/MCatNLO/srcHerwig} and {\tt MYPROC/MCatNLO/srcPythia},
and write their names in {\tt MYPROC/MCatNLO/MCatNLO\_MadFKS.inputs}.
In the case of \PYs, the routines {\tt UPINIT}, {\tt UPEVNT}, {\tt PDFSET}, 
{\tt STRUCTP}, and {\tt STRUCTM} need also be commented out
of the source code.}.
On the other hand, the modern C++ PSMC (\PYe\ and \HWpp) must be installed
locally prior to running \aNLO. The paths to their executables/libraries
have to be included in the setup file {\tt mg5\_configuration.txt}
(see appendix~\ref{sec:depen}).

When steering of the PSMC by \aNLO\ one will obtain, at the end of 
the PSMC run, the file {\tt MC.output} mentioned in item 2 at the 
beginning of this section. By default (i.e., if the user does not 
write his/her own analysis) this file will contain the full event
record of each showered event (in {\tt StdHEP} format for Fortran MCs,
and in {\tt HepMC} format for C++ PSMCs), and it will 
be named as follows:

\noindent
{\tt ~MYPROC/Events/run\_*/events\_MCTYPE\_ll.hep.gz}

\noindent
Here, the {\tt run\_*} directory is one of those introduced before;
{\tt MC.output} is moved to the same directory where
one finds the hard-subprocess event file fed to the PSMC during
the course of the run. {\tt MCTYPE} is a string equal to the name
of the PSMC used, and {\tt ll} is an integer, chosen by \aNLO, that
allows one to distinguish different files obtained by showering the
same hard-event file multiple times (for example by changing the
shower parameters). Any other output created by the PSMC (such as
its standard output, which would be printed on the screen during an
interactive run) will be found in the directory:

\noindent
{\tt ~MYPROC/MCatNLO/RUN\_MCTYPE\_pp}

\noindent
with {\tt pp} an integer that enumerates the various runs\footnote{Note
that in general {\tt pp}$\,\ne\,${\tt ll}: for example, a PSMC run may not
create an {\tt StdHEP} event record, in which case {\tt pp} would be 
increased, while {\tt ll} would not.}.

While {\tt StdHEP} full event records may be analysed off-line,
they have the disadvantage of using a very significant amount of
disk space. It is very often more convenient to analyse showered
events on-the-fly, using the event kinematics and weight to construct
observables and fill the corresponding histograms. This can be done
in a very flexible manner in the PSMC runs steered by \aNLO. The 
user's analysis can be stored in one of the directories\footnote{Which
contain several ready-to-run templates.}

\noindent
{\tt ~MYPROC/MCatNLO/XYZAnalyzer}

\noindent
where the string {\tt XYZ} depends on the PSMC adopted. The name of
such an analysis, and those of all its dependencies (be they in the 
form of either source codes or libraries) can be given in input to
\aNLO\ at runtime, as parameters in {\tt shower\_card.dat}.
It is clear that it is the user's analysis that determines the
format of {\tt MC.output}. Since \aNLO\ cannot know it beforehand,
it will treat {\tt MC.output} as any other standard file produced by 
the PSMC, which will thus be found in the directory {\tt RUN\_MCTYPE\_pp}.
An exception is that of the topdrawer format (which is human readable);
in this case, {\tt MC.output} will  be named as follows:

\noindent
{\tt ~MYPROC/Events/run\_*/plot\_MCTYPE\_kk\_*.top}

\noindent
with again {\tt kk} an integer that allows one to distinguish the
outputs relevant to different showering of the same hard-event file.

We conclude this section by mentioning the fact that when the {\tt launch}
command is executed with {\tt madspin=ON} and {\tt shower=ON} (i.e.,
both undecayed and decayed events are produced, the PSMC runs immediately 
follows the integration of the cross section, and it is steered by \aNLO)
only the decayed event file will be showered. In order to shower the
undecayed event file, or to perform other shower runs, one needs to use the
\aNLO\ shell command {\tt shower}. Please see appendix~\ref{sec:verbose}
for more details.

\subsubsection{f(N)LO results\label{sec:FO}}
As was discussed in sect.~\ref{sec:NLOgen}, fixed-order results
can be obtained with \aNLO\ by setting {\tt fixed\_order=ON}, with the 
perturbative order assigned according to the value of the switch 
{\tt order}. 

We remind the reader that an NLO computation not matched to
a parton shower cannot produce {\em un}weighted events, since the 
matrix elements are not bounded from above in the phase space;
unweighting events would require the introduction of unphysical 
cutoffs that would bias predictions, and must therefore be avoided.
The only thing one can do is that of considering weighted events,
namely parton-level kinematic configurations supplemented by a
weight (which is basically equal to the matrix element times
a phase-space factor). Such events, which are readily obtained when
sampling the phase space in the course of the integration of the
cross section, can be either stored in a file or used on-the-fly 
to predict physical observables. While we choose the latter option
as our default, we remark that the former can be easily implemented
by the interested user, as it will become clear in what follows.

The input parameters relevant to f(N)LO runs can be found in the same
cards that control NLO+PS runs, namely {\tt run\_card.dat} and 
{\tt param\_card.dat}. Since observables will be plotted on the
fly, the user is expected to write his/her own analysis (which
is a trivial task, in view of the rather small final-state 
multiplicities of parton-level computations). This analysis
must be written in Fortran (or at least it must include a Fortran 
front-end interface to the user's core analysis, written in a language
other than Fortran) and stored in the directory:

\noindent
{\tt ~MYPROC/FixedOrderAnalysis}

\noindent
which contains several templates, meant to be used as examples.
As the templates show, \aNLO\ at present supports two output formats
for user's analyses -- Root and topdrawer (see appendix.~\ref{sec:depen}
for more details about this). The analysis name and the format of 
its output must be given in input to \aNLO, which is done by including 
them in the input card:

\noindent
{\tt ~MYPROC/Cards/FO\_analyse\_card.dat}

\noindent
where the user will also be able to specify any other source
file or library needed by the analysis. Upon running, \aNLO\
will produce the final-result file:

\noindent
{\tt ~MYPROC/Events/run\_nn/MADatNLO.root}

\noindent
if the Root format has been chosen, and:

\noindent
{\tt ~MYPROC/Events/run\_nn/MADatNLO.top}

\noindent
for the topdrawer format; these will contain plots for the observables
defined by the user. As in the case of NLO+PS runs, the integer {\tt nn}
will be chosen by \aNLO\ in a non-ambiguous way (it should be obvious
that both (N)LO+PS and f(N)LO simulations can be performed any number of
times starting from a given generation, and {\tt nn} will allow one to
distinguish their results). By default, if the user does not write an 
analysis, \aNLO\ will produce a file which will contain the predictions 
for the total rate (possibly within the cuts, as specified in 
{\tt run\_card.dat} and {\tt cuts.f}).

In essence, the user's analysis will have to construct the desired
observables, given in input the pair composed of a kinematic configuration 
and its corresponding weight\footnote{Such a weight becomes an array of
weights when the user asks \aNLO\ to compute scale and PDF uncertainties.
See sect.~\ref{sec:errors} for more details.}, which are provided by \aNLO. 
These pairs can be read from eq.~(\ref{NLOdiff}): for each choice of the 
random variables
\mbox{$\{\meas_{Bj}^{(ij)},\meas_{n+1}^{(ij)}\}$} that scan the phase
space and Bjorken $x$'s, \aNLO\ will call the user's analysis three times,
with input pairs:
\beqn
&&\left(\confnpoE\,,\,
\frac{d\sigmaNLOE_{ij}}{d\meas_{Bj}^{(ij)}d\meas_{n+1}^{(ij)}}
\right),
\label{anaE}
\\
&&\left(\confnpoS\,,
\sum_{\alpha=S,C,SC} 
\frac{d\sigmaNLOa_{ij}|_{\rm non-Born}}{d\meas_{Bj}^{(ij)}d\meas_{n+1}^{(ij)}}
\right),
\label{anaS}
\\
&&\left(\confnpoS\,,
\frac{d\sigmaNLOS_{ij}|_{\rm Born}}{d\meas_{Bj}^{(ij)}d\meas_{n+1}^{(ij)}}
\right),
\label{anaB}
\eeqn
where
\beqn
d\sigmaNLOa_{ij}|_{\rm non-Born}&=&d\sigmaNLOa_{ij}
\;\;\;\;\;\;\;\;\alpha=C,SC\,,
\label{wgt2C}
\\
d\sigmaNLOS_{ij}|_{\rm Born}+d\sigmaNLOS_{ij}|_{\rm non-Born}&=&
d\sigmaNLOS_{ij}\,,
\label{wgt2S}
\eeqn
and $d\sigmaNLOS_{ij}|_{\rm Born}$ is the contribution of the Born
matrix elements to the soft-counterevent weight.
The user's analysis must treat eqs.~(\ref{anaE})--(\ref{anaB})
precisely in the same way: this is an essential condition which
guarantees that infrared safety is not spoiled.
In particular, to any given histogram all of these weights must contribute
(obviously, to the bins and subject to the cuts determined by the respective 
kinematics configurations, $\confnpoE$ and $\confnpoS$), if the NLO
accuracy is to be maintained. Note, also, that the weights
(\ref{anaE})--(\ref{anaB}) are correlated, and thus cannot be 
individually used in a statistical analysis as they would if they
had been the result e.g.~of an unweighted-event procedure.

Thanks to the fact that the Born weight, eq.~(\ref{anaB}), is
kept separate from the other contributions to the NLO cross section,
\aNLO\ allows the user to plot, during the course of the same run, a 
given observable both at the NLO accuracy (by using 
eqs.~(\ref{anaE})--(\ref{anaB})) and at the LO accuracy 
(by using eq.~(\ref{anaB}) only).
In order to allow the user's analysis to tell eq.~(\ref{anaB})
apart from eqs.~(\ref{anaE}) and~(\ref{anaS}), \aNLO\ will tag these
weights with an integer, equal to 3, 1, and 2 respectively. For
explicit examples, see one of the template analyses in
{\tt MYPROC/FixedOrderAnalysis}.

We conclude this section by pointing out that the writing of weighted
events can be seen as a special type of analysis. The flexibility
inherent in the structure sketched above should easily allow a user
to write and exploit such an analysis. In this case, we also note
that, rather than associating a single weight with each event, one
can use all of the scale- and PDF-independent ones defined in
ref.~\cite{Frederix:2011ss} (which are available as variables in
a common block) in order to be able to compute scale and PDF uncertainties
through reweighting. These uncertainties can obviously be included
when plotting observables on the fly -- see appendix~\ref{sec:errors}.

\subsection{Possibilities for LO simulations
\label{sec:LO}}
As was mentioned in sect.~\ref{sec:howto}, \aNLO\ offers two ways
to obtain LO results. One is through an LO-type generation, where 
when executing the command {\tt generate} one does not include
the keyword {\tt [QCD]}; this is completely equivalent to what one 
would do if using \MadGraphf. The other is accessed through an 
NLO-type generation, by using {\tt order=LO}.

We stress once again that all of the capabilities of 
\MadGraphf\ are inherited by \aNLO. 
Therefore, if one is interested {\em only} in LO physics, there is no reason 
to simulate it within an NLO-type generation (which in that case would 
simply be a waste of time).
This is also in view of the fact that many of the options available 
in the context of an LO-type generation are disabled when an LO process
is computed after an NLO-type generation.
The idea of LO runs in the context of NLO-type generation is that
of rendering the comparison between LO and NLO a completely trivial
affair. Note, for example, that the input cards relevant to LO- and
NLO-type generations are different: many of the cuts that are present
in the former would simply not make sense in the latter (owing to
the necessity of being compliant with infrared safety).
Also, one should bear in mind that the functional form of $\as$
used by \aNLO\ (i.e., at one or two loops) is determined by the set 
of PDFs employed in the runs. Therefore, when working with an NLO-type
generation, and assuming that fLO (or LO+PS) results are obtained during the
same run as their fNLO (or NLO+PS) counterparts, the
former will be based on NLO PDFs and two-loop $\as$ (which is
of course fully consistent with perturbation theory, and actually
the best option if one is interested in assessing the perturbative
behaviour of the partonic short-distance matrix elements).

In summary, both options for LO runs lead to
exactly the same physics. However, by running the code at the LO 
with its default inputs one generally does {\em not} obtain exactly
the same numbers, since the relevant input cards are tailored either to 
LO-type or to NLO-type generations, which have different necessities 
and emphases. Hence, we recommend to consider fLO and/or LO+PS results
obtained through NLO-type generation mainly in the context of studies 
that feature fNLO and/or NLO+PS simulations as well.

\section{Illustrative results\label{sec:res}}

\subsection{Total cross sections\label{sec:tot}}
In this section we present benchmark results for total rates
(possibly within the cuts which will be specified later), 
at both the LO and the NLO. These are fLO and fNLO results
respectively, with the former computed in the context of 
an NLO-type generation (see sect.~\ref{sec:LO}). 
On the one hand, the aim of this section is that of showing the 
extreme flexibility and reach of \aNLO: in keeping with our
general philosophy, no part of the code has been customised 
to the computation of any of the processes below. The code 
has been run as is extracted from the tarball (apart from the
occasional necessity of defining some final-state cuts in {\tt cuts.f},
in some cases relevant to ($b$-)jet production, which are
explicitly mentioned in the captions of 
tables~\ref{tab:results_vj}--\ref{tab:results_eettV}).
We stress that several of these cross sections have never 
been computed before to NLO accuracy.

We summarise here the main physics parameters used in our runs;
the complete list of inputs, in the form of the input cards of
\aNLO, can be found by visiting 
{\tt http://amcatnlo.cern.ch/cards\_paper.htm},
which should render it particularly
easy to reproduce the results that follow. 
\begin{itemize}
\item $m_H = 125$~GeV, $m_t = 173.2$~GeV.
\item MSTWnlo2008~\cite{Martin:2009iq}~PDFs 
with errors at 68\perc CL, $n_f=4,5$.
Note that these PDFs are used to obtain the fLO results as well,
and that they set the value of $\as(m_Z)$.
\item Central scale choice: $\mu_0=\Ht/2$, with $\Ht$ the scalar sum
of the transverse masses $\sqrt{\pt^2+m^2}$ of all final state particles.
\item Scale variations: independent, $1/2 \mu_0 < \muR, \muF < 2\mu_0$.
\item Diagonal CKM matrix.
\end{itemize}
Final-state objects are defined as follows:
\begin{itemize}
\item 
Jets: anti-$\kt$ algorithm~\cite{Cacciari:2008gp} with 
$R=0.5$, $\pt(j)>30$ GeV, $\abs{\eta(j)}<4$.
\item 
Photons: Frixione isolation~\cite{Frixione:1998jh} with 
$R=0.7$, $\pt(\gamma)>20$ GeV, $\abs{\eta(\gamma)}<2$.
\end{itemize}
While \aNLO\ can handle intermediate resonances by using the 
complex mass scheme, in this section we consider $W$'s, $Z$'s, and
top quarks as stable, and thus set their widths equal to zero.
Furthermore, although matrix elements for loop-induced processes (e.g.~such 
as $gg\to ZZ$), can be obtained with \MadLoop, they have not been considered 
in this section.  Apart from reporting the absolute values of the total cross
sections, the following tables also show (as percentages) scale and PDF
uncertainties. These are computed exactly but without re-running, by 
exploiting the reweighting method presented in ref.~\cite{Frederix:2011ss}.
Following ref.~\cite{Hirschi:2011pa}, Yukawa's are renormalised with
an on-shell scheme.

We tag the processes for which we are not aware of any fNLO result being 
available in the literature with an asterisk. We stress that, in those
cases where an fNLO prediction had been previously obtained, the 
corresponding matching to PSMCs might not necessarily have been achieved; this
is the typical situation for several of the high-multiplicity reactions. On
the other hand, for all processes presented in the tables below the
corresponding NLO+PS event samples can be obtained with \aNLO. Some
non-exhaustive examples will be given in sect.~\ref{sec:diff}.

The following results are organised into broad classes of processes
that share some defining characteristic. In hadroproduction (with a
c.m.~energy of $13$~TeV), we have
considered vector bosons plus up to three light jets 
(table~\ref{tab:results_vj}), vector boson pairs plus up to two
light jets (table~\ref{tab:results_vvj}), three vector bosons plus 
up to one light jet (table~\ref{tab:results_multiv}), four
vector bosons (table~\ref{tab:results_vvvv}), light jets, $b$-jets,
and top quarks, possibly in association with each other
(table~\ref{tab:results_qqj}), vector bosons in association with
top or bottom quarks and light jets (table~\ref{tab:results_tv}),
single top quarks in association with $b$ quarks (four-flavour 
scheme\footnote{Throughout this paper, we adopt standard definitions 
for the four- and five-flavour schemes. See e.g.~sect.~1 of 
ref.~\cite{Frederix:2012dh} for a recent short review.})
and vector bosons (table~\ref{tab:results_stop}), Higgs and double-Higgs
in association with (possibly multiple) light jets, vector bosons
and heavy quarks (tables~\ref{tab:results_h} and~\ref{tab:results_hh}).
In $e^+e^-$ collisions (with a
c.m.~energy of $1$~TeV), we have considered light jets, possibly in
association with heavy quarks (table~\ref{tab:results_eej}),
and top quark pairs in association with (possibly multiple) vector or 
Higgs bosons (table~\ref{tab:results_eettV}).


\begin{landscape}


\begin{table}
\begin{center}
\begin{small}
\begin{tabular}{l r@{$\,\to\,$}l lll}
\toprule
\multicolumn{3}{c}{Process~~~~~~~~~~~~~~~~~~~}& Syntax  & \multicolumn{2}{c}{Cross section (pb)}\\
\multicolumn{3}{c}{Vector boson +jets~~~~~~~~~~~}& &  \multicolumn{1}{c}{ LO 13 TeV} &  \multicolumn{1}{c}{ NLO 13 TeV}\\
\midrule
a.1 & $pp$ & $W^\pm$  & {\tt p p > wpm } &$ 1.375 \pm 0.002\, \cdot 10^{5} \quad {}^{+ 15.4 \% }_{- 16.6 \% } \,\, {}^{+  2.0 \% }_{-  1.6 \% }$& $\,  1.773 \pm 0.007\, \cdot 10^{5} \quad {}^{+  5.2 \% }_{-  9.4 \% } \,\, {}^{+  1.9 \% }_{-  1.6 \% }$\\
a.2 & $pp$ & $W^\pm j$    & {\tt p p > wpm j}  &$ 2.045 \pm 0.001\, \cdot 10^{4} \quad {}^{+ 19.7 \% }_{- 17.2 \% } \,\, {}^{+  1.4 \% }_{-  1.1 \% }$& $\,  2.843 \pm 0.010\, \cdot 10^{4} \quad {}^{+  5.9 \% }_{-  8.0 \% } \,\, {}^{+  1.3 \% }_{-  1.1 \% }$ \\
a.3 & $pp$ & $W^\pm jj$   & {\tt p p > wpm j j}  &$ 6.805 \pm 0.015\, \cdot 10^{3} \quad {}^{+ 24.5 \% }_{- 18.6 \% } \,\, {}^{+  0.8 \% }_{-  0.7 \% }$& $\,  7.786 \pm 0.030\, \cdot 10^{3} \quad {}^{+  2.4 \% }_{-  6.0 \% } \,\, {}^{+  0.9 \% }_{-  0.8 \% }$ \\
a.4 & $pp$ & $W^\pm jjj$   & {\tt p p > wpm j j j}  &$ 1.821 \pm 0.002\, \cdot 10^{3} \quad {}^{+ 41.0 \% }_{- 27.1 \% } \,\, {}^{+  0.5 \% }_{-  0.5 \% }$& $\,  2.005 \pm 0.008\, \cdot 10^{3} \quad {}^{+  0.9 \% }_{-  6.7 \% } \,\, {}^{+  0.6 \% }_{-  0.5 \% }$ \\
\midrule
a.5 & $pp$ & $Z$     & {\tt p p > z  }  &$ 4.248 \pm 0.005\, \cdot 10^{4} \quad {}^{+ 14.6 \% }_{- 15.8 \% } \,\, {}^{+  2.0 \% }_{-  1.6 \% }$& $\,  5.410 \pm 0.022\, \cdot 10^{4} \quad {}^{+  4.6 \% }_{-  8.6 \% } \,\, {}^{+  1.9 \% }_{-  1.5 \% }$ \\
a.6 & $pp$ & $Zj$   & {\tt p p > z j  } &$ 7.209 \pm 0.005\, \cdot 10^{3} \quad {}^{+ 19.3 \% }_{- 17.0 \% } \,\, {}^{+  1.2 \% }_{-  1.0 \% }$& $\,  9.742 \pm 0.035\, \cdot 10^{3} \quad {}^{+  5.8 \% }_{-  7.8 \% } \,\, {}^{+  1.2 \% }_{-  1.0 \% }$ \\
a.7 & $pp$ & $Zjj$  & {\tt p p > z j j } &$ 2.348 \pm 0.006\, \cdot 10^{3} \quad {}^{+ 24.3 \% }_{- 18.5 \% } \,\, {}^{+  0.6 \% }_{-  0.6 \% }$& $\,  2.665 \pm 0.010\, \cdot 10^{3} \quad {}^{+  2.5 \% }_{-  6.0 \% } \,\, {}^{+  0.7 \% }_{-  0.7 \% }$ \\
a.8 & $pp$ & $Zjjj$  & {\tt p p > z j j j} &$ 6.314 \pm 0.008\, \cdot 10^{2} \quad {}^{+ 40.8 \% }_{- 27.0 \% } \,\, {}^{+  0.5 \% }_{-  0.5 \% }$& $\,  6.996 \pm 0.028\, \cdot 10^{2} \quad {}^{+  1.1 \% }_{-  6.8 \% } \,\, {}^{+  0.5 \% }_{-  0.5 \% }$ \\
\midrule
a.9 & $pp$ & $\gamma j$  & {\tt p p > a j }  &$ 1.964 \pm 0.001\, \cdot 10^{4} \quad {}^{+ 31.2 \% }_{- 26.0 \% } \,\, {}^{+  1.7 \% }_{-  1.8 \% }$& $\,  5.218 \pm 0.025\, \cdot 10^{4} \quad {}^{+ 24.5 \% }_{- 21.4 \% } \,\, {}^{+  1.4 \% }_{-  1.6 \% }$ \\
a.10 & $pp$ & $\gamma jj$ & {\tt p p > a j j }  &$ 7.815 \pm 0.008\, \cdot 10^{3} \quad {}^{+ 32.8 \% }_{- 24.2 \% } \,\, {}^{+  0.9 \% }_{-  1.2 \% }$& $\,  1.004 \pm 0.004\, \cdot 10^{4} \quad {}^{+  5.9 \% }_{- 10.9 \% } \,\, {}^{+  0.8 \% }_{-  1.2 \% }$\\
\bottomrule
\end{tabular}
\end{small}
\end{center}
\caption{ \label{tab:results_vj} 
Sample of LO and NLO rates for vector-boson
production, possibly within cuts and in association with jets, at the 13-TeV 
LHC; we also report the integration errors, and the fractional scale
(left) and PDF (right) uncertainties.
Where relevant, the notation understands the sum of the $W^+$ and $W^-$ 
cross sections, and {\tt wpm} is a label that includes both $W^+$ and $W^-$, 
defined from the shell with {\tt define wpm = w+ w-}. All cross sections are 
calculated in the five-flavour scheme. Results at the NLO accuracy for
$W/Z$ plus jets are also available in {\sc MCFM} for up to two 
jets~\cite{Campbell:2002tg,Campbell:2003hd,Campbell:2013vha}, including 
heavy-flavour identification~\cite{Campbell:2008hh,Campbell:2003dd,
Campbell:2005zv,Campbell:2006cu,Caola:2011pz}, and in 
{\sc\small POWHEG}~\cite{Alioli:2008gx,Alioli:2010qp,Re:2012zi}.  
NLO cross sections for $W$ plus three jets have appeared in 
refs.~\cite{Ellis:2009zw,Melnikov:2009wh}.  The {\sc BlackHat+SHERPA} 
collaboration has provided samples and results for up to
$Z$ plus four jets and $W$ plus five jets at the NLO~\cite{Berger:2009zg,
Berger:2009ep,Berger:2010vm,Berger:2010zx,Ita:2011wn}. 
NLO+PS merged samples for $W$ plus up to three jets are also available 
in \Sherpa~\cite{Hoeche:2012ft}.
$\gamma$ plus up to three jets calculations have been presented in
refs.~\cite{Catani:2002ny,Bern:2011pa}. We do not show cross sections 
for EW-induced $V$ plus two jets processes with $V=\gamma,Z,W^\pm$, which are
available in {\sc VBFNLO}~\cite{Arnold:2011wj} and have been studied in 
ref.~\cite{Jager:2012xk}.
}
\end{table}


\begin{table}
\begin{center}
\begin{small}
\begin{tabular}{l r@{$\,\to\,$}l lll}
\toprule
\multicolumn{3}{c}{Process~~~~~~~~~~~~~~~~~}& Syntax  & \multicolumn{2}{c}{Cross section (pb)}\\
\multicolumn{3}{c}{Vector-boson pair +jets~~~~~~}& &  \multicolumn{1}{c}{ LO 13 TeV} &  \multicolumn{1}{c}{  NLO 13 TeV}\\
\midrule
b.1 & $pp$ & $W^+W^-$   (4f)   & {\tt p p > w+ w- } &$ 7.355 \pm 0.005\, \cdot 10^{1} \quad {}^{+  5.0 \% }_{-  6.1 \% } \,\, {}^{+  2.0 \% }_{-  1.5 \% }$& $\,  1.028 \pm 0.003\, \cdot 10^{2} \quad {}^{+  4.0 \% }_{-  4.5 \% } \,\, {}^{+  1.9 \% }_{-  1.4 \% }$\\
b.2 & $pp$ & $ZZ$      & {\tt p p > z z } &$ 1.097 \pm 0.002\, \cdot 10^{1} \quad {}^{+  4.5 \% }_{-  5.6 \% } \,\, {}^{+  1.9 \% }_{-  1.5 \% }$& $\,  1.415 \pm 0.005\, \cdot 10^{1} \quad {}^{+  3.1 \% }_{-  3.7 \% } \,\, {}^{+  1.8 \% }_{-  1.4 \% }$\\
b.3 & $pp$ & $ZW^\pm$    & {\tt p p >  z wpm } &$ 2.777 \pm 0.003\, \cdot 10^{1} \quad {}^{+  3.6 \% }_{-  4.7 \% } \,\, {}^{+  2.0 \% }_{-  1.5 \% }$& $\,  4.487 \pm 0.013\, \cdot 10^{1} \quad {}^{+  4.4 \% }_{-  4.4 \% } \,\, {}^{+  1.7 \% }_{-  1.3 \% }$\\
b.4 & $pp$ & $\gamma \gamma$ & {\tt p p > a a  } &$ 2.510 \pm 0.002\, \cdot 10^{1} \quad {}^{+ 22.1 \% }_{- 22.4 \% } \,\, {}^{+  2.4 \% }_{-  2.1 \% }$& $\,  6.593 \pm 0.021\, \cdot 10^{1} \quad {}^{+ 17.6 \% }_{- 18.8 \% } \,\, {}^{+  2.0 \% }_{-  1.9 \% }$\\
b.5 & $pp$ & $\gamma Z$      & {\tt p p > a z } &$ 2.523 \pm 0.004\, \cdot 10^{1} \quad {}^{+  9.9 \% }_{- 11.2 \% } \,\, {}^{+  2.0 \% }_{-  1.6 \% }$& $\,  3.695 \pm 0.013\, \cdot 10^{1} \quad {}^{+  5.4 \% }_{-  7.1 \% } \,\, {}^{+  1.8 \% }_{-  1.4 \% }$\\
b.6 & $pp$ & $\gamma W^\pm$  & {\tt p p > a wpm } &$ 2.954 \pm 0.005\, \cdot 10^{1} \quad {}^{+  9.5 \% }_{- 11.0 \% } \,\, {}^{+  2.0 \% }_{-  1.7 \% }$& $\,  7.124 \pm 0.026\, \cdot 10^{1} \quad {}^{+  9.7 \% }_{-  9.9 \% } \,\, {}^{+  1.5 \% }_{-  1.3 \% }$\\
\midrule
b.7 & $pp$ & $W^+W^-j$   (4f)   & {\tt p p > w+ w- j } &$ 2.865 \pm 0.003\, \cdot 10^{1} \quad {}^{+ 11.6 \% }_{- 10.0 \% } \,\, {}^{+  1.0 \% }_{-  0.8 \% }$& $\,  3.730 \pm 0.013\, \cdot 10^{1} \quad {}^{+  4.9 \% }_{-  4.9 \% } \,\, {}^{+  1.1 \% }_{-  0.8 \% }$\\
b.8 & $pp$ & $ZZj$      & {\tt p p > z z j} &$ 3.662 \pm 0.003\, \cdot 10^{0} \quad {}^{+ 10.9 \% }_{-  9.3 \% } \,\, {}^{+  1.0 \% }_{-  0.8 \% }$& $\,  4.830 \pm 0.016\, \cdot 10^{0} \quad {}^{+  5.0 \% }_{-  4.8 \% } \,\, {}^{+  1.1 \% }_{-  0.9 \% }$\\
b.9 & $pp$ & $ZW^\pm j $    & {\tt p p >  z wpm j } &$ 1.605 \pm 0.005\, \cdot 10^{1} \quad {}^{+ 11.6 \% }_{- 10.0 \% } \,\, {}^{+  0.9 \% }_{-  0.7 \% }$& $\,  2.086 \pm 0.007\, \cdot 10^{1} \quad {}^{+  4.9 \% }_{-  4.8 \% } \,\, {}^{+  0.9 \% }_{-  0.7 \% }$\\
b.10& $pp$ & $\gamma \gamma j $ & {\tt p p > a a  j } &$ 1.022 \pm 0.001\, \cdot 10^{1} \quad {}^{+ 20.3 \% }_{- 17.7 \% } \,\, {}^{+  1.2 \% }_{-  1.5 \% }$& $\,  2.292 \pm 0.010\, \cdot 10^{1} \quad {}^{+ 17.2 \% }_{- 15.1 \% } \,\, {}^{+  1.0 \% }_{-  1.4 \% }$\\
b.11${}^*$  & $pp$ & $\gamma Z j $      & {\tt p p > a z  j} &$ 8.310 \pm 0.017\, \cdot 10^{0} \quad {}^{+ 14.5 \% }_{- 12.8 \% } \,\, {}^{+  1.0 \% }_{-  1.0 \% }$& $\,  1.220 \pm 0.005\, \cdot 10^{1} \quad {}^{+  7.3 \% }_{-  7.4 \% } \,\, {}^{+  0.9 \% }_{-  0.9 \% }$\\
b.12${}^*$  & $pp$ & $\gamma W^\pm j $  & {\tt p p > a wpm  j} &$ 2.546 \pm 0.010\, \cdot 10^{1} \quad {}^{+ 13.7 \% }_{- 12.1 \% } \,\, {}^{+  0.9 \% }_{-  1.0 \% }$& $\,  3.713 \pm 0.015\, \cdot 10^{1} \quad {}^{+  7.2 \% }_{-  7.1 \% } \,\, {}^{+  0.9 \% }_{-  1.0 \% }$\\
\midrule
b.13 & $pp$ & $W^+W^+ jj$     & {\tt p p > w+ w+  j j} &$ 1.484 \pm 0.006\, \cdot 10^{-1} \quad {}^{+ 25.4 \% }_{- 18.9 \% } \,\, {}^{+  2.1 \% }_{-  1.5 \% }$& $\,  2.251 \pm 0.011\, \cdot 10^{-1} \quad {}^{+ 10.5 \% }_{- 10.6 \% } \,\, {}^{+  2.2 \% }_{-  1.6 \% }$\\
b.14 & $pp$ & $W^-W^- jj$     & {\tt p p > w- w-  j j} &$ 6.752 \pm 0.007\, \cdot 10^{-2} \quad {}^{+ 25.4 \% }_{- 18.9 \% } \,\, {}^{+  2.4 \% }_{-  1.7 \% }$& $\,  1.003 \pm 0.003\, \cdot 10^{-1} \quad {}^{+ 10.1 \% }_{- 10.4 \% } \,\, {}^{+  2.5 \% }_{-  1.8 \% }$\\
b.15 & $pp$ & $W^+W^-jj$  (4f)   & {\tt p p > w+ w-  j j} &$ 1.144 \pm 0.002\, \cdot 10^{1} \quad {}^{+ 27.2 \% }_{- 19.9 \% } \,\, {}^{+  0.7 \% }_{-  0.5 \% }$& $\,  1.396 \pm 0.005\, \cdot 10^{1} \quad {}^{+  5.0 \% }_{-  6.8 \% } \,\, {}^{+  0.7 \% }_{-  0.6 \% }$\\
b.16 & $pp$ & $ZZjj$      & {\tt p p > z z j j } &$ 1.344 \pm 0.002\, \cdot 10^{0} \quad {}^{+ 26.6 \% }_{- 19.6 \% } \,\, {}^{+  0.7 \% }_{-  0.6 \% }$& $\,  1.706 \pm 0.011\, \cdot 10^{0} \quad {}^{+  5.8 \% }_{-  7.2 \% } \,\, {}^{+  0.8 \% }_{-  0.6 \% }$\\
b.17 & $pp$ & $ZW^\pm jj$    & {\tt p p >  z wpm j j } &$ 8.038 \pm 0.009\, \cdot 10^{0} \quad {}^{+ 26.7 \% }_{- 19.7 \% } \,\, {}^{+  0.7 \% }_{-  0.5 \% }$& $\,  9.139 \pm 0.031\, \cdot 10^{0} \quad {}^{+  3.1 \% }_{-  5.1 \% } \,\, {}^{+  0.7 \% }_{-  0.5 \% }$\\
b.18 & $pp$ & $\gamma \gamma jj$ & {\tt p p > a a  j j } &$ 5.377 \pm 0.029\, \cdot 10^{0} \quad {}^{+ 26.2 \% }_{- 19.8 \% } \,\, {}^{+  0.6 \% }_{-  1.0 \% }$& $\,  7.501 \pm 0.032\, \cdot 10^{0} \quad {}^{+  8.8 \% }_{- 10.1 \% } \,\, {}^{+  0.6 \% }_{-  1.0 \% }$\\
b.19${}^*$ & $pp$ & $\gamma Z jj$      & {\tt p p > a z j j } &$ 3.260 \pm 0.009\, \cdot 10^{0} \quad {}^{+ 24.3 \% }_{- 18.4 \% } \,\, {}^{+  0.6 \% }_{-  0.6 \% }$& $\,  4.242 \pm 0.016\, \cdot 10^{0} \quad {}^{+  6.5 \% }_{-  7.3 \% } \,\, {}^{+  0.6 \% }_{-  0.6 \% }$\\
b.20${}^*$ & $pp$ & $\gamma W^\pm jj$  & {\tt p p > a wpm j j } &$ 1.233 \pm 0.002\, \cdot 10^{1} \quad {}^{+ 24.7 \% }_{- 18.6 \% } \,\, {}^{+  0.6 \% }_{-  0.6 \% }$& $\,  1.448 \pm 0.005\, \cdot 10^{1} \quad {}^{+  3.6 \% }_{-  5.4 \% } \,\, {}^{+  0.6 \% }_{-  0.7 \% }$\\
\bottomrule
\end{tabular}
\end{small}
\end{center}
\caption{\label{tab:results_vvj} 
Sample of LO and NLO rates for vector-boson pair production, possibly 
within cuts  and in association with jets, at the 13-TeV LHC; we also 
report the integration errors, and the fractional scale
(left) and PDF (right) uncertainties. See 
table~\ref{tab:results_vj} for the meaning of {\tt wpm}. All cross 
sections are calculated in the five-flavour scheme, except for
processes b.1, b.7, and b.15, which are obtained in the four-flavour 
scheme to avoid resonant-top contributions. NLO results for $VV$ production
have been known for some time~\cite{Mele:1990bq,Ohnemus:1990za,
Frixione:1992pj,Ohnemus:1991kk,Ohnemus:1991gb,Frixione:1993yp,
Campbell:1999ah,Dixon:1999di,DeFlorian:2000sg,Campbell:2011bn}, 
are publicly available in {\sc MCFM} and in {\sc VBFNLO}~\cite{Arnold:2011wj},
and are matched to parton showers in \mcatnlo~\cite{Frixione:2002ik}
and {\sc\small POWHEG}~\cite{Nason:2006hfa}.
NLO results for $VV$ with up to an extra jet have been
made available in {\sc POWHEG}~\cite{Melia:2011tj,Nason:2013ydw}.  
NLO corrections to $\gamma \gamma$ plus up to three jets are also 
known~\cite{DelDuca:2003uz,Gehrmann:2013bga,Badger:2013ava,Bern:2013bha,
Bern:2014vza}. Other available results are: 
$W^+W^-jj$~\cite{Greiner:2012im,Melia:2011dw},
$W^\pm W^\pm jj$~\cite{Melia:2010bm}, 
$W^\pm W^\pm jj$ (EW+QCD)~\cite{Campanario:2013gea}, 
$Z \gamma j$~\cite{Campbell:2012ft}, 
$W\gamma jj$~\cite{Campanario:2014dpa}, 
$WZjj$~\cite{Campanario:2013ysa}, 
$W\gamma j$~\cite{Campanario:2009um,Campanario:2010hv}, 
$WZj$~\cite{Campanario:2010hp}.
We do not show results for NLO corrections to EW-induced production of 
$VV$ plus two jets, such as $W^\pm W^\mp jj$~\cite{Jager:2013mu}, 
$WZ jj$~\cite{Schissler:2013nga}, and $ZZjj$~\cite{Jager:2013iza}, which 
can also be obtained with {\sc POWHEG} and {\sc VBFNLO}.  
}
\end{table}


\begin{table}
\begin{center}
\begin{small}
\begin{tabular}{l r@{$\,\to\,$}l lll}
\toprule
\multicolumn{3}{c}{Process~~~~~~~~~~~~~~~~~}& Syntax  & \multicolumn{2}{c}{Cross section (pb)}\\
\multicolumn{3}{c}{Three vector bosons +jet~}& &  \multicolumn{1}{c}{  LO 13 TeV}&  \multicolumn{1}{c}{  NLO 13 TeV}\\
\midrule
c.1 & $pp$ & $W^+ W^- W^\pm$ (4f)  & {\tt p p > w+ w- wpm  } &$ 1.307 \pm 0.003\, \cdot 10^{-1} \quad {}^{+  0.0 \% }_{-  0.3 \% } \,\, {}^{+  2.0 \% }_{-  1.5 \% }$& $\,  2.109 \pm 0.006\, \cdot 10^{-1} \quad {}^{+  5.1 \% }_{-  4.1 \% } \,\, {}^{+  1.6 \% }_{-  1.2 \% }$\\
c.2 & $pp$ & $Z W^+ W^-$ (4f)  & {\tt p p > z w+  w- } &$ 9.658 \pm 0.065\, \cdot 10^{-2} \quad {}^{+  0.8 \% }_{-  1.1 \% } \,\, {}^{+  2.1 \% }_{-  1.6 \% }$& $\,  1.679 \pm 0.005\, \cdot 10^{-1} \quad {}^{+  6.3 \% }_{-  5.1 \% } \,\, {}^{+  1.6 \% }_{-  1.2 \% }$\\
c.3 & $pp$ & $Z  Z W^\pm$  & {\tt p p > z z wpm  } &$ 2.996 \pm 0.016\, \cdot 10^{-2} \quad {}^{+  1.0 \% }_{-  1.4 \% } \,\, {}^{+  2.0 \% }_{-  1.6 \% }$& $\,  5.550 \pm 0.020\, \cdot 10^{-2} \quad {}^{+  6.8 \% }_{-  5.5 \% } \,\, {}^{+  1.5 \% }_{-  1.1 \% }$\\
c.4 & $pp$ & $Z  Z Z$      & {\tt p p > z z z } &$ 1.085 \pm 0.002\, \cdot 10^{-2} \quad {}^{+  0.0 \% }_{-  0.5 \% } \,\, {}^{+  1.9 \% }_{-  1.5 \% }$& $\,  1.417 \pm 0.005\, \cdot 10^{-2} \quad {}^{+  2.7 \% }_{-  2.1 \% } \,\, {}^{+  1.9 \% }_{-  1.5 \% }$\\
c.5 & $pp$ & $\gamma W^+ W^-$  (4f)    & {\tt p p > a w+ w- } &$ 1.427 \pm 0.011\, \cdot 10^{-1} \quad {}^{+  1.9 \% }_{-  2.6 \% } \,\, {}^{+  2.0 \% }_{-  1.5 \% }$& $\,  2.581 \pm 0.008\, \cdot 10^{-1} \quad {}^{+  5.4 \% }_{-  4.3 \% } \,\, {}^{+  1.4 \% }_{-  1.1 \% }$\\
c.6 & $pp$ & $\gamma \gamma W^\pm$    & {\tt p p > a a wpm } &$ 2.681 \pm 0.007\, \cdot 10^{-2} \quad {}^{+  4.4 \% }_{-  5.6 \% } \,\, {}^{+  1.9 \% }_{-  1.6 \% }$& $\,  8.251 \pm 0.032\, \cdot 10^{-2} \quad {}^{+  7.6 \% }_{-  7.0 \% } \,\, {}^{+  1.0 \% }_{-  1.0 \% }$\\
c.7 & $pp$ & $\gamma Z W^\pm$       & {\tt p p > a z wpm } &$ 4.994 \pm 0.011\, \cdot 10^{-2} \quad {}^{+  0.8 \% }_{-  1.4 \% } \,\, {}^{+  1.9 \% }_{-  1.6 \% }$& $\,  1.117 \pm 0.004\, \cdot 10^{-1} \quad {}^{+  7.2 \% }_{-  5.9 \% } \,\, {}^{+  1.2 \% }_{-  0.9 \% }$\\
c.8 & $pp$ & $\gamma Z Z$      & {\tt p p > a z z } &$ 2.320 \pm 0.005\, \cdot 10^{-2} \quad {}^{+  2.0 \% }_{-  2.9 \% } \,\, {}^{+  1.9 \% }_{-  1.5 \% }$& $\,  3.118 \pm 0.012\, \cdot 10^{-2} \quad {}^{+  2.8 \% }_{-  2.7 \% } \,\, {}^{+  1.8 \% }_{-  1.4 \% }$\\
c.9 & $pp$ & $\gamma \gamma Z$    & {\tt p p > a a z } &$ 3.078 \pm 0.007\, \cdot 10^{-2} \quad {}^{+  5.6 \% }_{-  6.8 \% } \,\, {}^{+  1.9 \% }_{-  1.6 \% }$& $\,  4.634 \pm 0.020\, \cdot 10^{-2} \quad {}^{+  4.5 \% }_{-  5.0 \% } \,\, {}^{+  1.7 \% }_{-  1.3 \% }$\\
c.10 & $pp$ & $\gamma \gamma \gamma$    & {\tt p p > a a a  } &$ 1.269 \pm 0.003\, \cdot 10^{-2} \quad {}^{+  9.8 \% }_{- 11.0 \% } \,\, {}^{+  2.0 \% }_{-  1.8 \% }$& $\,  3.441 \pm 0.012\, \cdot 10^{-2} \quad {}^{+ 11.8 \% }_{- 11.6 \% } \,\, {}^{+  1.4 \% }_{-  1.5 \% }$\\
\midrule
c.11 & $pp$ & $W^+ W^- W^\pm j$ (4f)  & {\tt p p > w+ w- wpm j } &$ 9.167 \pm 0.010\, \cdot 10^{-2} \quad {}^{+ 15.0 \% }_{- 12.2 \% } \,\, {}^{+  1.0 \% }_{-  0.7 \% }$& $\,  1.197 \pm 0.004\, \cdot 10^{-1} \quad {}^{+  5.2 \% }_{-  5.6 \% } \,\, {}^{+  1.0 \% }_{-  0.8 \% }$\\
c.12${}^*$ & $pp$ & $Z W^+ W^- j$ (4f)  & {\tt p p > z w+  w- j} &$ 8.340 \pm 0.010\, \cdot 10^{-2} \quad {}^{+ 15.6 \% }_{- 12.6 \% } \,\, {}^{+  1.0 \% }_{-  0.7 \% }$& $\,  1.066 \pm 0.003\, \cdot 10^{-1} \quad {}^{+  4.5 \% }_{-  5.3 \% } \,\, {}^{+  1.0 \% }_{-  0.7 \% }$\\
c.13${}^*$ & $pp$ & $Z  Z W^\pm j$  & {\tt p p > z z wpm j } &$ 2.810 \pm 0.004\, \cdot 10^{-2} \quad {}^{+ 16.1 \% }_{- 13.0 \% } \,\, {}^{+  1.0 \% }_{-  0.7 \% }$& $\,  3.660 \pm 0.013\, \cdot 10^{-2} \quad {}^{+  4.8 \% }_{-  5.6 \% } \,\, {}^{+  1.0 \% }_{-  0.7 \% }$\\
c.14${}^*$ & $pp$ & $Z  Z Z j$      & {\tt p p > z z z j} &$ 4.823 \pm 0.011\, \cdot 10^{-3} \quad {}^{+ 14.3 \% }_{- 11.8 \% } \,\, {}^{+  1.4 \% }_{-  1.0 \% }$& $\,  6.341 \pm 0.025\, \cdot 10^{-3} \quad {}^{+  4.9 \% }_{-  5.4 \% } \,\, {}^{+  1.4 \% }_{-  1.0 \% }$\\
c.15${}^*$ & $pp$ & $\gamma W^+ W^- j$  (4f)    & {\tt p p > a w+ w- j} &$ 1.182 \pm 0.004\, \cdot 10^{-1} \quad {}^{+ 13.4 \% }_{- 11.2 \% } \,\, {}^{+  0.8 \% }_{-  0.7 \% }$& $\,  1.233 \pm 0.004\, \cdot 10^{3} \quad {}^{+ 18.9 \% }_{- 19.9 \% } \,\, {}^{+  1.0 \% }_{-  1.5 \% }$\\
c.16 & $pp$ & $\gamma \gamma W^\pm j$    & {\tt p p > a a wpm j} &$ 4.107 \pm 0.015\, \cdot 10^{-2} \quad {}^{+ 11.8 \% }_{- 10.2 \% } \,\, {}^{+  0.6 \% }_{-  0.8 \% }$& $\,  5.807 \pm 0.023\, \cdot 10^{-2} \quad {}^{+  5.8 \% }_{-  5.5 \% } \,\, {}^{+  0.7 \% }_{-  0.7 \% }$\\
c.17${}^*$ & $pp$ & $\gamma Z W^\pm j$       & {\tt p p > a z wpm j} &$ 5.833 \pm 0.023\, \cdot 10^{-2} \quad {}^{+ 14.4 \% }_{- 12.0 \% } \,\, {}^{+  0.7 \% }_{-  0.6 \% }$& $\,  7.764 \pm 0.025\, \cdot 10^{-2} \quad {}^{+  5.1 \% }_{-  5.5 \% } \,\, {}^{+  0.8 \% }_{-  0.6 \% }$\\
c.18${}^*$ & $pp$ & $\gamma Z Z j$      & {\tt p p > a z z j} &$ 9.995 \pm 0.013\, \cdot 10^{-3} \quad {}^{+ 12.5 \% }_{- 10.6 \% } \,\, {}^{+  1.2 \% }_{-  0.9 \% }$& $\,  1.371 \pm 0.005\, \cdot 10^{-2} \quad {}^{+  5.6 \% }_{-  5.5 \% } \,\, {}^{+  1.2 \% }_{-  0.9 \% }$\\
c.19${}^*$ & $pp$ & $\gamma \gamma Z j$    & {\tt p p > a a z j} &$ 1.372 \pm 0.003\, \cdot 10^{-2} \quad {}^{+ 10.9 \% }_{-  9.4 \% } \,\, {}^{+  1.0 \% }_{-  0.9 \% }$& $\,  2.051 \pm 0.011\, \cdot 10^{-2} \quad {}^{+  7.0 \% }_{-  6.3 \% } \,\, {}^{+  1.0 \% }_{-  0.9 \% }$\\
c.20${}^*$ & $pp$ & $\gamma \gamma \gamma j$    & {\tt p p > a a a j } &$ 1.031 \pm 0.006\, \cdot 10^{-2} \quad {}^{+ 14.3 \% }_{- 12.6 \% } \,\, {}^{+  0.9 \% }_{-  1.2 \% }$& $\, 2.020 \pm 0.008\, \cdot 10^{-2} \quad {}^{+ 12.8 \% }_{- 11.0 \% } \,\, {}^{+  0.8 \% }_{-  1.2 \% }$\\
\bottomrule
\end{tabular}
\end{small}
\end{center}
\caption{\label{tab:results_multiv} 
Sample of LO and NLO rates for triple-vector-boson production, possibly 
within cuts  and in association with one jet, at the 13-TeV LHC; we also 
report the integration errors, and the fractional scale
(left) and PDF (right) uncertainties. See 
table~\ref{tab:results_vj} for the meaning of {\tt wpm}. All cross sections are
calculated in the five-flavour scheme, except for processes with at least 
two $W$ bosons, where the four-flavour scheme is adopted to avoid
resonant-top contributions.
Triple-vector-boson cross sections at the NLO have been computed in:
$Z \gamma \gamma $~\cite{Campbell:2012ft,Bozzi:2011en}, 
$\gamma \gamma W^\pm$~\cite{Bozzi:2011wwa}, 
$\gamma Z W^\pm$~\cite{Bozzi:2010sj}, 
$WW \gamma$ and $ZZ \gamma$~\cite{Bozzi:2009ig}, 
$ZWW$~\cite{Binoth:2008kt}, 
$ZZW$ and $WWW$~\cite{Campanario:2008yg,Binoth:2008kt}, 
$\gamma\gamma\gamma$~\cite{Campbell:2014yka,Mandal:2014vpa}, 
$ZZZ$~\cite{Lazopoulos:2007ix,Binoth:2008kt}.  
The complete set of triple-vector-boson cross sections 
at the NLO is also available in {\sc VBFNLO}~\cite{Arnold:2011wj}.
Except for $\gamma \gamma W^\pm j$ and $W^+ W^- W^\pm j$ that have 
appeared in ref.~\cite{Campanario:2011ud} and ref.~\cite{Hoeche:2014rya}
respectively, $VVVj$ cross sections at the NLO have been computed 
here for the first time.  
}
\end{table}


\begin{table}
\begin{center}
\begin{small}
\begin{tabular}{l r@{$\,\to\,$}l lll}
\toprule
\multicolumn{3}{c}{Process~~~~~~~~~~~~~~~~~}& Syntax  & \multicolumn{2}{c}{Cross section (pb)}\\
\multicolumn{3}{c}{Four vector bosons~~~~~~}& &  \multicolumn{1}{c}{  LO 13 TeV}&  \multicolumn{1}{c}{  NLO 13 TeV}\\
\midrule
c.21${}^*$ & $pp$ & $W^+ W^- W^+ W^-$ (4f) & {\tt p p > w+ w- w+ w- } &$ 5.721 \pm 0.014\, \cdot 10^{-4} \quad {}^{+  3.7 \% }_{-  3.5 \% } \,\, {}^{+  2.3 \% }_{-  1.7 \% }$& $\,  9.959 \pm 0.035\, \cdot 10^{-4} \quad {}^{+  7.4 \% }_{-  6.0 \% } \,\, {}^{+  1.7 \% }_{-  1.2 \% }$\\
c.22${}^*$ & $pp$ & $W^+ W^- W^\pm Z$ (4f) & {\tt p p > w+ w- wpm z } &$ 6.391 \pm 0.076\, \cdot 10^{-4} \quad {}^{+  4.4 \% }_{-  4.1 \% } \,\, {}^{+  2.4 \% }_{-  1.8 \% }$& $\,  1.188 \pm 0.004\, \cdot 10^{-3} \quad {}^{+  8.4 \% }_{-  6.8 \% } \,\, {}^{+  1.7 \% }_{-  1.2 \% }$\\
c.23${}^*$ & $pp$ & $W^+ W^- W^\pm \gamma$ (4f) & {\tt p p > w+ w- wpm a } &$ 8.115 \pm 0.064\, \cdot 10^{-4} \quad {}^{+  2.5 \% }_{-  2.5 \% } \,\, {}^{+  2.2 \% }_{-  1.7 \% }$& $\,  1.546 \pm 0.005\, \cdot 10^{-3} \quad {}^{+  7.9 \% }_{-  6.3 \% } \,\, {}^{+  1.5 \% }_{-  1.1 \% }$\\
c.24${}^*$ & $pp$ & $W^+ W^- Z Z$ (4f) & {\tt p p > w+ w- z z } &$ 4.320 \pm 0.013\, \cdot 10^{-4} \quad {}^{+  4.4 \% }_{-  4.1 \% } \,\, {}^{+  2.4 \% }_{-  1.7 \% }$& $\,  7.107 \pm 0.020\, \cdot 10^{-4} \quad {}^{+  7.0 \% }_{-  5.7 \% } \,\, {}^{+  1.8 \% }_{-  1.3 \% }$\\
c.25${}^*$ & $pp$ & $W^+ W^- Z \gamma$ (4f) & {\tt p p > w+ w- z a } &$ 8.403 \pm 0.016\, \cdot 10^{-4} \quad {}^{+  3.0 \% }_{-  2.9 \% } \,\, {}^{+  2.3 \% }_{-  1.7 \% }$& $\,  1.483 \pm 0.004\, \cdot 10^{-3} \quad {}^{+  7.2 \% }_{-  5.8 \% } \,\, {}^{+  1.6 \% }_{-  1.2 \% }$\\
c.26${}^*$ & $pp$ & $W^+ W^- \gamma \gamma$ (4f) & {\tt p p > w+ w- a a } &$ 5.198 \pm 0.012\, \cdot 10^{-4} \quad {}^{+  0.6 \% }_{-  0.9 \% } \,\, {}^{+  2.1 \% }_{-  1.6 \% }$& $\,  9.381 \pm 0.032\, \cdot 10^{-4} \quad {}^{+  6.7 \% }_{-  5.3 \% } \,\, {}^{+  1.4 \% }_{-  1.1 \% }$\\
c.27${}^*$ & $pp$ & $W^\pm Z Z Z$ & {\tt p p > wpm z z z } &$ 5.862 \pm 0.010\, \cdot 10^{-5} \quad {}^{+  5.1 \% }_{-  4.7 \% } \,\, {}^{+  2.4 \% }_{-  1.8 \% }$& $\,  1.240 \pm 0.004\, \cdot 10^{-4} \quad {}^{+  9.9 \% }_{-  8.0 \% } \,\, {}^{+  1.7 \% }_{-  1.2 \% }$\\
c.28${}^*$ & $pp$ & $W^\pm Z Z \gamma $ & {\tt p p > wpm z z a } &$ 1.148 \pm 0.003\, \cdot 10^{-4} \quad {}^{+  3.6 \% }_{-  3.5 \% } \,\, {}^{+  2.2 \% }_{-  1.7 \% }$& $\,  2.945 \pm 0.008\, \cdot 10^{-4} \quad {}^{+ 10.8 \% }_{-  8.7 \% } \,\, {}^{+  1.3 \% }_{-  1.0 \% }$\\
c.29${}^*$ & $pp$ & $W^\pm Z \gamma \gamma$ & {\tt p p > wpm z a a } &$ 1.054 \pm 0.004\, \cdot 10^{-4} \quad {}^{+  1.7 \% }_{-  1.9 \% } \,\, {}^{+  2.1 \% }_{-  1.7 \% }$& $\,  3.033 \pm 0.010\, \cdot 10^{-4} \quad {}^{+ 10.6 \% }_{-  8.6 \% } \,\, {}^{+  1.1 \% }_{-  0.8 \% }$\\
c.30${}^*$ & $pp$ & $W^\pm \gamma \gamma \gamma$ & {\tt p p > wpm a a a } &$ 3.600 \pm 0.013\, \cdot 10^{-5} \quad {}^{+  0.4 \% }_{-  1.0 \% } \,\, {}^{+  2.0 \% }_{-  1.6 \% }$& $\,  1.246 \pm 0.005\, \cdot 10^{-4} \quad {}^{+  9.8 \% }_{-  8.1 \% } \,\, {}^{+  0.9 \% }_{-  0.8 \% }$\\
c.31${}^*$ & $pp$ & $Z Z Z Z$ & {\tt p p > z z z z } &$ 1.989 \pm 0.002\, \cdot 10^{-5} \quad {}^{+  3.8 \% }_{-  3.6 \% } \,\, {}^{+  2.2 \% }_{-  1.7 \% }$& $\,  2.629 \pm 0.008\, \cdot 10^{-5} \quad {}^{+  3.5 \% }_{-  3.0 \% } \,\, {}^{+  2.2 \% }_{-  1.7 \% }$\\
c.32${}^*$ & $pp$ & $Z Z Z \gamma$ & {\tt p p > z z z a } &$ 3.945 \pm 0.007\, \cdot 10^{-5} \quad {}^{+  1.9 \% }_{-  2.1 \% } \,\, {}^{+  2.1 \% }_{-  1.6 \% }$& $\,  5.224 \pm 0.016\, \cdot 10^{-5} \quad {}^{+  3.3 \% }_{-  2.7 \% } \,\, {}^{+  2.1 \% }_{-  1.6 \% }$\\
c.33${}^*$ & $pp$ & $Z Z \gamma \gamma$ & {\tt p p > z z a a } &$ 5.513 \pm 0.017\, \cdot 10^{-5} \quad {}^{+  0.0 \% }_{-  0.3 \% } \,\, {}^{+  2.1 \% }_{-  1.6 \% }$& $\,  7.518 \pm 0.032\, \cdot 10^{-5} \quad {}^{+  3.4 \% }_{-  2.6 \% } \,\, {}^{+  2.0 \% }_{-  1.5 \% }$\\
c.34${}^*$ & $pp$ & $Z \gamma \gamma \gamma$ & {\tt p p > z a a a } &$ 4.790 \pm 0.012\, \cdot 10^{-5} \quad {}^{+  2.3 \% }_{-  3.1 \% } \,\, {}^{+  2.0 \% }_{-  1.6 \% }$& $\,  7.103 \pm 0.026\, \cdot 10^{-5} \quad {}^{+  3.4 \% }_{-  3.2 \% } \,\, {}^{+  1.6 \% }_{-  1.5 \% }$\\
c.35${}^*$ & $pp$ & $\gamma \gamma \gamma \gamma$ & {\tt p p > a a a a } &$ 1.594 \pm 0.004\, \cdot 10^{-5} \quad {}^{+  4.7 \% }_{-  5.7 \% } \,\, {}^{+  1.9 \% }_{-  1.7 \% }$& $\,  3.389 \pm 0.012\, \cdot 10^{-5} \quad {}^{+  7.0 \% }_{-  6.7 \% } \,\, {}^{+  1.3 \% }_{-  1.3 \% }$\\
\bottomrule
\end{tabular}
\end{small}
\end{center}
\caption{\label{tab:results_vvvv} 
Sample of LO and NLO rates for quadruple-vector-boson production, possibly 
within cuts, at the 13-TeV LHC; we also 
report the integration errors, and the fractional scale
(left) and PDF (right) uncertainties. See table~\ref{tab:results_vj} for the 
meaning of {\tt wpm}. All cross sections are calculated in the five-flavour 
scheme, except the processes with at least two $W$ bosons, where the 
four-flavour scheme is adopted to avoid resonant-top contributions.
For all processes in this table NLO QCD corrections have
never been computed before.  }
\end{table}


\begin{table}
\begin{center}
\begin{small}
\begin{tabular}{l r@{$\,\to\,$}l lll}
\toprule
\multicolumn{3}{c}{Process~~~~~~~~~~~~~~~~~}& Syntax  & \multicolumn{2}{c}{Cross section (pb)}\\
\multicolumn{3}{c}{Heavy quarks and jets~~~~~~~~~}& &  \multicolumn{1}{c}{  LO 13 TeV}&  \multicolumn{1}{c}{  NLO 13 TeV}\\
\midrule
d.1 & $pp$ & $jj$          & {\tt p p > j j   } &$ 1.162 \pm 0.001\, \cdot 10^{6} \quad {}^{+ 24.9 \% }_{- 18.8 \% } \,\, {}^{+  0.8 \% }_{-  0.9 \% }$& $\,  1.580 \pm 0.007\, \cdot 10^{6} \quad {}^{+  8.4 \% }_{-  9.0 \% } \,\, {}^{+  0.7 \% }_{-  0.9 \% }$\\
d.2 & $pp$ & $jjj$       & {\tt p p >  j j j   } &$ 8.940 \pm 0.021\, \cdot 10^{4} \quad {}^{+ 43.8 \% }_{- 28.4 \% } \,\, {}^{+  1.2 \% }_{-  1.4 \% }$& $\,  7.791 \pm 0.037\, \cdot 10^{4} \quad {}^{+  2.1 \% }_{- 23.2 \% } \,\, {}^{+  1.1 \% }_{-  1.3 \% }$\\
\midrule
d.3 & $pp$ & $b\bar b$ (4f)    & {\tt p p > b b$\sim$ } &$ 3.743 \pm 0.004\, \cdot 10^{3} \quad {}^{+ 25.2 \% }_{- 18.9 \% } \,\, {}^{+  1.5 \% }_{-  1.8 \% }$& $\,  6.438 \pm 0.028\, \cdot 10^{3} \quad {}^{+ 15.9 \% }_{- 13.3 \% } \,\, {}^{+  1.5 \% }_{-  1.7 \% }$\\
d.4${}^*$ & $pp$ & $b \bar b j $ (4f)  & {\tt p p >  b b$\sim$ j } &$ 1.050 \pm 0.002\, \cdot 10^{3} \quad {}^{+ 44.1 \% }_{- 28.5 \% } \,\, {}^{+  1.6 \% }_{-  1.8 \% }$& $\,  1.327 \pm 0.007\, \cdot 10^{3} \quad {}^{+  6.8 \% }_{- 11.6 \% } \,\, {}^{+  1.5 \% }_{-  1.8 \% }$\\
d.5${}^*$& $pp$ & $b\bar{b}jj$ (4f)  & {\tt p p > b b$\sim$  j j  } &$ 1.852 \pm 0.006\, \cdot 10^{2} \quad {}^{+ 61.8 \% }_{- 35.6 \% } \,\, {}^{+  2.1 \% }_{-  2.4 \% }$& $\,  2.471 \pm 0.012\, \cdot 10^{2} \quad {}^{+  8.2 \% }_{- 16.4 \% } \,\, {}^{+  2.0 \% }_{-  2.3 \% }$\\
d.6 & $pp$ & $b \bar b b \bar b$ (4f) & {\tt p p > b b$\sim$ b b$\sim$ } &$ 5.050 \pm 0.007\, \cdot 10^{-1} \quad {}^{+ 61.7 \% }_{- 35.6 \% } \,\, {}^{+  2.9 \% }_{-  3.4 \% }$& $\,  8.736 \pm 0.034\, \cdot 10^{-1} \quad {}^{+ 20.9 \% }_{- 22.0 \% } \,\, {}^{+  2.9 \% }_{-  3.4 \% }$\\
\midrule
d.7 & $pp$ & $t\bar{t}$          & {\tt p p > t t$\sim$  } &$ 4.584 \pm 0.003\, \cdot 10^{2} \quad {}^{+ 29.0 \% }_{- 21.1 \% } \,\, {}^{+  1.8 \% }_{-  2.0 \% }$& $\,  6.741 \pm 0.023\, \cdot 10^{2} \quad {}^{+  9.8 \% }_{- 10.9 \% } \,\, {}^{+  1.8 \% }_{-  2.1 \% }$\\
d.8 & $pp$ & $t\bar{t}j$         & {\tt p p > t t$\sim$  j } &$ 3.135 \pm 0.002\, \cdot 10^{2} \quad {}^{+ 45.1 \% }_{- 29.0 \% } \,\, {}^{+  2.2 \% }_{-  2.5 \% }$& $\,  4.106 \pm 0.015\, \cdot 10^{2} \quad {}^{+  8.1 \% }_{- 12.2 \% } \,\, {}^{+  2.1 \% }_{-  2.5 \% }$\\
d.9 & $pp$ & $t\bar{t}jj$   & {\tt p p > t t$\sim$ j j  } &$ 1.361 \pm 0.001\, \cdot 10^{2} \quad {}^{+ 61.4 \% }_{- 35.6 \% } \,\, {}^{+  2.6 \% }_{-  3.0 \% }$& $\,  1.795 \pm 0.006\, \cdot 10^{2} \quad {}^{+  9.3 \% }_{- 16.1 \% } \,\, {}^{+  2.4 \% }_{-  2.9 \% }$\\
d.10 & $pp$ & $t\bar{t}t \bar {t}$   & {\tt p p > t t$\sim$  t t$\sim$  } &$ 4.505 \pm 0.005\, \cdot 10^{-3} \quad {}^{+ 63.8 \% }_{- 36.5 \% } \,\, {}^{+  5.4 \% }_{-  5.7 \% }$& $\,  9.201 \pm 0.028\, \cdot 10^{-3} \quad {}^{+ 30.8 \% }_{- 25.6 \% } \,\, {}^{+  5.5 \% }_{-  5.9 \% }$\\
\midrule
d.11 & $pp$ & $t\bar{t} b \bar {b}$ (4f) & {\tt p p > t t$\sim$  b b$\sim$  }&$ 6.119 \pm 0.004\, \cdot 10^{0} \quad {}^{+ 62.1 \% }_{- 35.7 \% } \,\, {}^{+  2.9 \% }_{-  3.5 \% }$& $\,  1.452 \pm 0.005\, \cdot 10^{1} \quad {}^{+ 37.6 \% }_{- 27.5 \% } \,\, {}^{+  2.9 \% }_{-  3.5 \% }$\\
\bottomrule
\end{tabular}
\end{small}
\end{center}
\caption{\label{tab:results_qqj} 
Sample of LO and NLO total rates for the production of heavy
quarks and/or jets, possibly within cuts, at the 13-TeV LHC; we also 
report the integration errors, and the fractional scale
(left) and PDF (right) uncertainties. Processes 
d.1 and d.2, as well as processes involving at least a top pair, are
computed in the five-flavour scheme. Processes that explicitly involve 
$b$-quarks in the final state are calculated in the four-flavour scheme. 
For processes d.3--d.6 we require 2 (or 4) $b$-jets in the 
final state with $|\eta|<2.5$.  For processes d.1--d.6, we require the 
($b$)-jets to have $\pt>80$~GeV, with at least one of them with $\pt>100$~GeV.
Calculations of cross sections at the NLO for this
class of processes are available in the literature as well as in public codes:
from the seminal results for the hadroproduction of a heavy quark 
pair~\cite{Nason:1987xz,Beenakker:1988bq,Nason:1989zy,Beenakker:1990maa,
Mangano:1991jk}, to their NLO+PS implementation in 
\mcatnlo~\cite{Frixione:2003ei}
and {\sc POWHEG}~\cite{Frixione:2007nw}, to 
$t\bar tj$~\cite{Dittmaier:2007wz} (also including top 
decays~\cite{Melia:2011tj,Melnikov:2011qx} and parton shower 
effects~\cite{Alioli:2011as,Kardos:2011qa}), to the computation of 
$t\bt jj$~\cite{Bevilacqua:2010ve}. 
Merged NLO+PS results for $t\bar t$ plus jets are also 
available~\cite{Frederix:2012ps,Schonherr:2013bpa,Hoeche:2014qda}.  
NLO results for three jets~\cite{Nagy:2003tz}, four jets~\cite{Bern:2011ep},
and up to five jets~\cite{Badger:2012pf,Badger:2013yda} have been published.  
Two- and three-jet event generation is available in 
{\sc POWHEG}~\cite{Alioli:2010xa,Kardos:2014dua}.  
Calculations for $b \bar b b \bar b$~\cite{Greiner:2011mp,
Bevilacqua:2013taa}, $t \bar t b \bar b$~\cite{Bevilacqua:2009zn,
Bredenstein:2009aj,Bredenstein:2010rs,Kardos:2013vxa}, and 
$t \bar t t \bar t$~\cite{Bevilacqua:2012em} production 
have appeared in the literature.  
}
\end{table}


\begin{table}
\begin{center}
\begin{small}
\begin{tabular}{l r@{$\,\to\,$}l lll}
\toprule
\multicolumn{3}{c}{Process~~~~~~~~~~~~~~~~~~~}& Syntax  & \multicolumn{2}{c}{Cross section (pb)}\\
\multicolumn{3}{c}{Heavy quarks+vector bosons~~~}& &  \multicolumn{1}{c}{  LO 13 TeV}&  \multicolumn{1}{c}{  NLO 13 TeV}\\
\midrule
e.1 & $pp$ & $W^\pm \,b\bar{b}$ (4f)  & {\tt p p > wpm  b b$\sim$ } &$ 3.074 \pm 0.002\, \cdot 10^{2} \quad {}^{+ 42.3 \% }_{- 29.2 \% } \,\, {}^{+  2.0 \% }_{-  1.6 \% }$& $\,  8.162 \pm 0.034\, \cdot 10^{2} \quad {}^{+ 29.8 \% }_{- 23.6 \% } \,\, {}^{+  1.5 \% }_{-  1.2 \% }$\\
e.2 & $pp$ & $ Z \, b\bar{b}$ (4f)  & {\tt p p >  z b b$\sim$    } &$ 6.993 \pm 0.003\, \cdot 10^{2} \quad {}^{+ 33.5 \% }_{- 24.4 \% } \,\, {}^{+  1.0 \% }_{-  1.4 \% }$& $\,  1.235 \pm 0.004\, \cdot 10^{3} \quad {}^{+ 19.9 \% }_{- 17.4 \% } \,\, {}^{+  1.0 \% }_{-  1.4 \% }$\\
e.3 & $pp$ & $ \gamma \,b\bar{b}$ (4f)  & {\tt p p > a b b$\sim$  } &$ 1.731 \pm 0.001\, \cdot 10^{3} \quad {}^{+ 51.9 \% }_{- 34.8 \% } \,\, {}^{+  1.6 \% }_{-  2.1 \% }$& $\,  4.171 \pm 0.015\, \cdot 10^{3} \quad {}^{+ 33.7 \% }_{- 27.1 \% } \,\, {}^{+  1.4 \% }_{-  1.9 \% }$\\
\midrule
e.4${}^*$ & $pp$ & $ W^\pm\, b\bar{b} \,j $ (4f) & {\tt p p > wpm b b$\sim$ j } &$ 1.861 \pm 0.003\, \cdot 10^{2} \quad {}^{+ 42.5 \% }_{- 27.7 \% } \,\, {}^{+  0.7 \% }_{-  0.7 \% }$& $\,  3.957 \pm 0.013\, \cdot 10^{2} \quad {}^{+ 27.0 \% }_{- 21.0 \% } \,\, {}^{+  0.7 \% }_{-  0.6 \% }$\\
e.5${}^*$ & $pp$ & $  Z \,b\bar{b} \,j$ (4f) & {\tt p p > z b b$\sim$  j} &$ 1.604 \pm 0.001\, \cdot 10^{2} \quad {}^{+ 42.4 \% }_{- 27.6 \% } \,\, {}^{+  0.9 \% }_{-  1.1 \% }$& $\,  2.805 \pm 0.009\, \cdot 10^{2} \quad {}^{+ 21.0 \% }_{- 17.6 \% } \,\, {}^{+  0.8 \% }_{-  1.0 \% }$\\
e.6${}^*$ & $pp$ & $ \gamma \,b\bar{b} \,j$ (4f)     & {\tt p p > a b b$\sim$  j} &$ 7.812 \pm 0.017\, \cdot 10^{2} \quad {}^{+ 51.2 \% }_{- 32.0 \% } \,\, {}^{+  1.0 \% }_{-  1.5 \% }$& $\,  1.233 \pm 0.004\, \cdot 10^{3} \quad {}^{+ 18.9 \% }_{- 19.9 \% } \,\, {}^{+  1.0 \% }_{-  1.5 \% }$\\
\midrule
e.7 & $pp$ & $ t\bar{t}\, W^\pm $       & {\tt p p > t t$\sim$ wpm  } &$ 3.777 \pm 0.003\, \cdot 10^{-1} \quad {}^{+ 23.9 \% }_{- 18.0 \% } \,\, {}^{+  2.1 \% }_{-  1.6 \% }$& $ 5.662 \pm 0.021\, \cdot 10^{-1} \quad {}^{+ 11.2 \% }_{- 10.6 \% } \,\, {}^{+  1.7 \% }_{-  1.3 \% }$\\
e.8 & $pp$ & $ t\bar{t}\,  Z $  & {\tt p p > t t$\sim$ z } &$ 5.273 \pm 0.004\, \cdot 10^{-1} \quad {}^{+ 30.5 \% }_{- 21.8 \% } \,\, {}^{+  1.8 \% }_{-  2.1 \% }$& $\,  7.598 \pm 0.026\, \cdot 10^{-1} \quad {}^{+  9.7 \% }_{- 11.1 \% } \,\, {}^{+  1.9 \% }_{-  2.2 \% }$\\
e.9 & $pp$ & $ t\bar{t}\,  \gamma $        & {\tt p p > t t$\sim$  a } &$ 1.204 \pm 0.001\, \cdot 10^{0} \quad {}^{+ 29.6 \% }_{- 21.3 \% } \,\, {}^{+  1.6 \% }_{-  1.8 \% }$& $\,  1.744 \pm 0.005\, \cdot 10^{0} \quad {}^{+  9.8 \% }_{- 11.0 \% } \,\, {}^{+  1.7 \% }_{-  2.0 \% }$\\ 
\midrule
e.10${}^*$& $pp$ & $ t\bar{t}\, W^\pm j $       & {\tt p p > t t$\sim$ wpm j   } &$ 2.352 \pm 0.002\, \cdot 10^{-1} \quad {}^{+ 40.9 \% }_{- 27.1 \% } \,\, {}^{+  1.3 \% }_{-  1.0 \% }$& $\,  3.404 \pm 0.011\, \cdot 10^{-1} \quad {}^{+ 11.2 \% }_{- 14.0 \% } \,\, {}^{+  1.2 \% }_{-  0.9 \% }$\\ 
e.11${}^*$ & $pp$ & $ t\bar{t}\,  Z j $    & {\tt p p > t t$\sim$ z j  } &$ 3.953 \pm 0.004\, \cdot 10^{-1} \quad {}^{+ 46.2 \% }_{- 29.5 \% } \,\, {}^{+  2.7 \% }_{-  3.0 \% }$& $\,  5.074 \pm 0.016\, \cdot 10^{-1} \quad {}^{+  7.0 \% }_{- 12.3 \% } \,\, {}^{+  2.5 \% }_{-  2.9 \% }$\\
e.12${}^*$ & $pp$ & $ t\bar{t}\,  \gamma j $   & {\tt p p > t t$\sim$  a j  } &$ 8.726 \pm 0.010\, \cdot 10^{-1} \quad {}^{+ 45.4 \% }_{- 29.1 \% } \,\, {}^{+  2.3 \% }_{-  2.6 \% }$& $\,  1.135 \pm 0.004\, \cdot 10^{0} \quad {}^{+  7.5 \% }_{- 12.2 \% } \,\, {}^{+  2.2 \% }_{-  2.5 \% }$\\  
\midrule
e.13${}^*$ & $pp$ & $t\bar{t} \,W^- W^+  $ (4f)  & {\tt p p > t t$\sim$ w+ w- } &$ 6.675 \pm 0.006\, \cdot 10^{-3} \quad {}^{+ 30.9 \% }_{- 21.9 \% } \,\, {}^{+  2.1 \% }_{-  2.0 \% }$& $\,  9.904 \pm 0.026\, \cdot 10^{-3} \quad {}^{+ 10.9 \% }_{- 11.8 \% } \,\, {}^{+  2.1 \% }_{-  2.1 \% }$\\
e.14${}^*$ & $pp$ & $t\bar{t} \,W^\pm Z  $    & {\tt p p > t t$\sim$ wpm z } &$ 2.404 \pm 0.002\, \cdot 10^{-3} \quad {}^{+ 26.6 \% }_{- 19.6 \% } \,\, {}^{+  2.5 \% }_{-  1.8 \% }$& $\,  3.525 \pm 0.010\, \cdot 10^{-3} \quad {}^{+ 10.6 \% }_{- 10.8 \% } \,\, {}^{+  2.3 \% }_{-  1.6 \% }$\\
e.15${}^*$ & $pp$ & $t\bar{t} \,W^\pm \gamma  $  & {\tt p p > t t$\sim$ wpm a  } &$ 2.718 \pm 0.003\, \cdot 10^{-3} \quad {}^{+ 25.4 \% }_{- 18.9 \% } \,\, {}^{+  2.3 \% }_{-  1.8 \% }$& $\,  3.927 \pm 0.013\, \cdot 10^{-3} \quad {}^{+ 10.3 \% }_{- 10.4 \% } \,\, {}^{+  2.0 \% }_{-  1.5 \% }$\\
e.16${}^*$ & $pp$ & $ t\bar{t} \,Z Z  $       & {\tt p p > t t$\sim$  z z } &$ 1.349 \pm 0.014\, \cdot 10^{-3} \quad {}^{+ 29.3 \% }_{- 21.1 \% } \,\, {}^{+  1.7 \% }_{-  1.5 \% }$& $\,  1.840 \pm 0.007\, \cdot 10^{-3} \quad {}^{+  7.9 \% }_{-  9.9 \% } \,\, {}^{+  1.7 \% }_{-  1.5 \% }$\\
e.17${}^*$ & $pp$ & $ t\bar{t} \,Z \gamma  $  & {\tt p p > t t$\sim$ z a  } &$ 2.548 \pm 0.003\, \cdot 10^{-3} \quad {}^{+ 30.1 \% }_{- 21.5 \% } \,\, {}^{+  1.7 \% }_{-  1.6 \% }$& $\,  3.656 \pm 0.012\, \cdot 10^{-3} \quad {}^{+  9.7 \% }_{- 11.0 \% } \,\, {}^{+  1.8 \% }_{-  1.9 \% }$\\
e.18${}^*$ & $pp$ & $t\bar{t} \, \gamma \gamma $  & {\tt p p > t t$\sim$  a a } &$ 3.272 \pm 0.006\, \cdot 10^{-3} \quad {}^{+ 28.4 \% }_{- 20.6 \% } \,\, {}^{+  1.3 \% }_{-  1.1 \% }$& $\,  4.402 \pm 0.015\, \cdot 10^{-3} \quad {}^{+  7.8 \% }_{-  9.7 \% } \,\, {}^{+  1.4 \% }_{-  1.4 \% }$\\
\bottomrule
\end{tabular}
\end{small}
\end{center}
\caption{\label{tab:results_tv} 
Sample of LO and NLO total rates for the production of heavy
quarks in association with vector bosons, possibly within cuts and
in association with jets, at the 13-TeV LHC; we also 
report the integration errors, and the fractional scale
(left) and PDF (right) uncertainties. Processes that explicitly 
involve $b$-quarks in the final state, and process e.13, are calculated 
in the four-flavour scheme, while all of the others are in the five-flavour 
scheme. Results are available in the literature for 
$Wb\bb$~\cite{Ellis:1998fv,Badger:2010mg,Hirschi:2011pa,Frederix:2011qg,
Oleari:2011ey}, $Zb\bb$~\cite{Campbell:2000bg,Hirschi:2011pa,Frederix:2011qg}, 
$t\bt\gamma$~\cite{Melnikov:2011ta}, $t\bt Z$~\cite{Lazopoulos:2008de,
Hirschi:2011pa,Garzelli:2011is,Kardos:2011na,Garzelli:2012bn}, 
$t\bt W$~\cite{Hirschi:2011pa,Campbell:2012dh,Garzelli:2012bn} production. 
For the majority of the processes in this table, NLO corrections are 
calculated in this work for the first time.
}
\end{table}


\begin{table}
\begin{center}
\begin{small}
\begin{tabular}{l r@{$\,\to\,$}l lll}
\toprule
\multicolumn{3}{c}{Process~~~~~~~~~~~~~~~~~~~}& Syntax  & \multicolumn{2}{c}{Cross section (pb)}\\
\multicolumn{3}{c}{Single-top~~~~~~~~~~~}& &  \multicolumn{1}{c}{  LO 13 TeV}&  \multicolumn{1}{c}{  NLO 13 TeV}\\
\midrule
f.1 & $pp$ & $t j$ (t-channel)    & {\tt p p > tt  j \$\$ w+ w-} &$ 1.520 \pm 0.001\, \cdot 10^{2} \quad {}^{+  9.4 \% }_{- 11.9 \% } \,\, {}^{+  0.4 \% }_{-  0.6 \% }$& $\,  1.563 \pm 0.005\, \cdot 10^{2} \quad {}^{+  1.4 \% }_{-  1.8 \% } \,\, {}^{+  0.4 \% }_{-  0.6 \% }$\\
f.2 & $pp$ & $t \gamma j $ (t-channel)    & {\tt p p >  tt  a j \$\$ w+ w-} &$ 9.956 \pm 0.014\, \cdot 10^{-1} \quad {}^{+  6.4 \% }_{-  8.8 \% } \,\, {}^{+  0.9 \% }_{-  1.0 \% }$& $\,  1.017 \pm 0.003\, \cdot 10^{0} \quad {}^{+  1.3 \% }_{-  1.2 \% } \,\, {}^{+  0.8 \% }_{-  0.9 \% }$\\
f.3 & $pp$ & $t  Z j $  (t-channel)   & {\tt p p > tt z j  \$\$ w+ w- } &$ 6.967 \pm 0.007\, \cdot 10^{-1} \quad {}^{+  3.5 \% }_{-  5.5 \% } \,\, {}^{+  0.9 \% }_{-  1.0 \% }$& $\,  6.993 \pm 0.021\, \cdot 10^{-1} \quad {}^{+  1.6 \% }_{-  1.1 \% } \,\, {}^{+  0.9 \% }_{-  1.0 \% }$\\
f.4 & $pp$ & $t b j$ ($t$-channel, 4f)    & {\tt p p > tt  bb j \$\$ w+ w-} &$ 1.003 \pm 0.000\, \cdot 10^{2} \quad {}^{+ 13.8 \% }_{- 11.5 \% } \,\, {}^{+  0.4 \% }_{-  0.5 \% }$& $\,  1.319 \pm 0.003\, \cdot 10^{2} \quad {}^{+  5.8 \% }_{-  5.2 \% } \,\, {}^{+  0.4 \% }_{-  0.5 \% }$\\
f.5${}^*$ & $pp$ & $t b j \gamma$ ($t$-channel, 4f)    & {\tt p p > tt  bb j a \$\$ w+ w-} &$ 6.293 \pm 0.006\, \cdot 10^{-1} \quad {}^{+ 16.8 \% }_{- 13.5 \% } \,\, {}^{+  0.8 \% }_{-  0.9 \% }$& $\, 8.612 \pm 0.025\, \cdot 10^{-1} \quad {}^{+  6.2 \% }_{-  6.6 \% } \,\, {}^{+  0.8 \% }_{-  0.9 \% }$\\
f.6${}^*$ & $pp$ & $t b j Z $ ($t$-channel, 4f)    & {\tt p p > tt  bb j z \$\$ w+ w-} &$ 3.934 \pm 0.002\, \cdot 10^{-1} \quad {}^{+ 18.7 \% }_{- 14.7 \% } \,\, {}^{+  1.0 \% }_{-  0.9 \% }$& $\,  5.657 \pm 0.014\, \cdot 10^{-1} \quad {}^{+  7.7 \% }_{-  7.9 \% } \,\, {}^{+  0.9 \% }_{-  0.9 \% }$\\
\midrule
f.7 & $pp$ & $t b $ ($s$-channel, 4f)    & {\scriptsize\tt p p > w+ > t b$\sim$, p p > w- > t$\sim$ b} &$ 7.489 \pm 0.007\, \cdot 10^{0} \quad {}^{+  3.5 \% }_{-  4.4 \% } \,\, {}^{+  1.9 \% }_{-  1.4 \% }$& $\,  1.001 \pm 0.004\, \cdot 10^{1} \quad {}^{+  3.7 \% }_{-  3.9 \% } \,\, {}^{+  1.9 \% }_{-  1.5 \% }$ \\
f.8${}^*$ & $pp$ & $t b \gamma   $ ($s$-channel, 4f)    & {\scriptsize\tt p p > w+ > t b$\sim$ a, p p > w- > t$\sim$ b a} &$ 1.490 \pm 0.001\, \cdot 10^{-2} \quad {}^{+  1.2 \% }_{-  1.8 \% } \,\, {}^{+  1.9 \% }_{-  1.5 \% }$& $\,  1.952 \pm 0.007\, \cdot 10^{-2} \quad {}^{+  2.6 \% }_{-  2.3 \% } \,\, {}^{+  1.7 \% }_{-  1.4 \% }$\\
f.9${}^*$ & $pp$ & $t  b Z  $  ($s$-channel, 4f)   & {\scriptsize\tt p p > w+ > t b$\sim$ z, p p > w- > t$\sim$ b z} &$ 1.072 \pm 0.001\, \cdot 10^{-2} \quad {}^{+  1.3 \% }_{-  1.5 \% } \,\, {}^{+  2.0 \% }_{-  1.6 \% }$& $\,  1.539 \pm 0.005\, \cdot 10^{-2} \quad {}^{+  3.9 \% }_{-  3.2 \% } \,\, {}^{+  1.9 \% }_{-  1.5 \% }$\\
\bottomrule
\end{tabular}
\end{small}
\end{center}
\caption{\label{tab:results_stop} 
Sample of LO and NLO total rates for the production of a single top,
possibly in association and within cuts, at the 13-TeV LHC; we also 
report the integration errors, and the fractional scale
(left) and PDF (right) uncertainties. 
The notation understands the sum of the $t$ and $\bt$ cross sections
for all processes, and {\tt tt} is a label that includes both $t$ and $\bt$, 
defined from the shell with {\tt define tt = t t\~{}} (and analogously 
for the label {\tt bb}). Processes that explicitly involve
$b$-quarks in the final state are calculated in the four-flavour scheme, while
all of the others are in the five-flavour scheme. Being an EW-induced 
process, single-top production requires special care for the 
\aNLO\ generation syntax: ${\tt \$\$}$ means excluding particles
in the $s$-channel, while the {\tt > w+ > } ({\tt > w- > }) forces a $W^+$
($W^-$) to be present in the $s$-channel (see appendix~\ref{sec:verbose}).
Total NLO cross sections for $t$- and $s$-channel single-top production 
have been known for some time~\cite{Stelzer:1995mi,Stelzer:1997ns}.  
All single-top channels are also available in MCFM~\cite{Campbell:2004ch,
Campbell:2009gj,Campbell:2009ss,Campbell:2005bb}, 
\mcatnlo~\cite{Frixione:2005vw,Frixione:2008yi}, and
{\sc POWHEG}~\cite{Alioli:2009je,Re:2010bp}. An NLO calculation for 
$tZj$ production has appeared in ref.~\cite{Campbell:2013yla}. 
}
\end{table}


\begin{table}
\begin{center}
\begin{small}
\begin{tabular}{l r@{$\,\to\,$}l lll}
\toprule
\multicolumn{3}{c}{Process~~~~~~~~~~~~~~~~~~~}& Syntax  & \multicolumn{2}{c}{Cross section (pb)}\\
\multicolumn{3}{c}{Single Higgs production~~~~~~~~~~~}& &  \multicolumn{1}{c}{  LO 13 TeV}&  \multicolumn{1}{c}{  NLO 13 TeV}\\
\midrule
g.1 & $pp$ & $H$   (HEFT)         & {\tt p p > h } &  $ 1.593 \pm 0.003\, \cdot 10^{1} \quad {}^{+ 34.8 \% }_{- 26.0 \% } \,\, {}^{+  1.2 \% }_{-  1.7 \% }$ & $ 3.261 \pm 0.010\, \cdot 10^{1} \quad {}^{+ 20.2 \% }_{- 17.9 \% } \,\, {}^{+  1.1 \% }_{-  1.6 \% }$\\
g.2 & $pp$ & $Hj$      (HEFT)     & {\tt p p > h j } &  $ 8.367 \pm 0.003\, \cdot 10^{0} \quad {}^{+ 39.4 \% }_{- 26.4 \% } \,\, {}^{+  1.2 \% }_{-  1.4 \% }$ & $ 1.422 \pm 0.006\, \cdot 10^{1} \quad {}^{+ 18.5 \% }_{- 16.6 \% } \,\, {}^{+  1.1 \% }_{-  1.4 \% }$\\
g.3 & $pp$ & $Hjj$   (HEFT)       & {\tt p p > h j j} & $ 3.020 \pm 0.002\, \cdot 10^{0} \quad {}^{+ 59.1 \% }_{- 34.7 \% } \,\, {}^{+  1.4 \% }_{-  1.7 \% }$ & $ 5.124 \pm 0.020\, \cdot 10^{0} \quad {}^{+ 20.7 \% }_{- 21.0 \% } \,\, {}^{+  1.3 \% }_{-  1.5 \% } $\\
\midrule
g.4 & $pp$ & $Hjj$ (VBF)         & {\tt p p > h j j \$\$ w+ w- z} & $ 1.987 \pm 0.002\, \cdot 10^{0} \quad {}^{+  1.7 \% }_{-  2.0 \% } \,\, {}^{+  1.9 \% }_{-  1.4 \% }$ & $ 1.900 \pm 0.006\, \cdot 10^{0} \quad {}^{+  0.8 \% }_{-  0.9 \% } \,\, {}^{+  2.0 \% }_{-  1.5 \% }$ \\
g.5 & $pp$ & $Hjjj$ (VBF)        & {\tt p p > h j j j \$\$ w+ w- z} &  $ 2.824 \pm 0.005\, \cdot 10^{-1} \quad {}^{+ 15.7 \% }_{- 12.7 \% } \,\, {}^{+  1.5 \% }_{-  1.0 \% }$ & $ 3.085 \pm 0.010\, \cdot 10^{-1} \quad {}^{+  2.0 \% }_{-  3.0 \% } \,\, {}^{+  1.5 \% }_{-  1.1 \% }  $\\
\midrule
g.6 & $pp$ & $HW^\pm$         & {\tt p p > h wpm  } &  $ 1.195 \pm 0.002\, \cdot 10^{0} \quad {}^{+  3.5 \% }_{-  4.5 \% } \,\, {}^{+  1.9 \% }_{-  1.5 \% }$ & $ 1.419 \pm 0.005\, \cdot 10^{0} \quad {}^{+  2.1 \% }_{-  2.6 \% } \,\, {}^{+  1.9 \% }_{-  1.4 \% } $\\
g.7 & $pp$ & $HW^\pm\,j$      & {\tt p p > h wpm j } & $ 4.018 \pm 0.003\, \cdot 10^{-1} \quad {}^{+ 10.7 \% }_{-  9.3 \% } \,\, {}^{+  1.2 \% }_{-  0.9 \% }$ & $ 4.842 \pm 0.017\, \cdot 10^{-1} \quad {}^{+  3.6 \% }_{-  3.7 \% } \,\, {}^{+  1.2 \% }_{-  1.0 \% } $\\
g.8${}^*$ & $pp$ & $HW^\pm\,jj$     & {\tt p p > h wpm j j  } &  $ 1.198 \pm 0.016\, \cdot 10^{-1} \quad {}^{+ 26.1 \% }_{- 19.4 \% } \,\, {}^{+  0.8 \% }_{-  0.6 \% }$ & $ 1.574 \pm 0.014\, \cdot 10^{-1} \quad {}^{+  5.0 \% }_{-  6.5 \% } \,\, {}^{+  0.9 \% }_{-  0.6 \% }$\\
\midrule
g.9 & $pp$ & $HZ$       & {\tt p p > h z   } &  $ 6.468 \pm 0.008\, \cdot 10^{-1} \quad {}^{+  3.5 \% }_{-  4.5 \% } \,\, {}^{+  1.9 \% }_{-  1.4 \% }$ & $ 7.674 \pm 0.027\, \cdot 10^{-1} \quad {}^{+  2.0 \% }_{-  2.5 \% } \,\, {}^{+  1.9 \% }_{-  1.4 \% } $ \\
g.10 & $pp$ & $HZ\,j$    & {\tt p p > h z j  } &  $ 2.225 \pm 0.001\, \cdot 10^{-1} \quad {}^{+ 10.6 \% }_{-  9.2 \% } \,\, {}^{+  1.1 \% }_{-  0.8 \% }$ & $ 2.667 \pm 0.010\, \cdot 10^{-1} \quad {}^{+  3.5 \% }_{-  3.6 \% } \,\, {}^{+  1.1 \% }_{-  0.9 \% } $\\
g.11${}^*$ & $pp$ & $HZ\,j j$  & {\tt p p > h z j j  } & $ 7.262 \pm 0.012\, \cdot 10^{-2} \quad {}^{+ 26.2 \% }_{- 19.4 \% } \,\, {}^{+  0.7 \% }_{-  0.6 \% }$ & $  8.753 \pm 0.037\, \cdot 10^{-2} \quad {}^{+  4.8 \% }_{-  6.3 \% } \,\, {}^{+  0.7 \% }_{-  0.6 \% }  $\\
\midrule
g.12${}^*$ & $pp$ & $HW^+W^-$ (4f)   & {\tt p p > h w+ w- } &  $ 8.325 \pm 0.139\, \cdot 10^{-3} \quad {}^{+  0.0 \% }_{-  0.3 \% } \,\, {}^{+  2.0 \% }_{-  1.6 \% }$ & $ 1.065 \pm 0.003\, \cdot 10^{-2} \quad {}^{+  2.5 \% }_{-  1.9 \% } \,\, {}^{+  2.0 \% }_{-  1.5 \% }  $\\
g.13${}^*$ & $pp$ & $HW^\pm\gamma$    & {\tt p p > h wpm a } & $ 2.518 \pm 0.006\, \cdot 10^{-3} \quad {}^{+  0.7 \% }_{-  1.4 \% } \,\, {}^{+  1.9 \% }_{-  1.5 \% }$ & $ 3.309 \pm 0.011\, \cdot 10^{-3} \quad {}^{+  2.7 \% }_{-  2.0 \% } \,\, {}^{+  1.7 \% }_{-  1.4 \% }  $\\
g.14${}^*$ & $pp$ & $HZW^\pm$    & {\tt p p > h z wpm  } &  $ 3.763 \pm 0.007\, \cdot 10^{-3} \quad {}^{+  1.1 \% }_{-  1.5 \% } \,\, {}^{+  2.0 \% }_{-  1.6 \% }$ & $ 5.292 \pm 0.015\, \cdot 10^{-3} \quad {}^{+  3.9 \% }_{-  3.1 \% } \,\, {}^{+  1.8 \% }_{-  1.4 \% }  $\\
g.15${}^*$ & $pp$ & $HZZ$       & {\tt p p > h z  z } &  $ 2.093 \pm 0.003\, \cdot 10^{-3} \quad {}^{+  0.1 \% }_{-  0.6 \% } \,\, {}^{+  1.9 \% }_{-  1.5 \% }$ & $ 2.538 \pm 0.007\, \cdot 10^{-3} \quad {}^{+  1.9 \% }_{-  1.4 \% } \,\, {}^{+  2.0 \% }_{-  1.5 \% } $\\
\midrule
g.16 & $pp$ & $Ht\bar{t}$   & {\tt p p > h t t$\sim$  } &  $ 3.579 \pm 0.003\, \cdot 10^{-1} \quad {}^{+ 30.0 \% }_{- 21.5 \% } \,\, {}^{+  1.7 \% }_{-  2.0 \% }$ & $ 4.608 \pm 0.016\, \cdot 10^{-1} \quad {}^{+  5.7 \% }_{-  9.0 \% } \,\, {}^{+  2.0 \% }_{-  2.3 \% }  $\\
g.17 & $pp$ & $Htj $    & {\tt p p > h tt j   } &  $ 4.994 \pm 0.005\, \cdot 10^{-2} \quad {}^{+  2.4 \% }_{-  4.2 \% } \,\, {}^{+  1.2 \% }_{-  1.3 \% }$ & $ 6.328 \pm 0.022\, \cdot 10^{-2} \quad {}^{+  2.9 \% }_{-  1.8 \% } \,\, {}^{+  1.5 \% }_{-  1.6 \% } $\\
g.18 & $pp$ & $Hb\bar{b}$ (4f) & {\tt p p > h b b$\sim$} &$ 4.983 \pm 0.002\, \cdot 10^{-1} \quad {}^{+ 28.1 \% }_{- 21.0 \% } \,\, {}^{+  1.5 \% }_{-  1.8 \% }$& $ 6.085 \pm 0.026\, \cdot 10^{-1} \quad {}^{+  7.3 \% }_{-  9.6 \% } \,\, {}^{+  1.6 \% }_{-  2.0 \% }$\\
\midrule
g.19 & $pp$ & $Ht\bar{t}j$   & {\tt p p > h t t$\sim$ j } &$ 2.674 \pm 0.041\, \cdot 10^{-1} \quad {}^{+ 45.6 \% }_{- 29.2 \% } \,\, {}^{+  2.6 \% }_{-  2.9 \% }$& $ 3.244 \pm 0.025\, \cdot 10^{-1} \quad {}^{+  3.5 \% }_{-  8.7 \% } \,\, {}^{+  2.5 \% }_{-  2.9 \% }  $\\
g.20${}^*$ & $pp$ & $Hb\bar{b}j$ (4f) & {\tt p p > h b b$\sim$ j } &$ 7.367 \pm 0.002\, \cdot 10^{-2} \quad {}^{+ 45.6 \% }_{- 29.1 \% } \,\, {}^{+  1.8 \% }_{-  2.1 \% }$& $ 9.034 \pm 0.032\, \cdot 10^{-2} \quad {}^{+  7.9 \% }_{- 11.0 \% } \,\, {}^{+  1.8 \% }_{-  2.2 \% }$\\
\bottomrule
\end{tabular}
\end{small}
\end{center}
\caption{\label{tab:results_h} 
Sample of LO and NLO total rates for the production of a single SM Higgs,
possibly in association and within cuts, at the 13-TeV LHC; we also 
report the integration errors, and the fractional scale
(left) and PDF (right) uncertainties. 
See table~\ref{tab:results_vj} for the meaning of {\tt wpm}, and
table~\ref{tab:results_stop}  for the meaning of {\tt tt}, {\tt bb}, 
and the generation syntax. Processes that explicitly involve $b$-quarks 
in the final state are calculated in the four-flavour scheme, while all of
the others are in the five-flavour scheme, except for g.12.  
A complete set of references relevant to NLO rates for Higgs production
can be found in refs.~\cite{Dittmaier:2011ti,Dittmaier:2012vm,
Heinemeyer:2013tqa,Cullen:2013saa,vanDeurzen:2013xla}. 
The $W$-boson width is set equal to 2.0476 GeV 
for process g.17. Cross sections at the NLO for $HVjj$ and $HVV$ 
production appear in this work for the first time. 
}
\end{table}
  \vfill\newpage


\begin{table}
\begin{center}
\begin{small}
\begin{tabular}{l r@{$\,\to\,$}l lll}
\toprule
\multicolumn{3}{c}{Process~~~~~~~~~~~~~~~~~~~}& Syntax  & \multicolumn{2}{c}{Cross section (pb)}\\
\multicolumn{3}{c}{Higgs pair production~~~~~~~~~~~}& &  \multicolumn{1}{c}{  LO 13 TeV}&  \multicolumn{1}{c}{  NLO 13 TeV}\\
\midrule
h.1 & $pp$ & $HH$   (Loop improved)    & {\tt p p > h h   } &  $ 1.772 \pm 0.006 \, \cdot 10^{-2} \quad {}^{+  29.5 \% }_{-  21.4 \% } \,\, {}^{+  2.1 \% }_{-  2.6 \% }  $ & $ 2.763 \pm 0.008 \, \cdot 10^{-2} \quad {}^{+  11.4 \% }_{-  11.8 \% } \,\, {}^{+  2.1 \% }_{-  2.6 \% }  $\\
h.2 & $pp$ & $HHjj$ (VBF)       & {\tt p p > h h j j \$\$ w+ w- z } &  $ 6.503 \pm 0.019\, \cdot 10^{-4} \quad {}^{+  7.2 \% }_{-  6.4 \% } \,\, {}^{+  2.3 \% }_{-  1.6 \% }$ & $ 6.820 \pm 0.026\, \cdot 10^{-4} \quad {}^{+  0.8 \% }_{-  1.0 \% } \,\, {}^{+  2.4 \% }_{-  1.7 \% }  $\\
h.3 & $pp$ & $HHW^\pm$        & {\tt p p > h h wpm  } &  $ 4.303 \pm 0.005\, \cdot 10^{-4} \quad {}^{+  0.9 \% }_{-  1.3 \% } \,\, {}^{+  2.0 \% }_{-  1.5 \% }$ & $ 5.002 \pm 0.014\, \cdot 10^{-4} \quad {}^{+  1.5 \% }_{-  1.2 \% } \,\, {}^{+  2.0 \% }_{-  1.6 \% }$\\
h.4${}^*$  & $pp$ & $HHW^\pm j$        & {\tt p p > h h wpm j } & $ 1.922 \pm 0.002\, \cdot 10^{-4} \quad {}^{+ 14.2 \% }_{- 11.7 \% } \,\, {}^{+  1.5 \% }_{-  1.1 \% }$  & $ 2.218 \pm 0.009\, \cdot 10^{-4} \quad {}^{+  2.7 \% }_{-  3.3 \% } \,\, {}^{+  1.6 \% }_{-  1.1 \% }$  \\
h.5${}^*$  & $pp$ & $HHW^\pm\gamma$         & {\tt p p > h h wpm a  } &   $ 1.952 \pm 0.004\, \cdot 10^{-6} \quad {}^{+  3.0 \% }_{-  3.0 \% } \,\, {}^{+  2.2 \% }_{-  1.6 \% }$  &$ 2.347 \pm 0.007\, \cdot 10^{-6} \quad {}^{+  2.4 \% }_{-  2.0 \% } \,\, {}^{+  2.1 \% }_{-  1.6 \% }$  \\
h.6 & $pp$ & $HHZ$         & {\tt p p > h h z   } &  $ 2.701 \pm 0.007\, \cdot 10^{-4} \quad {}^{+  0.9 \% }_{-  1.3 \% } \,\, {}^{+  2.0 \% }_{-  1.5 \% }$  &  $ 3.130 \pm 0.008\, \cdot 10^{-4} \quad {}^{+  1.6 \% }_{-  1.2 \% } \,\, {}^{+  2.0 \% }_{-  1.5 \% }$\\
h.7${}^*$  & $pp$ & $HHZj$         & {\tt p p > h h z j  } &   $ 1.211 \pm 0.001\, \cdot 10^{-4} \quad {}^{+ 14.1 \% }_{- 11.7 \% } \,\, {}^{+  1.4 \% }_{-  1.1 \% }$  &  $ 1.394 \pm 0.006\, \cdot 10^{-4} \quad {}^{+  2.7 \% }_{-  3.2 \% } \,\, {}^{+  1.5 \% }_{-  1.1 \% }$ \\
h.8${}^*$  & $pp$ & $HHZ\gamma$         & {\tt p p > h h z a  } & $ 1.397 \pm 0.003\, \cdot 10^{-6} \quad {}^{+  2.4 \% }_{-  2.5 \% } \,\, {}^{+  2.2 \% }_{-  1.7 \% }$   & $ 1.604 \pm 0.005\, \cdot 10^{-6} \quad {}^{+  1.7 \% }_{-  1.4 \% } \,\, {}^{+  2.3 \% }_{-  1.7 \% }$ \\
h.9${}^*$  & $pp$ & $HHZZ$         & {\tt p p > h h z z  } &   $ 2.309 \pm 0.005\, \cdot 10^{-6} \quad {}^{+  3.9 \% }_{-  3.8 \% } \,\, {}^{+  2.2 \% }_{-  1.7 \% }$  &  $ 2.754 \pm 0.009\, \cdot 10^{-6} \quad {}^{+  2.3 \% }_{-  2.0 \% } \,\, {}^{+  2.3 \% }_{-  1.7 \% }$ \\
h.10${}^*$ & $pp$ & $HHZW^\pm$         & {\tt p p > h h z wpm  } & $ 3.708 \pm 0.013\, \cdot 10^{-6} \quad {}^{+  4.8 \% }_{-  4.5 \% } \,\, {}^{+  2.3 \% }_{-  1.7 \% }$  &  $ 4.904 \pm 0.029\, \cdot 10^{-6} \quad {}^{+  3.7 \% }_{-  3.2 \% } \,\, {}^{+  2.2 \% }_{-  1.6 \% }$ \\
h.11${}^*$ & $pp$ & $HHW^+ W^-$ (4f)        & {\tt p p > h h w+ w-  } &  $ 7.524 \pm 0.070\, \cdot 10^{-6} \quad {}^{+  3.5 \% }_{-  3.4 \% } \,\, {}^{+  2.3 \% }_{-  1.7 \% }$   &  $ 9.268 \pm 0.030\, \cdot 10^{-6} \quad {}^{+  2.3 \% }_{-  2.1 \% } \,\, {}^{+  2.3 \% }_{-  1.7 \% }$ \\
h.12& $pp$ & $HHt\bar{t}$ & {\tt p p > h h t t$\sim$  } &  $ 6.756 \pm 0.007\, \cdot 10^{-4} \quad {}^{+ 30.2 \% }_{- 21.6 \% } \,\, {}^{+  1.8 \% }_{-  1.8 \% }$ & $  7.301 \pm 0.024\, \cdot 10^{-4} \quad {}^{+  1.4 \% }_{-  5.7 \% } \,\, {}^{+  2.2 \% }_{-  2.3 \% }  $\\
h.13& $pp$ & $HHtj$& {\tt p p > h h tt j   } &  $ 1.844 \pm 0.008\, \cdot 10^{-5} \quad {}^{+  0.0 \% }_{-  0.6 \% } \,\, {}^{+  1.8 \% }_{-  1.8 \% }$ & $  2.444 \pm 0.009\, \cdot 10^{-5} \quad {}^{+  4.5 \% }_{-  3.1 \% } \,\, {}^{+  2.8 \% }_{-  3.0 \% }  $\\
h.14${}^*$& $pp$ & $HHb\bar{b}$ & {\tt p p > h h b b$\sim$  } & $ 7.849 \pm 0.022\, \cdot 10^{-8} \quad {}^{+ 34.3 \% }_{- 23.9 \% } \,\, {}^{+  3.1\%}_{-  3.7 \% }$  & $ 1.084 \pm 0.012\, \cdot 10^{-7} \quad {}^{+  7.4 \% }_{- 10.8 \% } \,\, {}^{+  3.1 \% }_{-  3.7 \% }$  \\
\bottomrule
\end{tabular}
\end{small}
\end{center}
\caption{\label{tab:results_hh} 
Sample of LO and NLO total rates for Higgs-pair production,
possibly in association and within cuts, at the 13-TeV LHC; we also 
report the integration errors, and the fractional scale
(left) and PDF (right) uncertainties. 
See table~\ref{tab:results_vj} for the meaning of {\tt wpm}, and
table~\ref{tab:results_stop}  for the meaning of {\tt tt}. 
All cross sections are calculated in the five-flavour
scheme, except for process h.11 which is obtained in the four-flavour 
scheme to avoid resonant-top contributions. Processes h.1, h.2, h.3, h.6, 
h.12, and h.13 have appeared in ref.~\cite{Frederix:2014hta} as NLO+PS
results; some of these were already known at the NLO~\cite{Baglio:2012np}. 
The $W$-boson width is set equal to 2.0476 GeV for process h.13. 
Previous to the release of \aNLO, the only available public code for
Higgs pair production was {\sc HPAIR}~\cite{Plehn:1996wb,Dawson:1998py},
relevant to process h.1 (see ref.~\cite{Frederix:2014hta} for more details
on the different approach adopted by \aNLO). Process h.2 has been recently 
added to {\sc\small VBFNLO}.
}
\end{table}




\begin{table}
\begin{center}
\begin{small}
\begin{tabular}{l r@{$\,\to\,$}l lll}
\toprule
\multicolumn{3}{c}{Process~~~~~~~~~~~~~~~~~~~}& Syntax  & \multicolumn{2}{c}{Cross section (pb)}\\
\multicolumn{3}{c}{Heavy quarks and jets~~~~~~~~~}& &  \multicolumn{1}{c}{ LO 1 TeV} &  \multicolumn{1}{c}{ NLO 1 TeV}\\
\midrule
i.1 & $e^+e^-$ & $jj$  & {\tt e+ e- > j j } &$ 6.223 \pm 0.005\, \cdot 10^{-1} \quad {}^{+  0.0 \% }_{-  0.0 \% }$& $ 6.389 \pm 0.013\, \cdot 10^{-1} \quad {}^{+  0.2 \% }_{-  0.2 \% }$\\
i.2 & $e^+e^-$ & $jjj$    & {\tt e+ e- > j j j}  & $3.401 \pm 0.002\, \cdot 10^{-1} \quad {}^{+  9.6 \% }_{-  8.0 \% }$&$ 3.166 \pm 0.019\, \cdot 10^{-1} \quad {}^{+  0.2 \% }_{-  2.1 \% }$ \\
i.3 & $e^+e^-$ & $jjjj$   & {\tt e+ e- > j j j j}  & $1.047 \pm 0.001\, \cdot 10^{-1} \quad {}^{+ 20.0 \% }_{- 15.3 \% }$ &$ 1.090 \pm 0.006\, \cdot 10^{-1} \quad {}^{+  0.0 \% }_{-  2.8 \% }$\\
i.4 & $e^+e^-$ & $jjjjj$   & {\tt e+ e- > j j j j j}  & $2.211 \pm 0.006\, \cdot 10^{-2} \quad {}^{+ 31.4 \% }_{- 22.0 \% }$ &$2.771 \pm 0.021\, \cdot 10^{-2} \quad {}^{+  4.4 \% }_{-  8.6 \% }$\\
\midrule
i.5 & $e^+e^-$ & $t\bar t$     & {\tt e+ e- > t t$\sim$ }  &$ 1.662 \pm 0.002\, \cdot 10^{-1} \quad {}^{+  0.0 \% }_{-  0.0 \% }$& $1.745 \pm 0.006\, \cdot 10^{-1} \quad {}^{+  0.4 \% }_{-  0.4 \% }$ \\
i.6 & $e^+e^-$ & $t\bar tj$     & {\tt e+ e- > t t$\sim$ j }  &$ 4.813 \pm 0.005\, \cdot 10^{-2} \quad {}^{+  9.3 \% }_{-  7.8 \% }$& $5.276 \pm 0.022\, \cdot 10^{-2} \quad {}^{+  1.3 \% }_{-  2.1 \% }$ \\
i.7${}^*$ & $e^+e^-$ & $t\bar tjj$     & {\tt e+ e- > t t$\sim$ j j}  &$ 8.614 \pm 0.009\, \cdot 10^{-3} \quad {}^{+ 19.4 \% }_{- 15.0 \% }$& $1.094 \pm 0.005\, \cdot 10^{-2} \quad {}^{+  5.0 \% }_{-  6.3 \% }$ \\
i.8${}^*$ & $e^+e^-$ & $t\bar tjjj$     & {\tt e+ e- > t t$\sim$ j j j}  &$1.044 \pm 0.002\, \cdot 10^{-3} \quad {}^{+ 30.5 \% }_{- 21.6 \% }$& $1.546 \pm 0.010\, \cdot 10^{-3} \quad {}^{+ 10.6 \% }_{- 11.6 \% }$ \\
i.9${}^*$ & $e^+e^-$ & $t\bar tt\bar t$     & {\tt e+ e- > t t$\sim$ t t$\sim$}  &$6.456 \pm 0.016\, \cdot 10^{-7} \quad {}^{+ 19.1 \% }_{- 14.8 \% }$& $1.221 \pm 0.005\, \cdot 10^{-6} \quad {}^{+ 13.2 \% }_{- 11.2 \% }$ \\
i.10${}^*$ & $e^+e^-$ & $t\bar tt\bar t j$     & {\tt e+ e- > t t$\sim$ t t$\sim$ j}  &$2.719 \pm 0.005\, \cdot 10^{-8} \quad {}^{+ 29.9 \% }_{- 21.3 \% }$& $5.338 \pm 0.027\, \cdot 10^{-8} \quad {}^{+ 18.3 \% }_{- 15.4 \% }$ \\
\midrule
i.11 & $e^+e^-$ & $b\bar b$ (4f)  & {\tt e+ e- > b b$\sim$ }  &$9.198 \pm 0.004\, \cdot 10^{-2} \quad {}^{+  0.0 \% }_{-  0.0 \% }$& $9.282 \pm 0.031\, \cdot 10^{-2} \quad {}^{+  0.0 \% }_{-  0.0 \% }$ \\
i.12 & $e^+e^-$ & $b\bar bj$ (4f)   & {\tt e+ e- > b b$\sim$ j }  &$5.029 \pm 0.003\, \cdot 10^{-2} \quad {}^{+  9.5 \% }_{-  8.0 \% }$& $4.826 \pm 0.026\, \cdot 10^{-2} \quad {}^{+  0.5 \% }_{-  2.5 \% }$ \\
i.13${}^*$ & $e^+e^-$ & $b\bar bjj$ (4f)     & {\tt e+ e- > b b$\sim$ j j}  &$1.621 \pm 0.001\, \cdot 10^{-2} \quad {}^{+ 20.0 \% }_{- 15.3 \% }$& $1.817 \pm 0.009\, \cdot 10^{-2} \quad {}^{+  0.0 \% }_{-  3.1 \% }$ \\
i.14${}^*$ & $e^+e^-$ & $b\bar bjjj$ (4f)     & {\tt e+ e- > b b$\sim$ j j j}  &$3.641 \pm 0.009\, \cdot 10^{-3} \quad {}^{+ 31.4 \% }_{- 22.1 \% }$& $4.936 \pm 0.038\, \cdot 10^{-3} \quad {}^{+  4.8 \% }_{-  8.9 \% }$ \\
i.15${}^*$ & $e^+e^-$ & $b\bar bb\bar b$ (4f)     & {\tt e+ e- > b b$\sim$ b b$\sim$}  &$1.644 \pm 0.003\, \cdot 10^{-4} \quad {}^{+ 19.9 \% }_{- 15.3 \% }$& $3.601 \pm 0.017\, \cdot 10^{-4} \quad {}^{+ 15.2 \% }_{- 12.5 \% }$ \\
i.16${}^*$ & $e^+e^-$ & $b\bar bb\bar b j$ (4f)     & {\tt e+ e- > b b$\sim$ b b$\sim$ j}  &$7.660 \pm 0.022\, \cdot 10^{-5} \quad {}^{+ 31.3 \% }_{- 22.0 \% }$& $1.537 \pm 0.011\, \cdot 10^{-4} \quad {}^{+ 17.9 \% }_{- 15.3 \% }$ \\
\midrule
i.17${}^*$ & $e^+e^-$ & $t\bar tb\bar b$ (4f)     & {\tt e+ e- > t t$\sim$ b b$\sim$ }  &$1.819 \pm 0.003\, \cdot 10^{-4} \quad {}^{+ 19.5 \% }_{- 15.0 \% }$& $2.923 \pm 0.011\, \cdot 10^{-4} \quad {}^{+  9.2 \% }_{-  8.9 \% }$ \\
i.18${}^*$ & $e^+e^-$ & $t\bar tb\bar bj$ (4f)     & {\tt e+ e- > t t$\sim$ b b$\sim$ j}  &$4.045 \pm 0.011\, \cdot 10^{-5} \quad {}^{+ 30.5 \% }_{- 21.6 \% }$& $7.049 \pm 0.052\, \cdot 10^{-5} \quad {}^{+ 13.7 \% }_{- 13.1 \% }$ \\
\bottomrule
\end{tabular}
\end{small}
\end{center}
\caption{ \label{tab:results_eej} 
Sample of LO and NLO rates for the production of light jets in
association with heavy quarks, possibly within cuts, at a 1-TeV $e^+e^-$
collider; we also report the integration errors, and the fractional scale
uncertainties. Cross sections for processes i.1--i.10 are calculated in the
five-flavour scheme. For processes i.11--i.18 we use the four-flavour
scheme, and require the presence of at least two (four in i.15--i.16)
$b$-jets in the final state. $b$-jets are clustered with the same parameters
as light jets. Results at NLO accuracy for up to seven light jets can be 
found in refs.~\cite{Ellis:1980wv,Signer:1996bf,Signer:1997dm,Nagy:1997mf,
Nagy:1997yn,Frederix:2010ne,Becker:2011vg},
and for a heavy-quark-pair plus up to one jet in refs.~\cite{Bilenky:1994ad,
Schmidt:1995mr,Oleari:1997az,Nason:1997nw,Bernreuther:1997jn,
Brandenburg:1997pu,Brandenburg:1999gm}. All other processes are computed 
here for the first time at the NLO.
}
\end{table}


\begin{table}
\begin{center}
\begin{small}
\begin{tabular}{l r@{$\,\to\,$}l lll}
\toprule
\multicolumn{3}{c}{Process~~~~~~~~~~~~~~~~~}& Syntax  & \multicolumn{2}{c}{Cross section (pb)}\\
\multicolumn{3}{c}{Top quarks +bosons~~~~~~}& &  \multicolumn{1}{c}{ LO 1 TeV} &  \multicolumn{1}{c}{  NLO 1 TeV}\\
\midrule
j.1 & $e^+e^-$ & $t\bar t H$     & {\tt e+ e- > t t$\sim$ h}  &$2.018 \pm 0.003\, \cdot 10^{-3} \quad {}^{+  0.0 \% }_{-  0.0 \% }$& $1.911 \pm 0.006\, \cdot 10^{-3} \quad {}^{+  0.4 \% }_{-  0.5 \% }$ \\
j.2${}^*$ & $e^+e^-$ & $t\bar t H j$     & {\tt e+ e- > t t$\sim$ h j}  &$2.533 \pm 0.003\, \cdot 10^{-4} \quad {}^{+  9.2 \% }_{-  7.8 \% }$& $2.658 \pm 0.009\, \cdot 10^{-4} \quad {}^{+  0.5 \% }_{-  1.5 \% }$ \\
j.3${}^*$ & $e^+e^-$ & $t\bar t H jj$     & {\tt e+ e- > t t$\sim$ h j j}  &$2.663 \pm 0.004\, \cdot 10^{-5} \quad {}^{+ 19.3 \% }_{- 14.9 \% }$& $3.278 \pm 0.017\, \cdot 10^{-5} \quad {}^{+  4.0 \% }_{-  5.7 \% }$ \\
j.4${}^*$ & $e^+e^-$ & $t\bar t \gamma$     & {\tt e+ e- > t t$\sim$ a}  &$1.270 \pm 0.002\, \cdot 10^{-2} \quad {}^{+  0.0 \% }_{-  0.0 \% }$& $1.335 \pm 0.004\, \cdot 10^{-2} \quad {}^{+  0.5 \% }_{-  0.4 \% }$ \\
j.5${}^*$ & $e^+e^-$ & $t\bar t \gamma j$     & {\tt e+ e- > t t$\sim$ a j}  &$2.355 \pm 0.002\, \cdot 10^{-3} \quad {}^{+  9.3 \% }_{-  7.9 \% }$& $2.617 \pm 0.010\, \cdot 10^{-3} \quad {}^{+  1.6 \% }_{-  2.4 \% }$ \\
j.6${}^*$ & $e^+e^-$ & $t\bar t \gamma jj$     & {\tt e+ e- > t t$\sim$ a j j}  &$3.103 \pm 0.005\, \cdot 10^{-4} \quad {}^{+ 19.5 \% }_{- 15.0 \% }$& $4.002 \pm 0.021\, \cdot 10^{-4} \quad {}^{+  5.4 \% }_{-  6.6 \% }$ \\
j.7${}^*$ & $e^+e^-$ & $t\bar t Z$     & {\tt e+ e- > t t$\sim$ z}  &$4.642 \pm 0.006\, \cdot 10^{-3} \quad {}^{+  0.0 \% }_{-  0.0 \% }$& $4.949 \pm 0.014\, \cdot 10^{-3} \quad {}^{+  0.6 \% }_{-  0.5 \% }$ \\
j.8${}^*$ & $e^+e^-$ & $t\bar t Z j$     & {\tt e+ e- > t t$\sim$ z j}  &$6.059 \pm 0.006\, \cdot 10^{-4} \quad {}^{+  9.3 \% }_{-  7.8 \% }$& $6.940 \pm 0.028\, \cdot 10^{-4} \quad {}^{+  2.0 \% }_{-  2.6 \% }$ \\
j.9${}^*$ & $e^+e^-$ & $t\bar t Z jj$     & {\tt e+ e- > t t$\sim$ z j j}  &$6.351 \pm 0.028\, \cdot 10^{-5} \quad {}^{+ 19.4 \% }_{- 15.0 \% }$&$8.439 \pm 0.051\, \cdot 10^{-5} \quad {}^{+  5.8 \% }_{-  6.8 \% }$ \\
j.10${}^*$ & $e^+e^-$ & $t\bar t W^\pm jj$     & {\tt e+ e- > t t$\sim$ wpm j j}  &$2.400 \pm 0.004\, \cdot 10^{-7} \quad {}^{+ 19.3 \% }_{- 14.9 \% }$&$3.723 \pm 0.012\, \cdot 10^{-7} \quad {}^{+  9.6 \% }_{-  9.1 \% }$ \\

\midrule
j.11${}^*$ & $e^+e^-$ & $t\bar tHZ$     & {\tt e+ e- > t t$\sim$ h z}  &$3.600 \pm 0.006\, \cdot 10^{-5} \quad {}^{+  0.0 \% }_{-  0.0 \% }$& $3.579 \pm 0.013\, \cdot 10^{-5} \quad {}^{+  0.1 \% }_{-  0.0 \% }$ \\
j.12${}^*$ & $e^+e^-$ & $t\bar t\gamma Z$     & {\tt e+ e- > t t$\sim$ a z}  &$2.212 \pm 0.003\, \cdot 10^{-4} \quad {}^{+  0.0 \% }_{-  0.0 \% }$& $2.364 \pm 0.006\, \cdot 10^{-4} \quad {}^{+  0.6 \% }_{-  0.5 \% }$ \\
j.13${}^*$ & $e^+e^-$ & $t\bar t\gamma H$     & {\tt e+ e- > t t$\sim$ a h}  &$9.756 \pm 0.016\, \cdot 10^{-5} \quad {}^{+  0.0 \% }_{-  0.0 \% }$& $9.423 \pm 0.032\, \cdot 10^{-5} \quad {}^{+  0.3 \% }_{-  0.4 \% }$ \\
j.14${}^*$ & $e^+e^-$ & $t\bar t\gamma \gamma$     & {\tt e+ e- > t t$\sim$ a a}  &$3.650 \pm 0.008\, \cdot 10^{-4} \quad {}^{+  0.0 \% }_{-  0.0 \% }$& $3.833 \pm 0.013\, \cdot 10^{-4} \quad {}^{+  0.4 \% }_{-  0.4 \% }$ \\
j.15${}^*$ & $e^+e^-$ & $t\bar tZZ$     & {\tt e+ e- > t t$\sim$ z z}  &$3.788 \pm 0.004\, \cdot 10^{-5} \quad {}^{+  0.0 \% }_{-  0.0 \% }$& $4.007 \pm 0.013\, \cdot 10^{-5} \quad {}^{+  0.5 \% }_{-  0.5 \% } $ \\
j.16${}^*$ & $e^+e^-$ & $t\bar tHH$     & {\tt e+ e- > t t$\sim$ h h}  &$1.358 \pm 0.001\, \cdot 10^{-5} \quad {}^{+  0.0 \% }_{-  0.0 \% }$& $1.206 \pm 0.003\, \cdot 10^{-5} \quad {}^{+  0.9 \% }_{-  1.1 \% }$ \\
j.17${}^*$ & $e^+e^-$ & $t\bar tW^+W^-$     & {\tt e+ e- > t t$\sim$ w+ w-}  &$1.372 \pm 0.003\, \cdot 10^{-4} \quad {}^{+  0.0 \% }_{-  0.0 \% }$& $1.540 \pm 0.006\, \cdot 10^{-4} \quad {}^{+  1.0 \% }_{-  0.9 \% }$ \\
\bottomrule
\end{tabular}
\end{small}
\end{center}
\caption{
\label{tab:results_eettV} 
Sample of LO and NLO rates for the production of top quarks in
association with bosons, possibly within cuts and in association
with jets, at a 1-TeV $e^+e^-$ collider, and the fractional scale
uncertainties. Cross sections are 
calculated in the five-flavour scheme; see table~\ref{tab:results_vj} 
for the meaning of {\tt wpm}. Results at NLO accuracy 
for $t\bar tH$ production can be found in ref.~\cite{Dittmaier:1998dz}. 
All of the other processes are computed here for the first time at the NLO.
}
\end{table}


\end{landscape}

\subsection{Differential distributions\label{sec:diff}}
In this section we present sample results for differential observables
relevant to several processes, which we have simulated at the 8, 13,
or 14~TeV LHC. While some of these have never been computed before
at the NLO+PS accuracy (or even at fNLO; see sect.~\ref{sec:tot}),
and appear here for the first time, we do not aim to present
a series of phenomenological analyses, which would be out of the
scope of this work, but rather at showing yet again the flexibility
of \aNLO, and the type of results that one can obtain with it.
For this reason, the various PSMCs which we shall use have been run with 
their default parameters, and no underlying events have been generated.
Having said that, some of the predictions given here are motivated
by recent measurements by ATLAS and CMS. Furthermore, the present
section constitutes a complement in particular to sect.~\ref{sec:FxFx},
since we shall discuss, using explicit examples, several features
of the FxFx merging procedure which have been outlined before in a general
fashion. We shall be mainly concerned with (N)LO+PS results, but we 
shall also consider f(N)LO ones where necessary. As was the case 
for the total rates presented in sect.~\ref{sec:tot}, the computation of
scale and PDF uncertainties has been carried out by using the reweighting
procedure introduced in ref.~\cite{Frederix:2011ss} (see also
appendix~\ref{sec:errors}). NLO+PS results that have never appeared
in the literature are: six-lepton, $t\bt W^+W^-$, and SM Higgs 
in VBF$+1j$ production; furthermore, double-Higgs production in association 
with either a $t\bt$ pair or a $Z$ boson has been solely computed with 
\aNLO, in ref.~\cite{Frederix:2014hta}. Finally, FxFx-merged results
for $ZZ$ and $He^+\nu_e$ production are also presented here for the 
first time.

\vskip 0.4truecm
\noindent
{\bf $\blacklozenge$ Six-lepton production}

\noindent
We start by studying the (N)LO+PS production of six leptons:
\beq
pp\;\longrightarrow\;e^+e^-\mu^+\nu_\mu\tau^-\bar{\nu}_\tau\,,
\label{sixlep}
\eeq
which we have computed by using the complex mass scheme; the
$\tau^-$ lepton is set stable, and its mass is kept at the physical
value, while the electron and the muon are treated as massless. On top
of the computation carried out with the exact six-lepton matrix
elements of eq.~(\ref{sixlep}), we have also considered the
production of the $ZW^+W^-$ triplet, with the subsequent decays
of the vector bosons performed with either \Madspin\ or by the
PSMC (in this case, \HWs):
\beq
pp\;\longrightarrow\;Z(\to e^+e^-)\,W^+(\to\mu^+\nu_\mu)\,
W^-(\to\tau^-\bar{\nu}_\tau)\,.
\label{sixlepun}
\eeq
While \Madspin\ multiplies the undecayed matrix elements by the
branching ratios of the relevant decays, so that the rates resulting
from eq.~(\ref{sixlepun}) are in absolute value directly comparable
to those of eq.~(\ref{sixlep}), the PSMC does not; in that case, we
have therefore manually included such an overall factor.
To all samples, we have applied the following cut:
\beq
M(\ell^+\ell^{(\prime)-})>30~{\rm GeV}\,,
\label{Meecut}
\eeq
on all opposite-charged lepton pairs; given the lepton flavours
considered here, not surprisingly the vastly dominant effect of
such a cut is that due to the $e^+e^-$ pair. We shall call 
eq.~(\ref{Meecut}) the generation cut\footnote{Despite the fact
that it has been imposed at the analysis level, and the true
generation cut is marginally lower so as to avoid biases.}.
On top of eq.~(\ref{Meecut}), we have also imposed (also at
the analysis level):
\beqn
&&\abs{M(e^+e^-)-m_Z}<20~{\rm GeV}\,,
\label{Mllc1}
\\
&&\abs{M(\mu^+\nu_\mu)-m_W}<20~{\rm GeV}\,,
\label{Mllc2}
\\
&&\abs{M(\tau^-\bar{\nu}_\tau)-m_W}<20~{\rm GeV}\,,
\label{Mllc3}
\eeqn
which we shall call $V$-reco cuts.
Since eq.~(\ref{sixlepun}) features only 3-resonant contributions
(see sect.~\ref{sec:madspin} about the notation used here for
resonant and non-resonant diagrams),
the results of the \Madspin- and PSMC-decayed samples are basically
the same if one considers only the generation cut, or the generation
and $V$-reco cuts together; for this reason, we shall discuss only
the latter scenario. On the other hand, one of the reasons for comparing
eqs.~(\ref{sixlep}) and~(\ref{sixlepun}) is precisely that of assessing
the importance of non-3-resonant contributions to six-lepton matrix
elements; hence, in this case we shall present both the generation-cut-only 
and the generation-plus-$V$-reco cuts results.

\begin{figure}[h]
 \begin{center}
 \epsfig{file=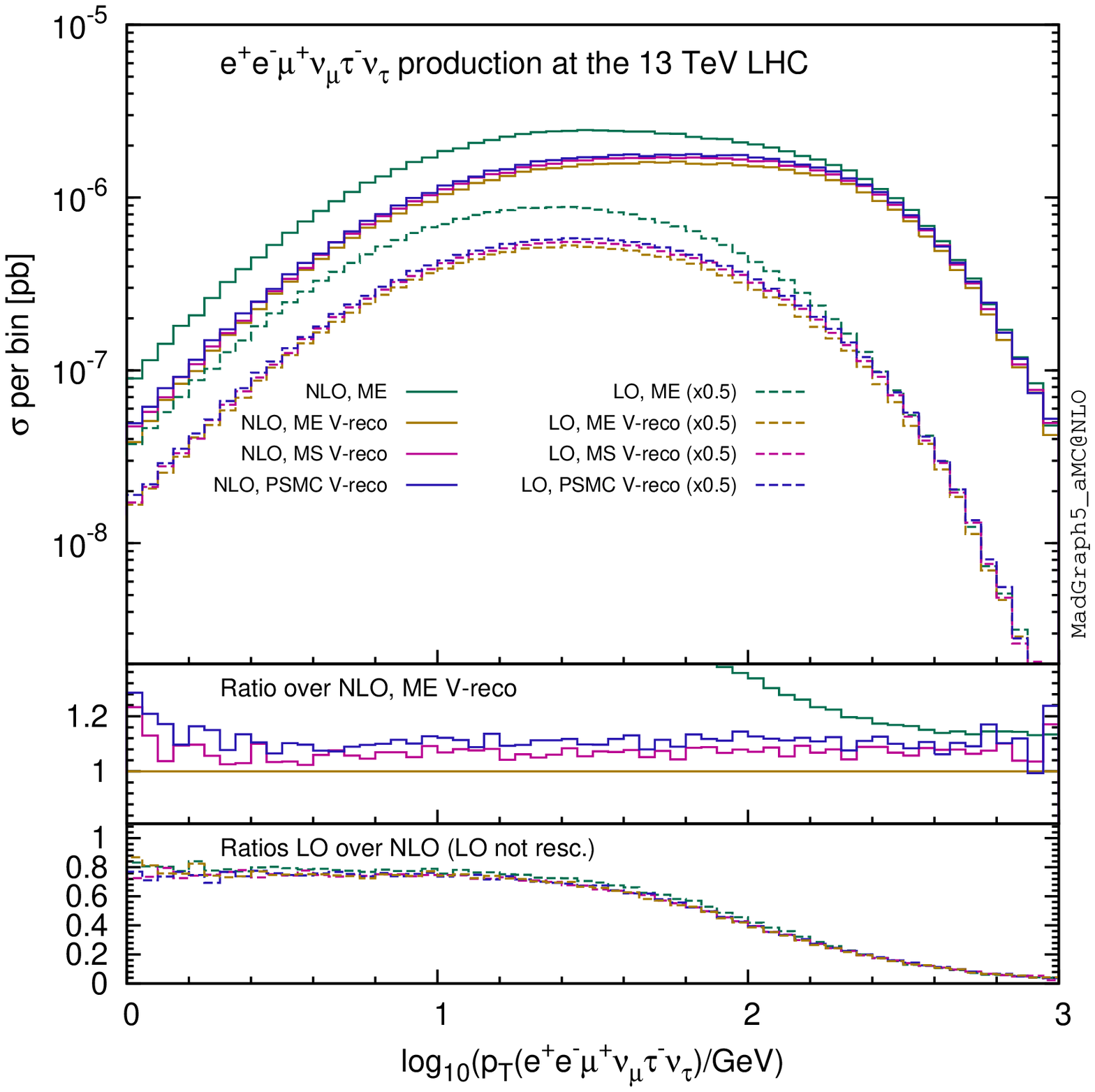, width=0.48\textwidth}
 \epsfig{file=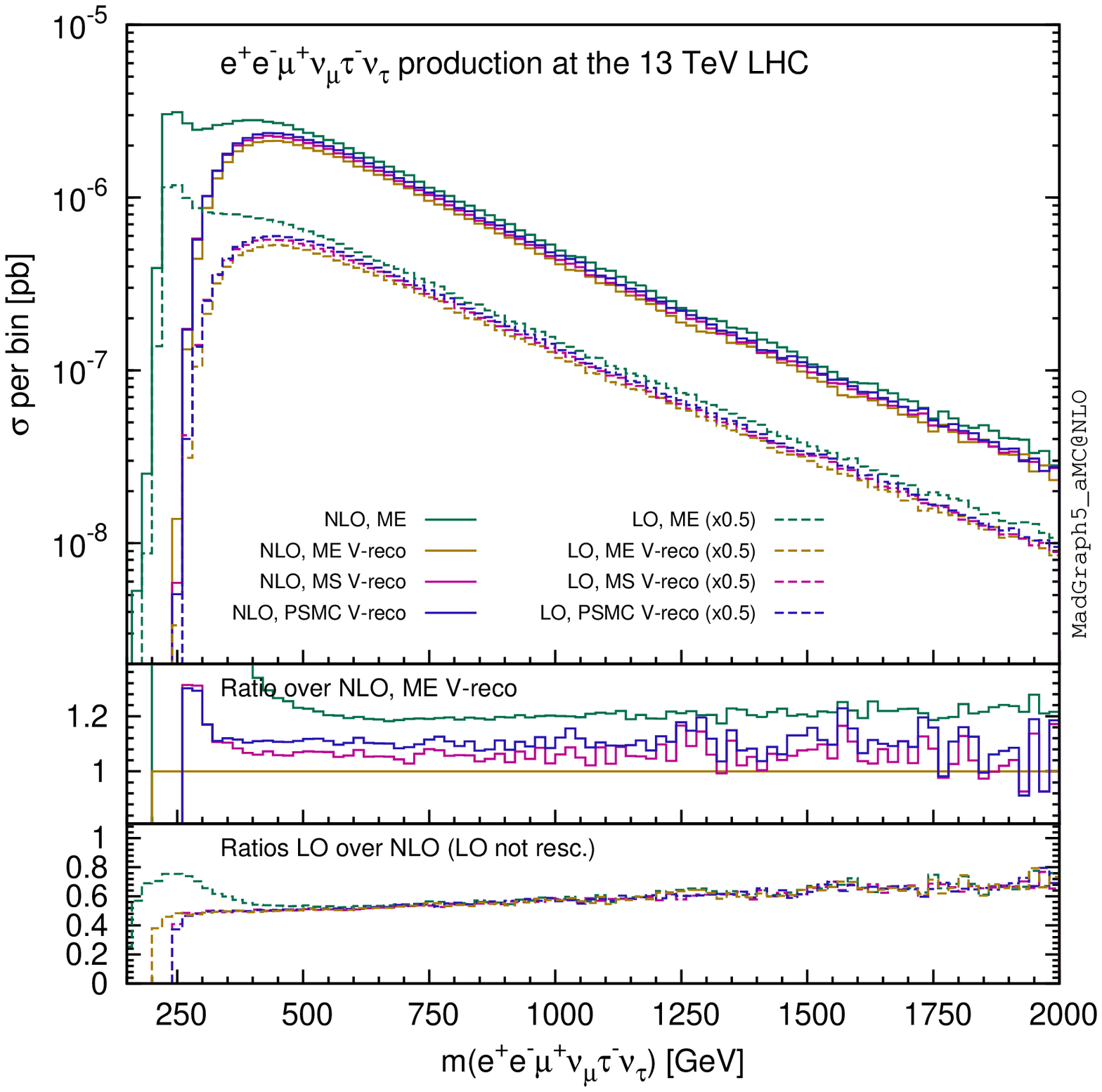, width=0.48\textwidth}
 \end{center}
\caption{Transverse momentum (left panel) and invariant mass (right panel)
of the six-lepton system, for the processes of eqs.~(\ref{sixlep})
and~(\ref{sixlepun}). See the text for details.}
\label{fig:VVV1}
\end{figure}
\begin{figure}[h]
 \begin{center}
 \epsfig{file=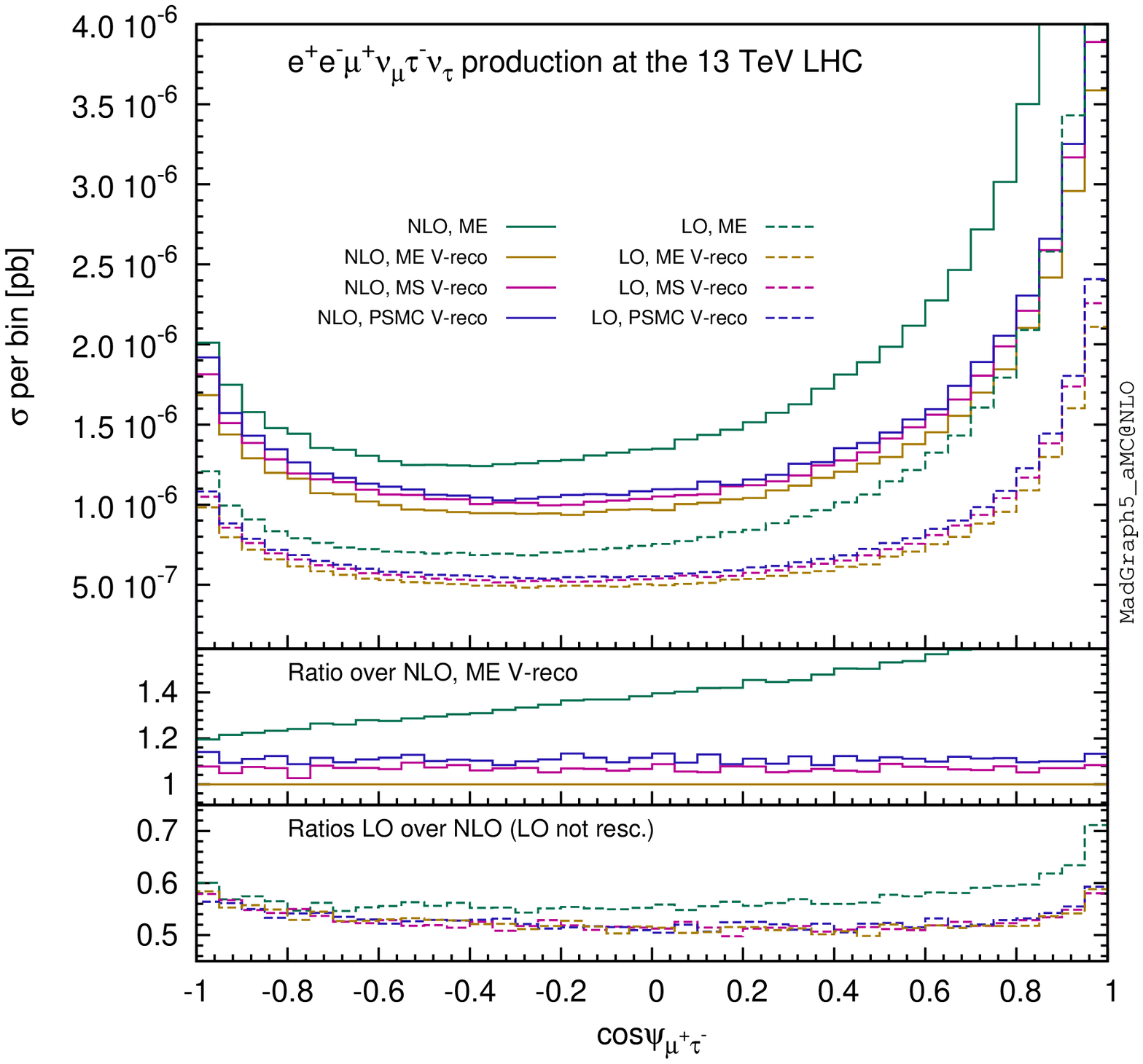, width=0.48\textwidth}
 \epsfig{file=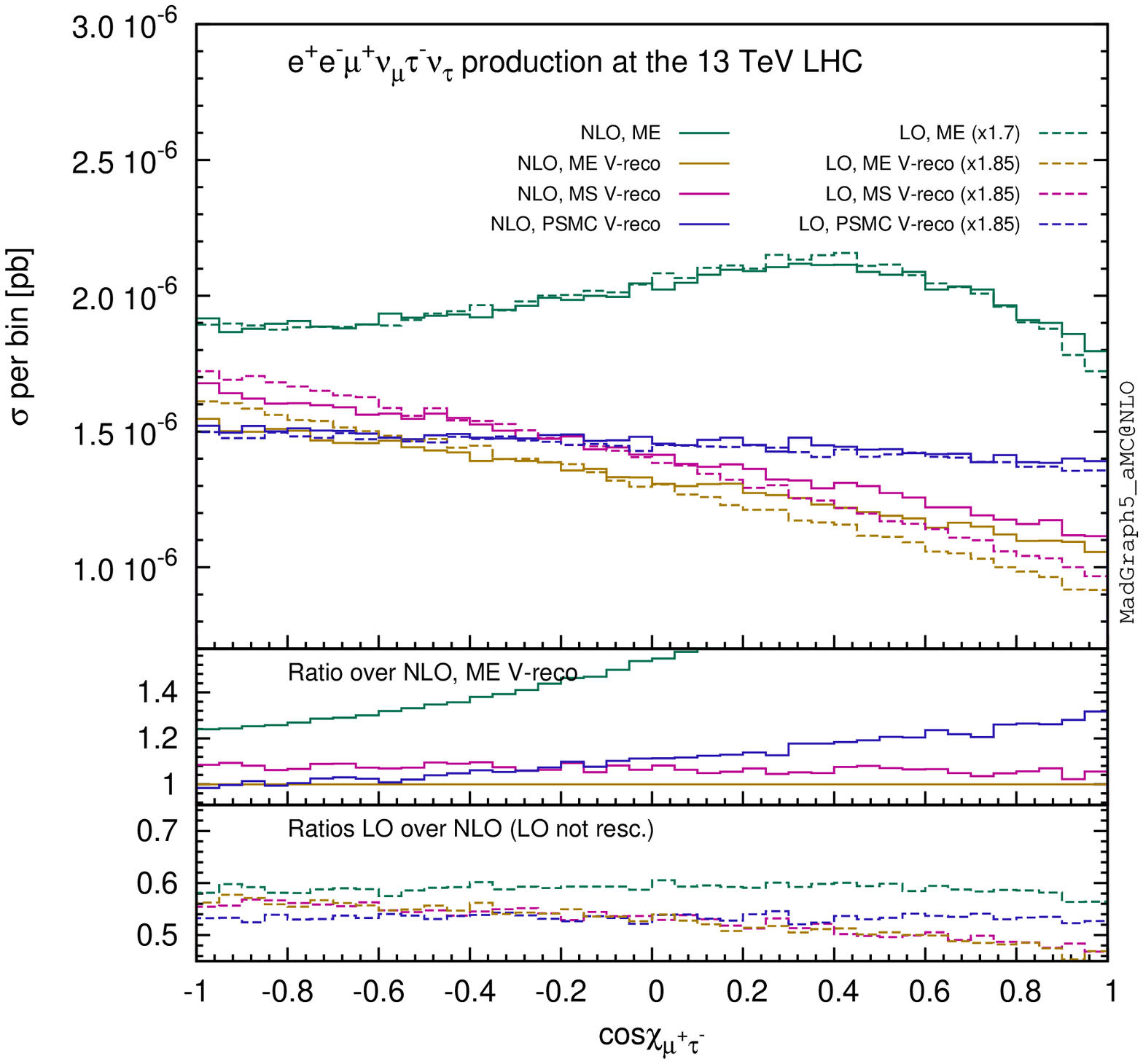, width=0.48\textwidth}
 \end{center}
\caption{Cosine of correlation angles for the $\mu^+\tau^-$ pair,
in six-lepton production, eqs.~(\ref{sixlep}) and~(\ref{sixlepun}). 
See the text for details.}
\label{fig:VVV2}
\end{figure}
In figure~\ref{fig:VVV1} we show observables relevant to the
six-lepton system, i.e.~obtained by summing the four-momenta
of the leptons: the transverse momentum (left panel) and
the invariant mass (right panel). Both the NLO+PS (solid histograms)
and LO+PS (dashed histograms, rescaled as indicated in order to fit 
into the layout) are displayed. The green histograms are the results of
eq.~(\ref{sixlep}) with only the generation cut (denoted by ``(N)LO ME''); 
the results for the generation-plus-$V$-reco cuts are shown as yellow 
(eq.~(\ref{sixlep}), denoted by ``(N)LO ME $V$-reco''), red 
(eq.~(\ref{sixlepun}) with \Madspin, denoted by ``(N)LO MS $V$-reco''), 
and blue (eq.~(\ref{sixlepun}) with PSMC decays, denoted by ``(N)LO PSMC 
$V$-reco'') histograms respectively. In the middle insets, the ratios
of all the NLO results over the NLO ME $V$-reco ones are presented.
Finally, in the lower insets each LO prediction is divided by 
its NLO counterpart (so these are essentially the 
inverse of the $K$ factors). The plots show clearly the large impact
of non-3-resonant contributions, which induce dramatic shape modifications
for $M(6\ell)<500$~GeV and $10<\pt(6\ell)<250$~GeV (the very small $\pt$
region being dominated by PSMC radiation effects). On the other hand,
by imposing $V$-reco cuts the three predictions agree rather well with
each other, which is the signal that spin correlations are unimportant
for these observables (and, more importantly in view of the aim of this
paper, that all is fine from a technical viewpoint, in the context of 
a very involved production process). We have performed similar comparisons
for a large number of observables; here, we limit ourselves to reporting
the results for the cosine of the angle defined by the directions of 
flight of the $\mu^+$ and $\tau^-$ leptons, which we denote by
$\psi_{\mu^+\tau^-}$ (left panel of fig.~\ref{fig:VVV2}) when it
is computed in the laboratory frame, and by $\chi_{\mu^+\tau^-}$ 
(right panel of fig.~\ref{fig:VVV2}) when it is computed by first
boosting the four-momentum of the $\mu^+$ and $\tau^-$ leptons to
the rest frame of the $\mu^+\nu_\mu$ and $\tau^-\bar{\nu}_\tau$
systems respectively (i.e., to the virtual-$W^+$ and $W^-$ rest
frames in the case of resonant contributions); the latter observable
is known to be particularly suitable for the study of spin-correlation
effects. The same conclusions as for the observables of fig.~\ref{fig:VVV1}
apply here, bar for the $\chi_{\mu^+\tau^-}$ NLO PSMC $V$-reco one that
is fairly different from both the NLO ME $V$-reco and NLO MS $V$-reco 
predictions, which in turn agree with each other quite well. As it was
expected, this is a manifestation of the importance of spin correlations
for such an observable, and a direct validation of the \Madspin\ procedure.

The overall messages that one can obtain from the present study are
the following. Firstly, we did verify that the conclusions reached above
are not qualitatively modified if one replaces the (N)LO ME results
with those obtained by imposing the generation cut and eq.~(\ref{Mllc1})
only -- in other words, it is the simultaneous action of the three
cuts of eqs.~(\ref{Mllc1})--(\ref{Mllc3}) that brings the predictions
for eq.~(\ref{sixlep}) in agreement with those for eq.~(\ref{sixlepun})
and \Madspin; this is obviously because it is important that all three
vector bosons be near their respective mass shell. Secondly, the effects
of NLO corrections are non negligible, in both rate and shape; however,
the patterns of comparison among the various calculations are to a large 
extent independent of the perturbative accuracy of the latter. Thirdly, 
production spin correlations are present, that can be properly described
only by the full six-lepton computation, and by \Madspin\ as well {\em if} 
one limits oneself to the 3-resonant region. It is clear that the cuts
of eqs.~(\ref{Mllc2}) and~(\ref{Mllc3}), and the definition of the
observables considered here, bar $\psi_{\mu^+\tau^-}$, cannot be achieved 
experimentally, owing to the presence of the neutrino four-momenta.
However, they have helped us reach conclusions which have a general
validity, and in particular in the case of a fully realistic analysis:
namely, that non-resonant effects in six-lepton production may be quite
large and that, for all those cuts that render the 3-resonant contributions
dominant, the undecayed-plus-\Madspin\ simulation provides one with a 
very good approximation of the exact calculation.

\vskip 0.4truecm
\noindent
{\bf $\blacklozenge$ $t\bt W^- W^+$ production}

\begin{figure}[h]
 \begin{center}
 \epsfig{file=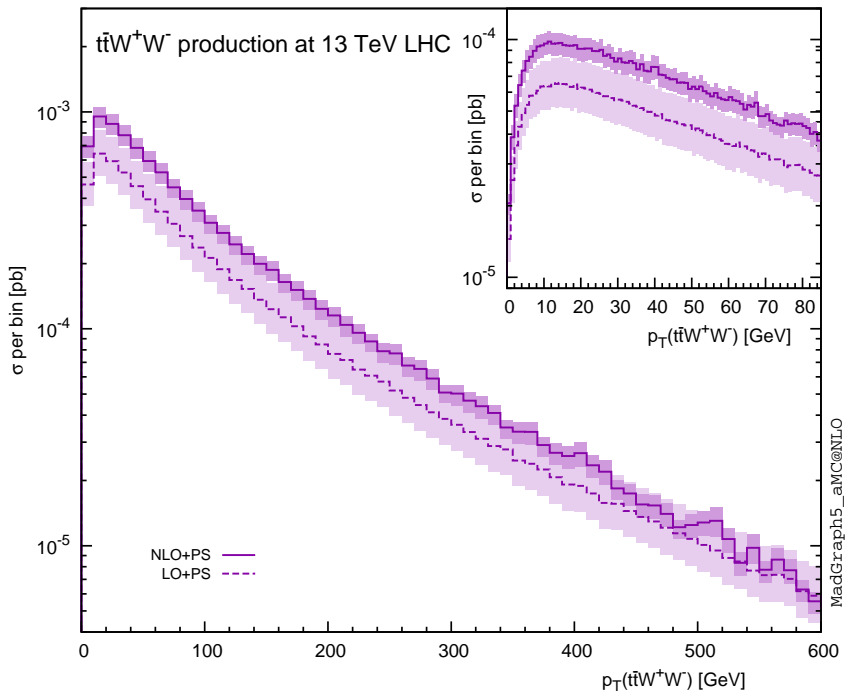, width=0.75\textwidth}
 \end{center}
\caption{Transverse momentum of the system of the four primary final-state
particles in $t\bt W^+W^-$ production.}
\label{fig:ttWW1}
\end{figure}

\noindent
We now turn to considering the process:
\beq
pp\;\longrightarrow\;t(\to e^+\nu_e b)\,\bt(\to e^-\bar{\nu}_e\bb)\,
W^-(\to\mu^-\bar{\nu}_\mu)\,W^+(\to\mu^+\nu_\mu)\,,
\label{ttWW}
\eeq
which we have simulated at the (N)LO+PS accuracy, by only considering
the undecayed matrix elements with $t\bt W^+W^-$ final states, and by using 
\PYe\ as PSMC and either \Madspin\ or the internal \PYe\ routine (which 
correctly accounts for decay spin correlations) for the decays of
the top quarks and $W$ bosons. In fig.~\ref{fig:ttWW1} we present
the transverse momentum of the $t\bt W^+W^-$ system, which is
the typical observable whose small-$\pt$ behaviour is dominated
by MC effects (whose systematics will not be studied here), 
and which is thus unreliable if computed at fNLO
accuracy. Both the NLO+PS (solid histograms) and LO+PS (dashed
histograms) results are displayed, with the respective scale-uncertainty
bands (in dark and light shades respectively). The very significant
reduction of such theoretical systematics when higher-order corrections
are included is evident in the whole range considered (see also
entry e.13 in table~\ref{tab:results_tv} for its total-rate
counterpart). While for asymptotically-large transverse momenta
one expects the NLO+PS scale dependence to be of LO type (because in
that region the computation is dominated by tree-level contributions),
for such a massive system these $\pt$'s are confortably in
the TeV-range.
\begin{figure}[h]
 \begin{center}
 \epsfig{file=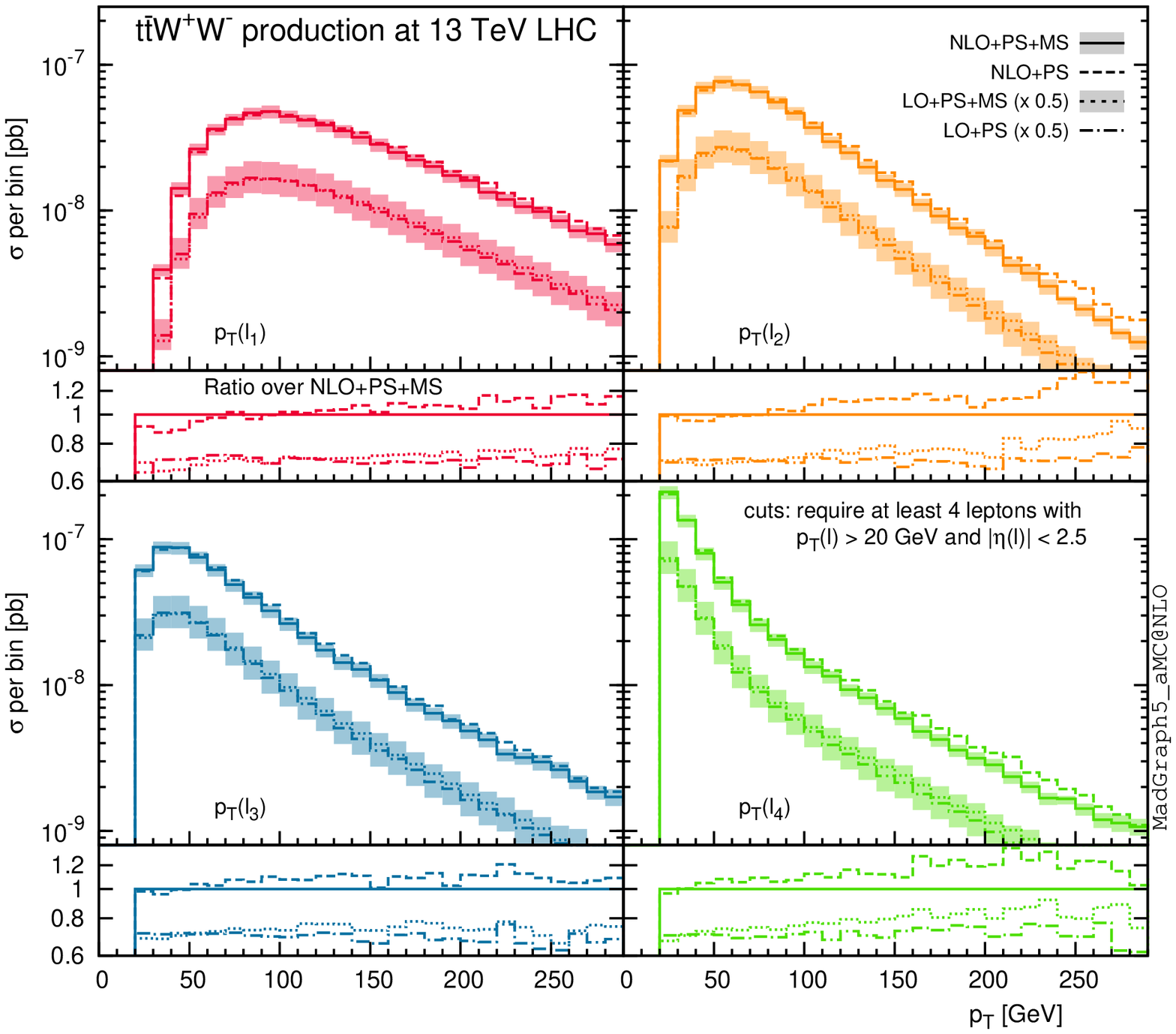, width=0.9\textwidth}
 \end{center}
\caption{Transverse momentum of the hardest four charged leptons in 
$t\bt W^+W^-$ production -- see eq.~(\ref{ttWW}). The LO+PS results
that appear in the insets have not been rescaled.}
\label{fig:ttWW2}
\end{figure}
In fig.~\ref{fig:ttWW2} we show the transverse momenta of the four-hardest
charged leptons in the events; the leptons are required to have 
$\pt(\ell)>20$~GeV
and $\abs{\eta(\ell)}<2.5$. At variance with those of fig.~\ref{fig:ttWW1},
these plots include the branching ratios of the decays reported 
in eq.~(\ref{ttWW}). The LO+PS results in the main frames are rescaled 
so as to fit into the layout. Each plot displays four histograms, that
correspond to NLO+PS (solid, with \Madspin\ decays; dashed, with
\PYe\ decays) and to LO+PS (dot-dashed, with \Madspin\ decays; dotted, 
with \PYe\ decays) results. The ratios of these predictions over the
NLO+PS, \Madspin-decayed ones are shown in the insets. The radiative 
corrections are large, but relatively flat in the $\pt$ ranges considered
here; as in the case of the $\pt$ of the system, their inclusion reduces
the scale uncertainty in a dramatic manner. Production spin correlations
are sizable, so much so that \Madspin- and \PYe-decayed results have shapes
which are barely within, or slightly outside of, the theoretical systematics
bands. This is true at the NLO; at the LO, the two predictions are 
compatible within uncertainties, but this is solely due to the fact that
the LO scale dependence is rather large: in fact, the pattern of the inclusion 
of production spin correlations is basically independent of the perturbative
order at which one is working. This is nothing but another manifestation
of the benefits inherent to the increased predictive power of simulations
that include both NLO and production spin correlation effects.

\vskip 0.4truecm
\noindent
{\bf $\blacklozenge$ Double-Higgs production}

\begin{figure}[h]
 \begin{center}
 \epsfig{file=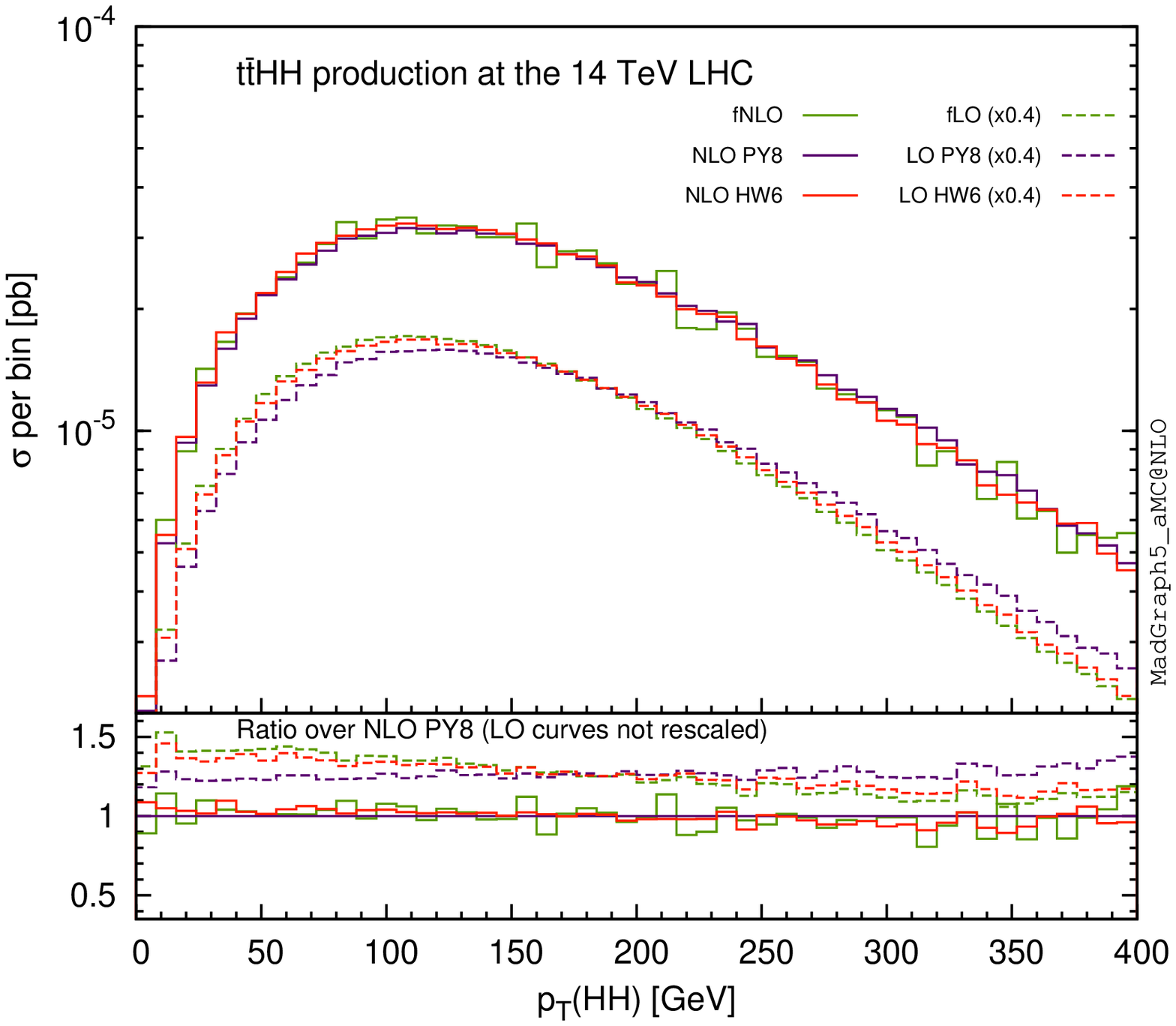, width=0.48\textwidth}
 \epsfig{file=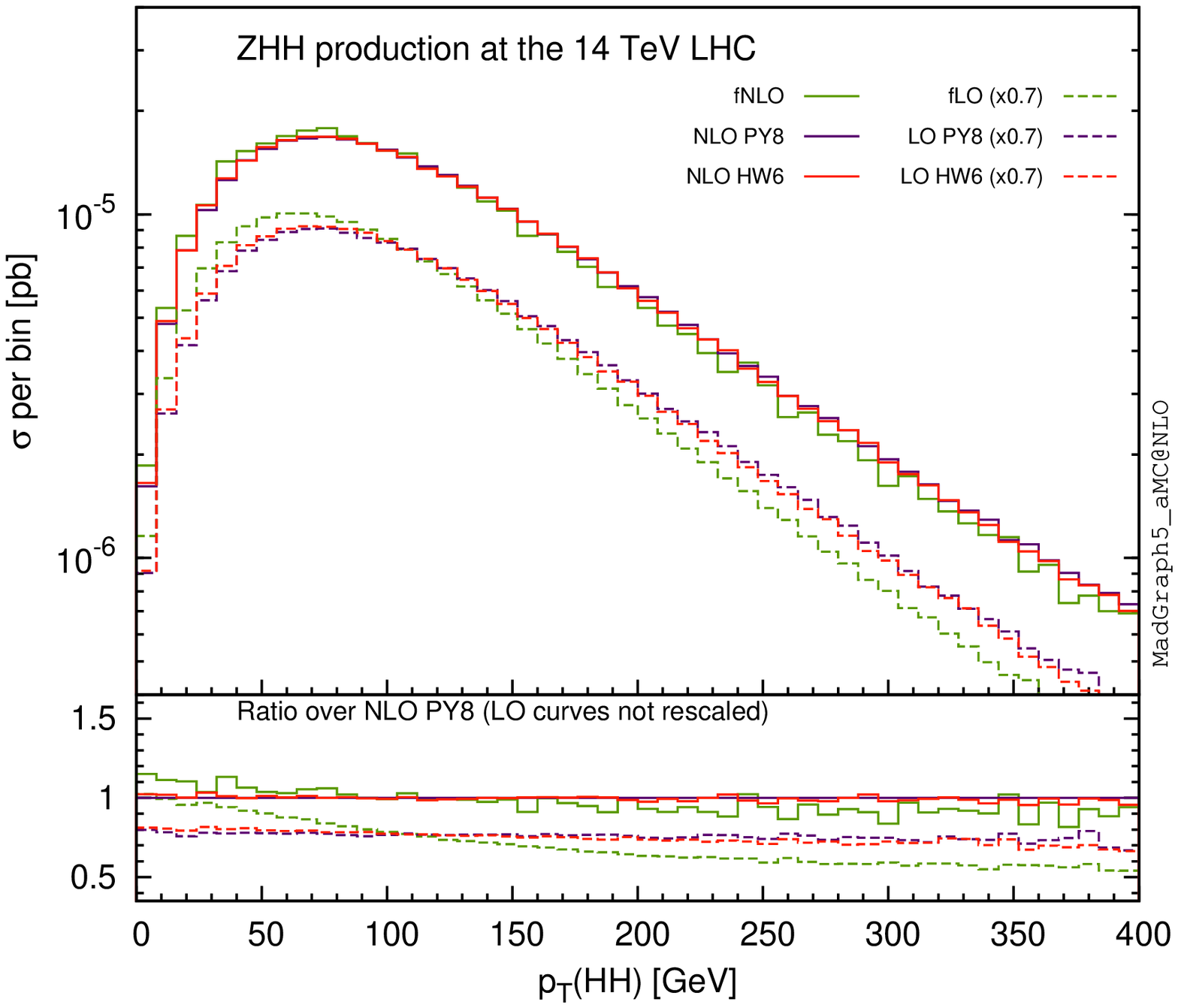, width=0.48\textwidth}
 \end{center}
\caption{Transverse momentum of the two-Higgs system in $t\bt HH$
(left panel) and $ZHH$ (right panel) production. (N)LO+PS and
f(N)LO results are shown. We have used \PYe\ and \HWs.
}
\label{fig:HH}
\end{figure}

\noindent
We now consider double-Higgs production in the SM at the 14~TeV LHC.
This process has been investigated recently with \aNLO\ in 
ref.~\cite{Frederix:2014hta}, where
all the six dominant channels at the LHC have been computed up to NLO+PS 
accuracy, some of them for the first time. For all channels, the results
of ref.~\cite{Frederix:2014hta} have improved what was available in
the literature in at least one respect. In ref.~\cite{Frederix:2014hta}
we have only presented (N)LO+PS distributions. Here, we amend this
by showing also f(N)LO spectra; we use the transverse momentum of the
Higgs pair, and the $t\bt HH$ and $ZHH$ channels, as a definite example;
as PSMCs, we adopt \PYe\ and \HWs.
The results are shown in fig.~\ref{fig:HH}, as solid (for NLO-accurate)
and dashed (for LO-accurate) histograms. The main frames display the
NLO predictions in absolute value, while the LO ones are rescaled in
order to fit into the layout; the $K$ factors can be read from the
insets, where we present the ratios of all results over those at the 
NLO+PS obtained with \PYe. The common feature of the two plots is
that NLO results are mutually closer than the corresponding LO ones;
the two NLO+PS predictions are extremely similar for both processes, while 
the fNLO spectrum in $ZHH$ production is only marginally softer (an overall
effect smaller than 20\%). It is interesting to see that this 
stabilisation due to the inclusion of higher-order corrections follows
different patterns for the two channels considered here. $ZHH$ is
predominantly a $q\bq$-initiated process: therefore, the difference
between the two standalone PSMCs (i.e., LO+PS) is expected to be smaller than 
in the case of $t\bt HH$ production, which mainly proceeds through $gg$ 
fusion. This is precisely what we see in fig.~\ref{fig:HH} (compare the
purple and red histograms in the insets). On the other hand, at fLO
the Higgs pair recoils against a $Z$ boson and a $t\bt$-system in $ZHH$ 
and $t\bt HH$ production respectively; the kinematics of the $t\bt$
pair being non-trivial (at variance with that of a single $Z$) implies
that the fLO prediction for $\pt(HH)$ is farther away from the corresponding
LO+PS ones in the $ZHH$ channel than in the $t\bt HH$ channel (see the
green dashed histograms in the insets, and compare the position of the
peaks of the fLO and fNLO results).

\vskip 0.4truecm
\noindent
{\bf $\blacklozenge$ Single-Higgs production}

\begin{figure}[h]
 \begin{center}
 \epsfig{file=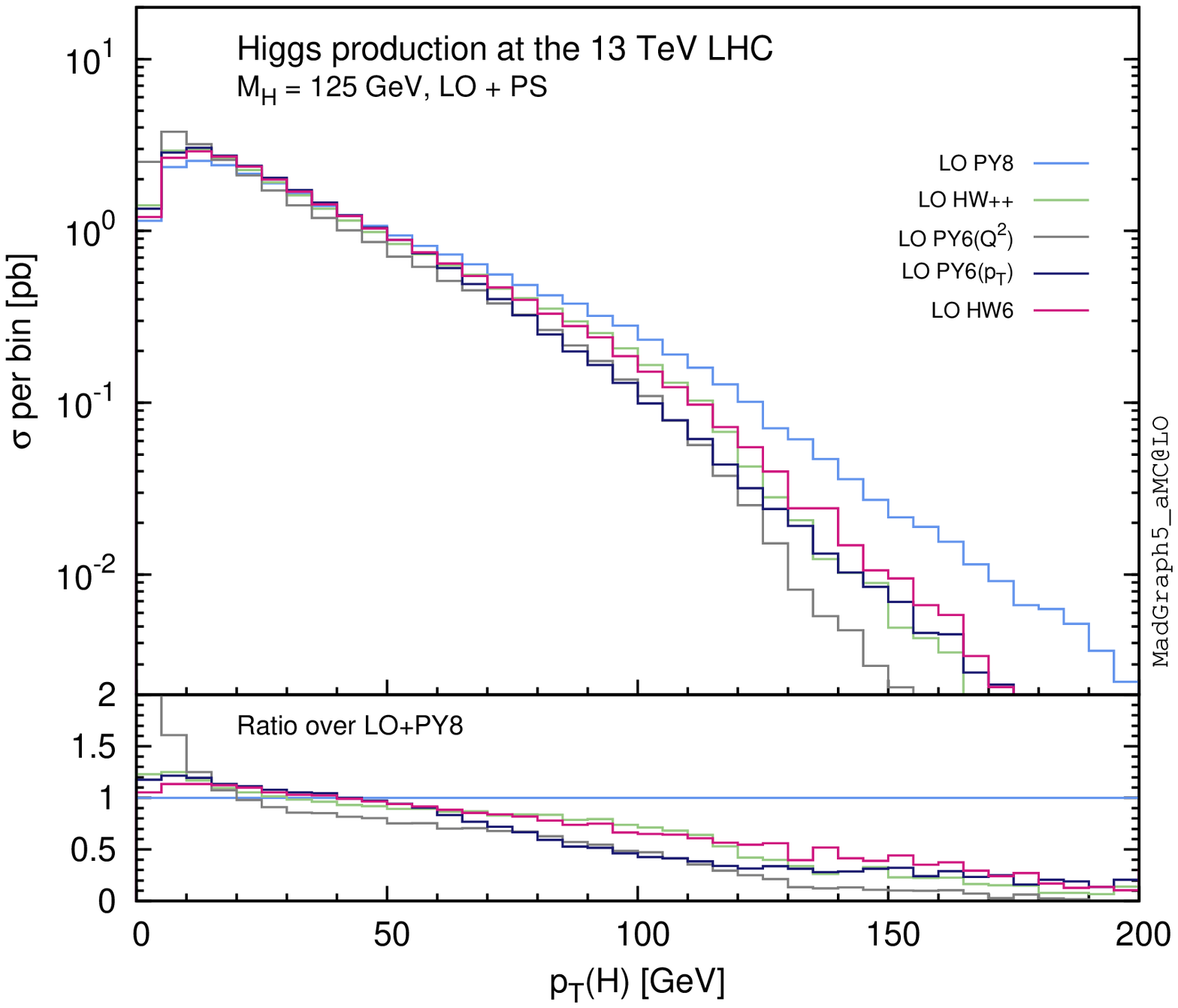, width=0.48\textwidth}
 \epsfig{file=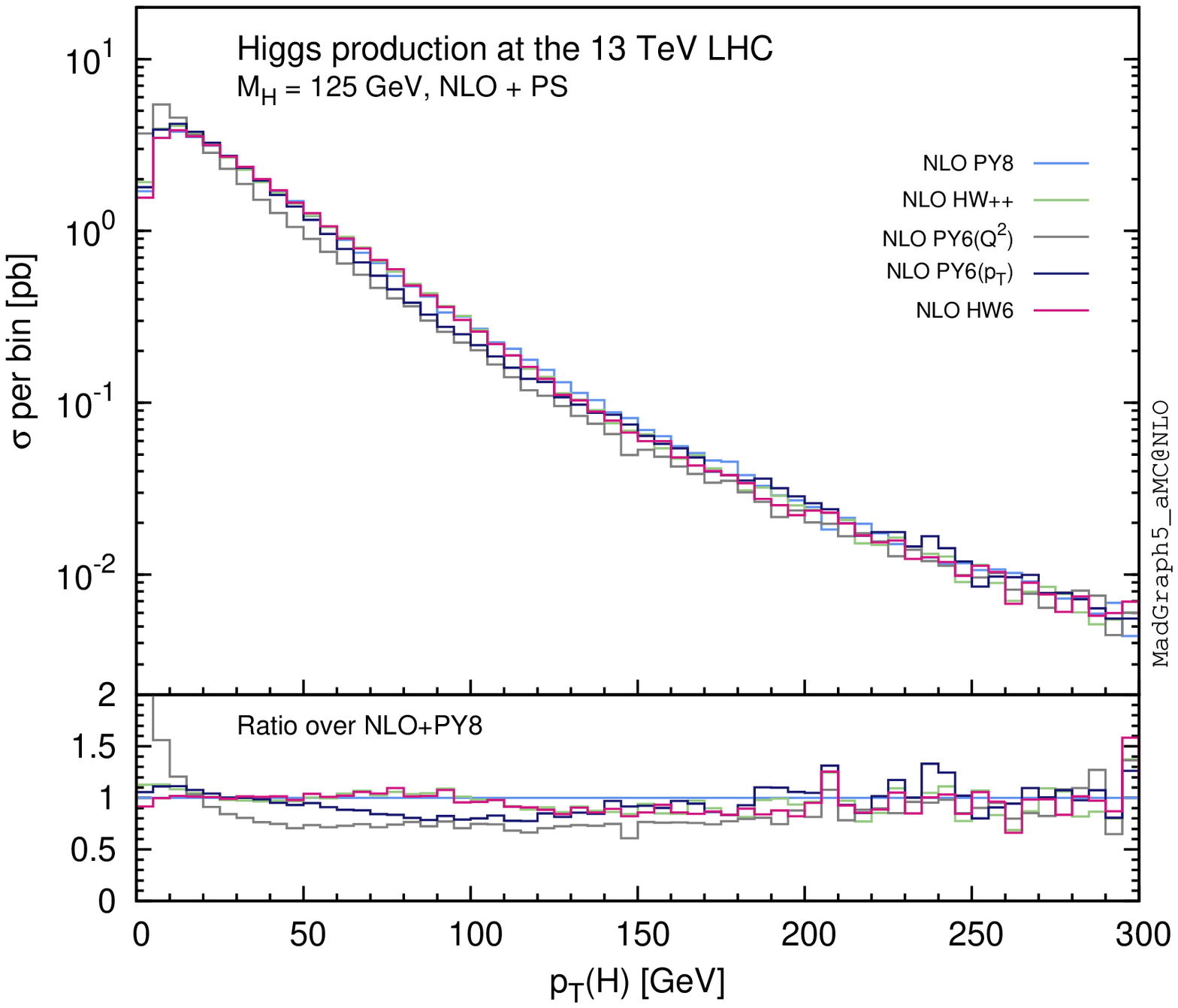, width=0.48\textwidth}
 \end{center}
\caption{Higgs $\pt$ spectrum in $gg$ fusion (HEFT), for various PSMCs,
at the LO+PS (left panel) and NLO+PS (right panel) accuracy. Note the
larger $\pt$ range in the right panel.
}
\label{fig:ptggH}
\end{figure}

\begin{figure}[h]
 \begin{center}
 \epsfig{file=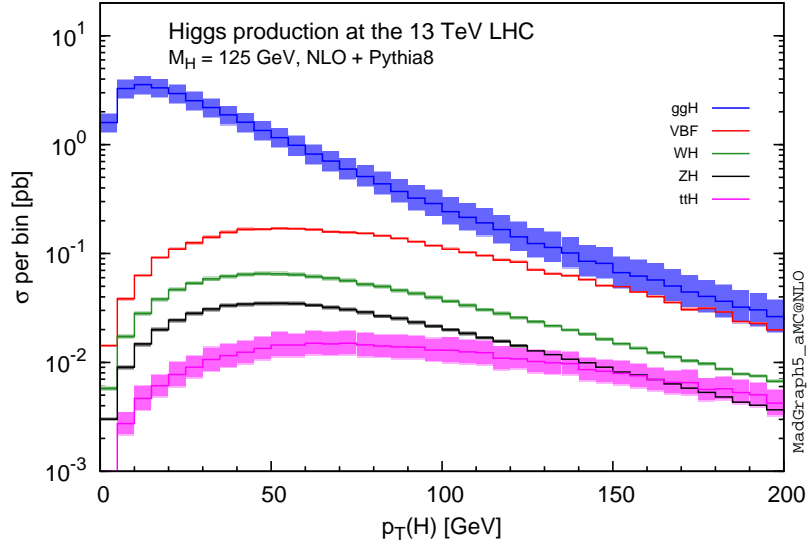, width=0.7\textwidth}
 \end{center}
\caption{Higgs $\pt$ spectrum for the five dominant production
channels at the LHC, at the NLO+PS accuracy with \PYe.
}
\label{fig:ptHall}
\end{figure}

\noindent
We now turn to discussing the production of a single SM Higgs
at the 13~TeV LHC. The aims of fig.~\ref{fig:ptggH}, where we
show the Higgs transverse momentum, are twofold.
Firstly, we compare LO+PS with NLO+PS predictions; secondly,
we present results for all PSMCs which are matched to NLO 
calculations in \aNLO\footnote{We remind the reader that
\PYs($\pt$) is available for ISR-only processes.}. The comparison
between the two panels of the figure shows the two expected behaviours:
at large $\pt$'s, all NLO+PS predictions coincide (and are just on 
top of the corresponding fNLO result, not displayed here), while the LO+PS
are vastly different; on the other hand, at small $\pt$'s, the relative 
behaviour of the various PSMCs is the same, regardless of the perturbative
order of the underlying matrix-element computations. 
We point out that all PSMCs are  treated on equal footing, 
i.e.~they are given the same numerical values as shower-scales 
parameters (such scales are equal to $m_H$ at the LO, and controlled
by the $D$ function at the NLO); so while different scales for different
PSMCs could bring them in better agreement at the LO, this is actually
a negative implication of the loss of predictivity at this perturbative
order, and it is unnecessary when NLO corrections are included in
the simulations; a further example of this pattern will be given below,
in the study of Higgs production through VBF. We conclude this part
by showing, in fig.~\ref{fig:ptHall}, the Higgs $\pt$ that results
from the five dominant production channels at the 13~TeV
LHC (whose total rates are reported in lines g.1, g.4, g.6, g.9,
and g.16 of table~\ref{tab:results_h}, where one can also find the
\aNLO\ shell commands relevant to their generation); 
the thickness of the bands represent the combined scale and
PDF uncertainties; all the predictions are
obtained at the NLO+PS, with the use of \PYe. This plot is another
demonstration of the flexibility of the \aNLO\ framework.

\newpage
\noindent
{\bf $\blacklozenge$ Higgs production in VBF}

\begin{figure}[h]
 \begin{center}
 \epsfig{file=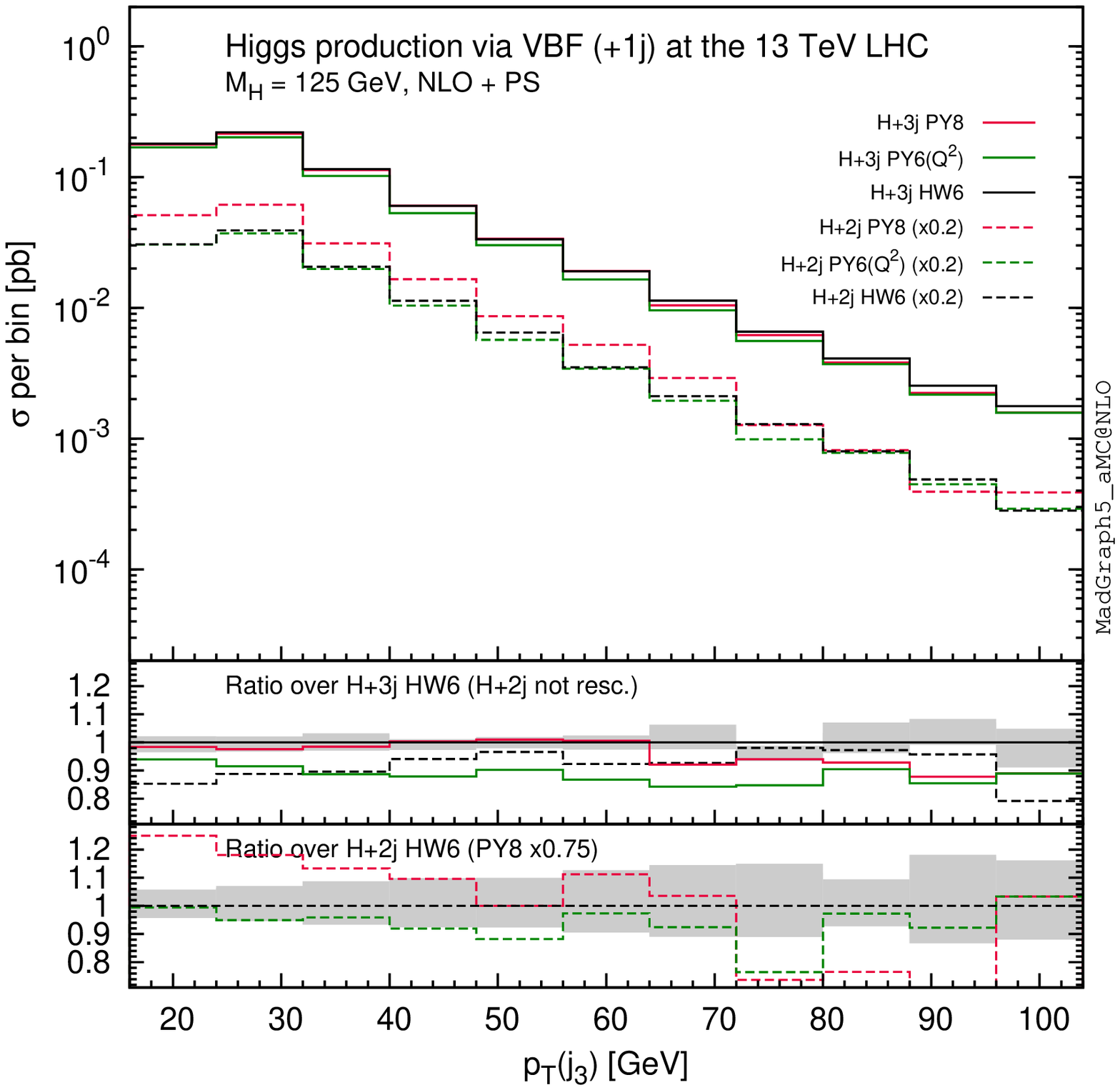, width=0.48\textwidth}
 \epsfig{file=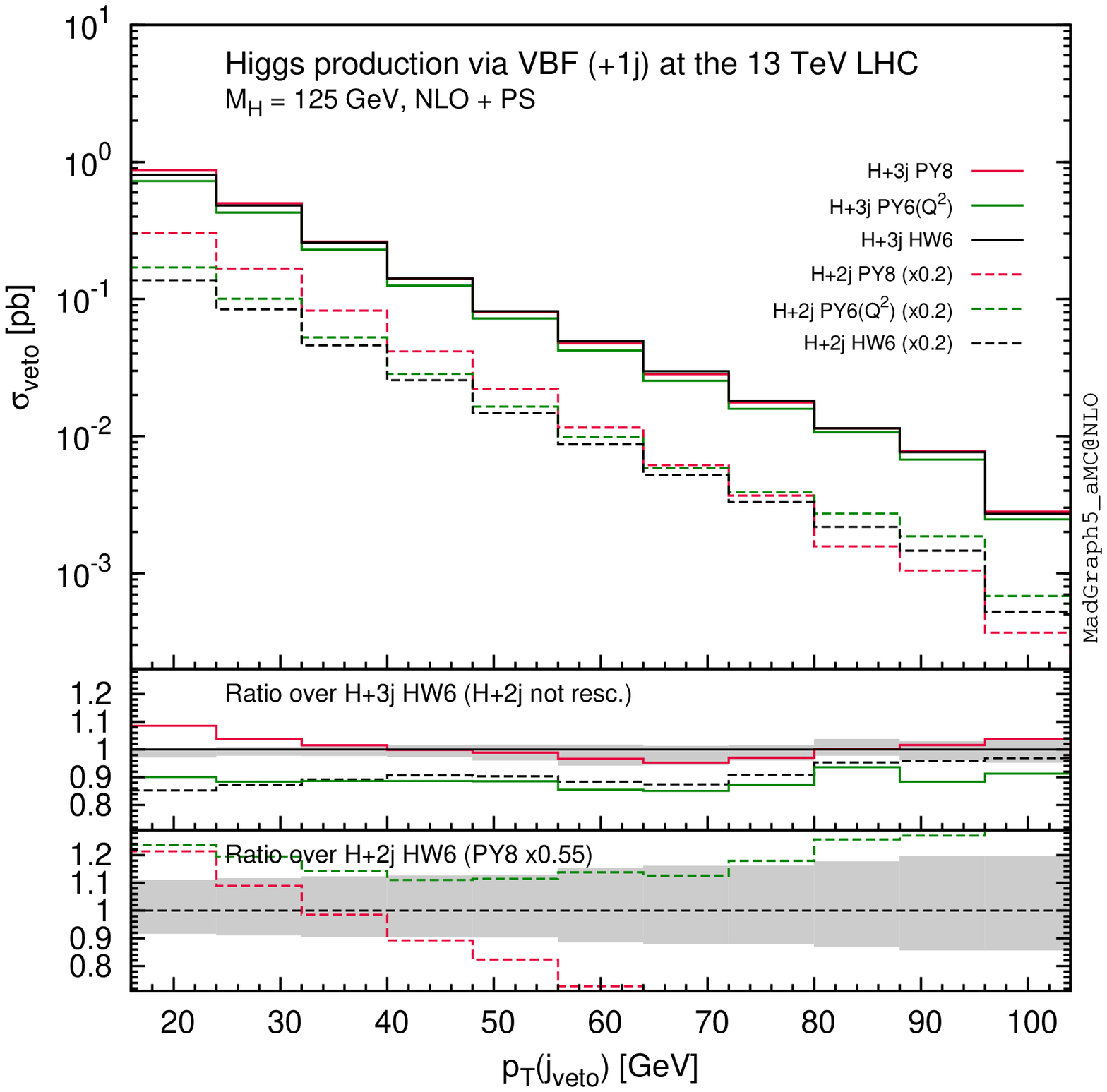, width=0.48\textwidth}
 \end{center}
\caption{Third-jet observables in Higgs VBF$+0j$ and VBF$+1j$ production, 
both at the NLO+PS, with \PYe, \HWs, and \PYs($Q^2$).
}
\label{fig:VBFj}
\end{figure}

\noindent
As a further example of the stabilisation of the predictions
that result from including higher-order matrix elements (which
in turn serves as a validation exercise for the whole NLO+PS
machinery in \aNLO), we consider Higgs production in VBF,
which we compute in two ways: by considering the process whose Born is 
of ${\cal O}(\aem^3)$ (which we denote by VBF$+0j$~\cite{Frixione:2013mta},
and whose final state at the Born level thus features a Higgs-plus-two-parton
system), and the process whose Born is of ${\cal O}(\aem^3\as)$ (which we 
denote by VBF$+1j$, and whose final state at the Born level thus features 
a Higgs-plus-three-parton system).
These are the processes reported in lines g.4 and g.5 of 
table~\ref{tab:results_h}, where the interested reader can
also find the \aNLO\ shell commands that one must use in
order to generate them. The only analysis cuts we impose here are
on the transverse momenta of the (anti-$\kt$, $R=0.5$) jets, by requiring 
that $\pt(j)>20$~GeV. The only observables for which a direct
comparison between VBF$+0j$ and VBF$+1j$ is sensible, and allow
one to assess the impact of perturbative corrections, are those
related to the third jet, where one expects to have an effective
LO and NLO description respectively. In fig.~\ref{fig:VBFj} we
present predictions for the transverse momentum spectrum of the
third-hardest jet, and for the rate as a function of the transverse
momentum of the veto jet:
\beq
\sigma_{veto}\left(\pt(j_{veto})\right)=\int_{\pt(j_{veto})}^\infty d\pt\,
\frac{d\sigma}{d\pt}\,.
\label{sigveto}
\eeq
The veto jet is the hardest jet which is not one of the two tagging
jet (which are defined to be the two hardest ones overall, and which we denote 
by $j_1$ and $j_2$ respectively), and whose rapidity obeys the condition:
\beq
\min\left(y_{j_1},y_{j_2}\right)\le y(j_{veto})\le
\max\left(y_{j_1},y_{j_2}\right)\,.
\eeq
The quantity defined in eq.~(\ref{sigveto}) is related to $P_{veto}$, 
defined e.g.~in eq.~(41) of ref.~\cite{Heinemeyer:2013tqa}, by a simple 
normalisation factor, \mbox{$\sigma_{veto}=\sigma_{{\rm NLO}}\,P_{veto}$}.
For both VBF$+0j$ (dashed histograms) and VBF$+1j$ (solid histograms) 
the results for three PSMCs (\PYe\ (red), \PYs($Q^2$) (green), and 
\HWs\ (black)) are displayed, with the VBF$+0j$ ones 
rescaled by a factor $1/5$ in
order for them to fit into the layout. The ratios of the \PY\ predictions
over the \HWs\ ones are presented in the insets. In the inset
relevant to VBF$+1j$, we also report the ratio of the \HWs\ VBF$+0j$
result over the VBF$+1j$ one (black dashed histogram), which is related to
the inverse of the $K$ factor -- for both observables, the latter is of 
the order of 1.05--1.15. In the insets we also display the scale 
uncertainties as gray bands: it is evident that the inclusion
of the contributions of relative ${\cal O}(\as)$ in VBF$+1j$ significantly
reduces the theoretical systematics. Apart from this, the most striking
consequence of such an inclusion is the fact that the three PSMCs in
VBF$+1j$ are fairly close to each other; this is not the case in
VBF$+0j$, where \PYe\ has a much softer shape than either \PYs\ or \HWs.
We point out that this is a feature of quantities related to the third jet:
other observables which have an NLO nature in VBF$+0j$ are much better
behaved, with the three PSMCs in good agreement already at 
${\cal O}(\aem^3(1+\as))$.
Therefore, although one could possibly find settings for the PSMCs that
would bring the predictions for the $\pt(j_3)$ and $\sigma_{veto}$
VBF$+0j$ spectra in better agreement than in fig.~\ref{fig:VBFj},
this would simply be the signal of an unsatisfactory predictive capability,
which is restored by considering these observables in VBF$+1j$ production.

\vskip 0.4truecm
\noindent
{\bf $\blacklozenge$ Top-pair production}

\begin{figure}[h]
 \begin{center}
 \epsfig{file=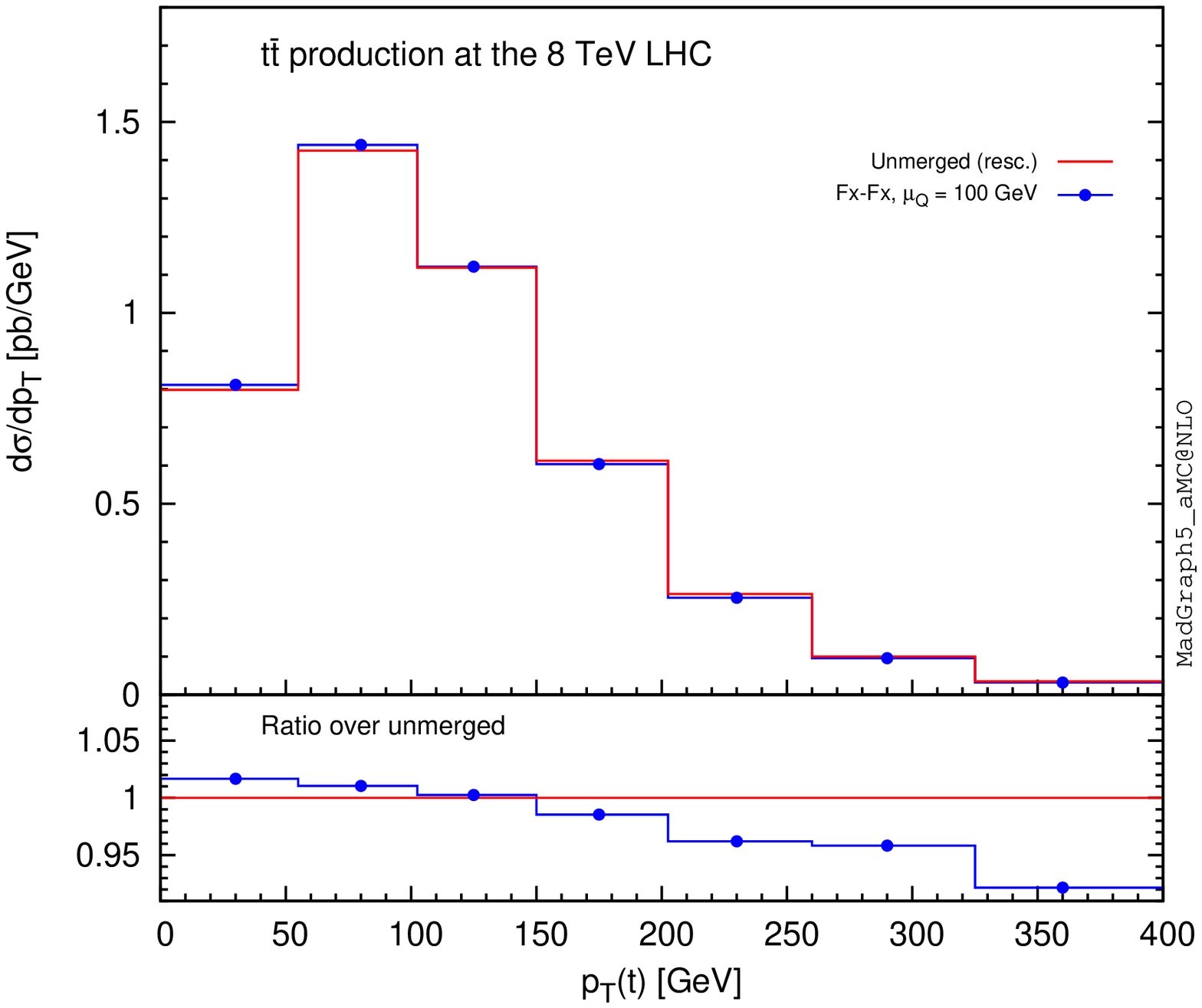, width=0.48\textwidth}
 \epsfig{file=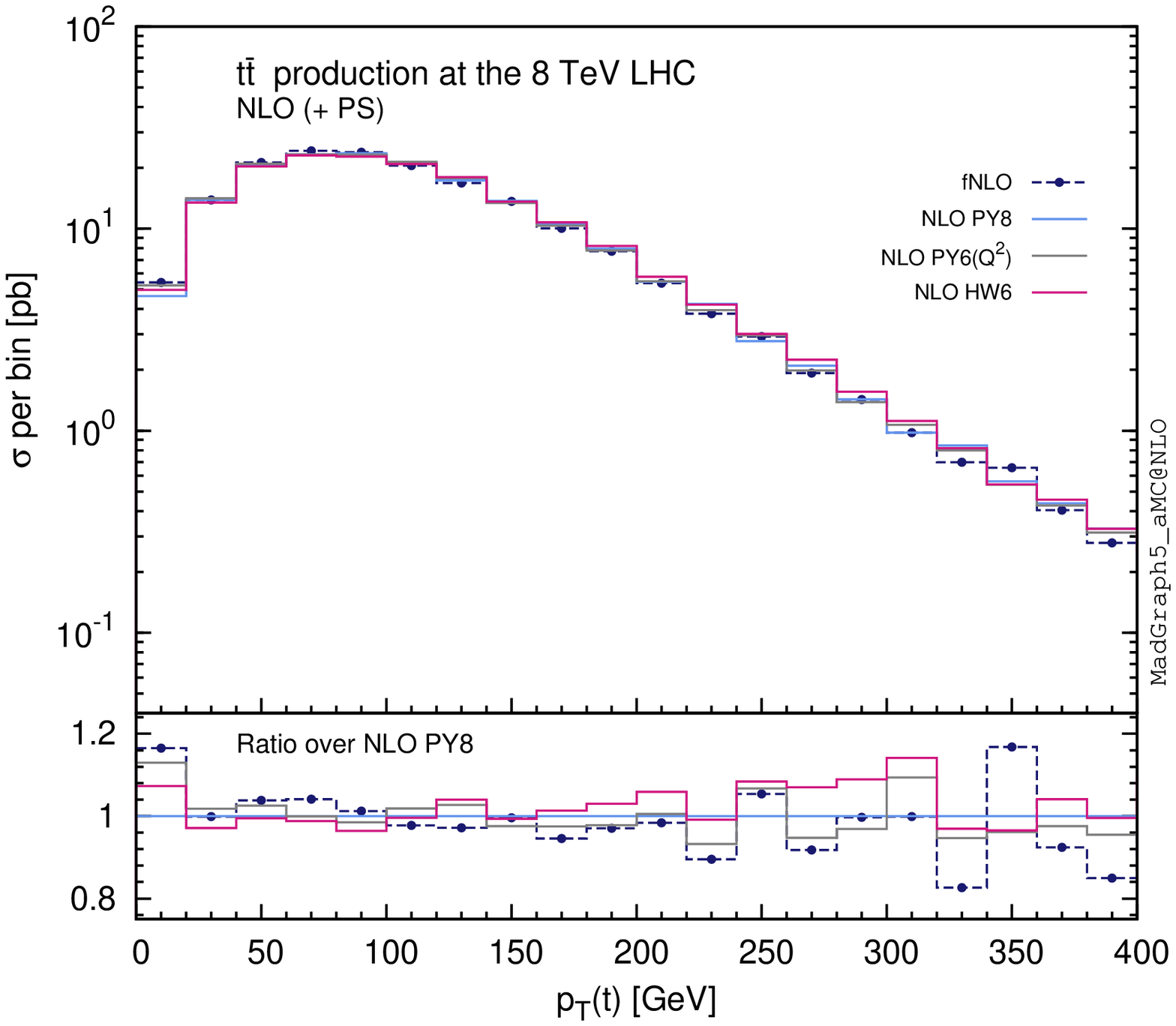, width=0.48\textwidth}
 \end{center}
\caption{Transverse momentum of the top quark in $t\bt$ production.
Left panel: comparison between FxFx-merged (blue) and unmerged (red) 
predictions; the binning is the same as that of ref.~\cite{CMSpttop27}.
Right panel: NLO+PS predictions obtained with different PSMCs,
compared to the fNLO result.}
\label{fig:pttop}
\end{figure}

\noindent
While differential distributions relevant to $t\bt$ production
measured at the LHC by the ATLAS (at 7~TeV in lepton+jets
events~\cite{ATLASpttop}) and the CMS (at 7 and 8~TeV, in lepton+jets
and dilepton events~\cite{Chatrchyan:2012saa,CMSpttop27,CMSpttop28})
collaborations are generally in very good agreement with theoretical 
predictions, the CMS data for the reconstructed transverse momentum of 
the top quark ($\pt(t)$) are visibly softer than NLO+PS predictions, 
and in disagreement with those of ATLAS for $\pt(t)<200$~GeV (ATLAS 
data are harder). Given this inconsistency between measurements
it is premature to speculate on the origin of a possible discrepancy
between data and theory; it is however of some interest to discuss
the theoretical systematics that affect the NLO+PS spectrum.
Among these, those due to scale, PDFs, and choice of top-quark mass 
have been studied by the experimental collaborations, and shown to 
be smaller than the disagreement between data and 
theory~\cite{Chatrchyan:2012saa}. Here, we
therefore concentrate on other sources of systematics. One of these
is due to missing higher orders, since the NLO+PS predictions used
by the experiments include only up to ${\cal O}(\as^3)$ terms, namely
$t\bt+0j$ samples at the NLO. While the impact of missing higher orders
in perturbation theory is estimated by scale variations, an important
and independent check of this assessment may be obtained by considering
NLO-merged prections. In the left panel of fig.~\ref{fig:pttop} we thus
compare the unmerged $t\bt+0j$ prediction with the FxFx one,
where the $t\bt+0j$ and $t\bt+1j$ samples are combined with $\mu_Q=100$~GeV.
Both merged and unmerged results have been obtained with \HWs,
by setting the collider energy equal to 8~TeV;
the latter curve has been rescaled\footnote{If visually that may not
seem to be the case, it is because the bin widths are not equal: note
that the cross section is differential.} in order for its visible integral to
coincide with that of the former (since in this case we are specifically
interested in a shape comparison: in absolute values, the two cross 
sections differ however by only 2.5\%). As one
can gather from the plot, the two predictions are close to
each other; the FxFx prediction is sightly softer than the unmerged
one, but this does not appear to be sufficient to bring it in
agreement with the CMS measurement. The variation
of the merging scale in a large range ($30$--$155$~GeV) does not induce
any significant change. It therefore appears that the systematics due
to higher-order corrections are fully under control, since scale variations
and NLO-merging give consistent results for this observable, and we thus
confirm the previous findings that it cannot explain the discrepancy
between theory and CMS data.
Another source of theoretical systematics in NLO+PS predictions is
that due to the choice of the PSMC. This is presented in the right
panel of fig.~\ref{fig:pttop}, where we compare the ($t\bt+0j$ unmerged)
predictions obtained with \PYe\ (cyan), \HWs\ (red), and \PYs($Q^2$)
(grey); the lower inset presents the bin-by-bin ratios of the latter
two predictions over that of \PYe. It is clear from the plot that
the three PSMCs are amply consistent with each other; this 
is expected, since the pure-NLO result for $\pt(t)$ must not
be dramatically modified by shower effects; such an expectation is
confirmed by the fNLO prediction, also reported in the right panel
of fig.~\ref{fig:pttop} as the dashed histogram overlaid with full circles,
which is in fair agreement with all the other curves. We therefore conclude
that it does not seem possible to get NLO+PS predictions to agree
with CMS data by changing the PSMC used in the simulations.
We also point out that this statement by no means implies that
acceptance corrections, which we do not compute here, are PSMC-independent;
a careful investigation of these may be necessary should the discrepancy
discussed here persist, given that $\pt(t)$ is not a quantity that
can be measured directly.

Inclusive quantities that stem from either or both tops in $t\bt$
events are an ideal testing ground for NLO+PS predictions, which should
give a good description of the data for absolute normalisation as well
as for shapes. On the other hand, $t\bt$ events are characterised by
large c.m.~energies which imply a large amount of QCD radiation
(only a tiny fraction of which originate from the top quarks, owing
to their being very massive). The study of such radiation and of its
topological properties is an interesting subject, particularly in view
of the large statistical accuracy that can be obtained at the LHC.
At variance with the case of inclusive observables, there is no reason
to expect that {\em all} radiation-related observables will be
well described by NLO+PS $t\bt+0j$ predictions. In particular, those
for which large jet multiplicities are important should be sensibly 
compared only to merged results, or at least to unmerged samples whose
underlying matrix elements feature a sufficiently large number of
hard partons -- the most obvious example is that of $N_{jet}$.
On the other hand, for radiation-related observables which are also
inclusive enough, NLO+PS simulations should do relatively well.
One such case is that of the so-called gap fractions, which have
been measured in the dilepton channel by both ATLAS 
(at 7 TeV~\cite{ATLAS:2012al}) and CMS (at 7~\cite{CMSgf7tev}
and 8 TeV~\cite{CMSgf8tev}), and compared to various theoretical
predictions -- \mcatnlo\ with \HWs, POWHEG with both \HWs\ and \PY,
\MadGraph\ with \PY, \Alpgen\ with \HWs, and \Sherpa. There are different
levels of agreement among generators and with the data, whose discussion
is outside the scope of this work; here, we concentrate on \mcatnlo, since
the relevant formalism is the same as that used in NLO+PS simulations in
\aNLO. Before going into the details, let us define the main quantity
that we shall study, namely the gap fraction for the $\pt$ of the 
hardest jet. In order to be definite, we shall use the same setup
as in ref.~\cite{CMSgf8tev} (which is CMS's at the 8~TeV LHC): jets are 
reconstructed with the anti-$\kt$ algorithm with $R=0.5$, and the 
following cuts are imposed:
\beq
\pt(\ell)\ge 20~{\rm GeV}\,,\;\;\;\;
\abs{\eta(\ell)}\le 2.4\,,\;\;\;\;
\pt(j_b)\ge 30~{\rm GeV}\,,\;\;\;\;
\abs{\eta(j_b)}\le 2.4\,,
\label{CMSgfcuts}
\eeq
on both charged leptons, and on the two hardest $b$-jets. 
Other type of cuts (e.g.~lepton isolation) are seen to be unimportant, 
and are not imposed here. We then define:
\beq
{\rm GF}_{\pt(j_1)}(Q)=\frac{1}{\sigma}\int d\Phi\,
\stepf\left(Q-\hat{p}_{{\sss T}}(j_1)\right)\frac{d\sigma}{d\Phi}\,,
\label{GFdef}
\eeq
where $\sigma$ is the cross section within the cuts of
eq.~(\ref{CMSgfcuts}), and for the notation of the argument 
of the gap fraction we adopt one similar to that of ref.~\cite{ATLAS:2012al}, 
which is not liable to generate confusion. We have also introduced:
\beqn
\hat{p}_{{\sss T}}(j_1)=\left\{
\begin{array}{ll}
\pt(j_1)        &\phantom{aaaa} \pt(j_1)\ge 30~{\rm GeV}~~{\rm and}~~
                           \eta_{\min}\le\abs{\eta(j_1)}\le\eta_{\max}\,,\\
0               &\phantom{aaaa} {\rm otherwise}\,,\\
\end{array}
\right.
\label{phtdef}
\eeqn
with $j_1$ the hardest jet which is neither of the two $b$-jets on which
the cuts of eq.~(\ref{CMSgfcuts}) are applied. Note that if there is
no jet harder than $30$~GeV in the pseudorapidity interval
\mbox{$(\eta_{\min},\eta_{\max})$}, eq.~(\ref{phtdef}) implies that
the $\stepf$ function in eq.~(\ref{GFdef}) is identically equal to one. 
Therefore, ${\rm GF}_{\pt(j_1)}(Q)$ is a constant for $Q<30$~GeV, equal 
to the fraction of events which do not have any jet harder than $30$~GeV in 
the relevant pseudorapidity interval: for this reason, gap fractions are 
not displayed in this range. The quantity \mbox{$\hat{p}_{{\sss T}}(j_1)$}
in eq.~(\ref{GFdef}) can be replaced by a function, defined analogously to
what is done in eq.~(\ref{phtdef}) in terms of any observable ${\cal O}$
with mass dimension equal to one; in this way, one constructs a different
type of gap fraction, ${\rm GF}_{{\cal O}}(Q)$. The transverse momentum
of the second-hardest jet~\cite{CMSgf8tev}, and 
$\Ht$~\cite{ATLAS:2012al,CMSgf8tev} have been considered in the literature;
they are rather strongly correlated with $\pt(j_1)$, and will not be
investigated any further here.

\begin{figure}[h]
 \begin{center}
 \epsfig{file=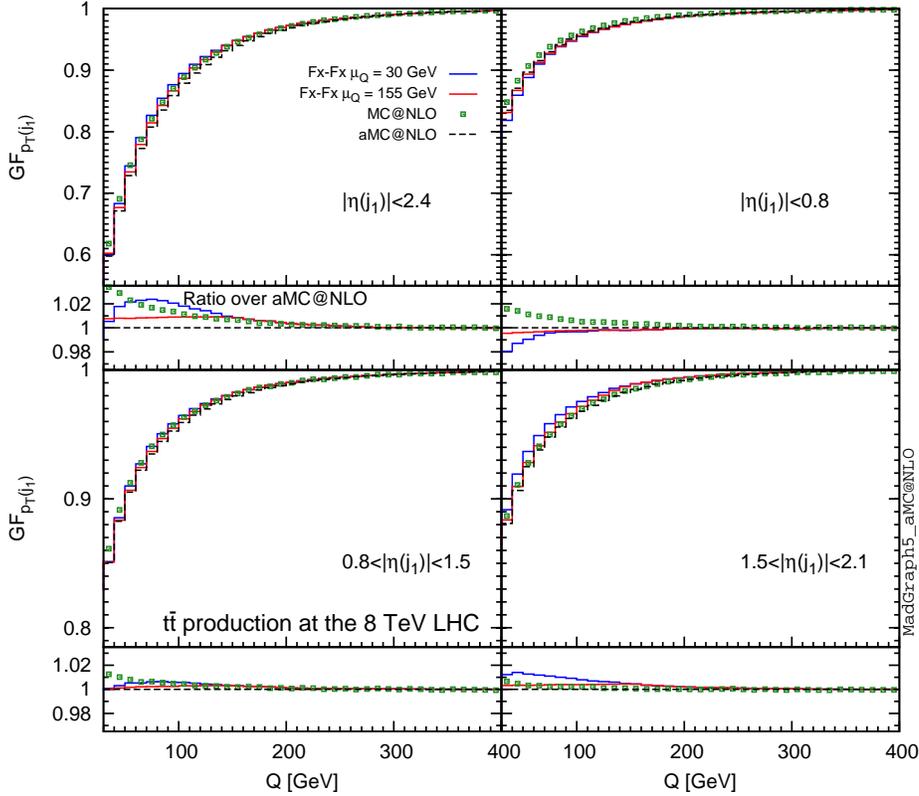, width=0.8\textwidth}
 \end{center}
  \caption{Gap fraction for the $\pt$ of the hardest jet, in four
different pseudorapidity intervals, as predicted by FxFx-merged, 
unmerged (labelled \amcatnlo), and \mcatnlo\ simulations. 
The setup follows closely that of ref.~\cite{CMSgf8tev}. 
See the text for details.
}
\label{fig:g1j}
\end{figure}
We now aim to compare the NLO+PS predictions of the \mcatnlo\ program (v4.09)
with those obtained with \aNLO\ (both NLO+PS and FxFx-merged), which is 
interesting in two respects. Firstly, the unmerged NLO+PS results
of the two codes would be identical, were it not
for the fact that in the latter we have included the effect of the function
$D$ (see eq.~(\ref{Ddef})); therefore, any difference between the two
is the signal of matching systematics\footnote{There may be other 
very tiny differences between \mcatnlo\ and \aNLO\ NLO+PS predictions,
owing to possibly non-identical choices of parameters,
which are negligible for all purposes.}. Secondly, the comparison 
between unmerged $t\bt+0j$ and FxFx simulations helps assess the 
impact of the inclusion of matrix elements of higher order in the latter.
We have adopted \HWs\ as PSMC, and the FxFx-merged simulations have
been performed for two extreme choices of the merging scale,
$\mu_Q=30$ and 155~GeV. The results for the hardest-jet gap fraction 
are presented in fig.~\ref{fig:g1j}, for the same four pseudorapidity
intervals \mbox{$(\eta_{\min},\eta_{\max})$}$=$\mbox{$(0,2.4)$}, 
\mbox{$(0,0.8)$}, \mbox{$(0.8,1.5)$}, and \mbox{$(1.5,2.1)$}
as in ref.~\cite{CMSgf8tev}; the insets display the ratios of the
\mcatnlo\ v4.09 and FxFx-merged results over the unmerged \aNLO\ ones.
We first observe that all predictions
are quite close to each other, the largest deviation being about 3\%.
Interestingly, some of the largest relative differences are between the
two unmerged predictions, which implies that the matching systematics
is not negligible. In general, \aNLO\ predicts more jet activity (i.e., a
lower curve) at NLO+PS than \mcatnlo\ v4.09 in the central and widest 
pseudorapidity regions,
this difference decreasing with increasing $\eta_{\min}$. This appears
to be fairly consistent with what is seen in the 8-TeV CMS data  
(compare the upper left corner of fig.~\ref{fig:g1j} with fig.~7 in
ref.~\cite{CMSgf8tev}, and the other panels of fig.~\ref{fig:g1j} 
with the upper row of fig.~8 in ref.~\cite{CMSgf8tev}), whose analysis
we have followed here. A similar trend as NLO+PS of \aNLO\ 
is seen in the FxFx-merged
result with $\mu_Q=155$~GeV, while that obtained with $\mu_Q=30$~GeV
follows a slightly different pattern. Although these findings are
encouraging, it is certaintly too early to draw any firm conclusions;
preliminarly, we can observe that the inclusion of more matrix-elements
results into matched predictions seems to be beneficial (be either through
an NLO-merging procedure, or because the $D$ function limits the impact
of the \HWs\ shower to smaller scales than it happens in \mcatnlo).
Furthermore, even by choosing an extremely large range for the merging
scale, the FxFx systematics is smaller than that of the
MLM-type merging in \aNLO\ (see the results labelled \MadGraph\ in
ref.~\cite{CMSgf8tev}).
We shall in fact see below another clear example of the pattern of the 
reduction of the merging systematics when going from the LO to the NLO.

\vskip 0.4truecm
\noindent
{\bf $\blacklozenge$ Multi-parton merged predictions}

\noindent
We now turn to illustrate some results of the FxFx merging which 
are directly relevant to the unitarity and the merging-scale-choice
arguments which have been discussed in general at the end of
sect.~\ref{sec:FxFx}. We shall do so by using an example which
one expects to be critical from these viewpoints, namely Higgs
production in gluon-gluon fusion (in HEFT) at the 8~TeV LHC, since 
such a process is characterised by very large higher-order corrections 
and by a very significant amount of radiation in PSMCs. As was done
in ref.~\cite{Frederix:2012ps}, we shall also compare to the
predictions obtained with \Alpgen, which we shall take as a benchmark
for the typical behaviour of LO merging procedures; we have used 
\HWs\ as PSMC.
\begin{table}
\begin{center}
\begin{tabular}{c|c|cccc}
\toprule
 &  & {$\mu_Q=20$~GeV} & {$\mu_Q=30$~GeV} & 
      {$\mu_Q=50$~GeV} & {$\mu_Q=70$~GeV} \\\hline
{no cuts} & 
      \begin{tabular}{c} {FxFx}\\
                         {\Alpgen}
      \end{tabular} 
    & \begin{tabular}{c} {14.47($-$0.6\%)}\\
                         { 8.84($-$0.9\%)}
      \end{tabular} 
    & \begin{tabular}{c} {14.56}\\
                         { 8.92}
      \end{tabular} 
    & \begin{tabular}{c} {14.77($+$1.5\%)}\\
                         { 9.08($+$1.8\%)}
      \end{tabular} 
    & \begin{tabular}{c} {14.78($+$1.5\%)}\\
                         { 9.07($+$1.7\%)}
      \end{tabular} 
    \\\hline
{two jets}  & 
      \begin{tabular}{c} {FxFx}\\
                         {\Alpgen}
      \end{tabular} 
    & \begin{tabular}{c} {1.65($+$0.8\%)}\\
                         {1.27($+$13.2\%)}
      \end{tabular} 
    & \begin{tabular}{c} {1.63}\\
                         {1.12}
      \end{tabular} 
    & \begin{tabular}{c} {1.60($-$2.4\%)}\\
                         {1.01($-$9.5\%)}
      \end{tabular} 
    & \begin{tabular}{c} {1.55($-$5.4\%)}\\
                         {0.92($-$18.4\%)}
      \end{tabular} 
    \\
\bottomrule
\end{tabular}
\end{center}
\caption{\label{tab:ggHres}
Total rates (in pb) for single-Higgs production in gluon-gluon fusion 
in HEFT, resulting from FxFx-merged (with up to two extra partons at 
the NLO) and \Alpgen\ (with up to three extra partons at the LO) samples.
Fractional differences w.r.t.~the corresponding results obtained with 
$\mu_Q=30$~GeV are also reported.
}
\end{table}
We start by presenting in table~\ref{tab:ggHres} the results
for the fully inclusive rates, both in the absence of cuts
(upper two rows), and by imposing the presence of at least two 
jets (lower two rows): the latter are defined by means of the
anti-$\kt$ algorithm with $R=0.4$, and have to obey the following 
conditions:
\beq
\pt(j)\ge 25~{\rm GeV}\,,\;\;\;\;\;\;
\abs{\eta(j)}\le 5\,.
\label{ggHcuts}
\eeq
The rates have been obtained by considering four different values
for the merging scale, which cover the very large range 
\mbox{$\mu_Q\in (20,70)$~GeV}. In order to be definite, we shall
take $\mu_Q=30$~GeV as our central value; in table~\ref{tab:ggHres},
we report in parenthesis the fractional difference of all results obtained 
with $\mu_Q\ne 30$~GeV w.r.t.~those obtained with $\mu_Q=30$~GeV that
appear in the same row. The rates in absence of jet cuts are seen
to be extremely stable against merging-scale variations. LO results
have an only marginally-larger $\mu_Q$ dependence, which is in any
case much smaller than the scale uncertainty (not shown here); the
same applies to NLO predictions. The NLO fully-inclusive rate for
the unmerged $H+0j$ sample is $13.40$~pb: it is therefore from
8\% to 10\% lower than the FxFx-merged results. Thus, despite the
fact that FxFx does not impose any unitarity condition, the merged
predictions are ``naturally'' quite close to the unmerged one.
\begin{figure}[h]
 \begin{center}
 \epsfig{file=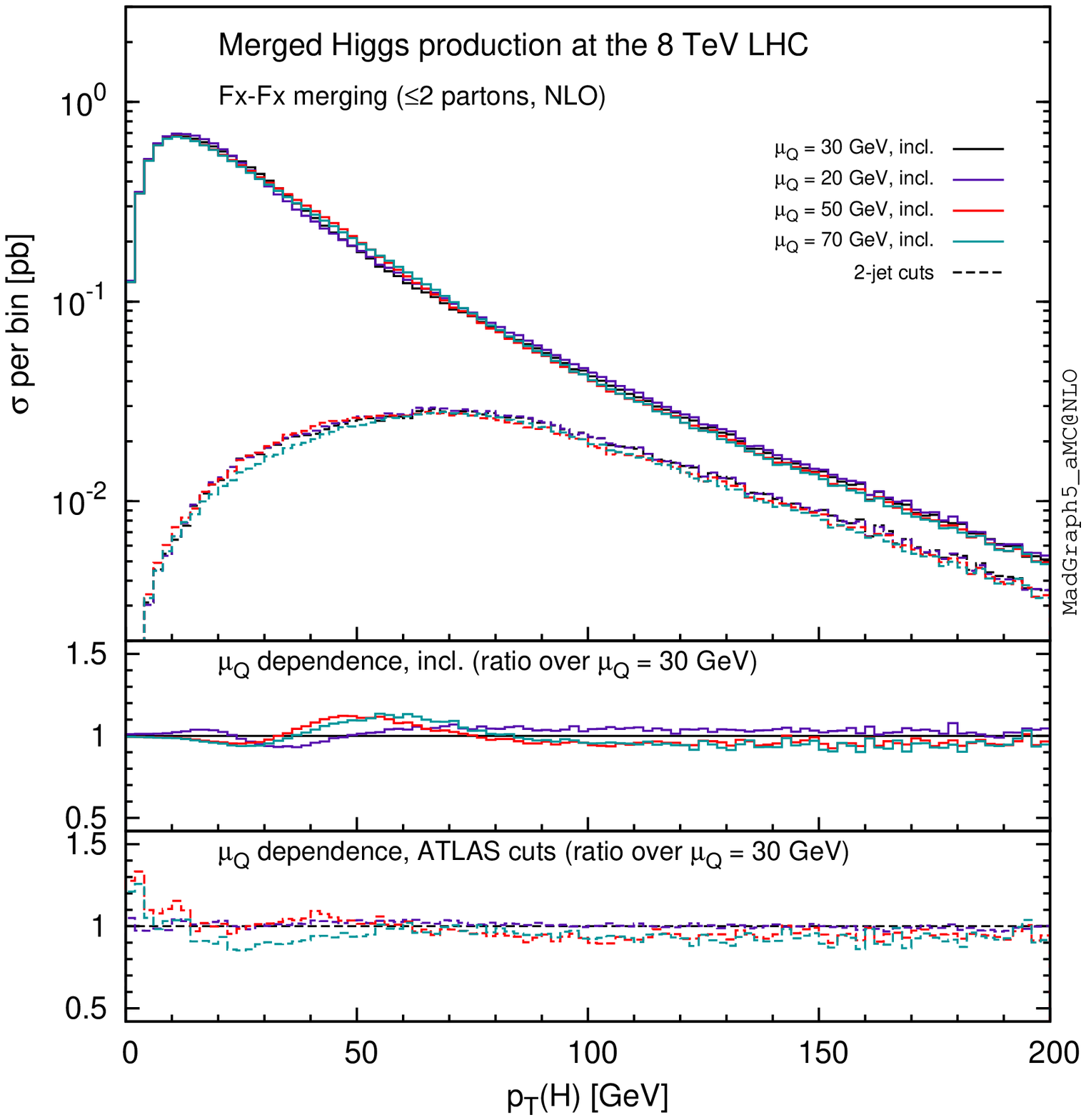, width=0.48\textwidth}
 \epsfig{file=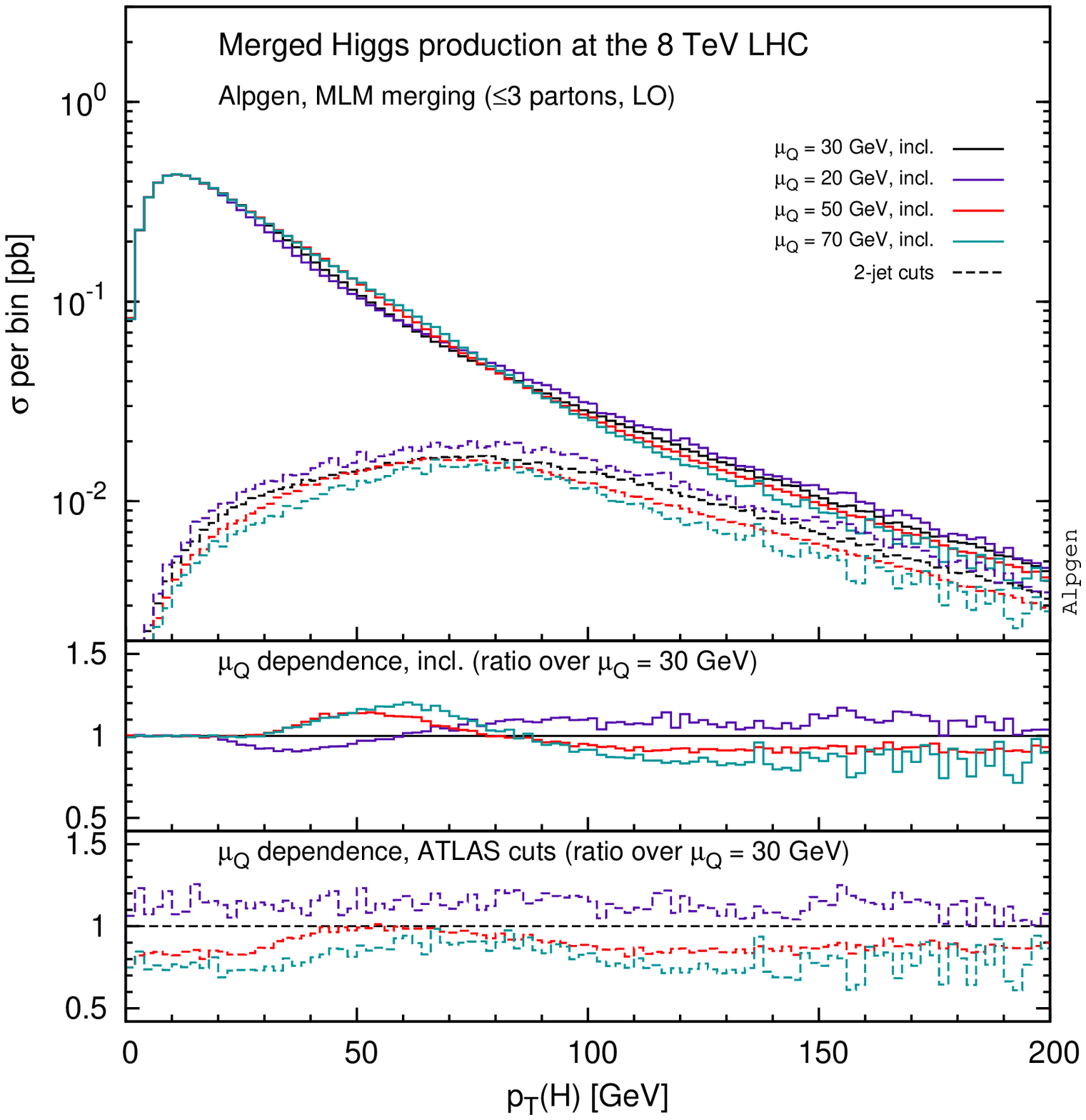, width=0.48\textwidth}
 \end{center}
  \caption{Higgs transverse momentum in single-Higgs production 
(gluon-gluon fusion in HEFT), as predicted by FxFx (left panel)
and \Alpgen\ (right panel), for various choices of the merging scale $\mu_Q$.
}
\label{fig:FxFxalp}
\end{figure}
They are not equal, nor they should be -- at the end of the day,
we are including here contributions up to ${\cal O}(\as^5)$.
The increase in the cross section when passing from unmerged
to merged results is rather consistent with expectations based
on perturbative scaling, and the known large NNLO/NLO $K$ factor that
characterises the Higgs-production mechanism we study here. We stress
that this feature is not an accident of this process, since we have 
observed it in all cases studied so far (see later for further examples).
When considering rates obtained within the cuts of eq.~(\ref{ggHcuts}),
we see that the NLO results are still quite stable, while the 
merging-scale dependence of the LO ones becomes sizable. In order
to further investigate this matter, we present in fig.~\ref{fig:FxFxalp}
the Higgs transverse momentum spectra, obtained with FxFx and \Alpgen\
for the same four merging scales as before; the insets display the
ratios of the various curves over the $\mu_Q=30$~GeV corresponding ones.
Obviously, the $\mu_Q$-dependence pattern reflects that of 
table~\ref{tab:ggHres}. However, it is interesting to see that
when no jet cuts are applied there is a significant compensation
in terms of rates in the \Alpgen\ curves (this can be clearly seen in
the ratio plot, with the presence of a crossing point at a $\pt(H)$ 
of about 70~GeV) -- in other words, the very small $\mu_Q$ dependence
of the LO total rates is partly an artifact, since locally in the phase 
space the various predictions differ by a larger amount. To some extent, 
the same is true for the FxFx-merged results, but the effects are much
more modest there. Note that, when jet cuts are applied, this phenomenon
does not occur any longer, and local and global merging-scale dependences
are quite similar: in particular, one sees how all the NLO curves are
close to each other also within these cuts. The significance of this
(in)dependence is tightly related to the range chosen for the merging
scale variation. The function \mbox{$\log(\mu_Q/m_H)$}, which one may
take as an indicator of the typical quantity relevant when the merging
scale is varied, changes by a factor of $3.16$ in the range 
\mbox{$(20,70)$~GeV} considered here; we believe that
this is a sufficiently-large range to give a sensible
indication of merging-scale systematics. As the results presented here
clearly show, the supposed spoiling of some underlying NLO accuracy
that occurs when ``large'' values of $\mu_Q$ are adopted is simply not
an issue if NLO and MC predictions are properly merged, and are reasonably
consistent with each other. The latter is in fact a key point: we 
have observed that, by imposing VBF-type cuts, e.g.:
\beq
M_{j_1j_2}\ge 400~{\rm GeV}\,,\;\;\;\;\;\;
\abs{\Delta y_{j_1j_2}}\ge 2.8\,,
\label{ggHVBF}
\eeq
the mild dependences shown in table~\ref{tab:ggHres} become huge
(of the order of 80\% and 70\% at the LO and NLO respectively).
It is clear that the invariant-mass cut of eq.~(\ref{ggHVBF}) introduces
a third scale in the game which renders its treatment a complicated
matter. Given that such a large merging-scale dependence is basically
driven by the largest $\mu_Q$'s, the problem is likely due to intrinsic
differences between the PSMC and matrix-element descriptions of
the VBF region. However, we are able to immediately notice
this {\em only} because we have considered a relatively large range
of $\mu_Q$, which does not give any issues for sufficiently inclusive
quantities, but it does when VBF cuts are applied. We conclude by
pointing out that the uncovering of this issue by means of merging-scale
systematics does not imply its most naive solution, which would be
that of restricting, to small values, the range of $\mu_Q$ in this
kinematic region, thus relying on a matrix-element-dominated description: 
in fact, such a description is not necessarily better than a PSMC one 
in the context of the multi-scale dynamics induced by eq.~(\ref{ggHcuts})
and~(\ref{ggHVBF}).

\begin{table}
\begin{center}
\begin{tabular}{c|ccc|c}
\toprule
 &  {$\mu_Q^{(\downarrow)}$} & {$\mu_Q^{(c)}$} & 
    {$\mu_Q^{(\uparrow)}$}   & unmerged \\\hline
$pp\to e^+\nu_e$ 
    & 7059($-$0.9\%)
    & 7121
    & 7160($+$0.5\%)
    & 7067($-$0.8\%)
    \\
$pp\to ZZ$ 
    & 7.383($-$0.01\%)
    & 7.384
    & 7.387($+$0.04\%)
    & 7.355($-$0.4\%)
    \\
$pp\to He^+\nu_e$ 
    & 0.05180 ($+$0.9\%)
    & 0.05131
    & 0.05117($-$0.2\%)
    & 0.05066($-$1.3\%)
    \\
\bottomrule
\end{tabular}
\end{center}
\caption{\label{tab:resFxFx}
Total rates (in pb) for three processes, computed with FxFx by using
three different merging scales (whose values are process-dependent,
see eqs.~(\ref{sc1})--(\ref{sc3})), and with the unmerged 
lowest-multiplicity samples.
Relative differences w.r.t~the FxFx results obtained with the central
merging scales are also reported.
}
\end{table}
\begin{figure}[h]
 \begin{center}
 \epsfig{file=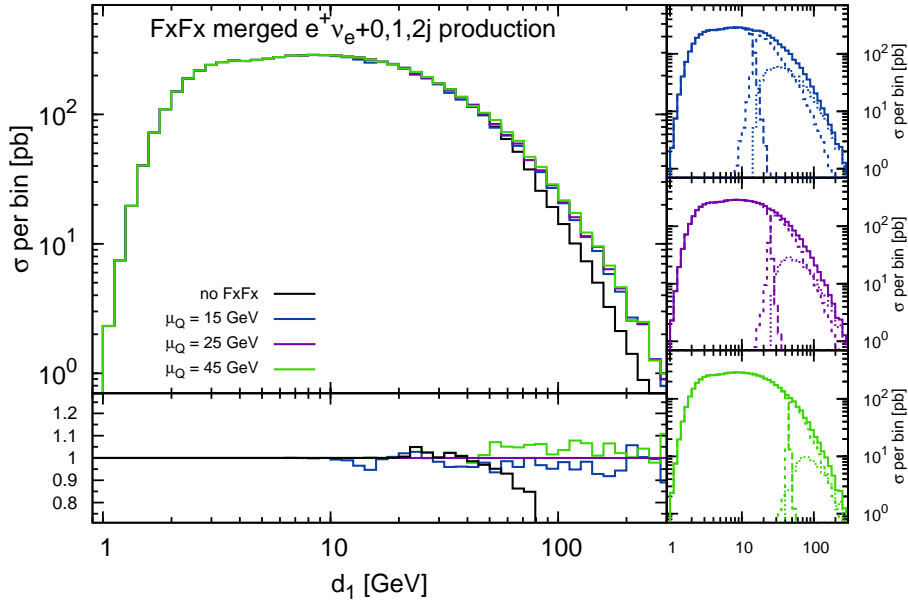, width=0.75\textwidth}
 \end{center}
\caption{One-jet rate in $e^+\nu_e$ production. The main frame presents
the three FxFx-merged predictions as well as the unmerged one. The lower
inset displays the ratios of these curves over the central FxFx-merged
one. The insets to the right show the separate contributions of the
unphysical $0$- (long-dashed), $1$- (dashed), and $2$-parton (dotted) 
samples, for the three merging scales.}
\label{fig:w012d01}
\end{figure}
\begin{figure}[h]
 \begin{center}
 \epsfig{file=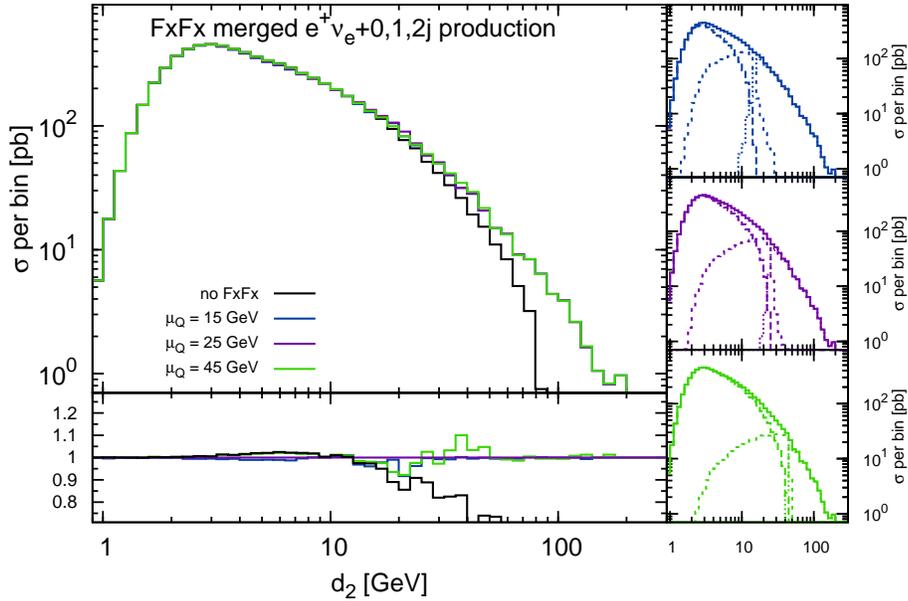, width=0.75\textwidth}
 \end{center}
\caption{As in fig.~\ref{fig:w012d01}, for the
two-jet rate in $e^+\nu_e$ production.}
\label{fig:w012d12}
\end{figure}
In order to further the previous arguments, and as a way to validate
the FxFx merging procedure with a special attention to cases where the 
construction of the Sudakovs that enter eqs.~(\ref{SudiS})--(\ref{Delsig})
is involved owing to the flavour structure of the hard process, we
consider here three different final states, namely $e^+\nu_e$, $ZZ$,
and $He^+\nu_e$, which we simulate at the 8~TeV LHC. In FxFx we
include the $0$-, $1$-, and $2$-parton samples for the former process,
and the $0$- and $1$-parton samples for the latter two. In all cases,
the unmerged $0$-parton results are also presented. All merged and unmerged
samples are showered with \HWs. We start with the fully-inclusive rates, 
reported in table~\ref{tab:resFxFx}; the values of the three merging 
scales are as follows:
\beqn
(\mu_Q^{(\downarrow)},\mu_Q^{(c)},\mu_Q^{(\uparrow)})
&=&(15,25,45)~{\rm GeV}\phantom{aaaaaaa}pp\to e^+\nu_e\,,
\label{sc1}
\\
&=&(45,65,105)~{\rm GeV}\phantom{aaaaaa}pp\to ZZ\,,
\label{sc2}
\\
&=&(50,75,100)~{\rm GeV}\phantom{aaaaaa}pp\to He^+\nu_e\,,
\label{sc3}
\eeqn
which cover quite wide ranges. The message emerging from the 
table is analogous to that relevant to Higgs production, which
we have discussed previously: the merging-scale dependence is
very small. Furthermore, the unmerged results are extremely
close to the merged one, in fact much closer than in the case
of $gg\to H$; this is a natural consequence of the relatively small
(compared to Higgs) $K$ factors for the present processes.
\begin{figure}[h]
 \begin{center}
 \epsfig{file=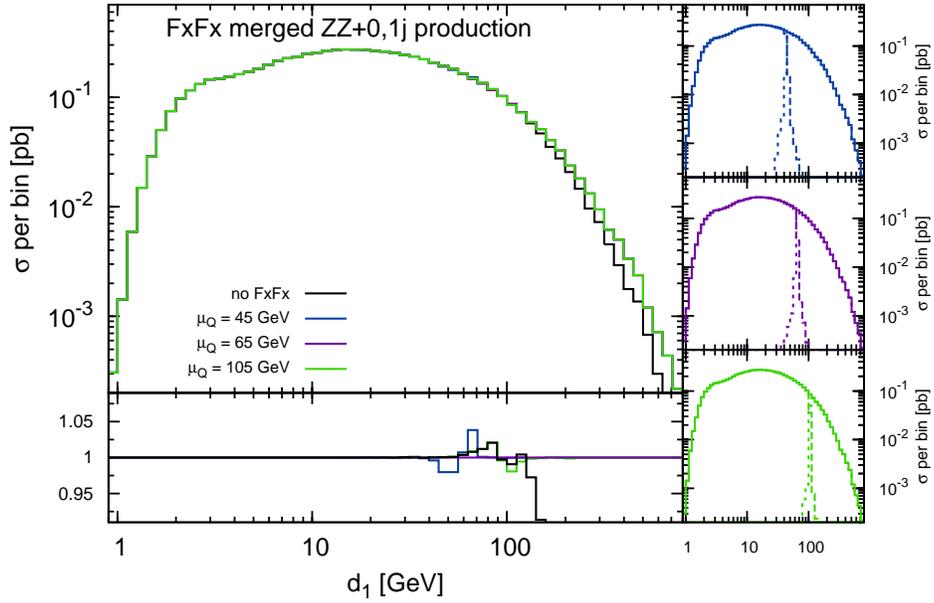, width=0.75\textwidth}
 \end{center}
\caption{As in fig.~\ref{fig:w012d01}, for the
one-jet rate in $ZZ$ production. Only the $0$- and $1$-parton
samples have been considered here.}
\label{fig:zz01d01}
\end{figure}
\begin{figure}[h]
 \begin{center}
 \epsfig{file=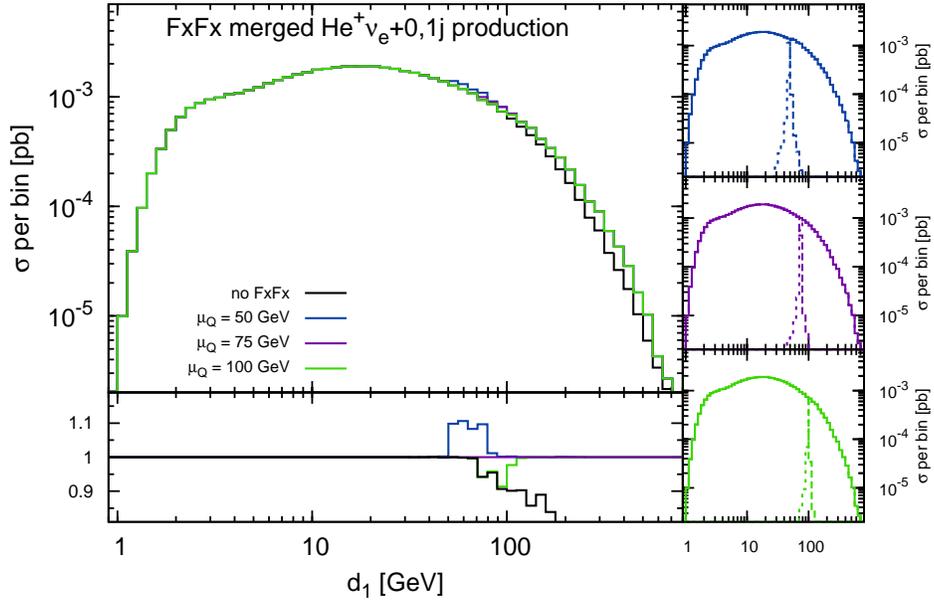, width=0.75\textwidth}
 \end{center}
\caption{As in fig.~\ref{fig:w012d01}, for the
one-jet rate in $He^+\nu_e$ production. Only the $0$- and $1$-parton
samples have been considered here.}
\label{fig:hw01d01}
\end{figure}
Again, this fact emerges naturally, without the need to impose
any unitarity condition in the merging. 
We conclude by showing some selected differential distributions,
and in particular the $j$-jet rates of the $\kt$ 
algorithm (denoted by $d_j$, see eq.~(\ref{dord}));
these quantities are known to be critical in the context of merging
procedures, since they are very sensitive to artefacts of the latter,
which show up as discontinuities in the spectra. Our results are
presented in figs.~\ref{fig:w012d01}--\ref{fig:hw01d01}, where
the main frames display the predictions in absolute values.
The lower insets display the ratios of the FxFx-merged and unmerged 
predictions over the FxFx one obtained with the central merging-scale
values. The insets at the right of the figures show the way in which
the $i$-parton samples combine in order to give the physical curves.
All results for all processes behave as expected.

\subsection{One-loop SM and BSM results: a look ahead\label{sec:next}}
All the results presented in sects.~\ref{sec:tot} and~\ref{sec:diff}
exploit the one-loop computations performed by the public version of \MLf, 
which amply demonstrates the reach and flexibility of this code. The aim of 
this section is on the other hand that of showing that the current,
still-private \MLf\ program has a much larger scope, being able to handle
computations of very high complexity also in the context of a mixed-coupling 
expansion (see sect.~\ref{sec:NLO}), and in theories other than the SM.
As was already discussed, while the corresponding capabilities
on the real-emission side have not yet been fully validated in \aNLO,
they do not pose any problem of principle, and only minor ones
from the technical point of view; therefore, the results presented
below constitute the proof that the major obstacles have been cleared
which prevent \aNLO\ from performing NLO computations in arbitrary
renormalisable theories.

We shall give here benchmark results for given phase-space
configurations. They will be presented in the form of the
coefficients $c_j$, $j=-2,-1,0$, introduced in eq.~(\ref{Vexpr})
for $V$, the colour- and helicity-summed virtual amplitude contracted
with the corresponding Born one (see eq.~(\ref{Vdef})). We shall
also denote by $a_0$ the Born amplitude squared:
\beq
a_0=\mathop{\sum_{\rm colour}}_{\rm spin}\abs{\ampnt}^2\,,
\eeq
where, similarly to $V$, the averages over initial-state colour and 
spin degrees of freedom are understood. Throughout this section, we 
have set $\mu\equiv Q=\muF=\muR=\sqrt{s}$, and all particles widths equal
to zero for simplicity; the leptons that circulate in the loops
are taken to be massless.
In order to maximize the numerical accuracy, the computations reported here
have been performed by using quadruple-precision arithmetics; the stability 
tests described in sect.~\ref{sec:OPP} have shown that these results are
numerically stable beyond the seventeen digits quoted below.
The coefficients $c_j$ are computed in the 't Hooft-Veltman scheme and,
in the case of the pole residues $c_{-2}$ and $c_{-1}$, compared
to their analytically-known forms (see e.g.~eq.~(B.2) of 
ref.~\cite{Frederix:2009yq}, whose generalisation to cases other 
than QCD is straighforward). Units for quantities of canonical dimension
equal to one are understood to be GeV. The CKM matrix is diagonal.
The integral-reduction OPP procedure has been adopted in all cases;
processes {\em C)} and {\em D)} have also been computed with TIR
(using IREGI), and perfect agreement with the OPP results has been
found. We point out that processes {\em A)} and {\em B)} feature
8- and 7-point rank-5 loop integrals respectively, and are therefore 
beyond the present capabilities of any TIR library.
Further technical details relevant to the calculations presented 
in this section are given in appendix~\ref{sec:MLperf}.
We emphasise that the one-loop results shown here have
never been presented in the literature, and can serve as
benchmarks for comparisons with future independent computations.

\vskip 0.4truecm
\noindent
{\em A)~~}{\bf High-multiplicity multi-scale QCD process:
$gg \to d \bar{d} b \bar{b} t \bar{t}$}

\noindent
This process involves up to 8-point loop diagrams with three external scales:
the top- and bottom-quark masses, and the partonic c.m.~energy. A total of
54614 loop diagrams contribute, and all pure-QCD UV renormalisation
counterterms are necessary, which makes it an excellent test case
for \MLf. The following parameters have been used:
\begin{center}
\begin{tabular}{cl|cl}\toprule
Parameter & value & Parameter & value
\\\midrule
$\as$ & \texttt{0.118} & $n_{lf}$ & \texttt{4}
\\
$m_{t}$ & \texttt{173.0}    & $m_{b}$ & \texttt{4.7}
\\\bottomrule
\end{tabular}
\end{center}
The kinematic configuration considered is (we use an $(E,p_x,p_y,p_z)$
notation):
\begin{footnotesize}
  \begin{align*}
    p_g            &=\textrm{( 500}&&                           \textrm{, \phantom{-}0}                               &&\textrm{, \phantom{-}0}                               &&\textrm{, \phantom{-}500}                           &&\textrm{)}\\    
    p_g            &=\textrm{( 500}&&                           \textrm{, \phantom{-}0}                               &&\textrm{, \phantom{-}0}                               &&\textrm{, -500}                                            &&\textrm{)}\\
    p_d             &=\textrm{( 159.884957663500}&&\textrm{, -100.187853644511}&&\textrm{, \phantom{-}83.9823400815702}&&\textrm{, \phantom{-}92.0465111972672}                  &&\textrm{)}\\   
    p_{\bar{d}}  &=\textrm{( 203.546206153656}&&\textrm{, -154.329441032052}&&\textrm{, -0.512510195103158}&&\textrm{, \phantom{-}132.714803257139}                  &&\textrm{)}\\
    p_b             &=\textrm{( 81.9036633616240}&&\textrm{, \phantom{-}4.56741073895954}&&\textrm{, -80.4386221767117}&&\textrm{, \phantom{-}13.9601895942747}                  &&\textrm{)}\\   
    p_{\bar{b}}  &=\textrm{( 41.5312244194448}&&\textrm{, \phantom{-}6.99982274816896}&&\textrm{, \phantom{-}9.96034329509376}&&\textrm{, \phantom{-}39.4277395334349}                  &&\textrm{)}\\
    p_t             &=\textrm{( 239.961310957973}&&\textrm{, \phantom{-}84.0110736983121}&&\textrm{, \phantom{-}18.3862699981019}&&\textrm{, -142.325385396572}                  &&\textrm{)}\\    
    p_{\bar{t}}  &=\textrm{( 273.172637443802}&&\textrm{, \phantom{-}158.938987491122}&&\textrm{, -31.3778210029510}&&\textrm{, -135.823858185543}                  &&\textrm{)}
  \end{align*}
\end{footnotesize}
For the ${\cal O}(\as^6)$ Born matrix element and ${\cal O}(\as^7)$ $V$ 
coefficients we have obtained:
\begin{center}
\begin{tabular}{cl}\toprule
  $g g \to d\bar{d} b\bar{b}t\bar{t}$ & \multicolumn{1}{c}{}
  \\\midrule
$a_0$ & \texttt{\phantom{-}1.7614866952133752e-14}
\\
$c_{0}$ & \texttt{\phantom{-}7.1888721656398052e-14}
\\
$c_{-1}$ & \texttt{-3.8948541926529643e-15}
\\
$c_{-2}$ & \texttt{-2.8670389920110557e-15}
\\\bottomrule
\end{tabular}
\end{center}
The relevant \aNLO-shell generation command is:

\noindent
~\prompt\ {\tt ~generate g g > d d\~{} b b\~{} t t\~{} [virt=QCD]}

\noindent
More details on this syntax can be found in appendix~\ref{sec:verbose}.
After the generation phase, the usual {\tt output} and {\tt launch}
commands are executed. We point out, however, than in the present context
(i.e., when one computes one-loop matrix elements pointwise), {\tt launch}
allows the user to specify the kinematic configuration for which the said
matrix elements are to be computed.

\vskip 0.4truecm
\noindent
{\em B)~~}{\bf Mixed-coupling expansion: 
$u \bar{d} \rightarrow d \bar{d} W^{+} Z H$}

\noindent
The Born matrix elements for this process receive contributions
at ${\cal O}(\as^2\aem^3)$, ${\cal O}(\as\aem^4)$, and ${\cal O}(\aem^5)$,
resulting from ${\cal O}(\gs^2 e^3)$ and ${\cal O}(e^5)$ amplitudes.
In the notation of sect.~\ref{sec:NLO}, this corresponds to
$k_0=5$, $c_s(k_0)=0$, $c(k_0)=3$, and $\Delta(k_0)=2$. We have
considered the full set of NLO corrections, thus obtaining four
terms of ${\cal O}(\as^n\aem^m)$, with \mbox{$0\le n\le 3$} and $m=6-n$
(see eqs.~(\ref{Delkpo}) and~(\ref{taylor4})). This process features
essentially all complications one faces in the case of a mixed QCD-EW
expansion, and in particular it tests fully the UV- and $R_2$-related
machinery, described in sect.~\ref{sec:OPP}, beyond the pure-QCD
cases considered so far. From a technical viewpoint, the major
challenges are represented by the fact that genuine EW corrections
(as opposed to QCD corrections to processes that may feature EW
external particles) significantly complicate the flavour structure
of the diagrams (whose number therefore grows and which always pose
a multi-scale problem), and by the necessity of keeping separate
track of the $\Sigma_{k_0+1,q}$ coefficients of eq.~(\ref{taylor4}).
We have performed our computation in the Feynman gauge, with
an \mbox{$(\aem(m_Z),m_Z,m_W)$} input scheme for the EW parameters,
and adopted the $\aem(m_Z)$ renormalisation scheme~\cite{Dittmaier:2001ay}
(we point out that the $G_{\mu}$ scheme\cite{Dittmaier:2001ay,Denner:1991kt}
is also available in \MLf). The full set of inputs is thus:
\begin{center}
\begin{tabular}{cl|cl}\toprule
Parameter & value & Parameter & value
\\\midrule
$\as$ & \texttt{0.118} & $n_{lf}$ & \texttt{5}
\\
$m_{t}$ & \texttt{173.0}    & $y_{t}$ & \texttt{173.0}
\\
$m_{W}$ & \texttt{80.419}    & $m_{Z}$ & \texttt{91.188}
\\
$m_{H}$ & \texttt{125.0}    & $\alpha^{-1}$ & \texttt{132.507}
\\\bottomrule
\end{tabular}
\end{center}
The kinematical configuration is:
\begin{footnotesize}
  \begin{align*}
    p_u            &=\textrm{( 500}&&                           \textrm{, \phantom{-}0}                               &&\textrm{, \phantom{-}0}                               &&\textrm{, \phantom{-}500}                           &&\textrm{)}\\    
    p_{\bar{d}} &=\textrm{( 500}&&                           \textrm{, \phantom{-}0}                               &&\textrm{, \phantom{-}0}                               &&\textrm{, -500}                                            &&\textrm{)}\\
    p_d             &=\textrm{( 77.3867935143263}&&\textrm{, -13.6335837243927}&&\textrm{, \phantom{-}33.7255664483738}                               &&\textrm{, -68.3039338032245}&&\textrm{)}\\    
    p_{\bar{d}}  &=\textrm{( 251.029839835656}&&\textrm{, -74.4940380485791}                 &&\textrm{, -235.871950829717}&&\textrm{, \phantom{-}42.7906718212678}                  &&\textrm{)}\\
    p_{W^{+}}             &=\textrm{( 139.739680522225}&&\textrm{, -81.0565319364851}&&\textrm{, -74.5408139008771}                               &&\textrm{, \phantom{-}30.5527158347332}&&\textrm{)}\\    
    p_{Z}  &=\textrm{( 382.164100735946}&&\textrm{, \phantom{-}208.038848497860}                 &&\textrm{, \phantom{-}298.200182616267}&&\textrm{, -74.3682536477996}                  &&\textrm{)}\\
    p_{H}             &=\textrm{( 149.679585391847}&&\textrm{, -38.8546947884028}&&\textrm{, -21.5129843340470}                               &&\textrm{, \phantom{-}69.3287997950232}&&\textrm{)}
  \end{align*}
\end{footnotesize}
The corresponding Born results are:
\begin{center}
\begin{tabular}{cl}\toprule
  $u \bar{d} \rightarrow d \bar{d} W^{+} Z H$ & \multicolumn{1}{c}{$a_0$} 
  \\\midrule
  $\mathcal{O}(\as^{2}\aem^{3})$ & \texttt{\phantom{-}2.8791434190645365e-16} 
  \\
  $\mathcal{O}(\as\aem^{4})$ & \texttt{-4.2378807039987007e-17} 
  \\
  $\mathcal{O}(\aem^{5})$ & \texttt{\phantom{-}5.8013051661550053e-18} 
\\\bottomrule
\end{tabular}
\end{center}
where the ${\cal O}(\as\aem^4)$ interference term happens to be negative,
while at one loop we obtain:
\begin{center}
\begin{tabular}{cll}\toprule
  $u \bar{d} \rightarrow d \bar{d} W^{+} Z H$ & \multicolumn{1}{c}{$\mathcal{O}(\as^{3}\aem^{3})$} & \multicolumn{1}{c}{$\mathcal{O}(\as^{2}\aem^{4})$} 
  \\\midrule
$c_{0}$ & \texttt{-4.9670212643498834e-17} & \texttt{\phantom{-}3.5197577360529166e-18} 
\\
$c_{-1}$ & \texttt{-1.0437771535958436e-16} & \texttt{\phantom{-}1.5619709675879874e-17} 
\\
$c_{-2}$ & \texttt{-2.8837935481452971e-17} & \texttt{\phantom{-}3.9757576347989499e-18}  
  \\\midrule
  & \multicolumn{1}{c}{$\mathcal{O}(\as\aem^{5})$} & \multicolumn{1}{c}{$\mathcal{O}(\aem^{6})$} 
  \\\midrule
$c_{0}$ & \texttt{\phantom{-}2.3220780285374270e-18} & \texttt{-1.4592469761033279e-18} 
\\
$c_{-1}$ & \texttt{-1.8146075843176133e-18} & \texttt{-5.0799804067050324e-21} 
\\
$c_{-2}$ & \texttt{-5.4147748433007504e-19} & \texttt{-5.4195415714279579e-21}  
\\\bottomrule
\end{tabular}
\end{center}
The relevant \aNLO-shell generation command is:

\noindent
~\prompt\ {\tt ~generate u d\~{} > d d\~{} w+ z h QCD=99 QED=99 
[virt=QCD QED]}

\noindent
where the {\tt QCD=99 QED=99} bit instructs \MLf\ to consider the corrections
to all Born-level contributions, and not only to the leading (QCD) 
${\cal O}(\as^2\aem^3)$ one, while 
the {\tt [virt=QCD QED]} syntax implies that both QCD and QED/EW corrections
need to be included. More details on this extended syntax can be found in 
appendix~\ref{sec:verbose}.

\vskip 0.4truecm
\noindent
{\em C)~~}{\bf Mixed-coupling expansion: 
$u \bar{u} \rightarrow d \bar{d} t \bar{t}$}

\noindent
Although the one-loop corrections to this process feature a smaller
number of diagrams than those relevant to \mbox{$u\bar{d}\to d\bar{d}W^{+}ZH$}
studied above, owing to the fact that the corresponding Born amplitudes 
have three quark lines the mixed-coupling expansion ladder depicted
in fig.~\ref{fig:corr} is wider: we have in fact $\Delta(k_0)=4$,
with $k_0=4$, $c_s(k_0)=0$, $c(k_0)=0$. 
The input-parameter and scheme choices are the same as those adopted
in the case of \mbox{$u\bar{d}\to d\bar{d}W^{+}ZH$} production, while
the kinematic configuration is:

\begin{footnotesize}
  \begin{align*}
    p_u            &=\textrm{( 500}&&                           \textrm{, \phantom{-}0}                               &&\textrm{, \phantom{-}0}                               &&\textrm{, \phantom{-}500}                           &&\textrm{)}\\    
    p_{\bar{u}}            &=\textrm{( 500}&&                           \textrm{, \phantom{-}0}                               &&\textrm{, \phantom{-}0}                               &&\textrm{, -500}                                            &&\textrm{)}\\
    p_d             &=\textrm{( 77.6887158960956}&&\textrm{, -19.3895923374881}&&\textrm{, \phantom{-}35.1636848900680}&&\textrm{, -66.5063572263756}                  &&\textrm{)}\\    
    p_{\bar{d}}  &=\textrm{( 288.053156184158}&&\textrm{, -91.1103505191485}&&\textrm{, -264.895455921162}&&\textrm{, \phantom{-}67.1112676698377}                  &&\textrm{)}\\
    p_t             &=\textrm{( 218.623451637725}&&\textrm{, -92.8925122931906}&&\textrm{, -85.7235692614867}&&\textrm{, \phantom{-}43.4702707482150}                  &&\textrm{)}\\   
    p_{\bar{t}}  &=\textrm{( 415.634676282022}&&\textrm{, \phantom{-}203.392455149827}&&\textrm{, \phantom{-}315.455340292580}&&\textrm{, -44.0751811916771}                  &&\textrm{)}
  \end{align*}
\end{footnotesize}
We thus obtain at the Born level:
\begin{center}
\begin{tabular}{cl}\toprule
  $u \bar{u} \rightarrow d \bar{d} t \bar{t}$ & \multicolumn{1}{c}{$a_0$} 
  \\\midrule
  $\mathcal{O}(\alpha_{s}^{4})$ & \texttt{\phantom{-}8.0443110796911884e-10} 
  \\
  $\mathcal{O}(\alpha_{s}^{3}\alpha)$ & \texttt{-4.1964024114099949e-11} 
  \\
  $\mathcal{O}(\alpha_{s}^{2}\alpha^{2})$ & \texttt{\phantom{-}3.2368049995513863e-11} 
  \\
  $\mathcal{O}(\alpha_{s}\alpha^{3})$ & \texttt{-7.9030872133243511e-13} 
  \\
  $\mathcal{O}(\alpha^{4})$ & \texttt{\phantom{-}1.8667390802029741e-13} 
\\\bottomrule
\end{tabular}
\end{center}
while the one-loop coefficients turn out to be:
\begin{center}
\begin{tabular}{cll}\toprule
  $u \bar{u} \rightarrow d \bar{d} t \bar{t}$ & \multicolumn{1}{c}{$\mathcal{O}(\alpha_{s}^{5})$} & \multicolumn{1}{c}{$\mathcal{O}(\alpha_{s}^{4}\alpha)$} 
  \\\midrule
$c_{0}$ & \texttt{\phantom{-}2.7744575300036875e-10} & \texttt{-6.1309409133299879e-11} 
\\
$c_{-1}$ & \texttt{-2.4891722473717473e-10} & \texttt{\phantom{-}5.1973614496390480e-12} 
\\
$c_{-2}$ & \texttt{-8.0573035150936874e-11} & \texttt{\phantom{-}3.1296167547367972e-12}  
  \\\midrule
  & \multicolumn{1}{c}{$\mathcal{O}(\alpha_{s}^{3}\alpha^{2})$} & \multicolumn{1}{c}{$\mathcal{O}(\alpha_{s}^{2}\alpha^{3})$} 
  \\\midrule
$c_{0}$ & \texttt{\phantom{-}1.2122291790182845e-11} & \texttt{-4.0611498141889722e-12} 
\\
$c_{-1}$ & \texttt{-8.6161115635612362e-12} & \texttt{\phantom{-}4.3209683736654367e-15} 
\\
$c_{-2}$ & \texttt{-3.1860291930204890e-12} & \texttt{\phantom{-}3.5961341456741816e-14}  
  \\\midrule
  & \multicolumn{1}{c}{$\mathcal{O}(\alpha_{s}\alpha^{4})$} & \multicolumn{1}{c}{$\mathcal{O}(\alpha^{5})$} 
  \\\midrule
$c_{0}$ & \texttt{-3.8642357648130340e-14} & \texttt{-1.1866388556893426e-14} 
\\
$c_{-1}$ & \texttt{-3.6050223887148020e-14} & \texttt{-4.7983631557836333e-16} 
\\
$c_{-2}$ & \texttt{-1.7642824564621470e-14} & \texttt{-2.4912793041300221e-16}
\\\bottomrule
\end{tabular}
\end{center}
The relevant \aNLO-shell generation command is:

\noindent
~\prompt\ {\tt ~generate u u\~{} > d d\~{} t t\~{} QCD=99 QED=99 
[virt=QCD QED]}

\vskip 0.4truecm
\noindent
{\em D)~~}{\bf A BSM case study: QCD corrections to 
$gg\rightarrow\tilde{t}_1\tilde{t}_1^\star g$}

\noindent
While \ML\ has been used to compute one-loop corrections to a very
large number of processes in the SM, its applications to other theories
have been pretty limited so far. Here, we present the first \MLf\ results
in a fully-fledged BSM model; namely, we compute QCD corrections
to $\tilde{t}_1\tilde{t}_1^\star g$ production in the MSSM.
From the technical point of view, the MSSM \UFO\ model at the NLO
is immensely complicated, and its writing by hand (which has been 
the procedure adopted for the SM) is inconceivable: we have therefore
obtained it by using a development version of \FeynRules.
All massive modes are subtracted at zero momentum, following
the same strategy as e.g. in ref.~\cite{Beenakker:2002nc}. It should
be pointed out that some of the elementary expressions and structures 
of such a model (related e.g.~to the presence of Majorana fermions)
are not featured in any other \UFO\ model employed so far. For this
reason, we have double checked our \MLf\ results against a completely
independent calculation performed with {\sc\small Mathematica}. This 
guarantees that the elementary building blocks are correct, thus rendering
analogous tests more and more irrelevant in the future.
We have chosen the input parameters as follows:
\begin{center}
\begin{tabular}{cl|cl}\toprule
Parameter & value & Parameter & value
\\\midrule
$\as$ & \texttt{0.118} & $n_{lf}$ & \texttt{4}
\\
$m_{b}$ & \texttt{4.75}    & $m_{t}$ & \texttt{175}
\\
$m_{W}$ & \texttt{79.82901}    & $m_{Z}$ & \texttt{91.1876}
\\
$m_{\tilde{g}}$ & \texttt{607.7137}    & $\tan\beta$ & $9.748624$
\\
$m_{\tilde{u}_1}$ & \texttt{561.119} & $m_{\tilde{u}_2}$ & \texttt{549.2593}
\\
$m_{\tilde{c}_1}$ & \texttt{561.119} & $m_{\tilde{c}_2}$ & \texttt{549.2593}
\\
$m_{\tilde{t}_1}$ & \texttt{399.6685} & $m_{\tilde{t}_2}$ & \texttt{585.7858}
\\
$m_{\tilde{d}_1}$ & \texttt{568.4411} & $m_{\tilde{d}_2}$ & \texttt{545.2285}
\\
$m_{\tilde{s}_1}$ & \texttt{568.4411} & $m_{\tilde{s}_2}$ & \texttt{545.2285}
\\
$m_{\tilde{b}_1}$ & \texttt{513.0652} & $m_{\tilde{b}_2}$ & \texttt{543.7267}
\\\bottomrule
\end{tabular}
\end{center}
with a diagonal squark-mixing matrix.
By using the following kinematic configuration:
\begin{footnotesize}
  \begin{align*}
    p_g            &=\textrm{( 500}&&                           \textrm{, \phantom{-}0}                               &&\textrm{, \phantom{-}0}                               &&\textrm{, \phantom{-}500}                           &&\textrm{)}\\    
    p_g            &=\textrm{( 500}&&                           \textrm{, \phantom{-}0}                               &&\textrm{, \phantom{-}0}                               &&\textrm{, -500}                                            &&\textrm{)}\\
    p_{\tilde{t}_1}  &=\textrm{( 465.457552338590}&&\textrm{, \phantom{-}88.1561012782457}&&\textrm{, \phantom{-}197.510478842819}&&\textrm{, -100.667451003198}                  &&\textrm{)}\\
    p_{\tilde{t}_1^\star}  &=\textrm{( 442.275748385439}&&\textrm{, -9.53590501776566}&&\textrm{, -180.889189039748}&&\textrm{, \phantom{-}55.3271680251616}                  &&\textrm{)}\\   
    p_g            &=\textrm{( 92.2666992759711}&&                           \textrm{, -78.6201962604800}                               &&\textrm{, -16.6212898030706}                               &&\textrm{,  \phantom{-}45.3402829780365}                                            &&\textrm{)}
  \end{align*}
\end{footnotesize}
we obtain:
\begin{center}
\begin{tabular}{cl}\toprule
  $g g \rightarrow \tilde{t}_1 \tilde{t}_1^\star g$ & \multicolumn{1}{c}{}
  \\\midrule
$a_0$ & \texttt{\phantom{-}2.839872059757065e-4}
\\
$c_{0}$ & \texttt{-2.081163174420354e-5}
\\
$c_{-1}$ & \texttt{-1.550338075591894e-4}
\\
$c_{-2}$ & \texttt{-4.800024159745521e-5}
\\\bottomrule
\end{tabular}
\end{center}
The relevant \aNLO-shell commands are:

\noindent
~\prompt\ {\tt ~import model loop\_MSSM}

\noindent
~\prompt\ {\tt ~generate g g > t1 t1\~{} g [virt=QCD]}

\noindent
We point out that, similarly to the version of \MLf\ used to
derive the results presented in this section, the {\tt loop\_MSSM}
model used here is not yet public.

\section{Conclusions and outlook\label{sec:conc}}
The motivation for pursuing this project stems from the fact
that all aspects of the computations of tree-level and NLO cross sections,
including their matching to parton shower simulations, are well 
understood, in a manner which is fully independent of the process 
under consideration. Therefore, the best way to make use of
this understanding is that of the full automation of such computations.
Automation has indeed already proven to be a very successful strategy
for obtaining tree-level results, as documented by the theoretical and
experimental activities based on, and spurred by, \MadGraph.
In this paper we have presented the successor of \MadGraphf, a code that
we have named \aNLO, which extends the capabilities of the former by 
giving the user the possibility of computing NLO QCD corrections, 
if desired in association with parton-shower matching. \aNLO\ is 
indeed the successor, and not just a plugin, of \MadGraphf, its 
main virtue being that of treating tree-level and NLO QCD 
computations on the very same footing -- as far as the user
is concerned, the difference between them is a switch in input.
In particular, {\em irrespective} of the perturbative order of the
computation, \aNLO\ features the following characteristics:
very lean dependencies, simplicity of use, and
flexibility. The information that the user has
to provide is only related to physics, such as values of 
masses, couplings, and scales, and the definitions of observables,
as well as the hard process one wants to generate.

In the current public version of \aNLO\ the inclusion of higher-order
effects is restricted to QCD corrections to SM processes. Such a limitation
is mostly due to the fact that, in the context of one-loop computations 
(performed by \MadLoop\ or by any other one-loop provider), one needs 
to take care of the UV renormalisation procedure and (typically, but
not always necessarily) of that related to the so-called $R_2$
counterterms, which are in any case just a simpler version of the former.
Both procedures are expressed as a set of rules that can be worked out directly
from the Lagrangian, an operation that has to be done only once for a given
theory, and that so far has been performed by means of analytical
computations. However, all obstacles preventing the automatic computation of
the UV and $R_2$ rules have now essentially been cleared, as hinted by the
results we have presented in sect.~\ref{sec:next}. Given that all remaining
obstacles are minor and of technical nature, this will allow \aNLO\ to
evaluate, in the near future, any type of NLO corrections, starting from the
same user-defined Lagrangians that are used in tree-level calculations.

This work shows clearly that automated techniques at the NLO 
are well past the developmental phase, and are indeed fully
established, as was already the case for their tree-level counterparts; 
we believe that this is amply demonstrated by the results of
sect.~\ref{sec:res}. For the non-trivial cases now of relevance
to collider phenomenology, automated computations are more robust, 
faster by orders of magnitude, and less error-prone than analytical, 
process-by-process traditional approaches. Furthermore, an increase
in complexity generally only requires more CPU power but no 
conceptually new solutions, one example of such a situation
being that of the computation of EW or QCD corrections to 
supersymmetric processes. As a counterexample, one may mention
the calculation of cross sections that feature final states with 
a very large number of QCD partons, for which dedicated optimisations
(such as recursive relations and colour reorganisation, which are being
investigated by us) will be needed in order to go beyond what is currently 
feasible by \aNLO.

The ready availability in \aNLO\ of perturbatively-accurate and realistic 
predictions for an extremely large range of processes of significant 
complexity should be seen as both solving a few problems, and opening up
new and exciting possibilities. Among the former, we would like to
mention explicitly the fact that automated tools help free experts in 
perturbative calculations from spending their resources in the increasingly
involved computations necessary to the experiments, thus allowing them
to concentrate on obtaining other, cutting-edge results (such as, to
limit oneself to perturbative QCD, calculations of NNLO accuracy with
universal subtraction methods, improvements to parton showers,
and so forth). As far as future possibilities are concerned, one
important characteristic of \aNLO\ to bear in mind is its modularity:
we shall be happy to support and help those interested in improving
parts of the code and the underpinning physics strategies (such as recursive
relations, alternative matching and merging schemes, integral-reduction
techniques and libraries). From a phenomenological viewpoint, many
different applications of \aNLO\ can be foreseen. The capability
of the code to assess systematically and in an easy manner the theoretical 
uncertainties due to scales, PDFs, and matching and merging methods should 
be routinely exploited by both theorists and experimentalists. The fact of 
having a basically unlimited set of processes predicted at the NLO accuracy 
has two immediate consequences: it gives one the chance of extracting 
the PDFs by using a much wider set of observables than employed at
present, at the same time possibly including EW and parton-shower
effects; and that of finally achieving PSMC tunings that properly include
NLO results. Exploratory studies at future colliders, such as circular
or linear $e^+e^-$ ones, or very-high-energy hadron machines, can also be
performed without the need of a dedicated effort. Finally, it is hard
to predict the kind of applications that will be relevant to BSM physics.
In any case, extending  the current flexibility of \aNLO\ for SM processes 
to new-physics models, be they renormalisable or effective, and thus 
being able to readily investigate the implications of any theory, will 
certainly be crucial in current and future analyses.

\section*{Acknowledgements}
We are grateful to Roberto Pittau for his support of \CutTools, to 
Stefan Prestel for his dedicated help with the NLO matching to \PYe\
and for discussions on related topics, to Matteo Cacciari for his help 
and support with \FJ\ {\sc\small (core)}, to Andreas Papaefstathiou
for his dedicated help with \HWpp, to Simon Badger, Daniel Maitre, and 
Markus Schulze for clarifications concerning bootstrap and $D$-dimensional 
unitarity methods, to Markus Cristinziani and Roberto Chierici
for information on top-physics measurements by ATLAS and CMS,
to Adam Alloul, Neil Christensen, Celine Degrande, Claude Duhr and 
Benjamin Fuks for their collaboration to BSM developments and support 
of \FeynRules, to Frank Krauss for explanations
on the use of the MC@NLO method in \Sherpa, to Simon de Visscher
for his help with LO merging, to Pierre Artoisenet and Robbert Rietkerk
for their work on \Madspin, to Valery Yundin and Andreas van Hameren
for their support of \PJF\ and {\sc\small OneLoop}, to Diogo Buarque 
Franzosi for his collaboration on the implementation of the complex mass
scheme, to Alexis Kalogeropoulos for his contributions in developing and 
testing large-scale event production, to Kentarou Mawatari for many fruitful 
collaborations, to Stefano Pozzorini, Peter Skands
and Torbj\"orn Sj\"ostrand for useful clarifications, 
to Josh Bendavid, Vitaliano Ciulli, and Sanjay Padhi
for the very useful feedback on the use of the code, and
to Michelangelo Mangano for his support and insight. We are
thankful to our cluster management team, Jer\^ome de Favereau, 
Pavel Demin, and Larry Nelson, for their help and collaboration over 
the years. FM and TS are particularly thankful to Kaoru Hagiwara for the
constant support and many insightful suggestions since the very beginning 
of the \MadGraph\ project.
SF wishes to thank Bryan Webber for his collaboration, over the course
of many years, on several of the topics discussed in this paper.
The authors not based at CERN are grateful to the PH/TH Unit for
the support and hospitality in many different occasions.
This work has been supported in part by and performed in the
framework of the ERC grant 291377 ``LHCtheory: Theoretical predictions 
and analyses of LHC physics: advancing the precision frontier", 
supported in part by the Swiss National Science Foundation
(SNF) under contract 200020-149517, by the European Commission through
the ``LHCPhenoNet" Initial Training Network PITN-GA-2010-264564,
by the Research Executive Agency (REA) of the European Union under 
Grant Agreement number PITN-GA-2012-315877 (MCNet). 
The work of FM and OM is supported by the IISN ``MadGraph'' convention
4.4511.10, by the IISN ``Fundamental interactions'' convention 4.4517.08, 
and in part by the Belgian Federal Science Policy Office through the 
Interuniversity Attraction Pole P7/37. OM is 
``Chercheur scientifique logistique postdoctoral F.R.S.-FNRS",
and wishes to thank the IPPP Durham for the kind hospitality
extended to him during his stay there.
The work of FM has been supported by the Francqui Foundation via a Francqui
Research Professorship. 
The work of MZ has been partly supported by the ILP LABEX (ANR-10-LABX-63),
in turn supported by French state funds managed by the ANR within the
``Investissements d'Avenir'' programme under reference ANR-11-IDEX-0004-02.
VH is supported by the SNF with grant PBELP2\_146525.
This material is based upon work supported in part by the US 
National Science Foundation under Grant No. PHY-1068326.

\appendix
\section{Technical prerequisites, setup, and structure\label{sec:depen}}
\aNLO\ is being developed and is routinely run on a variety of
Linux platforms and on Mac OS-X systems. The basic requirements 
for running the code
are the following:
\begin{itemize}
\item A {\tt bash} shell;
\item {\tt perl 5.8} or higher;
\item {\tt Python 2.6} or higher, but lower than {\tt 3.0};
\item {\tt gfortran/gcc 4.6} or higher; any other modern Fortran/C++ compiler
 should work, provided it supports computations in quadruple precision.
\end{itemize}
After downloading the tarball, and upacking it in what will be called
the {\em main directory} (which will contain several sub-directories,
such as {\tt aloha}, {\tt apidoc}, {\tt bin}, and so forth), the
code is ready to run. No installation of external packages is
mandatory, thanks to the fact that the tarball includes copies
of the third-party codes listed in appendix~\ref{sec:tpc}.

\newpage
\noindent
{\bf Setup}

\noindent
From a terminal shell in the main directory, type:

\noindent
{\tt ~./bin/mg5\_aMC}

\noindent
At this point, one has entered the \aNLO\ shell, which is made
evident by the fact that the prompt now reads as follows:

\noindent
~\prompt\

\noindent
A minimal setup phase may take place here, before generating and 
running the first process. This phase, that must be done at most once
(i.e., it does not have to be repeated before the generation of any
process after the first), consists essentially in defining configuration
variables. For example, in the case where a local installation of
\FJ\ were available\footnote{We point out that \FJ\ {\sc\small (core)}
is part of the \aNLO\ tarball.}, the path to it must be known by \aNLO: 
this is achieved by executing the following command:

\noindent
~\prompt\ {\tt set fastjet /<PATH\_TO\_FASTJET>/bin/fastjet-config}

\noindent
Likewise, for the local installation of LHAPDF~\cite{Whalley:2005nh} 
to be found, one should execute the command:

\noindent
~\prompt\ {\tt set lhapdf /<PATH\_TO\_LHAPDF>/bin/lhapdf-config}

\noindent
Each of these commands associates the given value with a variable 
in the file:

\noindent
{\tt ~input/mg5\_configuration.txt}

\noindent
The user can
find a list of all possible configuration variables by visiting
that file (or by auto-completion with the {\tt <TAB>} key
after typing {\tt set} in the \aNLO\ 
shell). We point out that each of these variables can be directly
edited in the file, as an alternative to executing the {\tt set}
command as shown above. More advanced setup options are described
in appendix~\ref{sec:verbose}.

\vskip 0.4truecm
\noindent
{\bf Structure}

\noindent
The various subdirectories of the main directory will be of no
interest to the regular user. The only possible exceptions are

\noindent
{\tt ~Template/NLO}

\noindent
{\tt ~Template/LO}

\noindent
a copy (with minor differences) of which is created in the
subdirectory 

\noindent
{\tt ~MYPROC}

\noindent 
of the main directory upon executing the command:

\noindent
~\prompt\ {\tt ~output MYPROC}

\noindent
after having executed one of the two following commands:

\noindent
~\prompt\ {\tt ~generate a b > c\_1...c\_n [QCD]} 

\noindent
~\prompt\ {\tt ~generate a b > c\_1...c\_n}

\noindent 
for the NLO and LO case respectively (see sect.~\ref{sec:howto}).
The minor differences in the copy alluded to before are due to the
fact that, after the {\tt generate} command has been issued, the program
knows e.g.~the number of final-state particles (equal to {\tt n} in
the examples given here), which is thus explicitly written in some
include files in the directory tree of {\tt MYPROC}. In any case, 
these include files and their analogues must not be modified by the user.

The directory {\tt MYPROC} will be called the {\em current-process directory},
and all the operations relevant to the process whose generation gave
rise to it are performed somewhere in its directory tree. Such
operations can roughly be arranged in two classes: input-type,
to be performed by the user before the {\tt launch} command,
and output-type, performed by \aNLO\ after the {\tt launch} command.

\noindent
The user may consider input-type operations in the following subdirectory:

\noindent
{\tt ~MYPROC/Cards}

\noindent
which contains the input cards (these are in plain text format, and liberally 
commented) that steer the \aNLO\ run (e.g., {\tt run\_card.dat}) or control 
the physical parameters of the theory (e.g., {\tt param\_card.dat}).
Contrary to other input-type operations, which will be mentioned below, 
the values of the entries
in the input cards can not only be modified by the user by visiting the
appropriate cards before the {\tt launch} command, but can also be accessed 
{\em after} the {\tt launch} command through an interactive talk-to
within the \aNLO\ shell. Both accessing modes can be used in the same 
run; note that the values of the inputs used in the actual run will be 
those stored in the input cards at the end of the talk-to phase.

Further input-type operations are specific to NLO-type generations,
require a minimal knowledge of Fortran, and must be completed 
before the {\tt launch} command is issued.
They will involve editing files in the following subdirectories:

\noindent
{\tt ~MYPROC/FixedOrderAnalysis}

\noindent
{\tt ~MYPROC/MCatNLO}

\noindent
{\tt ~MYPROC/SubProcesses}

\noindent
The {\tt FixedOrderAnalysis} subdirectory will need to contain the 
user's fixed-order analysis file(s) relevant to fLO or fNLO runs.
The {\tt MCatNLO} subdirectory is used only in the case when the user
chooses to steer the shower phase within the \aNLO\ framework; when
the LHE files produced by \aNLO\ are showered externally, such
a subdirectory is ignored. If also the showering is steered by \aNLO,
the user will be able to access the drivers of the various event generators,
and to write his/her own analysis inside the {\tt MCatNLO} directory tree
(whose structure is analogous to that of the \mcatnlo\ package,
for those familiar with it).
Finally, the subdirectory {\tt SubProcesses} contains the codes necessary 
for the computation of the cross section proper, and specific to the
process that has been constructed in the generation step. Typically, the
user will not need to modify any of these files; exceptions are those
of {\tt setscales.f}, where one defines the functional forms used for
dynamic-scale computations (see sect.~\ref{sec:defmu}), and of {\tt cuts.f},
where one sets any desired parton-level cuts (on top of those accessible
through {\tt run\_card.dat}); both of these files are amply commented.

As far as output-type operations are concerned, these are by definition
dealt with by \aNLO; we shall mainly describe in what follows those
relevant to an NLO-type generation. The relevant directories are

\noindent
{\tt ~MYPROC/Events/run\_*}

\noindent
There will be as many {\tt run\_*} subdirectories as number of 
runs\footnote{Runs may e.g.~differ by choices of input parameters,
or type of physics simulated, such as f(N)LO vs (N)LO+PS, or the 
inclusion of spin correlations as predicted by \MadSpin\ vs stable-particle
production.}; the string {\tt *} will feature a run-identification number.
These subdirectories will contain the final outputs of the corresponding
\aNLO\ runs, provided that such outputs are in one of the formats recognised
by the code. In particular, one will have:
\begin{enumerate}
\item For all types of runs: various plain-text files that summarise 
the inputs used and the results of the integration of the short-distance 
cross sections.
\item In (N)LO+PS runs: the Les Houches event file(s) that contain
hard-subprocess unweighted events which are to be showered.
\item In (N)LO+PS runs when \aNLO\ is used to steer the shower phase:
the final results after shower, provided that these are either: 
{\em a)} an {\tt StdHEP/HepMC} file that contains the event records; 
or {\em b)} a topdrawer file that contains histograms defined by the user
in his/her analysis.
\item In f(N)LO runs: the histograms defined by the user
in his/her analysis, provided that their format be either Root
or topdrawer.
\end{enumerate}
There is no problem if the format of the output of the user's analysis 
relevant to the shower phase (when \aNLO\ is used to steer the shower) 
is not compliant with either 3.{\em a} or 3.{\em b}. Simply, such an 
output will not be moved into {\tt ~MYPROC/Events/run\_*}, but will
be kept in the directory where the PSMC run has actually 
been performed. This directory will be named:

\noindent
{\tt ~MYPROC/MCatNLO/RUN\_MCTYPE\_nn}

\noindent
with {\tt MCTYPE}$=${\tt PYTHIA8} and so forth (depending on the
PSMC used), and {\tt nn} an integer number increased by one unity for 
each new PSMC run.

On the other hand, the use of formats other than Root or topdrawer
in f(N)LO runs is deprecated, since it implies some manual operations
and the writing of code by the user. The reason is the following:
cross sections are integrated by \aNLO\ through multi-channel
techniques -- this ensures optimal convergence and high degree of
parallelisation, but each channel is non-physical (only their sum is).
Analysis routines (regardless of the output format) are used by
individual channels; hence, their outputs are to be summed\footnote{This
assumes the output is a set of histograms. In case of n-tuples, these
will need to be combined, possibly after having rescaled their 
weights.}. Summing the results of the individual channels is performed
automatically by \aNLO\ for Root and topdrawer outputs, via dedicated
auxiliary codes; any new format would thus require the user to write
a new such code, and a script that finds all single-channel outputs
and feed them to his/her summing code.

We conclude this section by stressing again that the source codes which 
are compiled after executing the command {\tt launch}, and the input cards 
used throughout the run, are those in the current-process directory tree,
and not in the {\tt Template} directory tree. Therefore, any modifications
to files in the {\tt Template} directory tree will have no effect on
the current run. However, they will affect {\em all} subsequent process
generations, since as clarified at the beginning of this section it is 
the files in {\tt Template} that form the core of the contents of each
current-process directory. Hence, this procedure is reserved to the
very experienced users, and we strongly deprecate it.

\section{Advanced usage\label{sec:use}}
This section reports on some of the features of \aNLO\ whose understanding
allows the user to exploit the full physics potential of the code.
It is not meant to be a usage manual, but only to briefly expand
on some of the subjects which have been only touched upon in the
main text.

\subsection{Models, basic commands, and extended options\label{sec:verbose}}
We start with the following general comment: within the \aNLO\ shell,
the {\tt <TAB>} key plays the same role as in a normal terminal
shell: when hitting it, a list of possible completions (e.g.,
commands relevant to the current context, or completion of the
command syntax) is printed on the screen. Note that the shell commands
{\tt help} and {\tt tutorial} can be used as well, and will provide
the user with some minimal guidance.

\vskip 0.4truecm
\noindent
{\bf Models}

\noindent
As was explained in sect.~\ref{sec:method}, \aNLO\ needs a model
in order to generate a process. When one enters the \aNLO\ shell,
the default is that of assuming the SM: however one can choose to 
work in another theory by loading a new model,
by simply executing the command:

\noindent
~\prompt\ {\tt ~import model}~{\em ModelName}

\noindent 
where {\em ModelName} is the name of the desired model. The list
of available models (to which as usual the user can add his/her own)
can be obtained by hitting the {\tt <TAB>} key after {\tt import model}.
Each of these models is associated with a directory (under the main
directory):

\noindent
{\tt ~models/}{\em ModelClass}

\noindent 
In the directory {\em ModelClass}, one collects the definition of all
those models which are tightly connected with each other, for example
by having the same Lagrangian and differing by the choice of some
fundamental parameter. To give an explicit example: the default model
for the SM in \aNLO\ assumes the charm quark to be massless, but there
is a model where the charm quark is massive. For both of these, we have:

\noindent
{\tt ~models/}{\em ModelClass}\;~$\equiv$~\;{\tt models/sm}

\noindent 
The massless-charm or massive-charm SM is explicitly loaded by typing:

\noindent
~\prompt\ {\tt ~import model sm}

\noindent
~\prompt\ {\tt ~import model sm-c\_mass}

\noindent 
respectively. Technically, these two commands are in one-to-one 
correspondence with the two files:

\noindent
{\tt ~models/sm/restrict\_default.dat}

\noindent
{\tt ~models/sm/restrict\_c\_mass.dat}

\noindent 
The user interested in some non-extensive modification of the SM 
can thus simply create his/her own file {\tt models/sm/restrict\_XXX.dat},
which may be eventually loaded by executing {\tt import model sm-XXX}.
For more details on these matters, please visit:\\
{\tt https://cp3.irmp.ucl.ac.be/projects/madgraph/wiki/Models/USERMOD}.

We finally stress again that not all models support the computation
of NLO corrections. In the first public version of \aNLO\ such corrections 
are restricted to QCD to SM processes. The relevant models are all found
in the directory:

\noindent
{\tt ~models/loop\_sm}

\noindent 
Note that, by default, when performing an NLO-type generation 
(i.e., by using the {\tt [QCD]} keyword) the code switches automatically
from the LO-type model to the corresponding NLO-type one (i.e., in the
SM it switches from {\tt sm} to {\tt loop\_sm}). If the latter is not
available, a warning is issued and the code proceeds no further.

\vskip 0.4truecm
\noindent
{\bf Setup}

\noindent
A short description of the setup procedure has been already given
at the beginning of appendix~\ref{sec:depen}. Here, we wish to point
out that, on top of the environment variables found in 
{\tt input/mg5\_configuration.txt}, the user can also control
other options, which for example affect the physics schemes
used by \aNLO\ during the various computations. All such
options can be listed with the following command:

\noindent
~\prompt\ {\tt ~display options}

\noindent 
which will display all options and their current values. In order
to change the latter, one executes the {\tt set} command, whose
general syntax is:

\noindent
~\prompt\ {\tt ~set} {\em Option} {\em Value}

\noindent 
For example, \aNLO\ by default does not use the complex mass 
scheme in its computations. In order to change this,
it is sufficient to execute:

\noindent
~\prompt\ {\tt ~set complex\_mass\_scheme True}

\noindent 
Other examples relevant to environment variables have already been
given in appendix~\ref{sec:depen}.

The operation of computing the widths of the unstable particles
present in the imported model can be seen as part of the setup 
procedure, being a complement to the model itself, independent of 
the generation procedure, and mandatorily performed before the
running phase (see sect.~\ref{sec:madwidth}). Such an operation is 
carried out by issuing the shell command:

\noindent
~\prompt\ {\tt ~compute\_widths} [{\em \{Options\}}]

\noindent
which in turn executes the \madwidth\ module; the possible options
of the above command can be as usual explored by hitting the
{\tt <TAB>} key. We remind the reader that \madwidth\ works
at tree level and in the narrow-width approximation (in other 
words, the manual setting of widths in the context of an NLO
simulation may be necessary).

Finally, another setup-type operation is the diagonalisation of the
mass matrix. This is only available in a restricted class of \UFO\ models 
which include the \asperge~\cite{Alloul:2013fw} module, whose inputs
are accessible to the user during the interactive talk-to phase.

\vskip 0.4truecm
\noindent
{\bf Generation}

\noindent
The most general form of the {\tt generate} command is the following:

\noindent
~\prompt\ {\tt ~generate} {\em Process \{AmpOrders\}} 
[{\em \{\{Mode\} Couplings\}}]

\noindent 
the only mandatory option being {\em Process}, i.e. the actual process
one needs to generate. We shall now 
comment on the four options above in turn.

The option {\em Process} is simply the list of initial- and
final-state particles, separated by the conventional {\tt >} sign:

\noindent
{\em ~~Process}~$\equiv$~~~~~{\tt a b > c\_1...c\_n}

\noindent
One can further
refine the syntax above in order to include in the computation
only some of the contributions that one would normally obtain.
Such refinements are reported in table~\ref{tab:LOsyntax},
\begin{table}
\begin{center}
\begin{tabular}{llc}
\hline
syntax & example &  meaning \\
\hline
{\tt x, x>}~~~~~  & {\tt p p > z j, z > b b\~{}} & {\em s.1} \\
{\tt \$ x}     & {\tt p p > e+ e- \$ z} & {\em s.2} \\
{\tt / x}     & {\tt p p > e+ e- / z} & {\em s.3} \\
{\tt > x >}     & {\tt p p > z > e+ e- } & {\em s.4} \\
{\tt \$\$ x}     & {\tt p p > e+ e- \$\$ z} & {\em s.5} \\
\hline
\end{tabular}
\end{center}
\caption{\label{tab:LOsyntax}
Process-generation syntax refinements, also exemplified in the
case of various processes that involve a $Z$ boson. See the text 
for the explanation of the keywords {\em s.1}--{\em s.5}.
Only syntaxes {\em s.3}--{\em s.5} are supported for
NLO-type generations.
}
\end{table}
and have the following meaning:
\begin{itemize}
\item[{\em s.1}] A production process is generated that features {\tt x}
in the final state, with {\tt x} subsequently decaying into the list
of particles that follow the ``{\tt x >}'' string; more in general, there 
may be $p$ primary particles that play the same role as {\tt x}.
Only $p$-resonant diagrams (see sect.~\ref{sec:madspin}) 
are included in the computation. In the example of 
table~\ref{tab:LOsyntax}, one has the associated production of a $Z$ 
and a jet, with the $Z$ further decayed into a $b\bar{b}$ pair. 
Spin correlations and {\tt x} off-shell effects are taken into 
account exactly, but the virtuality $m_{\tt x}^\star$ of {\tt x} is 
forced to be in the following range:
\beq
\abs{m_{\tt x}^\star-m_{\tt x}}\le {\tt bwcutoff}\,\Gamma_{\tt x}\,,
\label{syns1}
\eeq
where $m_{\tt x}$ is the pole mass of {\tt x}, $\Gamma_{\tt x}$
its width, and {\tt bwcutoff} is a parameter controlled by the user 
(through {\tt run\_card.dat}). Syntax {\em s.1} thus loosely 
imposes an on-shell condition; it is called {\bf decay-chain syntax},
and can be iterated: any decay product can be decayed itself by using
this syntax (e.g. {\tt x > y z, y > w s}).
\item[{\em s.2}] If {\tt x} appears as an intermediate particle
in the generated process, its virtuality is forced to be
in the range:
\beq
\abs{m_{\tt x}^\star-m_{\tt x}}> {\tt bwcutoff}\,\Gamma_{\tt x}\,,
\label{syns2}
\eeq
which is the region complementary to that of eq.~(\ref{syns1}), and thus
loosely imposes an off-shell condition. All diagrams are kept. In the 
example of table~\ref{tab:LOsyntax}, one has Drell-Yan production with 
the invariant mass of the $e^+e^-$ pair larger than or smaller than
the $Z$ mass by at least {\tt bwcutoff}$\,\Gamma_Z$.
A consequence of the complementarity mentioned above
is that, while cross sections generated with either {\em s.1} or 
{\em s.2} are {\tt bwcutoff}-dependent, their sum is not (up to
interference terms, which are neglected by the process of discarding
non-resonant diagrams in {\em s.1}), and corresponds to the process 
generated with the simplest syntax. For example:
\beq
\frac{d\sigma}{dO}({\tt p~p~>~z})\simeq
\frac{d\sigma}{dO}({\tt p~p~>~z,~z~>~e+~e-})+
\frac{d\sigma}{dO}({\tt p~p~>~e+~e-~\$~z})\,,
\eeq
for any observable $O$. 
\item[{\em s.3}] All diagrams that feature (anywhere) the particle {\tt x} 
are discarded.
\item[{\em s.4}] The process is generated by demanding that at least
one particle of type {\tt x} be in an $s$-channel.
\item[{\em s.5}] All diagrams that feature the particle {\tt x} in
an $s$-channel are discarded.
\end{itemize}
We stress that all syntaxes but {\em s.2} produce in general results 
which are non physical, because gauge invariance might be violated 
(although there are exceptions: see e.g.~ref.~\cite{Papanastasiou:2013dta}), 
and have therefore to be used with extreme caution. The situation becomes 
more involved at the NLO, so even more care is required. Syntaxes
{\em s.3}, {\em s.4}, and {\em s.5} are supported; more refined 
selections of individual loop diagrams in \MLf\ can be imposed by 
editing the function {\tt user\_filter} in the source code 
{\tt loop\_diagram\_generation.py}.

The option {\em AmpOrders} allows the user to specify the
upper bounds on the powers of the coupling constants that enter
the scattering {\em amplitudes} (i.e., not the amplitudes squared);
therefore, the coefficients $\Sigma_{k_0+p,q}$ are selected only
in an indirect manner -- see sect.~\ref{sec:NLO}.
Furthermore, in the case of an NLO-type generation,
such amplitudes are the Born ones: the couplings of the one-loop
and real-emission amplitudes are then automatically determined
according to the type of corrections to be included.
The syntax for this option is the following:

\noindent
{\em ~~AmpOrders}~$\equiv$~~~~~{\tt coupling}$_1=p_1$~~
{\tt coupling}$_2=p_2$~~$\ldots$~~{\tt coupling}$_n=p_n$

\noindent
where {\tt coupling}$_i$ is the name of the $i^{th}$ coupling in the
model currently used, and $p_1$ is an integer which represents the
upper bound mentioned above. It should be obvious that {\tt coupling}$_i$
is an arbitrary name, chosen by the author of the model in use. In
order to see the list of all coupling names, one has just to type
(after having imported the model):

\noindent
~\prompt\ {\tt ~display coupling\_order}

\noindent
For example, by executing this command in the context of the SM,
one obtains what follows:

\noindent
{\tt ~QCD : weight = 1}

\noindent
{\tt ~QED : weight = 2}

\noindent
which implies that the internal names of the SM couplings $\gs$ and $\gW$
is {\tt QCD} and {\tt QED} respectively, with the latter being hierarchically
suppressed w.r.t.~the former. 
As an explicit example of the use of the {\em AmpOrders} option in
the SM, let us consider the case of:
\beq
p+p\;\longrightarrow\;W^+ + J_b + J_{light}\,,
\eeq
where $J_b$ and $J_{light}$ are a $b$- and a light-jet respectively,
and the five-flavour scheme is adopted.
Numerically, the dominant contributions to such a process are due
to diagrams whose corresponding amplitudes factorise the couplings
$\gs^2\gW$. However, amplitudes of order $\gW^3$ are more interesting,
since they feature diagrams with one top-propagator exchange, and are 
thus identified with single-top production (although of course at the
same order one has non-resonant diagrams as well). By executing
the command (after including the $b$-quark in the proton)

\noindent
~\prompt\ {\tt ~generate p p > w+ b j}

\noindent
one would obtain only the amplitudes of ${\cal O}(\gs^2\gW)$, the choice
being made by the code automatically according to the hierarchy shown 
above. In order to study single-top production, one can execute what 
follows instead~\cite{Papanastasiou:2013dta}:

\noindent
~\prompt\ {\tt ~generate p p > w+ b j QED=3 QCD=0}

\noindent
which will force the code to consider only ${\cal O}(\gW^3)$ amplitudes.
Note that, by entering {\tt QED=3 QCD=2}, one will generate {\em both} 
${\cal O}(\gs^2\gW)$ and ${\cal O}(\gW^3$) amplitudes. 
In future version of \aNLO, the syntax of the option {\em AmpOrders} 
will be extended, so as to give the user the possibility of selecting
directly cross-section-level quantities ($\Sigma_{k_0+p,q}$'s), at both
the leading and the next-to-leading order.

The option {\em Mode}\footnote{The use of which requires the use of the
option {\em Couplings} as well. The opposite is not true.}
allows the user to select which contributions 
to an NLO cross section (on top of that due to the Born, which is
always present) will be included in the computation. The possible
settings are the following:

\noindent
{\em ~~Mode}~$\equiv$~~{\tt all=}
$\phantom{aaaaaa}\Longrightarrow$\phantom{aa}both one-loop and 
FKS-subtracted real-emission

\noindent
{\em ~~Mode}~$\equiv$~~{\tt real=}
$\phantom{aaaaa}\Longrightarrow$\phantom{aa}only FKS-subtracted real-emission

\noindent
{\em ~~Mode}~$\equiv$~~{\tt virt=}
$\phantom{aaaaa}\Longrightarrow$\phantom{aa}only one-loop

\noindent
The setting {\tt all=} is equivalent to omitting the option 
{\em Mode} altogether, and should be the only one considered by the non-expert 
user, being the only one that leads to physical results. The setting 
{\tt real=}
instructs \aNLO\ to {\em not} generate the part of the code relevant
to virtual matrix elements with \MadLoop. In this way, the cross section
can still be dealt with as explained in sect.~\ref{sec:howto}, but
the results will be non-physical, {\em unless} an external one-loop
provider is linked to \aNLO\ (such an external OLP will thus effectively
play the same role as \MadLoop). Finally, the setting {\tt virt=} 
corresponds to the \MadLoop\ standalone mode. In such a mode, the
commands {\tt output} and {\tt launch} will behave differently
w.r.t.~what is described in sect.~\ref{sec:howto}, the idea being that
of using the code so generated in order to obtain the pole residues
and finite part of the virtual corrections for user-defined kinematic
configurations -- see sect.~\ref{sec:next} for explicit examples.

The option {\em Couplings} allows the user to specify which kind
of NLO corrections \aNLO\ will compute. The general syntax for this 
option is the following:

\noindent
{\em ~~Couplings}~$\equiv$~~~~~{\tt coupling}$_1$~~
{\tt coupling}$_2$~~$\ldots$~~{\tt coupling}$_n$

\noindent
However, in the current version only QCD corrections can be
computed, and therefore the only valid option read as follows:

\noindent
{\em ~~Couplings}~$\equiv$~~~~~{\tt QCD}

\noindent
as already mentioned several times in this paper. For examples of
the more general syntax that can be used in a still-private \aNLO\
version, see sect.~\ref{sec:next}.

\vskip 0.4truecm
\noindent
{\bf Output}

\noindent
The most general form of the {\tt output} command is the following:

\noindent
~\prompt\ {\tt ~output} [{\em OutputForm}] [{\tt MYPROC}]

\noindent
As was already mentioned in sect.~\ref{sec:howto} (see in particular
footnote~\ref{ft:ft}) the target-directory name {\tt MYPROC} may be
omitted, in which case \aNLO\ will choose automatically a name
(and print it out on the screen for the user to know). As the full
syntax above implies, however, there are a few names that are 
reserved because, if used, the code will interpret them as one of
the (optional) {\em OutputForm} keywords. These essentially serve
to create executables (or standalone libraries) which are not those
typically used for the integration of the cross sections and the
unweighting of the events. They are described in appendix~\ref{sec:MG5out}.

\vskip 0.4truecm
\noindent
{\bf Running}

\noindent
The most general form of the {\tt launch} command is the following:

\noindent
~\prompt\ {\tt ~launch} {\em \{ProcDir\}} {\em \{RunMode\}} {\em \{Options\}} 

\noindent 
We cannot possibly describe in this paper all possibilities implied
by the syntax above, which we plan to do in a forthcoming user manual; 
we urge the interested reader to explore them by either using the on-line 
tutorial, or executing the {\tt help launch} command, or exploiting 
the {\tt <TAB>} key.

Here, we limit ourselves to point out that {\em ProcDir}, if present,
must coincide with one of the current-process directories previously
generated. The implication is that, after the generation and output
phase, a user may not immediately run the process, but rather generate
a second one, or also quit the \aNLO\ shell. Being saved on disk, the full
information of a generated and outputed process can be retrieved
at any later time. In order to do this, one needs simply to execute:

\noindent
~\prompt\ {\tt ~launch MYPROC -i}

\noindent 
where {\tt MYPROC} is the name of the current-process directory
used throughout this paper. It should be clear that, upon executing 
the command above, one is again dealing with a specific current process.
For this reason, many early-stage commands (such as {\tt generate})
are disabled. In order to help the user remind this fact, the
prompt itself is actually changed, and reads:

\noindent
~{\tt MYPROC>}

\noindent
In short, we shall call the environment accessed by executing
the {\tt launch -i} command the {\em running mode}. One can re-enter
the running mode of a given process an unlimited amount of times.
Again, we shall not attempt to give here a full description of
the various implications of this fact. However, it is interesting
to discuss a specific usage in connection to what has been described
in sect.~\ref{sec:shower}, and namely how to shower hard-subprocess
event files previously generated. For example, we have discussed
in sect.~\ref{sec:shower} how, in the case of a process which features
particles whose decay products are dealt with \MadSpin, one obtains
(at least) one undecayed and one decayed hard-event files, for example:

\noindent
{\tt ~MYPROC/Events/run\_01/events.lhe.gz}

\noindent
{\tt ~MYPROC/Events/run\_01\_decayed\_1/events.lhe.gz}

\noindent
only the latter of which is showered by default when \aNLO\ steers 
the shower. In order to shower the former, one simply has to execute:

\noindent
~\prompt\ {\tt ~launch MYPROC -i}

\noindent
~{\tt MYPROC> ~shower run\_01}

\noindent
Namely, to access the running mode of the relevant process
directory, and then execute the command {\tt shower} followed
by the name of the subdirectory of the {\tt Events} directory
where the events to be showered are stored.
We point out that the shower command can be executed any number
of times with the same argument, e.g. if one desires to change
the seeds or the parameters in the PSMC.

\subsection{Setting the hard scales at the NLO\label{sec:defmu}}
The short distance NLO(+PS) cross sections in \aNLO\ feature three
hard scales: the renormalisation ($\muR$), factorisation ($\muF$),
and Ellis-Sexton ($\QES$) scales. The latter appears only at
the NLO and, at variance with the former two,
its variations do not induce any changes in the cross sections.
For this reason, $\QES$ is only useful in the context of validation
studies, performed by developers; regular users are recommended
to always set it equal to the factorisation scale. In view of the
standard way of performing scale variations, it is convenient
to write the hard scales as follows:
\beq
\muR=\ffR\muRz\,,\;\;\;\;\;\;
\muF=\ffF\muFz\,,\;\;\;\;\;\;
\QES=\ffQES\QESz,
\label{scales}
\eeq
where $\muRz$, $\muFz$, and $\QESz$ are called {\em reference} scales.

Hard scales must be set by the user prior to compiling and running
the code. They can be organised into two categories, typically
called ``fixed" and ``dynamical"; the scales belonging to the former
have constant values, independent of the event kinematics, while
those belonging to the latter depend on the four-momenta of the
final-state particles, and hence have values that change event-by-event
during the course of the run. Whether a scale is fixed or dynamical
is determined by an input parameter. For example, in the case of
$\muR$, the following setting in {\tt run\_card.dat}:

\noindent
{\tt ~~T        = fixed\_ren\_scale  ! if .true. use fixed ren scale}

\noindent
will instruct the code to use a fixed renormalisation scale. Such
a fixed value is also given by the user in input by setting
the pre-factor $\ffR$ and the reference scale $\muRz$ that appear
in eq.~(\ref{scales}). For example, the following entries
in {\tt run\_card.dat}:

\noindent
{\tt  ~~91.188~= muR\_ref\_fixed    ! fixed ren reference scale}

\noindent
{\tt  ~~2~~~~~~= muR\_over\_ref     ! ratio of current muR over reference muR}

\noindent
set $\muRz=91.188$~GeV and $\ffR=2$ respectively, and hence
$\muR=182.376$~GeV.

We now address the case of dynamical scales, again using the renormalisation
scale to give definite examples. A dynamical $\muR$ is used when:

\noindent
{\tt ~~F        = fixed\_ren\_scale  ! if .true. use fixed ren scale}

\noindent
With this setting, one still has $\ffR=${\tt muR\_over\_ref}. On the other
hand, the input {\tt muR\_ref\_fixed} is ignored, and the reference scale
is defined as follows:
\beq
\muRz={\tt muR\_ref\_dynamic}\,.
\eeq
The quantity {\tt muR\_ref\_dynamic} is a function of the four-momenta
of the final-state particles; its body is found in the file
{\tt MYPROC/SubProcesses/setscales.f}, where the user may 
include his/her definition of the dynamical scale best suited to
the process of interest. Although a few examples of typical dynamical
scales are given in the version of {\tt setscales.f} which is included
in the tarball of the package, we urge the user to study the structure
of that file (which is amply commented -- note, in particular, the role
of the variable {\tt temp\_scale\_id} in that file, which helps keep track
of the functional forms used), and to change it if need be.
It should be clear that {\tt setscales.f} follows the same rules
as all the other core files of the package (see sect.~\ref{sec:depen}).
Hence, in order for any modifications to it to be taken into account
in the current run, one must edit the file before the {\tt launch} 
command is issued.

What said above for the renormalisation scale applies without changes
to the factorisation and Ellis-Sexton scales; the relevant parameters
and functions in {\tt run\_card.dat} and {\tt setscales.f} have
self-explanatory names. Note that, in the case of the factorisation 
scale, one can in principle assign two different values to the two
incoming hadrons -- hence, in the input cards one can find the variables
{\tt *\_ref\_fixed} and {\tt *\_over\_ref}, with {\tt *}$=${\tt muF1}
and {\tt muF2}, which are respectively relevant to the hadron coming
from the left and the right. At the LO, 
this is unambiguous -- the scales only enter the relevant PDFs. 
At the NLO, however, there are several ways to write the logarithmic
terms whose arguments are ratios of scales. In order to avoid complications,
at the NLO the two factorisation scales must always be chosen to be equal;
when this is not the case, the code stops.

We point out that the structure outlined above will allow the user to
choose one scale to be fixed and another one to be dynamical; however,
in the vast majority of cases one will want the scales to be either all
fixed (not necessarily to the same value), or all dynamical. In the
latter case, the functions in {\tt setscales.f} are such that different
functional forms can be adopted for different scales. Again, this is
a somehow infrequent situation. Having this in mind, the default version
of {\tt setscales.f} sets all {\tt *\_ref\_dynamic} functions
(with {\tt *}$=${\tt muR}, {\tt muF}, and {\tt QES}) equal to the
same function {\tt scale\_global\_reference}, and one can limit
oneself to modifying the latter for a standard usage.

We conclude this section by summarising schematically what has been
described above.
\begin{itemize}
\item[1.] Decide whether to use fixed ({\tt fixed\_*\_scale}$=${\tt T})
or dynamical ({\tt fixed\_*\_scale}$=${\tt F}) scales. This is done
at runtime, either by editing {\tt run\_card.dat} before executing
the {\tt launch} command, or directly at the prompt after having 
executed it.
\item[2a.] If fixed scales are chosen: the relevant input parameters
are {\tt *\_over\_ref} (these are the $f_*$ factors in eq.~(\ref{scales}))
and {\tt *\_ref\_fixed} (these are the reference scales $\mu_{*0}$
in eq.~(\ref{scales})). Both can be set at runtime in the same
way as {\tt fixed\_*\_scale}.
\item[2b.] If dynamical scales are chosen: the relevant quantities 
are the input parameters {\tt *\_over\_ref} as in case 2a., and the
functions responsible for defining the reference scales, to be found
in {\tt setscales.f}. Modifications to the latter file must be carried
out before executing the {\tt launch} command (i.e., no modifications
are possible at runtime); by default, the reference dynamical scales 
are set equal to $\Ht/2$.
\end{itemize}
Note that the situation of FxFx-merged simulations is somewhat different,
owing to the specific prescriptions for the settings of the scales which
are inherent to the method; for more details, the user is encouraged
to check {\tt http://amcatnlo.cern.ch/FxFx\_merging.htm}.

\subsection{Scale and PDF uncertainties: the NLO case\label{sec:errors}}
Among all the dependencies of a cross section, those due to hard
scales and PDFs are special, since it is always possible to write
\beq
\sigma=\sum_i w_i b_i\,,
\label{decomp}
\eeq
where the coefficients $w_i$ (typically called ``weights") are independent
of both scales and PDFs, while the ``basis" members $b_i$ contain all
the information on scales and PDFs in simple forms such as:
\beq
b_i = f_{H_1}^{(i)}f_{H_2}^{(i)}\as^{k_i}
\Big\{1,\log\frac{\muR}{\QES},\log\frac{\muF}{\QES}\Big\}
\,.
\label{basis}
\eeq
The key point is that, while the weights might be very expensive to
compute CPU-wise, the basis members are straightforward. A convenient
strategy is therefore that of first evaluating the $w_i$'s, and then
of using them in eq.~(\ref{decomp}) for all the desired choices of
scales and PDFs; each of these will therefore result in a basically
instantaneous evaluation of the corresponding $\sigma$.
In practice, what is done in \aNLO\ is to compute $\sigma$ for a given
choice of scales and PDFs (the ``central" or ``default" choice), while
at the same time storing the values of the $w_i$'s (if this is required
by the user in input -- see later), in order to re-use them at a later
stage, typically for the assessment of the theoretical uncertainties.

We stress that eq.~(\ref{decomp}) is exact\footnote{The dependence on
PDFs of the (parton-shower) Sudakovs cannot be accounted for by
eq.~(\ref{decomp}). However, this is expected to be rather small,
and particularly so when computing PDF uncertainties. See 
ref.~\cite{Frederix:2011ss} for more details.}, and therefore so 
is the computation of a cross section starting from the weights for any
given scale and PDF choice. A complete discussion, which includes all
the relevant definitions of $w_i$ and $b_i$ for both fixed-order and
MC@NLO cross sections, is given in ref.~\cite{Frederix:2011ss} and 
will not be repeated here. The aim of this appendix is rather that
of giving some details on the way in which eq.~(\ref{decomp}) is
exploited when computing scale and PDF uncertainties in the context
of (N)LO+PS and f(N)LO simulations. 

The basic idea relevant to (N)LO+PS is the following. 
In the LHE file and event-by-event, the values of 
$\sigma$ will be stored that correspond to all the combinations 
of scales and PDFs selected by the user in input (we denote 
the numbers of these combinations by 
$N_\mu$ and $N_{\rm PDF}$ respectively). These $\sigma$'s will have 
to be treated in the same way as the cross section associated
with the central scales and PDFs (which is part of the standard LHE
information, and corresponds to the variable {\tt XWGTUP}); 
namely, each of them will 
constitute an entry in a histogram associated with that particular
combination of scales and PDFs. In other words, for each observable
of interest, one will have to fill $1+N_\mu+N_{\rm PDF}$ histograms
(the ``1" being for the central choices). At the end of the run, and
for each observable, the envelope of the $1+N_\mu$ histograms will
give the scale uncertainty affecting that observable, and the envelope
of the $1+N_{\rm PDF}$ histograms will be the PDF uncertainty.
The precise definition of these two envelopes is the user's
responsibility. In the case of the scales, one will typically want
to consider the largest and smallest cross sections bin-by-bin,
possibly excluding from the computation of such extremes some
of the $(\muR,\muF)$ combinations (see e.g.~\cite{Cacciari:2008zb}
for a discussion on this point). In the case of the PDFs, the envelope
must be defined following the prescription of the PDF authors.

We now show how the user can choose in input the scales and PDFs
that will be used in the calculation of the uncertainties; we start
from the former. In the present version of \aNLO\ we have fixed
$N_\mu=8$, which corresponds to the following combinations:
\beqn
&&
(\ffRd,\ffFd)\,,\;\;\;\;\;\;
(\ffRc,\ffFd)\,,\;\;\;\;\;\;
(\ffRu,\ffFd)\,,
\label{ffone}
\\*&&
(\ffRd,\ffFc)\,,\;\;\;\;\;\;
\phantom{(\ffRc,\ffFc)\,,}\;\;\;\;\;\;
(\ffRu,\ffFc)\,,
\label{fftwo}
\\*&&
(\ffRd,\ffFu)\,,\;\;\;\;\;\;
(\ffRc,\ffFu)\,,\;\;\;\;\;\;
(\ffRu,\ffFu)\,,
\label{ffthree}
\eeqn
these being the pre-factors introduced in eq.~(\ref{scales}) to
define the renormalisation and factorisation scales given the
reference scales:
\beq
\left(\muR^\alpha,\muF^\beta\right)=
\left(\ffR^\alpha\muRz,\ffF^\beta\muFz\right)\,,
\;\;\;\;\;\;\;\;\;
\{\alpha,\beta\}\in\{\downarrow,c,\uparrow\}\,.
\label{scalevar}
\eeq
The combination \mbox{$(\ffRc,\ffFc)$} missing in 
eqs.~(\ref{ffone})--(\ref{ffthree}) is obviously that corresponding
to the central scales, which is always computed. Equation~(\ref{scalevar})
implies that scale variations are defined by choosing the reference 
renormalisation and factorisation scales (which is done as explained 
in sect.~\ref{sec:defmu}), and by varying the pre-factors in front of them.
Such prefactors are defined by means of some input parameters found
in {\tt run\_card.dat}. More specifically, we have:
\beqn
\ffRd&=&{\tt muR\_over\_ref}\,\times\,{\tt rw\_Rscale\_down}\,,
\label{ffRF1}
\\
\ffRc&=&{\tt muR\_over\_ref}\,,
\\
\ffRu&=&{\tt muR\_over\_ref}\,\times\,{\tt rw\_Rscale\_up}\,,
\\
\ffFd&=&{\tt muF\_over\_ref}\,\times\,{\tt rw\_Fscale\_down}\,,
\\
\ffFc&=&{\tt muF\_over\_ref}\,,
\\
\ffFu&=&{\tt muF\_over\_ref}\,\times\,{\tt rw\_Fscale\_up}\,.
\label{ffRF5}
\eeqn
Given these inputs, \aNLO\ will consider all the combinations
given in eqs.~(\ref{ffone})--(\ref{ffthree}), compute the corresponding
$b_i$'s of eq.~(\ref{basis}) {\em using the central PDFs}, and combine 
them with the weights according to eq.~(\ref{decomp}). The resulting 
$\sigma$ (which could be denoted by $\sigma^{(\alpha,\beta)}$ for
consistency with eq.~(\ref{scalevar})) will be stored in the LHE file,
provided the user sets:

\noindent
{\tt ~.true.   = reweight\_scale   ! reweight to get scale dependence}

\noindent
in input.

As far as PDF uncertainties are concerned, we have assumed that one 
will use LHAPDF, where all members of an error set are 
identified by adjacent integer numbers. Using NNPDF 2.0~\cite{Ball:2010de} 
as an example to be definite, one will have the following inputs:

\noindent
{\tt ~lhapdf   = pdlabel   ! PDF set}

\noindent
{\tt ~90800    = lhaid     ! if pdlabel=lhapdf, this is the lhapdf number}

\noindent
{\tt ~90801   = PDF\_set\_min      ! First of the error PDF sets}

\noindent
{\tt ~90900   = PDF\_set\_max      ! Last of the error PDF sets}

\noindent
The first two lines instruct the code to use LHAPDF and to choose
the central NNPDF 2.0 set as the default. The last two lines will
set $N_{\rm PDF}=100$; for each of these hundred NNPDF 2.0 error
sets, \aNLO\ will compute the $b_i$'s of eq.~(\ref{basis}) 
{\em using the central scales}, and combine  them with the weights 
according to eq.~(\ref{decomp}). The resulting $\sigma$ will be stored 
in the LHE file, provided the user sets:

\noindent
{\tt ~ .true.  = reweight\_PDF     ! reweight to get PDF uncertainty}

\noindent
in input.

For what concerns the storage in the LHE file of the $\sigma$'s computed
as described above, we use a format which is fully compatible with
the LHA v2.0~\cite{Butterworth:2010ym}, and which is now officially
adopted as v3.0~\cite{LHEv30}. In the file header, there will 
be a part (which we call the reweight section) where a description is 
given of the meaning of the weights that will appear in each event.
Its structure may read as follows:

\noindent
{\tt
~<initrwgt>

\noindent
~~<weight id='1'> This is the central weight </weight>

\noindent
~~<weightgroup type='scale variation' combine='envelope'>

\noindent
~~~<weight id='2'> muR=0.5 muF=0.5 </weight>

\noindent
~~~<weight id='3'> muR=0.5 muF=1.0 </weight>

\noindent
~~~...

\noindent
~~~<weight id='9'> muR=2.0 muF=2.0 </weight>

\noindent
~~</weightgroup>

\noindent
~~<weightgroup type='NNPDF20' combine='gaussian'>

\noindent
~~~<weight id='10'> set001 </weight>

\noindent
~~~...

\noindent
~~~<weight id='109'> set100 </weight>

\noindent
~~</weightgroup>

\noindent
~</initrwgt>
}

\noindent
Thus, each choice of scales and of PDFs is uniquely identified
by an ID number. For example, {\tt id='1'} will correspond to
the central scale and PDF choices, \mbox{$(\ffR,\ffF)=(\ffRc,\ffFc)$}
and PDF number 90800 in the examples given above (this is the
same as {\tt XWGTUP}, and hence is redundant, but it is convenient
to include it in the reweight section as well). As far as {\tt id='2'} 
to {\tt id='9'} are concerned, these will correspond to the $N_\mu$ 
combinations given in eqs.~(\ref{ffone})--(\ref{ffthree}), the numerical 
values reported in the header next to {\tt muR} and {\tt muF} being those
of $\ffR^\alpha$ and $\ffF^\beta$ respectively. Finally, {\tt id='10'} 
to {\tt id='109'} correspond to the $N_{\rm PDF}$ error sets
$90801$--$90900$.

In keeping with the reweight section in the LHE file header, there
will be a reweight section event-by-event, which contains the actual
numbers to be used to fill the histograms as described previously.
Its structure will read as follows:

\noindent
{\tt
~<event>

\noindent
~...

\noindent
~~<rwgt>

\noindent
~~~<weight id='1'> 3.905e+01</wgt>

\noindent
~~~<weight id='2'> 4.142e+01</wgt>

\noindent
~~~...

\noindent
~~~<weight id='109'> 3.876e+01</wgt>

\noindent
~~</rwgt>

\noindent
~</event>

}

\noindent
The presence of the ID numbers in the reweight section of each event
facilitates debugging, does not require the weights to be always in the
same order, and is especially convenient for on-the-fly manipulations
(such as discarding some contributions, sorting them, and so forth).

We conclude this section with some general comments. The condition
$N_\mu=8$ can be trivially relaxed (which requires the inclusion
of $N_\mu$ itself in the list of the inputs); we have refrained from
doing so in the first public version of the code since three choices
for each of the hard scales typically give a good estimate of the 
uncertainties. A more extended set of input parameters would also
give one the possibility of performing scale variations by changing
the functional forms of the reference scales w.r.t.~those adopted for
the central results. Finally, the complete generality could be achieved
by storing in the LHE file the weights $w_i$'s. This would require,
however, that additional information be stored as well (such as
the values of the Bjorken $x$'s -- more details can be
found in ref.~\cite{Frederix:2011ss}). While this can be trivially
done, it would render the computation of the basis members $b_i$'s
of eq.~(\ref{basis}), and their subsequent combination with weights
according to eq.~(\ref{decomp}), a much more involved operation
than the present one if performed by an external user\footnote{This
{\em not} being the case at the LO, in LO-type generations we can
adopt a different strategy -- see appendix~\ref{sec:LOerrors}.}. 
Such a possibility is left open for future developments.

In the case of f(N)LO runs, the information on scale and PDF
variations is also available on an event-by-event basis. In keeping
with the fact that here one cannot have unweighted events, 
\aNLO\ will associate to each kinematic configuration an array
of weights (rather than a single weight), each of which gives
eq.~(\ref{decomp}) with the basis elements of eq.~(\ref{basis})
recomputed for all desired scale and PDF choices. This implies
that the second elements of the pairs in eqs.~(\ref{anaE})--(\ref{anaB})
will be turned into arrays, with dimensionality:
\beqn
1+(1+N_\mu)+N_{\rm PDF}\;\;\;\;&&
{\tt reweight\_scale=.true.}~~~~
{\tt reweight\_PDF=.true.}~~
\\
1+(1+N_\mu)\;\;\;\;&&
{\tt reweight\_scale=.true.}~~~~
{\tt reweight\_PDF=.false.}~~
\\
1+N_{\rm PDF}\;\;\;\;&&
{\tt reweight\_scale=.false.}~~~~
{\tt reweight\_PDF=.true.}~~~~
\eeqn
The first entry thus always gives the value of the weight associated
with central parameters. In the case of scale variations, this weight
is present twice (the ``1'' in \mbox{$(1+N_\mu)$}) essentially for
reasons of backward compatibility which should not concern the user.
The information of which weight is which is returned as the array
of strings {\tt weights\_info} and made available to the user's
initialisation routine {\tt analysis\_begin} -- this is equivalent
to the {\tt id} number discussed before in the context of LHE files.

\subsection{Scale and PDF uncertainties: the LO case\label{sec:LOerrors}}
At the LO, the structure of the cross section, eq.~(\ref{decomp}),
is trivial -- the sum contains one term $i=1$ (see however
footnote~\ref{ft:one}), which corresponds to 
the single PDF-and-coupling combination (i.e., to a single basis 
member, eq.~(\ref{basis})) relevant to this perturbative order:
this is the reason for the simplicity of eq.~(\ref{rff}).
Thus here, at variance with the NLO case, from the user's viewpoint 
it is therefore as simple to handle $w_1$ as is to handle $\sigma$.
This suggests the following strategy: when unweighted events are
produced, they are stored in an intermediate LHE file with a non-standard 
format, that features the weights $w_1$. {\em After} the end of the run, 
this LHE file is read by a standalone module, dubbed \syscalc, that converts 
the weights $w_1$ into the corresponding cross sections $\sigma$, and stores
them in a new LHE file, which has this time the standard format already
described in appendix~\ref{sec:errors}. This procedure is advantageous
because it allows one to run \syscalc\ as many times as desired using
the same intermediate LHE file obtained at generation time; this implies that
the type of scale or PDF variations need not be chosen before the
event-generation phase, as is the case at the NLO.
The \syscalc\ module can be installed\footnote{\syscalc\ requires 
that LHAPDF be installed as well.} with:

\noindent
~\prompt\ {\tt ~install syscalc}

\noindent
from the \aNLO\ shell. At runtime, \aNLO\ is instructed to save
the weights $w_1$ in the intermediate LHE file by setting

\noindent
{\tt ~~T        = use\_syst   ! Enable systematics studies}

\noindent
in {\tt run\_card.dat}. When this is done, \syscalc\ is automatically
called at the end of the run (although, as was said before, it can also
be run independently afterwards). The type of scale and PDF
variations considered are determined by the following entries
in {\tt run\_card.dat}:

\begin{verbatim}
0.5 1 2 = sys_scalefact  # Central scale factors
-1     = sys_scalecorrelation # for renormalization/scale variate
                              # -1: make all combination 
                              # -2: only correlated variation  
0.5 1 2 = sys_alpsfact  # \alpha_s emission scale factors
30 50 = sys_matchscale # variation of merging scale
# PDF sets and number of members (0 or none for all members).
CT10nlo.LHgrid = sys_pdf # 
\end{verbatim}

\noindent
The list of values to the left of {\tt sys\_scalefact} collects
the multiplicative factors in front of the reference (fixed or 
dynamic\footnote{Note that the {\tt setscales.f} code relevant to 
LO calculations is different w.r.t~that used in NLO ones.})
scales, and are thus analogous to the quantities $\ffR^\alpha$ and 
$\ffF^\beta$ that appear in eq.~(\ref{scalevar}); at variance with
the current NLO implementation, such a list can contain more than
three numbers. We stress that, when {\tt use\_syst=T}, the value
of {\tt scalefact} in {\tt run\_card.dat} is ignored.
The flag {\tt sys\_scalecorrelation} allows one to choose which of the
renormalisation/factorisation scale combinations are considered.
When ``-1' is entered, then all $\muR$ and $\muF$ values determined
by the list {\tt sys\_scalefact} are taken into account, while
with ``-2'' one restricts to code to dealing with $\muR=\muF$.
More sophisticated options are also available (allowing one to
select only some of the possible $(\muR,\muF)$ combinations), which
will be documented elsewhere. The entry {\tt sys\_pdf} is associated
with the study of PDF systematics; if set equal to a PDF error set,
all PDF members in that set will be considered. Alternatively, one
can specify the individual PDF members to be taken into account.
Finally, through \syscalc\ one can also investigate the systematics
relevant to tree-level merging (see sect.~\ref{sec:CKKW}), with 
{\tt sys\_alpsfact} and {\tt sys\_matchscale} corresponding to
$\as$-argument and $Q_{match}$ variations respectively. More
details on \syscalc\ can be found at:\\
$\phantom{a}${\tt https://cp3.irmp.ucl.ac.be/projects/madgraph/wiki/SysCalc}

\subsection{Other LO reweighting applications\label{sec:appLO}}
In this appendix we comment in the briefest of manners on some 
technicalities relevant to the matrix-element reweighting 
(eq.~(\ref{rwgtME})) and the matrix-element method (eq.~(\ref{rwgtLI2})) 
discussed in sect.~\ref{sec:rwgt}.

As was already mentioned in sect.~\ref{sec:LOgen}, the former procedure
is straightforwardly accessed through the \aNLO\ shell, by simply
setting {\tt reweight=ON} in the interactive talk-to phase. 
By doing so, \aNLO\ will use the file {\tt reweight\_card.dat}
to modify the parameters (found in {\tt param\_card.dat}) used
for the benchmark computation (that essentially corresponds to
the denominator of eq.~(\ref{rwgtME})). One may enter any number of
such modifications, each of which may contain any number of parameter
changes, to be done through the {\tt set} command (see the comments
inside the file {\tt reweight\_card.dat}). There are currently two
main limitations to this procedure. Firstly, the changes must occur
within one given model (i.e., one cannot reweight to matrix elements
computed in a model different w.r.t.~that adopted in the benchmark
calculation). Secondly, the accessible kinematical region in the
``new'' hypothesis must be equal to or smaller than the original
one. The interested reader can find more information by visiting:\\
$\phantom{a}${\tt https://cp3.irmp.ucl.ac.be/projects/madgraph/wiki/Reweight}

The matrix-element method is handled by \madweight. As was discussed
in sect.~\ref{sec:LOgen}, the current version embedded in \aNLO\ has vastly 
increased the speed of the previous version~\cite{Artoisenet:2010cn}.
Such an increase is mainly due to a better combination of subprocesses,
to a Monte-Carlo-type sum over jet-parton assigments (as opposed to
an exact sum), and to the possibility of performing the simultaneous
computation of $P({\bf q}|\alpha)$ in the case of multiple choices
of the transfer function. The \madweight\ executable specific
to the process generated by the user simply corresponds to employing
one of the reserved output keywords (see appendix~\ref{sec:MG5out}),
namely to executing, after the generation phase, the command:

\noindent
~\prompt\ {\tt ~output madweight MYPROC}

\noindent
After that, one may continue with using the \aNLO\ shell interface
by executing:

\noindent
~\prompt\ {\tt ~launch MYPROC}

\noindent
Further details can be found at:\\
$\phantom{a}${\tt https://cp3.irmp.ucl.ac.be/projects/madgraph/wiki/MadWeight}

\subsection{Output formats and standalone libraries\label{sec:MG5out}}

The main purpose of \aNLO\ is that of providing a self-contained framework 
where to compute cross sections and generate events at the desired level of 
perturbative accuracy, in both the SM and new-physics theories.  
On the other hand, the code may also be used to provide one with
only a given ingredient of a calculation (e.g., a matrix element)
which is to be performed elsewhere. This has been one of the defining
characteristic of \MadGraph, and has been wholly inherited by \aNLO.
Another example is the computation of a quantity which is not 
a cross section; a case in point is the likelihood dealt with
by \madweight\ (see sect.~\ref{sec:rwgt} and appendix~\ref{sec:appLO}).

All possible executables, libraries, or more elementary objects
that can be produced by \aNLO\ can simply be seen as {\em outputs}
of the (meta-)code; they are in fact in one-to-one correspondence with
the optional first keyword of the shell command:

\noindent
~\prompt\ {\tt ~output} [{\em OutputForm}] [{\tt MYPROC}]

\noindent
The possible choices of {\em OutputForm} can be readily obtained
by using the autocompletion {\tt <TAB>} key in the shell after having
typed {\tt output} (or with {\tt help output}). 
By doing so, one will notice that they all apply, 
bar one case ({\tt standalone}, see below), to LO-type generations.
Here we shall not list them all, but limit ourselves to commenting on 
those which are the most useful from the user's point of view. These are:
\begin{itemize}
\item {\tt standalone}: self-contained Fortran77 library for the
  computation of either tree-level matrix elements (after an LO-type 
  generation), or one-loop matrix elements (after an NLO-type generation
  with the keyword {\tt virt=coupling$_1$ ...} -- see 
  appendix~\ref{sec:verbose}).
  The directory structure thus created contains a simple program
  ({\tt check\_sa.f}) which allows one to evaluate the 
  matrix elements pointwise for test purposes.
\item {\tt standalone\_cpp}: the same as above, but in C++ rather
 than in Fortran77. Works only for tree-level matrix elements. 
\item {\tt pythia8}: self-contained library for the computation of 
  tree-level matrix elements, in a format which can be directly used in
  the \PYe\ PSMC. It includes a simple driver that one can employ to
  steer \PYe\ sample runs, and a {\tt Processes\_*} directory that 
  contains the said matrix elements. More details can be found
  in sect.~3.2 of ref.~\cite{Alwall:2011uj}. 
\item {\tt madweight}: this output keyword allows one to set up
  a computation with \madweight\ (see appendix~\ref{sec:appLO}).
\end{itemize}
We conclude this appendix by stressing that the above list 
(supplemented by that of the other output keywords not explicitly given 
here) is by no means all-inclusive. The reader who is interested in
a particular output format specifically suited to his/her need
is encouraged to contact us in order for that to be developed 
and included in future versions of \aNLO. For instance, dedicated 
outputs for matrix elements to be used in the 
{\sc\small MatchBox}~\cite{Platzer:2011bc} framework and by 
{\sc\small EventDeconstruction}~\cite{Soper:2012pb,Soper:2014rya}
are being developed.

\section{Features of one-loop computations\label{sec:OLtech}}

\subsection{TIR and IREGI\label{sec:IREGI}}
The aim of this section is that of presenting the basic procedures
used in TIR, and in particular by the program \IREGI\footnote{The acronym
stands for ``Integral REduction with General positive propagator
Indices''.}; more details on the latter will be given 
elsewhere~\cite{ShaoIREGI}.
\IREGI\ computes the integral that appears as first element in
the set of eq.~(\ref{ML5inputs}); owing to the fact that we apply
TIR to a given loop topology (see eq.~(\ref{redML5})), we
can exploit eq.~(\ref{topdef}) to simplify the notation here and
drop the dependence on $l_t$. On the other hand, at variance with
what was done in sect.~\ref{sec:OPP}, in order to make things more
explicit it is convenient to insert in the notation the dependence
on the external (four-dimensional, owing to the 't~Hooft-Veltman scheme)
four-momenta $p_i$ and on the masses $m_i$ that circulate in the loop 
(see eq.~(\ref{Dbardef})), so that the integral we are 
interested in reads as follows:
\beq
I^{\mu_1\ldots\mu_r}(\{p_i\},\{m_i\})=
\int d^d \bqloop\,\frac{\qloop^{\mu_1}\ldots\qloop^{\mu_r}}
{\prod_{i=0}^{m-1}\db{i}}\,.
\label{iregi1}
\eeq
Lorentz-covariance then guarantees that eq.~(\ref{iregi1})
can be re-written as follows:
\beqn
&&I^{\mu_1\ldots\mu_r}(\{p_i\},\{m_i\})=
\label{eq:step11}
\\*&&\phantom{aaaa}
\sum_{2j+i_0+i_1+\cdots+i_{m-1}=r}{\{[g]^{j}[p_0]^{i_0}\cdots 
[p_{m-1}]^{i_{m-1}}\}^{\mu_1\ldots\mu_r}
I_{ji_0\ldots i_{m-1}}(\{p_i\},\{m_i\})}\,,
\nonumber
\eeqn
for certain scalar integrals $I_{ji_0\ldots i_{m-1}}$, and where
the symmetric tensor form\\ \mbox{$\{[{g}]^{j}[p_0]^{i_0}\ldots 
[p_{m-1}]^{i_{m-1}}\}^{\mu_1\ldots\mu_r}$} is defined in such a way 
that all non-equivalent permutations of the Lorentz indices 
\mbox{$\mu_1,\ldots\mu_r$} on $j$ metric tensors ${g}$ and $i_s$ 
external momenta $p_s$ contribute with weight one. For example:
\beqn
\{[{g}]^2[p_0]^0[p_1]^0\}^{\mu_1\ldots\mu_4}&=&
{g}^{\mu_1\mu_2}{g}^{\mu_3\mu_4}+{g}^{\mu_1\mu_3}{g}^{\mu_2\mu_4}+
{g}^{\mu_1\mu_4}{g}^{\mu_2\mu_3},
\nonumber\\
\{[{g}]^1[p_0]^0[p_1]^2\}^{\mu_1\ldots\mu_4}&=&
{g}^{\mu_1\mu_2}p_1^{\mu_3}p_1^{\mu_4}+
{g}^{\mu_1\mu_3}p_1^{\mu_2}p_1^{\mu_4}+
{g}^{\mu_1\mu_4}p_1^{\mu_2}p_1^{\mu_3},
\nonumber\\
\{[{g}]^0[p_0]^0[p_1]^4\}^{\mu_1\ldots\mu_4}&=&
p_1^{\mu_1}p_1^{\mu_2}p_1^{\mu_3}p_1^{\mu_4},
\nonumber\\
\{[{g}]^0[p_0]^1[p_1]^2\}^{\mu_1\mu_2\mu_3}&=&
p_0^{\mu_1}p_1^{\mu_2}p_1^{\mu_3}+
p_0^{\mu_2}p_1^{\mu_1}p_1^{\mu_3}+
p_0^{\mu_3}p_1^{\mu_1}p_1^{\mu_2}.
\eeqn
Given eq.~(\ref{eq:step11}), the computation of the original tensor 
integral of eq.~(\ref{iregi1}) is reduced to that of the scalar
integrals $I_{ji_0\ldots i_{m-1}}$. One starts by observing that the
latter are independent of the number of dimensions used in the
numerator of eq.~(\ref{iregi1}), because the decomposition
of eq.~(\ref{eq:step11}) would hold, with the formal replacement
$g\to\bar{g}$, if one had replaced $\qloop\to\bqloop$ in eq.~(\ref{iregi1}).
Therefore, since working with the same number of dimensions in all
parts of a computation is algebraically convenient, one determines
the $I_{ji_0\ldots i_{m-1}}$ directly in $d$ dimensions.
This is done by recursively expressing such scalar integrals in
terms of lower-point ones, till only integrals that cannot be
further reduced are obtained -- these are just a few and well
known: \IREGI\ makes use of those tabulated in 
{\sc\small OneLoop}~\cite{vanHameren:2010cp} and 
{\sc\small QCDloop}~\cite{Ellis:2007qk}. 

One way of performing such recursive reduction stems from the pioneering 
work of Passarino and Veltman~\cite{Passarino:1978jh}: by contracting
both sides of eq.~(\ref{iregi1}) (in $d$-dimensions) with metric
tensors and external four-momenta, one relates the $I_{ji_0\ldots i_{m-1}}$ 
integrals to lower-rank tensor or lower-point scalar integrals. This 
lowering is due to the fact that, when contracting, one obtains the scalar
products $\bqloop^2$ and $\bqloop\mydot p_i$, which are then
re-expressed as follows:
\beqn
\bqloop^2&=&\db{0}+m_0^2\,,
\\
\bqloop\mydot p_i&=&(\db{i}-\db{0}+m_i^2-m_0^2)/2\,,
\eeqn
thus either cancelling some of the denominators or simplifying the
dependence on $\bqloop$ in the numerator. The procedure is algebraic,
and one ends up with a system of equations where the unknowns are the
scalar integrals, and the coefficients known functions of the kinematic
variables. The solution of such a system is non-trivial, owing for example
to the presence of special kinematic configurations; \IREGI\ implements
to a large extent the strategies proposed in ref.~\cite{Denner:2005nn}.

An alternative and independent way for performing the recursive
reduction follows the work of Davydychev~\cite{Davydychev:1991va}.
One introduces the generalised loop-tensor and basic-scalar integrals:
\beqn
\bar{I}^{\mu_1\ldots\mu_r}(d,\{\nu_i\},\{p_i\},\{m_i\})&=&
\frac{(\mu^2)^{2-d/2}}{(2\pi)^d}
\int{d^d\bqloop \frac{\bqloop^{\mu_1}\ldots \bqloop^{\mu_r}}
{\prod_{i=0}^{m-1}{\bar{D}_i^{\nu_i}}}}\,,
\label{eq:genI}
\\
I_0(d,\{\nu_i\},\{p_i\},\{m_i\})&=&
\frac{(\mu^2)^{2-d/2}}{(2\pi)^d}
\int{d^d\bqloop \frac{1}{\prod_{i=0}^{m-1}{\bar{D}_i^{\nu_i}}}}\,,
\label{eq:genI0}
\eeqn
where the scale-dependent prefactor is conventional, and the
indices $\nu_0,\nu_1,\cdots,\nu_{m-1}$ are positive integers.
By using a Feynman-parameter representation and the $d$-dimensional
analogue of eq.~(\ref{eq:step11}):
\beqn
\bar{I}^{\mu_1\ldots\mu_r}(d,\{\nu_i\},\{p_i\},\{m_i\})&=&
\frac{i}{(4\pi)}(4\pi\mu^2)^{2-d/2}
\label{eq:feynparam0}
\\*
&\times&\sum_{2j+i_0+\ldots+i_{m-1}=r}{\{[\bar{g}]^{j}[p_0]^{i_0}\ldots
[p_{m-1}]^{i_{m-1}}\}^{\mu_1\ldots\mu_r}}
\nonumber\\*
&\times&(-1)^{\sum_{i=0}^{m-1}{\nu_i}+r-j}
\frac{\Gamma(\sum_{i=0}^{m-1}{\nu_i}-d/2-j)}{2^j\prod_{i}^{m-1}{\Gamma(\nu_i
)}}\int^{1}_{0}
{\prod_{i=0}^{m-1}{dy_iy_i^{\nu_i+i_i-1}}}
\nonumber\\*
&\times&\delta(\sum_{i=0}^{m-1}{y_i}-1)
\left[-\sum_{i<j}{y_iy_j(p_i-p_j)^2+\sum_{i=0}^{m-1}{y_im_i^2}}
\right]^{j+d/2-\sum_{i=0}^{m-1}{\nu_i}},
\nonumber\\
I_0(d,\{\nu_i\},\{p_i\},\{m_i\})&=&
\frac{i}{(4\pi)}(4\pi\mu^2)^{2-d/2}
\frac{\Gamma(\sum_{i=0}^{m-1}{\nu_i}-d/2)}{\prod^{m-1}_{i=0}
{\Gamma(\nu_i)}}(-1)^{\sum^{m-1}_{i=0}{\nu_i}}
\nonumber\\*
&\times&\int_{0}^{1}{\prod_{i=0}^{m-1}{dy_iy_i^{\nu_i-1}}
\delta(\sum_{i=0}^{m-1}{y_i}-1)}
\nonumber\\*
&\times&
\left[-\sum_{i<j}{y_iy_j(p_i-p_j)^2+\sum^{m-1}_{i=0}{y_im_i^2}}
\right]^{d/2-\sum_{i=0}^{m-1}{\nu_i}},
\label{eq:feynparam}
\eeqn
one arrives at:
\beqn
\bar{I}^{\mu_1\ldots\mu_r}(d,\{\nu_i\},\{p_i\},\{m_i\})&=&
\sum_{2j+i_0+\cdots+i_{m-1}=r}
{\{[\bar{g}]^{j}[p_0]^{i_0}\cdots [p_{m-1}]^{i_{m-1}}\}^{\mu_1\cdots\mu_r}}
\nonumber\\*
&\times&
\frac{(4\pi\mu^2)^{r-j}}{(-2)^j}
\left(\prod^{m-1}_{i=0}{\frac{\Gamma(\nu_i+i_i)}{\Gamma(\nu_i)}}\right)
\nonumber\\*
&\times&
I_0(d+2(r-j),\{\nu_i+i_i\},\{p_i\},\{m_i\})\,.
\label{eq:IbarDavy}
\eeqn
By using eqs.~(\ref{eq:step11}) (in $d$ dimensions) and~(\ref{eq:IbarDavy}),
one finally obtains the relation:
\beqn
I_{ji_0\ldots i_{m-1}}(\{p_i\},\{m_i\})&=&
\left[\frac{(\mu^2)^{2-d/2}}{(2\pi)^d}\right]^{-1}
\frac{(4\pi\mu^2)^{r-j}}{(-2)^j}
\left(\prod_{i=0}^{m-1}{\frac{\Gamma(1+i_i)}{\Gamma(1)}}\right)
\nonumber\\*
&\times&
\left.I_0(d+2(r-j),\{1+i_i\},\{p_i\},\{m_i\})\right|_{d=4-2\epsilon}\,.
\label{eq:sfrelation}
\eeqn
With eq.~(\ref{eq:feynparam}) one is also able to 
derive relationships among scalar integrals in different dimensions. 
For instance, eq.~(6) of ref.~\cite{Davydychev:1991va} can be
easily obtained: 
\beq
I_0(d-2,\{\nu_i\},\{p_i\},\{m_i\})=
-4\pi\mu^2\,\sum_{s=0}^{m-1}{\nu_s\,I_0(d,\{\nu_i+\delta_{is}\},
\{p_i\},\{m_i\})}\,,
\label{iregi2}
\eeq
where by $\delta_{is}$ we have denoted the Kronecker symbol.
Furthermore, other recursion relations for the scalar integrals 
$I_0(d,\{\nu_i\},\{p_i\},\{m_i\})$ can be obtained with the help of 
the integration-by-parts method~\cite{Tarasov:1996br,Duplancic:2003tv}, 
which exploits the fact that integrals are translation-invariant in 
dimensional regularization. The practical implementation in \IREGI\ 
of the method discussed here follows ref.~\cite{Duplancic:2003tv}.

\IREGI\ can use either of the two methods presented above for the
recursive reduction, the actual choice being made by the calling 
code (in our case, \MLf). The current default in \aNLO\ is
the use of Passarino and Veltman; this is straightforward to change,
since it is simply controlled by a parameter in an input card.
\IREGI\ has a minimal internal stability control: should 
e.g.~Passarino and Veltman procedure be flagged unstable, the code
will turn to using Davydychev's. We stress that this by no means
replaces the stability control performed by \MLf, described
in sect.~\ref{sec:OPP}.

\subsection{Quantitative profile of MadLoop performances\label{sec:MLperf}}
We have already stressed that the results presented in sect.~\ref{sec:next}
can be used as benchmarks for the validation of other codes. However, they 
reveal only one aspect of the performances of \MLf, which we complement 
in this appendix by reviewing some quantitative characteristics of the
handling of the scattering processes considered before.
A summary of such characteristics is reported in table~\ref{tab:MLperf}.
\begin{table}
\begin{center}
\begin{tabular}{l|c|c|c|c}\toprule
   & $g g \rightarrow d \bar{d} b \bar{b} t \bar{t}$
   & $u \bar{u} \rightarrow d \bar{d} t \bar{t}$
   & $u \bar{d} \rightarrow d \bar{d} W^{+} Z H$
   & $g g \rightarrow \tilde{t}_1 \tilde{t}_1^\star g$
  \\\midrule
\# Feynman diagrams & 54614 & 10947 & 187138 & 3952
\\
\# topologies & 8190 & 811 & 8098 & 437
\\
$\Delta(k_0)$ & 0 & 4 & 2 & 0
\\
\# non-zero hel. configs. & 128 & 16 & 27 & 8
\\
Generation time & 15h & 28min & 25h & 1min 38s
\\
Running time & 18.6s (19\%) & 895ms (72\%) & 26.4s (32\%) & 83.6ms (68\%)
\\
Output code size & 600 Mb & 20 Mb & 700 Mb & 6 Mb
\\
Runtime RAM usage & 3.6 Gb & 152 Mb & 8.3 Gb & 81 Mb
\\
Stability & $2\cdot 10^{-8}$ & $1\cdot 10^{-7}$ & $4\cdot 10^{-7}$ & $1\cdot 10^{-7}$ 
\\\bottomrule
\end{tabular}
\end{center}
\caption{\label{tab:MLperf}
Performances of \MLf\ in the context of the computations
presented in sect.~\ref{sec:next}. See the text for details.
}
\end{table}

The number of topologies is the upper bound of the sum over the
index $t$ in eq.~(\ref{redML5}), and it corresponds to the number 
of independent loop reductions (i.e., of evaluations of the $\RED[\,]$ 
operator introduced in eq.~(\ref{REDdef})) for one kinematic
configuration. As one can see from the table, such a number is much 
smaller than the number of Feynman diagrams, which emphasises the 
importance of the optimization induced by eq.~(\ref{topdef}).
The quantity $\Delta(k_0)$ is equal to the number of coupling-constant
combinations, minus one, at the Born level, which is larger than zero in the 
case of a mixed-coupling expansion. See the beginning of sect.~\ref{sec:NLO},
and eqs.~(\ref{Deltadef}) and~(\ref{Delkpo}) in particular, for more details.
The generation time includes the compilation of the source code. The running
time corresponds to the time taken by the code output by \MLf\ to compute
the one-loop squared matrix element summed over colours but for a single
helicity configuration\footnote{The timing indicated is for a single core of a
  2.7 GHz i7 CPU, with the gfortran compiler v4.8.1 without optimization flags
  (which have been shown to have a negligible impact).}. The percentage in
parenthesis specifies the fraction of the running time spent in the loop
reduction which, we remind the reader, is independent of the number of 
non-zero helicity 
configurations considered. The complementary fraction of the time is spent in
the computation of the coefficients $C^{(r)}_{\mu_1\ldots\mu_r;h,l}$ of
eq.~(\ref{ML5inputs}), and scales linearly with the number of
helicity combinations. The size of the output code includes external 
data files (that essentially contain the colour coefficients $\Lambda_{lb}$)
loaded by the library which is in itself much lighter. The RAM measure 
reported is the runtime peak of residential memory allocated.
The figures given in the last row 
are the relative accuracies estimated by the \MLf\ internal stability tests 
in the context of double-precision computations that use the kinematic 
configurations considered in sect.~\ref{sec:next} (however, we stress 
again that the matrix-element results have been obtained in quadruple 
precision); obviously, these pointwise accuracies have only an indicative 
value, since the real figure of merit necessitates averaging over a large 
statistical sample of independent kinematic configurations.

We conclude this appendix by stressing that the data reported in
table~\ref{tab:MLperf} can be obtained by the user by using
the command {\tt check profile} in the \aNLO\ shell. For example,
one would have

\noindent
~\prompt\ {\tt ~check profile g g > t t\~{} z [virt=QCD]}

\noindent
in the case of the virtual corrections to $t\bt Z$ production
(i.e., process e.8 of table~\ref{tab:results_tv}).

\subsection{Computation of the integrand polynomial coefficients
\label{sec:pol}}
We have discussed in section sect.~\ref{sec:OPP} how the use of the
loop-integrand representation of eq.~(\ref{Nhlexp}) increases the speed of
OPP-based integral reductions, as well as giving one the possibility of using
TIR methods. This appendix elaborates on the techniques adopted in \MLf\ for
the computation of the coefficients $C^{(r)}_{\mu_1\ldots\mu_r;h,l}$.  The key
fact is that such coefficients are fully symmetric tensors of
rank $r$ with only $\binom{3+r}{r}$ independent entries. 
In renormalisable theories and in the Feynman gauge the number of 
loop propagators sets the maximal rank in $\qloop^\mu$ of that 
loop-integrand numerator, so
that the total number of coefficients necessary to express the numerator of
any loop of a, say, $2 \to 6$ process, is at most
$N_{coeff}(r_{\max}=8)\equiv\sum_{r=0}^{r_{\max}=8} \binom{3+r}{r}=495$ which
is well within the reach of modern computers.
We start by rewriting the analogue of eq.~(\ref{Nhlexp}) for any polynomial
$P^{(r_{\max})}(\qloop^\mu)$ of maximal rank $r_{\max}$ with the 
following shorthand and symbolic notation:
\beqn 
P^{(r_{\max})}(\qloop^\mu) =  
C^{(r_{\max})}_{\dot{k}}\qloop^{\dot{k}}\,,
\label{eq:polyshorthand}
\eeqn
where $\dot{k}$ takes values between 1 and $N_{coeff}(r_{\max})$,
and effectively defines a map between the sets of Lorentz indices
$\mu_i$ and integer numbers. The choice of such a map is arbitrary, 
and we use what follows:
\beqn
&&C^{(r)}_{\mu_1,\cdots,\mu_{r}}\,,\;\;\;\;
\mu_1 \leq \cdots \leq \mu_r\;\;\;\;
\rightarrow\;\;\;\; 
C^{(r_{\max})}_{\dot{k}({\mu_1,\cdots,\mu_{r}})}\,,
\\ 
&&\dot{k}({\mu_1,\cdots,\mu_{r}}) = N_{coeff}(r-1) +
   \sum\limits_{i=1}^{r}(1-\delta_{0\mu_i})\frac{(\mu_i+i-1)!}{i!(\mu_i-1)!}\,,
 \label{eq:dotdef}
 \eeqn
where we define $N_{coeff}(-1)=0$ (relevant to $r=0$). To make things
more explicit with one example, eq.~(\ref{eq:polyshorthand}) reads,
with $r_{\max}=2$:
 \beqn
 C^{(2)}_{\dot{k}} \qloop^{\dot{k}} &\equiv &\; C^{(2)}_{0}+C^{(2)}_{1}
 \qloop^{0}+C^{(2)}_{2} \qloop^{1}+C^{(2)}_{3} \qloop^{2}+C^{(4)}_{4}
 \qloop^{3} 
\nonumber\\ 
&+& C^{(2)}_{5} \qloop^{0}\qloop^{0} + C^{(2)}_{6}
 \qloop^{0}\qloop^{1} + C^{(2)}_{7} \qloop^{1}\qloop^{1} + C^{(2)}_{8}
 \qloop^{0}\qloop^{2} + C^{(2)}_{9} \qloop^{1}\qloop^{2} 
\nonumber\\ 
&+& C^{(2)}_{10}
 \qloop^{2}\qloop^{2} + C^{(2)}_{11} \qloop^{0}\qloop^{3} + C^{(2)}_{12}
 \qloop^{1}\qloop^{3} + C^{(2)}_{13} \qloop^{2}\qloop^{3} + C^{(2)}_{14}
 \qloop^{3}\qloop^{3}\,.
 \eeqn
The computation of the coefficients $C^{(r_{\max})}_{\dot{k}}$ with \MLf\ is
entirely numerical and follows the \MadGraph\ procedure for evaluating Feynman
diagrams. At tree level, the internal currents of a given Feynman diagram are
denoted by $w^{(n)}_{j}$, with the integer $n$ labeling them and the index $j$
spanning the representation of the particle associated with the current.
In \MLf\, the \emph{loop} currents are promoted to more general objects
embedding polynomials in $\qloop^\mu$ and they are denoted by
$W^{(n)}_{i,r,\dot{k},j}$, with the additional index $i$ specifying the choice
of the external current for the first \Lcut\ particle (which is non-physical
and whose only constraint is to reproduce the Lorentz trace when summed
over). The index $\dot{k}$ labels the coefficients of the polynomial of
maximal rank $r$ according to the convention of eq.~(\ref{eq:dotdef}).
\begin{figure}[h]
  \begin{center}
    \epsfig{figure=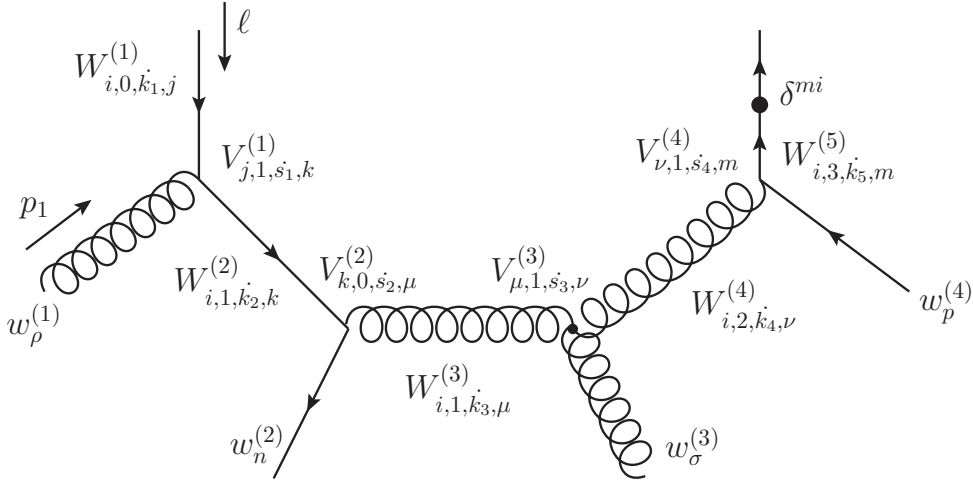,width=0.9\textwidth}
\caption{\label{fig:openloopex}An example of the \MLf\ construction of the
  coefficients $C^{(r_{\max})}_{\dot{k}}$. $W$ denotes the loop currents and
  $V$ the vertex polynomials (see the text). The figure depicts an \Lcut\ 
  diagram; the original box loop is obtained by sewing the two fermion lines at 
  the top.}
\end{center}
\end{figure}
Figure~\ref{fig:openloopex} shows a complete example of the different numerical
objects manipulated by \MLf\ in order to construct the polynomial coefficients
for the integrand of a box loop diagram . Note that since the currents
$w^{(\#)}$ attached to the loop are independent of the loop momentum
$\qloop^\mu$, it is irrelevant to know whether they are external
currents or originate from larger trees.  The starting loop current
$W^{(1)}_{i,0,\dot{k_1},j}$ is a polynomial of rank 0 since it does not have
any loop momentum dependence; its index $\dot{k_1}$ can therefore only take
the value 0. For all loop diagrams, the starting current is always
$W^{(1)}_{i,0,\dot{k_1},j}=\delta_{ij}\delta_{\dot{k_1}0}$.

The objects $V^{(\#)}_{a,r,\dot{s},b}$ are the representations (polynomials 
of rank $r$ in the loop momentum)  of the vertex plus propagator structures, 
with $a$ and $b$ the incoming and outgoing loop current indices respectively. 
For example, the explicit expression of $V^{(1)}_{j,1,\dot{s_1},k}$ for a 
massless quark in fig.~\ref{fig:openloopex} reads:
\beqn 
V^{(1)}_{j,1,\dot{s_1},k} \qloop^{\dot{s_1}} = \underbrace{\imath g_s
  \gamma^\rho_{ji} w^{(1)}_\rho}_{\mbox{vertex}}
\underbrace{\gamma_{ik}^\mu(\qloop_\mu+p_{1\mu})}_{\mbox{propagator}}
\label{eq:V1def}\,,
\eeqn
where the denominator of the propagator is removed since it is already
accounted for in the integral-reduction procedure. To be more definite, 
we show here the expression of each coefficient of the vertex polynomial of
eq.~(\ref{eq:V1def}):
\beqn 
V^{(1)}_{j,1,0,k} = \imath g_s
(\slashed{w}^{(1)}\slashed{p_1})_{jk}\,,\;\;\;\;V^{(1)}_{j,1,\dot{s_1},k} =
\imath g_s (\slashed{w}^{(1)}\gamma^{(\dot{s_1}-1)})_{jk}\;\;\;\;
{\rm for}\;\;\;\;\dot{s_1}=1,\dots,4\,.
\label{eq:V1defprecise}
\eeqn
In renormalisable theories, the rank of the vertex polynomials is maximally
equal to one as only one power of the loop momentum can arise from either the
propagator or the vertex itself (but not from both). This constraint in
not enforced by \MLf, since it would make it impossible to compute 
one-loop diagrams in effectives theories (such as HEFT). 
The numerical routines for the evaluation of
vertex polynomials are generated automatically by \ALOHA\ for each process,
using the \UFO\ model specifications; this is what renders the
optimisation discussed in this appendix applicable to any model.

Each subsequent loop current $W^{(n+1)}$ is obtained from the previous one
$W^{(n)}$ and the vertex polynomial $V^{(n)}$ placed in between via the
defining implicit relation:
\beqn
W^{(n+1)}_{i,r_1+r_2,\dot{k_1},j}\qloop^{\dot{k_1}}=
(W^{(n)}_{i,r_1,\dot{k_2},m}\qloop^{\dot{k_2}})(V^{(n)}_{m,r_2,\dot{s},j}
\qloop^{\dot{s}})
\label{eq:lwfupdate}.
\eeqn
The r.h.s.~of eq.~(\ref{eq:lwfupdate}) is a multiplication of two polynomials
and each coefficient of $W^{(n+1)}$ is obtained by summing the corresponding
terms in the expanded product. This implies that a symmetrisation of the
coefficients is performed after each loop vertex and this step is crucial in
order to limit their proliferation and the resulting computing time. To
illustrate this, eq.~(\ref{eq:lwfupdate}) is rewritten here for the case
$r_1=r_2=1$ with $\dot{k_1}=9$, corresponding to the term multiplying
$\qloop^1\qloop^2$ which can come either from $\dot{k_2}=2$, $\dot{s}=3$ or
$\dot{k_2}=3$, $\dot{s}=2$:
\beqn
W^{(n+1)}_{i,2,9,j}=W^{(n)}_{i,1,2,m}V^{(n)}_{m,1,3,j}+
W^{(n)}_{i,1,3,m}V^{(n)}_{m,1,2,j}\,.
\label{eq:lwfupdateex}
\eeqn
Note the implicit summation on the index $m$ which spans the representation 
of the particle circulating in the loop just before the vertex $V^{(n)}$.
Eq.~(\ref{eq:lwfupdate}) is then iteratively used to compute all
loop currents, until the second \Lcut\ leg, which includes the last
vertex and the \Lcut\ propagator, is reached. The coefficients 
$C_{\dot{k}}$ of an $n$-point loop diagram are then simply obtained 
by \emph{closing} the Lorentz trace:
\beqn
C_{\dot{k}}=W^{(n+1)}_{i,r,\dot{k},i}.
\eeqn
In the example of fig.~\ref{fig:openloopex}, this translates into:
\beqn
C_{\dot{k}}=W^{(5)}_{i,3,\dot{k},m}\delta^{mi}=
\sum\limits_{i=1}^4 W^{(5)}_{i,3,\dot{k},i}\,,
\eeqn
with the sum written explicitly for clarity. It is clear that many loop diagrams
can share some of their constituting loop currents; for example the \Lcut\ box
diagram of fig.~\ref{fig:openloopex} can be prolonged to form loops with more
propagators. \MLf\ takes advantage of this and computes each loop current only
once, for the first loop diagram in which it appears. An algorithm analogous
to the one described in ref.~\cite{Cascioli:2011va} is used for choosing the
\Lcut\ location in order to maximise the number of loop currents recycled.

We conclude this appendix by stressing that all of the optimisations inherited
from the integrand representation of eq.~(\ref{Nhlexp}) can be turned off in
\aNLO\ via the option {\tt{loop\_optimized\_output}} of the interactive
interface and that, when this is done, the structure of the code output 
by \MLf\ is completely different. For this reason and despite being 
significantly slower, the non optimized output mode provides a powerful 
self-consistency check and is useful for debugging purposes.

\section{Third-party codes included in \aNLO\label{sec:tpc}}
The tarball of \aNLO\ is self-contained, and ready-to-run.
This is also thanks to the fact that several third-party
codes are included into it. We list all of them here:
\ALOHA~\cite{deAquino:2011ub}, \CutTools~\cite{Ossola:2007ax},
\FJ\ {\sc\small (core)}~\cite{Cacciari:2011ma},
\HELAS~\cite{Murayama:1992gi}, \HWs~\cite{Corcella:2000bw},
{\sc\small MINT}~\cite{Nason:2007vt}, 
{\sc\small OneLoop}~\cite{vanHameren:2010cp},
\PYs~\cite{Sjostrand:2006za},
{\sc\small QCDloop}~\cite{Ellis:2007qk},
{\sc\small RAMBO}~\cite{Kleiss:1985gy},
{\tt StdHEP}~\cite{StdHEP},
{\sc\small Vegas}~\cite{Lepage:1977sw,Lepage:1980dq}.

In the case of \FJ, what is included in \aNLO\ is a stripped version of 
that code (visit {\tt www.fastjet.fr} for more details). For more extended 
jet-reconstruction capabilities, the user might want to install \FJ\ proper 
(thus including all the relevant plugins).

\bibliographystyle{UTPstyle}
\bibliography{amcatnlo5_bib}

\providecommand{\href}[2]{#2}\begingroup\raggedright\begin{thebibliography}{10%
0}

\bibitem{Gross:1973id}
D.~J. Gross and F.~Wilczek, {\it {Ultraviolet Behavior of Nonabelian Gauge
  Theories}},  {\em Phys.Rev.Lett.} {\bf 30} (1973) 1343--1346.

\bibitem{Politzer:1973fx}
H.~D. Politzer, {\it {Reliable Perturbative Results for Strong Interactions?}},
   {\em Phys.Rev.Lett.} {\bf 30} (1973) 1346--1349.

\bibitem{Appelquist:1973uz}
T.~Appelquist and H.~Georgi, {\it {$e^+e^-$ annihilation in gauge theories of
  strong interactions}},  {\em Phys.Rev.} {\bf D8} (1973) 4000--4002.

\bibitem{Sterman:1977wj}
G.~F. Sterman and S.~Weinberg, {\it {Jets from Quantum Chromodynamics}},  {\em
  Phys.Rev.Lett.} {\bf 39} (1977) 1436.

\bibitem{Buras:1977qg}
A.~Buras, E.~Floratos, D.~Ross, and C.~T. Sachrajda, {\it {Asymptotic Freedom
  Beyond the Leading Order}},  {\em Nucl.Phys.} {\bf B131} (1977) 308.

\bibitem{Bardeen:1978yd}
W.~A. Bardeen, A.~Buras, D.~Duke, and T.~Muta, {\it {Deep Inelastic Scattering
  Beyond the Leading Order in Asymptotically Free Gauge Theories}},  {\em
  Phys.Rev.} {\bf D18} (1978) 3998.

\bibitem{Altarelli:1978id}
G.~Altarelli, R.~K. Ellis, and G.~Martinelli, {\it {Leptoproduction and
  Drell-Yan Processes Beyond the Leading Approximation in Chromodynamics}},
  {\em Nucl.Phys.} {\bf B143} (1978) 521.

\bibitem{Celmaster:1980ji}
W.~Celmaster and R.~J. Gonsalves, {\it {Fourth Order QCD Contributions to the
  $e^+ e^-$ Annihilation Cross-Section}},  {\em Phys.Rev.} {\bf D21} (1980)
  3112.

\bibitem{Ellis:1980wv}
R.~K. Ellis, D.~Ross, and A.~Terrano, {\it {The Perturbative Calculation of Jet
  Structure in $e^+ e^-$ Annihilation}},  {\em Nucl.Phys.} {\bf B178} (1981)
  421.

\bibitem{Frixione:1995ms}
S.~Frixione, Z.~Kunszt, and A.~Signer, {\it {Three jet cross-sections to
  next-to-leading order}},  {\em Nucl.Phys.} {\bf B467} (1996) 399--442,
  [\href{http://xxx.lanl.gov/abs/hep-ph/9512328}{{\tt hep-ph/9512328}}].

\bibitem{Catani:1996vz}
S.~Catani and M.~Seymour, {\it {A General algorithm for calculating jet
  cross-sections in NLO QCD}},  {\em Nucl.Phys.} {\bf B485} (1997) 291--419,
  [\href{http://xxx.lanl.gov/abs/hep-ph/9605323}{{\tt hep-ph/9605323}}].

\bibitem{Frixione:1997np}
S.~Frixione, {\it {A General approach to jet cross-sections in QCD}},  {\em
  Nucl.Phys.} {\bf B507} (1997) 295--314,
  [\href{http://xxx.lanl.gov/abs/hep-ph/9706545}{{\tt hep-ph/9706545}}].

\bibitem{Kosower:1997zr}
D.~A. Kosower, {\it {Antenna factorization of gauge theory amplitudes}},  {\em
  Phys.Rev.} {\bf D57} (1998) 5410--5416,
  [\href{http://xxx.lanl.gov/abs/hep-ph/9710213}{{\tt hep-ph/9710213}}].

\bibitem{Campbell:1998nn}
J.~M. Campbell, M.~Cullen, and E.~N. Glover, {\it {Four jet event shapes in
  electron--positron annihilation}},  {\em Eur.Phys.J.} {\bf C9} (1999)
  245--265, [\href{http://xxx.lanl.gov/abs/hep-ph/9809429}{{\tt
  hep-ph/9809429}}].

\bibitem{Bern:1994zx}
Z.~Bern, L.~J. Dixon, D.~C. Dunbar, and D.~A. Kosower, {\it {One-loop n-point
  gauge theory amplitudes, unitarity and collinear limits}},  {\em Nucl.Phys.}
  {\bf B425} (1994) 217--260,
  [\href{http://xxx.lanl.gov/abs/hep-ph/9403226}{{\tt hep-ph/9403226}}].

\bibitem{delAguila:2004nf}
F.~del Aguila and R.~Pittau, {\it {Recursive numerical calculus of one-loop
  tensor integrals}},  {\em JHEP} {\bf 0407} (2004) 017,
  [\href{http://xxx.lanl.gov/abs/hep-ph/0404120}{{\tt hep-ph/0404120}}].

\bibitem{Bern:2005cq}
Z.~Bern, L.~J. Dixon, and D.~A. Kosower, {\it {Bootstrapping multi-parton loop
  amplitudes in QCD}},  {\em Phys.Rev.} {\bf D73} (2006) 065013,
  [\href{http://xxx.lanl.gov/abs/hep-ph/0507005}{{\tt hep-ph/0507005}}].

\bibitem{Ossola:2006us}
G.~Ossola, C.~G. Papadopoulos, and R.~Pittau, {\it {Reducing full one-loop
  amplitudes to scalar integrals at the integrand level}},  {\em Nucl.Phys.}
  {\bf B763} (2007) 147--169,
  [\href{http://xxx.lanl.gov/abs/hep-ph/0609007}{{\tt hep-ph/0609007}}].

\bibitem{Anastasiou:2006jv}
C.~Anastasiou, R.~Britto, B.~Feng, Z.~Kunszt, and P.~Mastrolia, {\it
  {D-dimensional unitarity cut method}},  {\em Phys.Lett.} {\bf B645} (2007)
  213--216, [\href{http://xxx.lanl.gov/abs/hep-ph/0609191}{{\tt
  hep-ph/0609191}}].

\bibitem{Anastasiou:2006gt}
C.~Anastasiou, R.~Britto, B.~Feng, Z.~Kunszt, and P.~Mastrolia, {\it {Unitarity
  cuts and Reduction to master integrals in d dimensions for one-loop
  amplitudes}},  {\em JHEP} {\bf 0703} (2007) 111,
  [\href{http://xxx.lanl.gov/abs/hep-ph/0612277}{{\tt hep-ph/0612277}}].

\bibitem{Ellis:2007br}
R.~K. Ellis, W.~Giele, and Z.~Kunszt, {\it {A Numerical Unitarity Formalism for
  Evaluating One-Loop Amplitudes}},  {\em JHEP} {\bf 0803} (2008) 003,
  [\href{http://xxx.lanl.gov/abs/0708.2398}{{\tt arXiv:0708.2398}}].

\bibitem{Ellis:2008ir}
R.~K. Ellis, W.~T. Giele, Z.~Kunszt, and K.~Melnikov, {\it {Masses, fermions
  and generalized $D$-dimensional unitarity}},  {\em Nucl.Phys.} {\bf B822}
  (2009) 270--282, [\href{http://xxx.lanl.gov/abs/0806.3467}{{\tt
  arXiv:0806.3467}}].

\bibitem{Giele:2008ve}
W.~T. Giele, Z.~Kunszt, and K.~Melnikov, {\it {Full one-loop amplitudes from
  tree amplitudes}},  {\em JHEP} {\bf 0804} (2008) 049,
  [\href{http://xxx.lanl.gov/abs/0801.2237}{{\tt arXiv:0801.2237}}].

\bibitem{Cascioli:2011va}
F.~Cascioli, P.~Maierhofer, and S.~Pozzorini, {\it {Scattering Amplitudes with
  Open Loops}},  {\em Phys.Rev.Lett.} {\bf 108} (2012) 111601,
  [\href{http://xxx.lanl.gov/abs/1111.5206}{{\tt arXiv:1111.5206}}].

\bibitem{Mastrolia:2012bu}
P.~Mastrolia, E.~Mirabella, and T.~Peraro, {\it {Integrand reduction of
  one-loop scattering amplitudes through Laurent series expansion}},  {\em
  JHEP} {\bf 1206} (2012) 095, [\href{http://xxx.lanl.gov/abs/1203.0291}{{\tt
  arXiv:1203.0291}}].

\bibitem{Frixione:2002ik}
S.~Frixione and B.~R. Webber, {\it {Matching NLO QCD computations and parton
  shower simulations}},  {\em JHEP} {\bf 0206} (2002) 029,
  [\href{http://xxx.lanl.gov/abs/hep-ph/0204244}{{\tt hep-ph/0204244}}].

\bibitem{Nason:2004rx}
P.~Nason, {\it {A New method for combining NLO QCD with shower Monte Carlo
  algorithms}},  {\em JHEP} {\bf 0411} (2004) 040,
  [\href{http://xxx.lanl.gov/abs/hep-ph/0409146}{{\tt hep-ph/0409146}}].

\bibitem{Dobbs:2001gb}
M.~Dobbs, {\it {Incorporating next-to-leading order matrix elements for
  hadronic diboson production in showering event generators}},  {\em Phys.Rev.}
  {\bf D64} (2001) 034016, [\href{http://xxx.lanl.gov/abs/hep-ph/0103174}{{\tt
  hep-ph/0103174}}].

\bibitem{Chen:2001nf}
Y.-j. Chen, J.~Collins, and X.-m. Zu, {\it {NLO corrections in MC event
  generator for angular distribution of Drell-Yan lepton pair production}},
  {\em JHEP} {\bf 0204} (2002) 041,
  [\href{http://xxx.lanl.gov/abs/hep-ph/0110257}{{\tt hep-ph/0110257}}].

\bibitem{Kurihara:2002ne}
Y.~Kurihara, J.~Fujimoto, T.~Ishikawa, K.~Kato, S.~Kawabata, et~al., {\it {QCD
  event generators with next-to-leading order matrix elements and parton
  showers}},  {\em Nucl.Phys.} {\bf B654} (2003) 301--319,
  [\href{http://xxx.lanl.gov/abs/hep-ph/0212216}{{\tt hep-ph/0212216}}].

\bibitem{Nagy:2005aa}
Z.~Nagy and D.~E. Soper, {\it {Matching parton showers to NLO computations}},
  {\em JHEP} {\bf 0510} (2005) 024,
  [\href{http://xxx.lanl.gov/abs/hep-ph/0503053}{{\tt hep-ph/0503053}}].

\bibitem{Bauer:2006mk}
C.~W. Bauer and M.~D. Schwartz, {\it {Event Generation from Effective Field
  Theory}},  {\em Phys.Rev.} {\bf D76} (2007) 074004,
  [\href{http://xxx.lanl.gov/abs/hep-ph/0607296}{{\tt hep-ph/0607296}}].

\bibitem{Nagy:2007ty}
Z.~Nagy and D.~E. Soper, {\it {Parton showers with quantum interference}},
  {\em JHEP} {\bf 0709} (2007) 114,
  [\href{http://xxx.lanl.gov/abs/0706.0017}{{\tt arXiv:0706.0017}}].

\bibitem{Giele:2007di}
W.~T. Giele, D.~A. Kosower, and P.~Z. Skands, {\it {A simple shower and
  matching algorithm}},  {\em Phys.Rev.} {\bf D78} (2008) 014026,
  [\href{http://xxx.lanl.gov/abs/0707.3652}{{\tt arXiv:0707.3652}}].

\bibitem{Bauer:2008qh}
C.~W. Bauer, F.~J. Tackmann, and J.~Thaler, {\it {GenEvA. I. A New framework
  for event generation}},  {\em JHEP} {\bf 0812} (2008) 010,
  [\href{http://xxx.lanl.gov/abs/0801.4026}{{\tt arXiv:0801.4026}}].

\bibitem{Hoeche:2011fd}
S.~Hoeche, F.~Krauss, M.~Schonherr, and F.~Siegert, {\it {A critical appraisal
  of NLO+PS matching methods}},  {\em JHEP} {\bf 1209} (2012) 049,
  [\href{http://xxx.lanl.gov/abs/1111.1220}{{\tt arXiv:1111.1220}}].

\bibitem{Hamilton:2013fea}
K.~Hamilton, P.~Nason, E.~Re, and G.~Zanderighi, {\it {NNLOPS simulation of
  Higgs boson production}},  {\em JHEP} {\bf 1310} (2013) 222,
  [\href{http://xxx.lanl.gov/abs/1309.0017}{{\tt arXiv:1309.0017}}].

\bibitem{Alwall:2011uj}
J.~Alwall, M.~Herquet, F.~Maltoni, O.~Mattelaer, and T.~Stelzer, {\it {MadGraph
  5 : Going Beyond}},  {\em JHEP} {\bf 1106} (2011) 128,
  [\href{http://xxx.lanl.gov/abs/1106.0522}{{\tt arXiv:1106.0522}}].

\bibitem{Stelzer:1994ta}
T.~Stelzer and W.~Long, {\it {Automatic generation of tree level helicity
  amplitudes}},  {\em Comput.Phys.Commun.} {\bf 81} (1994) 357--371,
  [\href{http://xxx.lanl.gov/abs/hep-ph/9401258}{{\tt hep-ph/9401258}}].

\bibitem{Caravaglios:1995cd}
F.~Caravaglios and M.~Moretti, {\it {An algorithm to compute Born scattering
  amplitudes without Feynman graphs}},  {\em Phys.Lett.} {\bf B358} (1995)
  332--338, [\href{http://xxx.lanl.gov/abs/hep-ph/9507237}{{\tt
  hep-ph/9507237}}].

\bibitem{Yuasa:1999rg}
F.~Yuasa, J.~Fujimoto, T.~Ishikawa, M.~Jimbo, T.~Kaneko, et~al., {\it
  {Automatic computation of cross-sections in HEP: Status of GRACE system}},
  {\em Prog.Theor.Phys.Suppl.} {\bf 138} (2000) 18--23,
  [\href{http://xxx.lanl.gov/abs/hep-ph/0007053}{{\tt hep-ph/0007053}}].

\bibitem{Kanaki:2000ey}
A.~Kanaki and C.~G. Papadopoulos, {\it {HELAC: A Package to compute electroweak
  helicity amplitudes}},  {\em Comput.Phys.Commun.} {\bf 132} (2000) 306--315,
  [\href{http://xxx.lanl.gov/abs/hep-ph/0002082}{{\tt hep-ph/0002082}}].

\bibitem{Moretti:2001zz}
M.~Moretti, T.~Ohl, and J.~Reuter, {\it {O'Mega: An Optimizing matrix element
  generator}},  \href{http://xxx.lanl.gov/abs/hep-ph/0102195}{{\tt
  hep-ph/0102195}}.

\bibitem{Krauss:2001iv}
F.~Krauss, R.~Kuhn, and G.~Soff, {\it {AMEGIC++ 1.0: A Matrix element generator
  in C++}},  {\em JHEP} {\bf 0202} (2002) 044,
  [\href{http://xxx.lanl.gov/abs/hep-ph/0109036}{{\tt hep-ph/0109036}}].

\bibitem{Mangano:2002ea}
M.~L. Mangano, M.~Moretti, F.~Piccinini, R.~Pittau, and A.~D. Polosa, {\it
  {ALPGEN, a generator for hard multiparton processes in hadronic collisions}},
   {\em JHEP} {\bf 0307} (2003) 001,
  [\href{http://xxx.lanl.gov/abs/hep-ph/0206293}{{\tt hep-ph/0206293}}].

\bibitem{Fujimoto:2002sj}
J.~Fujimoto, T.~Ishikawa, M.~Jimbo, T.~Kaneko, K.~Kato, et~al., {\it
  {GRACE/SUSY automatic generation of tree amplitudes in the minimal
  supersymmetric standard model}},  {\em Comput.Phys.Commun.} {\bf 153} (2003)
  106--134, [\href{http://xxx.lanl.gov/abs/hep-ph/0208036}{{\tt
  hep-ph/0208036}}].

\bibitem{Boos:2004kh}
{\bf CompHEP} Collaboration, E.~Boos et~al., {\it {CompHEP 4.4: Automatic
  computations from Lagrangians to events}},  {\em Nucl.Instrum.Meth.} {\bf
  A534} (2004) 250--259, [\href{http://xxx.lanl.gov/abs/hep-ph/0403113}{{\tt
  hep-ph/0403113}}].

\bibitem{Tsuno:2006cu}
S.~Tsuno, T.~Kaneko, Y.~Kurihara, S.~Odaka, and K.~Kato, {\it {GR@PPA 2.7 event
  generator for $pp$ / $p\bar{p}$ collisions}},  {\em Comput.Phys.Commun.} {\bf
  175} (2006) 665--677, [\href{http://xxx.lanl.gov/abs/hep-ph/0602213}{{\tt
  hep-ph/0602213}}].

\bibitem{Cafarella:2007pc}
A.~Cafarella, C.~G. Papadopoulos, and M.~Worek, {\it {Helac-Phegas: A Generator
  for all parton level processes}},  {\em Comput.Phys.Commun.} {\bf 180} (2009)
  1941--1955, [\href{http://xxx.lanl.gov/abs/0710.2427}{{\tt
  arXiv:0710.2427}}].

\bibitem{Kilian:2007gr}
W.~Kilian, T.~Ohl, and J.~Reuter, {\it {WHIZARD: Simulating Multi-Particle
  Processes at LHC and ILC}},  {\em Eur.Phys.J.} {\bf C71} (2011) 1742,
  [\href{http://xxx.lanl.gov/abs/0708.4233}{{\tt arXiv:0708.4233}}].

\bibitem{Alwall:2007st}
J.~Alwall, P.~Demin, S.~de~Visscher, R.~Frederix, M.~Herquet, et~al., {\it
  {MadGraph/MadEvent v4: The New Web Generation}},  {\em JHEP} {\bf 0709}
  (2007) 028, [\href{http://xxx.lanl.gov/abs/0706.2334}{{\tt
  arXiv:0706.2334}}].

\bibitem{Gleisberg:2008fv}
T.~Gleisberg and S.~Hoeche, {\it {Comix, a new matrix element generator}},
  {\em JHEP} {\bf 0812} (2008) 039,
  [\href{http://xxx.lanl.gov/abs/0808.3674}{{\tt arXiv:0808.3674}}].

\bibitem{Belyaev:2012qa}
A.~Belyaev, N.~D. Christensen, and A.~Pukhov, {\it {CalcHEP 3.4 for collider
  physics within and beyond the Standard Model}},  {\em Comput.Phys.Commun.}
  {\bf 184} (2013) 1729--1769, [\href{http://xxx.lanl.gov/abs/1207.6082}{{\tt
  arXiv:1207.6082}}].

\bibitem{Hahn:1998yk}
T.~Hahn and M.~Perez-Victoria, {\it {Automatized one loop calculations in
  four-dimensions and D-dimensions}},  {\em Comput.Phys.Commun.} {\bf 118}
  (1999) 153--165, [\href{http://xxx.lanl.gov/abs/hep-ph/9807565}{{\tt
  hep-ph/9807565}}].

\bibitem{Hahn:2000kx}
T.~Hahn, {\it {Generating Feynman diagrams and amplitudes with FeynArts 3}},
  {\em Comput.Phys.Commun.} {\bf 140} (2001) 418--431,
  [\href{http://xxx.lanl.gov/abs/hep-ph/0012260}{{\tt hep-ph/0012260}}].

\bibitem{Gleisberg:2007md}
T.~Gleisberg and F.~Krauss, {\it {Automating dipole subtraction for QCD NLO
  calculations}},  {\em Eur.Phys.J.} {\bf C53} (2008) 501--523,
  [\href{http://xxx.lanl.gov/abs/0709.2881}{{\tt arXiv:0709.2881}}].

\bibitem{Berger:2008sj}
C.~Berger, Z.~Bern, L.~Dixon, F.~Febres~Cordero, D.~Forde, et~al., {\it {An
  Automated Implementation of On-Shell Methods for One-Loop Amplitudes}},  {\em
  Phys.Rev.} {\bf D78} (2008) 036003,
  [\href{http://xxx.lanl.gov/abs/0803.4180}{{\tt arXiv:0803.4180}}].

\bibitem{Frederix:2008hu}
R.~Frederix, T.~Gehrmann, and N.~Greiner, {\it {Automation of the Dipole
  Subtraction Method in MadGraph/MadEvent}},  {\em JHEP} {\bf 0809} (2008) 122,
  [\href{http://xxx.lanl.gov/abs/0808.2128}{{\tt arXiv:0808.2128}}].

\bibitem{Giele:2008bc}
W.~Giele and G.~Zanderighi, {\it {On the Numerical Evaluation of One-Loop
  Amplitudes: The Gluonic Case}},  {\em JHEP} {\bf 0806} (2008) 038,
  [\href{http://xxx.lanl.gov/abs/0805.2152}{{\tt arXiv:0805.2152}}].

\bibitem{Czakon:2009ss}
M.~Czakon, C.~Papadopoulos, and M.~Worek, {\it {Polarizing the Dipoles}},  {\em
  JHEP} {\bf 0908} (2009) 085, [\href{http://xxx.lanl.gov/abs/0905.0883}{{\tt
  arXiv:0905.0883}}].

\bibitem{Frederix:2009yq}
R.~Frederix, S.~Frixione, F.~Maltoni, and T.~Stelzer, {\it {Automation of
  next-to-leading order computations in QCD: The FKS subtraction}},  {\em JHEP}
  {\bf 0910} (2009) 003, [\href{http://xxx.lanl.gov/abs/0908.4272}{{\tt
  arXiv:0908.4272}}].

\bibitem{Hasegawa:2009tx}
K.~Hasegawa, S.~Moch, and P.~Uwer, {\it {AutoDipole: Automated generation of
  dipole subtraction terms}},  {\em Comput.Phys.Commun.} {\bf 181} (2010)
  1802--1817, [\href{http://xxx.lanl.gov/abs/0911.4371}{{\tt
  arXiv:0911.4371}}].

\bibitem{Hoche:2010pf}
S.~Hoche, F.~Krauss, M.~Schonherr, and F.~Siegert, {\it {Automating the POWHEG
  method in Sherpa}},  {\em JHEP} {\bf 1104} (2011) 024,
  [\href{http://xxx.lanl.gov/abs/1008.5399}{{\tt arXiv:1008.5399}}].

\bibitem{Alioli:2010xd}
S.~Alioli, P.~Nason, C.~Oleari, and E.~Re, {\it {A general framework for
  implementing NLO calculations in shower Monte Carlo programs: the POWHEG
  BOX}},  {\em JHEP} {\bf 1006} (2010) 043,
  [\href{http://xxx.lanl.gov/abs/1002.2581}{{\tt arXiv:1002.2581}}].

\bibitem{Mastrolia:2010nb}
P.~Mastrolia, G.~Ossola, T.~Reiter, and F.~Tramontano, {\it {Scattering
  AMplitudes from Unitarity-based Reduction Algorithm at the Integrand-level}},
   {\em JHEP} {\bf 1008} (2010) 080,
  [\href{http://xxx.lanl.gov/abs/1006.0710}{{\tt arXiv:1006.0710}}].

\bibitem{Frederix:2010cj}
R.~Frederix, T.~Gehrmann, and N.~Greiner, {\it {Integrated dipoles with
  MadDipole in the MadGraph framework}},  {\em JHEP} {\bf 1006} (2010) 086,
  [\href{http://xxx.lanl.gov/abs/1004.2905}{{\tt arXiv:1004.2905}}].

\bibitem{Becker:2010ng}
S.~Becker, C.~Reuschle, and S.~Weinzierl, {\it {Numerical NLO QCD
  calculations}},  {\em JHEP} {\bf 1012} (2010) 013,
  [\href{http://xxx.lanl.gov/abs/1010.4187}{{\tt arXiv:1010.4187}}].

\bibitem{Hirschi:2011pa}
V.~Hirschi, R.~Frederix, S.~Frixione, M.~V. Garzelli, F.~Maltoni, et~al., {\it
  {Automation of one-loop QCD corrections}},  {\em JHEP} {\bf 1105} (2011) 044,
  [\href{http://xxx.lanl.gov/abs/1103.0621}{{\tt arXiv:1103.0621}}].

\bibitem{Bevilacqua:2011xh}
G.~Bevilacqua, M.~Czakon, M.~Garzelli, A.~van Hameren, A.~Kardos, et~al., {\it
  {HELAC-NLO}},  {\em Comput.Phys.Commun.} {\bf 184} (2013) 986--997,
  [\href{http://xxx.lanl.gov/abs/1110.1499}{{\tt arXiv:1110.1499}}].

\bibitem{Becker:2011vg}
S.~Becker, D.~Goetz, C.~Reuschle, C.~Schwan, and S.~Weinzierl, {\it {NLO
  results for five, six and seven jets in electron-positron annihilation}},
  {\em Phys.Rev.Lett.} {\bf 108} (2012) 032005,
  [\href{http://xxx.lanl.gov/abs/1111.1733}{{\tt arXiv:1111.1733}}].

\bibitem{Cullen:2011ac}
G.~Cullen, N.~Greiner, G.~Heinrich, G.~Luisoni, P.~Mastrolia, et~al., {\it
  {Automated One-Loop Calculations with GoSam}},  {\em Eur.Phys.J.} {\bf C72}
  (2012) 1889, [\href{http://xxx.lanl.gov/abs/1111.2034}{{\tt
  arXiv:1111.2034}}].

\bibitem{Binoth:2011xi}
T.~Binoth, D.~Goncalves~Netto, D.~Lopez-Val, K.~Mawatari, T.~Plehn, et~al.,
  {\it {Automized Squark-Neutralino Production to Next-to-Leading Order}},
  {\em Phys.Rev.} {\bf D84} (2011) 075005,
  [\href{http://xxx.lanl.gov/abs/1108.1250}{{\tt arXiv:1108.1250}}].

\bibitem{Agrawal:2011tm}
S.~Agrawal, T.~Hahn, and E.~Mirabella, {\it {FormCalc 7}},  {\em
  J.Phys.Conf.Ser.} {\bf 368} (2012) 012054,
  [\href{http://xxx.lanl.gov/abs/1112.0124}{{\tt arXiv:1112.0124}}].

\bibitem{Bern:2011ep}
Z.~Bern, G.~Diana, L.~Dixon, F.~Febres~Cordero, S.~Hoeche, et~al., {\it
  {Four-Jet Production at the Large Hadron Collider at Next-to-Leading Order in
  QCD}},  {\em Phys.Rev.Lett.} {\bf 109} (2012) 042001,
  [\href{http://xxx.lanl.gov/abs/1112.3940}{{\tt arXiv:1112.3940}}].

\bibitem{Actis:2012qn}
S.~Actis, A.~Denner, L.~Hofer, A.~Scharf, and S.~Uccirati, {\it {Recursive
  generation of one-loop amplitudes in the Standard Model}},  {\em JHEP} {\bf
  1304} (2013) 037, [\href{http://xxx.lanl.gov/abs/1211.6316}{{\tt
  arXiv:1211.6316}}].

\bibitem{Badger:2012pg}
S.~Badger, B.~Biedermann, P.~Uwer, and V.~Yundin, {\it {Numerical evaluation of
  virtual corrections to multi-jet production in massless QCD}},  {\em
  Comput.Phys.Commun.} {\bf 184} (2013) 1981--1998,
  [\href{http://xxx.lanl.gov/abs/1209.0100}{{\tt arXiv:1209.0100}}].

\bibitem{GoncalvesNetto:2012yt}
D.~Goncalves-Netto, D.~Lopez-Val, K.~Mawatari, T.~Plehn, and I.~Wigmore, {\it
  {Automated Squark and Gluino Production to Next-to-Leading Order}},  {\em
  Phys.Rev.} {\bf D87} (2013) 014002,
  [\href{http://xxx.lanl.gov/abs/1211.0286}{{\tt arXiv:1211.0286}}].

\bibitem{Badger:2013vpa}
S.~Badger, B.~Biedermann, P.~Uwer, and V.~Yundin, {\it {Computation of
  multi-leg amplitudes with NJet}},
  \href{http://xxx.lanl.gov/abs/1312.7140}{{\tt arXiv:1312.7140}}.

\bibitem{Bern:2013pya}
Z.~Bern, L.~Dixon, F.~F. Cordero, S.~Hoeche, H.~Ita, et~al., {\it {The BlackHat
  Library for One-Loop Amplitudes}},
  \href{http://xxx.lanl.gov/abs/1310.2808}{{\tt arXiv:1310.2808}}.

\bibitem{vanDeurzen:2013saa}
H.~van Deurzen, G.~Luisoni, P.~Mastrolia, E.~Mirabella, G.~Ossola, et~al., {\it
  {Multi-leg One-loop Massive Amplitudes from Integrand Reduction via Laurent
  Expansion}},  {\em JHEP} {\bf 1403} (2014) 115,
  [\href{http://xxx.lanl.gov/abs/1312.6678}{{\tt arXiv:1312.6678}}].

\bibitem{Cullen:2014yla}
G.~Cullen, H.~van Deurzen, N.~Greiner, G.~Heinrich, G.~Luisoni, et~al., {\it
  {GoSam-2.0: a tool for automated one-loop calculations within the Standard
  Model and beyond}},  \href{http://xxx.lanl.gov/abs/1404.7096}{{\tt
  arXiv:1404.7096}}.

\bibitem{Peraro:2014cba}
T.~Peraro, {\it {Ninja: Automated Integrand Reduction via Laurent Expansion for
  One-Loop Amplitudes}},  \href{http://xxx.lanl.gov/abs/1403.1229}{{\tt
  arXiv:1403.1229}}.

\bibitem{Byckling:1971vca}
E.~Byckling and K.~Kajantie, {\em {Particle Kinematics}}.
\newblock Wiley, 1971.

\bibitem{Christensen:2008py}
N.~D. Christensen and C.~Duhr, {\it {FeynRules - Feynman rules made easy}},
  {\em Comput.Phys.Commun.} {\bf 180} (2009) 1614--1641,
  [\href{http://xxx.lanl.gov/abs/0806.4194}{{\tt arXiv:0806.4194}}].

\bibitem{Christensen:2009jx}
N.~D. Christensen, P.~de~Aquino, C.~Degrande, C.~Duhr, B.~Fuks, et~al., {\it {A
  Comprehensive approach to new physics simulations}},  {\em Eur.Phys.J.} {\bf
  C71} (2011) 1541, [\href{http://xxx.lanl.gov/abs/0906.2474}{{\tt
  arXiv:0906.2474}}].

\bibitem{Christensen:2010wz}
N.~D. Christensen, C.~Duhr, B.~Fuks, J.~Reuter, and C.~Speckner, {\it
  {Introducing an interface between WHIZARD and FeynRules}},  {\em Eur.Phys.J.}
  {\bf C72} (2012) 1990, [\href{http://xxx.lanl.gov/abs/1010.3251}{{\tt
  arXiv:1010.3251}}].

\bibitem{Duhr:2011se}
C.~Duhr and B.~Fuks, {\it {A superspace module for the FeynRules package}},
  {\em Comput.Phys.Commun.} {\bf 182} (2011) 2404--2426,
  [\href{http://xxx.lanl.gov/abs/1102.4191}{{\tt arXiv:1102.4191}}].

\bibitem{Alloul:2013bka}
A.~Alloul, N.~D. Christensen, C.~Degrande, C.~Duhr, and B.~Fuks, {\it
  {FeynRules 2.0 - A complete toolbox for tree-level phenomenology}},
  \href{http://xxx.lanl.gov/abs/1310.1921}{{\tt arXiv:1310.1921}}.

\bibitem{Alloul:2013fw}
A.~Alloul, J.~D'Hondt, K.~De~Causmaecker, B.~Fuks, and M.~Rausch~de
  Traubenberg, {\it {Automated mass spectrum generation for new physics}},
  {\em Eur.Phys.J.} {\bf C73} (2013) 2325,
  [\href{http://xxx.lanl.gov/abs/1301.5932}{{\tt arXiv:1301.5932}}].

\bibitem{Degrande:nloct}
C.~Degrande, {\it {Automated computation of the $R_2$ rational terms and
  ultraviolet counterterms by NLOCT : An illustration on the 2HDM, in
  preparation.}}, .

\bibitem{Murayama:1992gi}
H.~Murayama, I.~Watanabe, and K.~Hagiwara, {\em {HELAS: Helicity amplitude
  subroutines for Feynman diagram evaluations}}.
\newblock KEK-91-11, 1992.

\bibitem{deAquino:2011ub}
P.~de~Aquino, W.~Link, F.~Maltoni, O.~Mattelaer, and T.~Stelzer, {\it {ALOHA:
  Automatic Libraries Of Helicity Amplitudes for Feynman Diagram
  Computations}},  {\em Comput.Phys.Commun.} {\bf 183} (2012) 2254--2263,
  [\href{http://xxx.lanl.gov/abs/1108.2041}{{\tt arXiv:1108.2041}}].

\bibitem{Degrande:2011ua}
C.~Degrande, C.~Duhr, B.~Fuks, D.~Grellscheid, O.~Mattelaer, et~al., {\it {UFO
  - The Universal FeynRules Output}},  {\em Comput.Phys.Commun.} {\bf 183}
  (2012) 1201--1214, [\href{http://xxx.lanl.gov/abs/1108.2040}{{\tt
  arXiv:1108.2040}}].

\bibitem{Alwall:2006yp}
J.~Alwall, A.~Ballestrero, P.~Bartalini, S.~Belov, E.~Boos, et~al., {\it {A
  Standard format for Les Houches event files}},  {\em Comput.Phys.Commun.}
  {\bf 176} (2007) 300--304,
  [\href{http://xxx.lanl.gov/abs/hep-ph/0609017}{{\tt hep-ph/0609017}}].

\bibitem{Butterworth:2010ym}
J.~Butterworth, A.~Arbey, L.~Basso, S.~Belov, A.~Bharucha, et~al., {\it {The
  Tools and Monte Carlo working group Summary Report}},
  \href{http://xxx.lanl.gov/abs/1003.1643}{{\tt arXiv:1003.1643}}.

\bibitem{Argyres:1995ym}
E.~N. Argyres, W.~Beenakker, G.~J. van Oldenborgh, A.~Denner, S.~Dittmaier,
  et~al., {\it {Stable calculations for unstable particles: Restoring gauge
  invariance}},  {\em Phys.Lett.} {\bf B358} (1995) 339--346,
  [\href{http://xxx.lanl.gov/abs/hep-ph/9507216}{{\tt hep-ph/9507216}}].

\bibitem{Beenakker:1996kn}
W.~Beenakker, G.~J. van Oldenborgh, A.~Denner, S.~Dittmaier, J.~Hoogland,
  et~al., {\it {The Fermion loop scheme for finite width effects in $e^{+}
  e^{-}$ annihilation into four fermions}},  {\em Nucl.Phys.} {\bf B500} (1997)
  255--298, [\href{http://xxx.lanl.gov/abs/hep-ph/9612260}{{\tt
  hep-ph/9612260}}].

\bibitem{Passarino:1999zh}
G.~Passarino, {\it {Unstable particles and nonconserved currents: A
  Generalization of the fermion loop scheme}},  {\em Nucl.Phys.} {\bf B574}
  (2000) 451--494, [\href{http://xxx.lanl.gov/abs/hep-ph/9911482}{{\tt
  hep-ph/9911482}}].

\bibitem{Beenakker:1999hi}
W.~Beenakker, F.~A. Berends, and A.~Chapovsky, {\it {An Effective Lagrangian
  approach for unstable particles}},  {\em Nucl.Phys.} {\bf B573} (2000)
  503--535, [\href{http://xxx.lanl.gov/abs/hep-ph/9909472}{{\tt
  hep-ph/9909472}}].

\bibitem{Beenakker:2003va}
W.~Beenakker, A.~Chapovsky, A.~Kanaki, C.~Papadopoulos, and R.~Pittau, {\it
  {Towards an effective Lagrangian approach to fermion loop corrections}},
  {\em Nucl.Phys.} {\bf B667} (2003) 359--393,
  [\href{http://xxx.lanl.gov/abs/hep-ph/0303105}{{\tt hep-ph/0303105}}].

\bibitem{Beneke:2003xh}
M.~Beneke, A.~Chapovsky, A.~Signer, and G.~Zanderighi, {\it {Effective theory
  approach to unstable particle production}},  {\em Phys.Rev.Lett.} {\bf 93}
  (2004) 011602, [\href{http://xxx.lanl.gov/abs/hep-ph/0312331}{{\tt
  hep-ph/0312331}}].

\bibitem{Denner:1999gp}
A.~Denner, S.~Dittmaier, M.~Roth, and D.~Wackeroth, {\it {Predictions for all
  processes $e^+e^- \to$ 4 fermions + $\gamma$}},  {\em Nucl.Phys.} {\bf B560}
  (1999) 33--65, [\href{http://xxx.lanl.gov/abs/hep-ph/9904472}{{\tt
  hep-ph/9904472}}].

\bibitem{Denner:2005fg}
A.~Denner, S.~Dittmaier, M.~Roth, and L.~Wieders, {\it {Electroweak corrections
  to charged-current $e^+ e^- \to$ 4 fermion processes: Technical details and
  further results}},  {\em Nucl.Phys.} {\bf B724} (2005) 247--294,
  [\href{http://xxx.lanl.gov/abs/hep-ph/0505042}{{\tt hep-ph/0505042}}].

\bibitem{Staub:2012pb}
F.~Staub, {\it {SARAH 3.2: Dirac Gauginos, UFO output, and more}},  {\em
  Computer Physics Communications} {\bf 184} (2013) pp. 1792--1809,
  [\href{http://xxx.lanl.gov/abs/1207.0906}{{\tt arXiv:1207.0906}}].

\bibitem{Christensen:2013aua}
N.~D. Christensen, P.~de~Aquino, N.~Deutschmann, C.~Duhr, B.~Fuks, et~al., {\it
  {Simulating spin-$ \frac{3}{2}$ particles at colliders}},  {\em Eur.Phys.J.}
  {\bf C73} (2013) 2580, [\href{http://xxx.lanl.gov/abs/1308.1668}{{\tt
  arXiv:1308.1668}}].

\bibitem{Alwall:2014bza}
J.~Alwall, C.~Duhr, B.~Fuks, O.~Mattelaer, D.~G. Ozturk, et~al., {\it
  {Computing decay rates for new physics theories with FeynRules and
  MadGraph5/aMC@NLO}},  \href{http://xxx.lanl.gov/abs/1402.1178}{{\tt
  arXiv:1402.1178}}.

\bibitem{Berends:1981rb}
F.~A. Berends, R.~Kleiss, P.~De~Causmaecker, R.~Gastmans, and T.~T. Wu, {\it
  {Single Bremsstrahlung Processes in Gauge Theories}},  {\em Phys.Lett.} {\bf
  B103} (1981) 124.

\bibitem{DeCausmaecker:1981bg}
P.~De~Causmaecker, R.~Gastmans, W.~Troost, and T.~T. Wu, {\it {Multiple
  Bremsstrahlung in Gauge Theories at High-Energies. 1. General Formalism for
  Quantum Electrodynamics}},  {\em Nucl.Phys.} {\bf B206} (1982) 53.

\bibitem{Kleiss:1985yh}
R.~Kleiss and W.~J. Stirling, {\it {Spinor Techniques for Calculating $p
  \bar{p} \to W^\pm / Z^0$ + Jets}},  {\em Nucl.Phys.} {\bf B262} (1985)
  235--262.

\bibitem{Gastmans:1990xh}
R.~Gastmans and T.~Wu, {\it {The Ubiquitous photon: Helicity method for QED and
  QCD}},  {\em Int.Ser.Monogr.Phys.} {\bf 80} (1990) 1--648.

\bibitem{Xu:1986xb}
Z.~Xu, D.-H. Zhang, and L.~Chang, {\it {Helicity Amplitudes for Multiple
  Bremsstrahlung in Massless Nonabelian Gauge Theories}},  {\em Nucl.Phys.}
  {\bf B291} (1987) 392.

\bibitem{Gunion:1985vca}
J.~Gunion and Z.~Kunszt, {\it {Improved Analytic Techniques for Tree Graph
  Calculations and the $ggq\bar{q}l\bar{l}$ Subprocess}},  {\em Phys.Lett.}
  {\bf B161} (1985) 333.

\bibitem{Hagiwara:1985yu}
K.~Hagiwara and D.~Zeppenfeld, {\it {Helicity Amplitudes for Heavy Lepton
  Production in $e^+e^-$ Annihilation}},  {\em Nucl.Phys.} {\bf B274} (1986) 1.

\bibitem{Mangano:1990by}
M.~L. Mangano and S.~J. Parke, {\it {Multiparton amplitudes in gauge
  theories}},  {\em Phys.Rept.} {\bf 200} (1991) 301--367,
  [\href{http://xxx.lanl.gov/abs/hep-th/0509223}{{\tt hep-th/0509223}}].

\bibitem{DelDuca:1999rs}
V.~Del~Duca, L.~J. Dixon, and F.~Maltoni, {\it {New color decompositions for
  gauge amplitudes at tree and loop level}},  {\em Nucl.Phys.} {\bf B571}
  (2000) 51--70, [\href{http://xxx.lanl.gov/abs/hep-ph/9910563}{{\tt
  hep-ph/9910563}}].

\bibitem{Maltoni:2002mq}
F.~Maltoni, K.~Paul, T.~Stelzer, and S.~Willenbrock, {\it {Color flow
  decomposition of QCD amplitudes}},  {\em Phys.Rev.} {\bf D67} (2003) 014026,
  [\href{http://xxx.lanl.gov/abs/hep-ph/0209271}{{\tt hep-ph/0209271}}].

\bibitem{Duhr:2006iq}
C.~Duhr, S.~Hoeche, and F.~Maltoni, {\it {Color-dressed recursive relations for
  multi-parton amplitudes}},  {\em JHEP} {\bf 0608} (2006) 062,
  [\href{http://xxx.lanl.gov/abs/hep-ph/0607057}{{\tt hep-ph/0607057}}].

\bibitem{Hagiwara:2008jb}
K.~Hagiwara, J.~Kanzaki, Q.~Li, and K.~Mawatari, {\it {HELAS and
  MadGraph/MadEvent with spin-2 particles}},  {\em Eur.Phys.J.} {\bf C56}
  (2008) 435--447, [\href{http://xxx.lanl.gov/abs/0805.2554}{{\tt
  arXiv:0805.2554}}].

\bibitem{Hagiwara:2009aq}
K.~Hagiwara, J.~Kanzaki, N.~Okamura, D.~Rainwater, and T.~Stelzer, {\it {Fast
  calculation of HELAS amplitudes using graphics processing unit (GPU)}},  {\em
  Eur.Phys.J.} {\bf C66} (2010) 477--492,
  [\href{http://xxx.lanl.gov/abs/0908.4403}{{\tt arXiv:0908.4403}}].

\bibitem{Hagiwara:2009cy}
K.~Hagiwara, J.~Kanzaki, N.~Okamura, D.~Rainwater, and T.~Stelzer, {\it
  {Calculation of HELAS amplitudes for QCD processes using graphics processing
  unit (GPU)}},  {\em Eur.Phys.J.} {\bf C70} (2010) 513--524,
  [\href{http://xxx.lanl.gov/abs/0909.5257}{{\tt arXiv:0909.5257}}].

\bibitem{Bahr:2008pv}
M.~Bahr, S.~Gieseke, M.~Gigg, D.~Grellscheid, K.~Hamilton, et~al., {\it
  {Herwig++ Physics and Manual}},  {\em Eur.Phys.J.} {\bf C58} (2008) 639--707,
  [\href{http://xxx.lanl.gov/abs/0803.0883}{{\tt arXiv:0803.0883}}].

\bibitem{Bellm:2013lba}
J.~Bellm, S.~Gieseke, D.~Grellscheid, A.~Papaefstathiou, S.~Platzer, et~al.,
  {\it {Herwig++ 2.7 Release Note}},
  \href{http://xxx.lanl.gov/abs/1310.6877}{{\tt arXiv:1310.6877}}.

\bibitem{Sjostrand:2007gs}
T.~Sjostrand, S.~Mrenna, and P.~Z. Skands, {\it {A Brief Introduction to PYTHIA
  8.1}},  {\em Comput.Phys.Commun.} {\bf 178} (2008) 852--867,
  [\href{http://xxx.lanl.gov/abs/0710.3820}{{\tt arXiv:0710.3820}}].

\bibitem{ArkaniHamed:1998rs}
N.~Arkani-Hamed, S.~Dimopoulos, and G.~Dvali, {\it {The Hierarchy problem and
  new dimensions at a millimeter}},  {\em Phys.Lett.} {\bf B429} (1998)
  263--272, [\href{http://xxx.lanl.gov/abs/hep-ph/9803315}{{\tt
  hep-ph/9803315}}].

\bibitem{Frederix:2011ss}
R.~Frederix, S.~Frixione, V.~Hirschi, F.~Maltoni, R.~Pittau, et~al., {\it
  {Four-lepton production at hadron colliders: aMC@NLO predictions with
  theoretical uncertainties}},  {\em JHEP} {\bf 1202} (2012) 099,
  [\href{http://xxx.lanl.gov/abs/1110.4738}{{\tt arXiv:1110.4738}}].

\bibitem{Alwall:2011cy}
J.~Alwall, Q.~Li, and F.~Maltoni, {\it {Matched predictions for Higgs
  production via heavy-quark loops in the SM and beyond}},  {\em Phys.Rev.}
  {\bf D85} (2012) 014031, [\href{http://xxx.lanl.gov/abs/1110.1728}{{\tt
  arXiv:1110.1728}}].

\bibitem{Frederix:2014hta}
R.~Frederix, S.~Frixione, V.~Hirschi, F.~Maltoni, O.~Mattelaer, et~al., {\it
  {Higgs pair production at the LHC with NLO and parton-shower effects}},
  \href{http://xxx.lanl.gov/abs/1401.7340}{{\tt arXiv:1401.7340}}.

\bibitem{Kondo:1988yd}
K.~Kondo, {\it {Dynamical Likelihood Method for Reconstruction of Events With
  Missing Momentum. 1: Method and Toy Models}},  {\em J.Phys.Soc.Jap.} {\bf 57}
  (1988) 4126--4140.

\bibitem{Dalitz:1991wa}
R.~Dalitz and G.~R. Goldstein, {\it {The decay and polarization properties of
  the top quark}},  {\em Phys.Rev.} {\bf D45} (1992) 1531--1543.

\bibitem{Kondo:2006ar}
K.~Kondo, {\it {Dynamical likelihood method and top quark mass measurement at
  CDF}},  {\em J.Phys.Conf.Ser.} {\bf 53} (2006) 202--213.

\bibitem{Gao:2010qx}
Y.~Gao, A.~V. Gritsan, Z.~Guo, K.~Melnikov, M.~Schulze, et~al., {\it {Spin
  determination of single-produced resonances at hadron colliders}},  {\em
  Phys.Rev.} {\bf D81} (2010) 075022,
  [\href{http://xxx.lanl.gov/abs/1001.3396}{{\tt arXiv:1001.3396}}].

\bibitem{Avery:2012um}
P.~Avery, D.~Bourilkov, M.~Chen, T.~Cheng, A.~Drozdetskiy, et~al., {\it
  {Precision studies of the Higgs boson decay channel $H\to ZZ\to 4\ell$ with
  MEKD}},  {\em Phys.Rev.} {\bf D87} (2013), no.~5 055006,
  [\href{http://xxx.lanl.gov/abs/1210.0896}{{\tt arXiv:1210.0896}}].

\bibitem{Campbell:2012cz}
J.~M. Campbell, W.~T. Giele, and C.~Williams, {\it {The Matrix Element Method
  at Next-to-Leading Order}},  {\em JHEP} {\bf 1211} (2012) 043,
  [\href{http://xxx.lanl.gov/abs/1204.4424}{{\tt arXiv:1204.4424}}].

\bibitem{Gainer:2013rxa}
J.~S. Gainer, J.~Lykken, K.~T. Matchev, S.~Mrenna, and M.~Park, {\it
  {Geolocating the Higgs Boson Candidate at the LHC}},  {\em Phys.Rev.Lett.}
  {\bf 111} (2013) 041801, [\href{http://xxx.lanl.gov/abs/1304.4936}{{\tt
  arXiv:1304.4936}}].

\bibitem{Plehn:2013paa}
T.~Plehn, P.~Schichtel, and D.~Wiegand, {\it {MadMax, or Where Boosted
  Significances Come From}},  \href{http://xxx.lanl.gov/abs/1311.2591}{{\tt
  arXiv:1311.2591}}.

\bibitem{Artoisenet:2010cn}
P.~Artoisenet, V.~Lemaitre, F.~Maltoni, and O.~Mattelaer, {\it {Automation of
  the matrix element reweighting method}},  {\em JHEP} {\bf 1012} (2010) 068,
  [\href{http://xxx.lanl.gov/abs/1007.3300}{{\tt arXiv:1007.3300}}].

\bibitem{Artoisenet:2014_MW}
P.~Artoisenet and O.~Mattelaer, {\it {MadWeight5.0, in preparation}}, .

\bibitem{Alwall:2010cq}
J.~Alwall, A.~Freitas, and O.~Mattelaer, {\it {The Matrix Element Method and
  QCD Radiation}},  {\em Phys.Rev.} {\bf D83} (2011) 074010,
  [\href{http://xxx.lanl.gov/abs/1010.2263}{{\tt arXiv:1010.2263}}].

\bibitem{Artoisenet:2013vfa}
P.~Artoisenet, P.~de~Aquino, F.~Maltoni, and O.~Mattelaer, {\it {Unravelling
  $t\overline{t}h$ via the Matrix Element Method}},  {\em Phys.Rev.Lett.} {\bf
  111} (2013), no.~9 091802, [\href{http://xxx.lanl.gov/abs/1304.6414}{{\tt
  arXiv:1304.6414}}].

\bibitem{Gleisberg:2003xi}
T.~Gleisberg, S.~Hoeche, F.~Krauss, A.~Schalicke, S.~Schumann, et~al., {\it
  {SHERPA 1. alpha: A Proof of concept version}},  {\em JHEP} {\bf 0402} (2004)
  056, [\href{http://xxx.lanl.gov/abs/hep-ph/0311263}{{\tt hep-ph/0311263}}].

\bibitem{Sjostrand:2006za}
T.~Sjostrand, S.~Mrenna, and P.~Z. Skands, {\it {PYTHIA 6.4 Physics and
  Manual}},  {\em JHEP} {\bf 0605} (2006) 026,
  [\href{http://xxx.lanl.gov/abs/hep-ph/0603175}{{\tt hep-ph/0603175}}].

\bibitem{Corcella:2000bw}
G.~Corcella, I.~Knowles, G.~Marchesini, S.~Moretti, K.~Odagiri, et~al., {\it
  {HERWIG 6: An Event generator for hadron emission reactions with interfering
  gluons (including supersymmetric processes)}},  {\em JHEP} {\bf 0101} (2001)
  010, [\href{http://xxx.lanl.gov/abs/hep-ph/0011363}{{\tt hep-ph/0011363}}].

\bibitem{Corcella:2002jc}
G.~Corcella, I.~Knowles, G.~Marchesini, S.~Moretti, K.~Odagiri, et~al., {\it
  {HERWIG 6.5 release note}},
  \href{http://xxx.lanl.gov/abs/hep-ph/0210213}{{\tt hep-ph/0210213}}.

\bibitem{Catani:1993hr}
S.~Catani, Y.~L. Dokshitzer, M.~Seymour, and B.~Webber, {\it {Longitudinally
  invariant $k_t$ clustering algorithms for hadron hadron collisions}},  {\em
  Nucl.Phys.} {\bf B406} (1993) 187--224.

\bibitem{Lonnblad:2001iq}
L.~Lonnblad, {\it {Correcting the color dipole cascade model with fixed order
  matrix elements}},  {\em JHEP} {\bf 0205} (2002) 046,
  [\href{http://xxx.lanl.gov/abs/hep-ph/0112284}{{\tt hep-ph/0112284}}].

\bibitem{Lonnblad:2011xx}
L.~Lonnblad and S.~Prestel, {\it {Matching Tree-Level Matrix Elements with
  Interleaved Showers}},  {\em JHEP} {\bf 1203} (2012) 019,
  [\href{http://xxx.lanl.gov/abs/1109.4829}{{\tt arXiv:1109.4829}}].

\bibitem{Lonnblad:2012ng}
L.~Lonnblad and S.~Prestel, {\it {Unitarising Matrix Element + Parton Shower
  merging}},  {\em JHEP} {\bf 1302} (2013) 094,
  [\href{http://xxx.lanl.gov/abs/1211.4827}{{\tt arXiv:1211.4827}}].

\bibitem{Alwall:2007fs}
J.~Alwall, S.~Hoche, F.~Krauss, N.~Lavesson, L.~Lonnblad, et~al., {\it
  {Comparative study of various algorithms for the merging of parton showers
  and matrix elements in hadronic collisions}},  {\em Eur.Phys.J.} {\bf C53}
  (2008) 473--500, [\href{http://xxx.lanl.gov/abs/0706.2569}{{\tt
  arXiv:0706.2569}}].

\bibitem{Alwall:2008qv}
J.~Alwall, S.~de~Visscher, and F.~Maltoni, {\it {QCD radiation in the
  production of heavy colored particles at the LHC}},  {\em JHEP} {\bf 0902}
  (2009) 017, [\href{http://xxx.lanl.gov/abs/0810.5350}{{\tt
  arXiv:0810.5350}}].

\bibitem{deAquino:2012ru}
P.~de~Aquino, F.~Maltoni, K.~Mawatari, and B.~Oexl, {\it {Light Gravitino
  Production in Association with Gluinos at the LHC}},  {\em JHEP} {\bf 1210}
  (2012) 008, [\href{http://xxx.lanl.gov/abs/1206.7098}{{\tt
  arXiv:1206.7098}}].

\bibitem{Artoisenet:2013puc}
P.~Artoisenet, P.~de~Aquino, F.~Demartin, R.~Frederix, S.~Frixione, et~al.,
  {\it {A framework for Higgs characterisation}},  {\em JHEP} {\bf 1311} (2013)
  043, [\href{http://xxx.lanl.gov/abs/1306.6464}{{\tt arXiv:1306.6464}}].

\bibitem{Maltoni:2002qb}
F.~Maltoni and T.~Stelzer, {\it {MadEvent: Automatic event generation with
  MadGraph}},  {\em JHEP} {\bf 0302} (2003) 027,
  [\href{http://xxx.lanl.gov/abs/hep-ph/0208156}{{\tt hep-ph/0208156}}].

\bibitem{Kleiss:1994qy}
R.~Kleiss and R.~Pittau, {\it {Weight optimization in multichannel Monte
  Carlo}},  {\em Comput.Phys.Commun.} {\bf 83} (1994) 141--146,
  [\href{http://xxx.lanl.gov/abs/hep-ph/9405257}{{\tt hep-ph/9405257}}].

\bibitem{Passarino:1978jh}
G.~Passarino and M.~Veltman, {\it {One Loop Corrections for $e^+e^-$
  Annihilation Into $\mu^+ \mu^-$ in the Weinberg Model}},  {\em Nucl.Phys.}
  {\bf B160} (1979) 151.

\bibitem{Davydychev:1991va}
A.~I. Davydychev, {\it {A Simple formula for reducing Feynman diagrams to
  scalar integrals}},  {\em Phys.Lett.} {\bf B263} (1991) 107--111.

\bibitem{Ossola:2007ax}
G.~Ossola, C.~G. Papadopoulos, and R.~Pittau, {\it {CutTools: A Program
  implementing the OPP reduction method to compute one-loop amplitudes}},  {\em
  JHEP} {\bf 0803} (2008) 042, [\href{http://xxx.lanl.gov/abs/0711.3596}{{\tt
  arXiv:0711.3596}}].

\bibitem{'tHooft:1972fi}
G.~'t~Hooft and M.~Veltman, {\it {Regularization and Renormalization of Gauge
  Fields}},  {\em Nucl.Phys.} {\bf B44} (1972) 189--213.

\bibitem{Ossola:2008xq}
G.~Ossola, C.~G. Papadopoulos, and R.~Pittau, {\it {On the Rational Terms of
  the one-loop amplitudes}},  {\em JHEP} {\bf 0805} (2008) 004,
  [\href{http://xxx.lanl.gov/abs/0802.1876}{{\tt arXiv:0802.1876}}].

\bibitem{Draggiotis:2009yb}
P.~Draggiotis, M.~Garzelli, C.~Papadopoulos, and R.~Pittau, {\it {Feynman Rules
  for the Rational Part of the QCD 1-loop amplitudes}},  {\em JHEP} {\bf 0904}
  (2009) 072, [\href{http://xxx.lanl.gov/abs/0903.0356}{{\tt
  arXiv:0903.0356}}].

\bibitem{Garzelli:2009is}
M.~Garzelli, I.~Malamos, and R.~Pittau, {\it {Feynman rules for the rational
  part of the Electroweak 1-loop amplitudes}},  {\em JHEP} {\bf 1001} (2010)
  040, [\href{http://xxx.lanl.gov/abs/0910.3130}{{\tt arXiv:0910.3130}}].

\bibitem{Garzelli:2010qm}
M.~Garzelli, I.~Malamos, and R.~Pittau, {\it {Feynman rules for the rational
  part of the Electroweak 1-loop amplitudes in the $R_\xi$ gauge and in the
  Unitary gauge}},  {\em JHEP} {\bf 1101} (2011) 029,
  [\href{http://xxx.lanl.gov/abs/1009.4302}{{\tt arXiv:1009.4302}}].

\bibitem{Shao:2011tg}
H.-S. Shao, Y.-J. Zhang, and K.-T. Chao, {\it {Feynman Rules for the Rational
  Part of the Standard Model One-loop Amplitudes in the 't Hooft-Veltman
  $\gamma_5$ Scheme}},  {\em JHEP} {\bf 09} (2011) 048,
  [\href{http://xxx.lanl.gov/abs/1106.5030}{{\tt arXiv:1106.5030}}].

\bibitem{Pittau:2011qp}
R.~Pittau, {\it {Primary Feynman rules to calculate the epsilon-dimensional
  integrand of any 1-loop amplitude}},  {\em JHEP} {\bf 1202} (2012) 029,
  [\href{http://xxx.lanl.gov/abs/1111.4965}{{\tt arXiv:1111.4965}}].

\bibitem{Shao:2012ja}
H.-S. Shao and Y.-J. Zhang, {\it {Feynman Rules for the Rational Part of
  One-loop QCD Corrections in the MSSM}},  {\em JHEP} {\bf 1206} (2012) 112,
  [\href{http://xxx.lanl.gov/abs/1205.1273}{{\tt arXiv:1205.1273}}].

\bibitem{Page:2013xla}
B.~Page and R.~Pittau, {\it {$R_{2}$ vertices for the effective $ggH$ theory}},
   {\em JHEP} {\bf 1309} (2013) 078,
  [\href{http://xxx.lanl.gov/abs/1307.6142}{{\tt arXiv:1307.6142}}].

\bibitem{Binoth:2006hk}
T.~Binoth, J.~P. Guillet, and G.~Heinrich, {\it {Algebraic evaluation of
  rational polynomials in one-loop amplitudes}},  {\em JHEP} {\bf 0702} (2007)
  013, [\href{http://xxx.lanl.gov/abs/hep-ph/0609054}{{\tt hep-ph/0609054}}].

\bibitem{Badger:2008cm}
S.~Badger, {\it {Direct Extraction Of One Loop Rational Terms}},  {\em JHEP}
  {\bf 0901} (2009) 049, [\href{http://xxx.lanl.gov/abs/0806.4600}{{\tt
  arXiv:0806.4600}}].

\bibitem{ShaoIREGI}
H.-S. Shao, {\it Iregi user manual, unpublished}, .

\bibitem{Fleischer:2012et}
J.~Fleischer, T.~Riemann, and V.~Yundin, {\it {New developments in PJFry}},
  {\em PoS} {\bf LL2012} (2012) 020,
  [\href{http://xxx.lanl.gov/abs/1210.4095}{{\tt arXiv:1210.4095}}].

\bibitem{YundinPhd:2012}
V.~Yundin, {\em Massive loop corrections for collider physics}.
\newblock PhD thesis, Humboldt-Universitat zu Berlin, 2012.

\bibitem{Nason:2007vt}
P.~Nason, {\it {MINT: A Computer program for adaptive Monte Carlo integration
  and generation of unweighted distributions}},
  \href{http://xxx.lanl.gov/abs/0709.2085}{{\tt arXiv:0709.2085}}.

\bibitem{Frixione:2003ei}
S.~Frixione, P.~Nason, and B.~R. Webber, {\it {Matching NLO QCD and parton
  showers in heavy flavor production}},  {\em JHEP} {\bf 0308} (2003) 007,
  [\href{http://xxx.lanl.gov/abs/hep-ph/0305252}{{\tt hep-ph/0305252}}].

\bibitem{Frixione:2005vw}
S.~Frixione, E.~Laenen, P.~Motylinski, and B.~R. Webber, {\it {Single-top
  production in MC@NLO}},  {\em JHEP} {\bf 0603} (2006) 092,
  [\href{http://xxx.lanl.gov/abs/hep-ph/0512250}{{\tt hep-ph/0512250}}].

\bibitem{Torrielli:2010aw}
P.~Torrielli and S.~Frixione, {\it {Matching NLO QCD computations with PYTHIA
  using MC@NLO}},  {\em JHEP} {\bf 1004} (2010) 110,
  [\href{http://xxx.lanl.gov/abs/1002.4293}{{\tt arXiv:1002.4293}}].

\bibitem{Frixione:2010ra}
S.~Frixione, F.~Stoeckli, P.~Torrielli, and B.~R. Webber, {\it {NLO QCD
  corrections in Herwig++ with MC@NLO}},  {\em JHEP} {\bf 1101} (2011) 053,
  [\href{http://xxx.lanl.gov/abs/1010.0568}{{\tt arXiv:1010.0568}}].

\bibitem{Lonnblad:1992tz}
L.~Lonnblad, {\it {ARIADNE version 4: A Program for simulation of QCD cascades
  implementing the color dipole model}},  {\em Comput.Phys.Commun.} {\bf 71}
  (1992) 15--31.

\bibitem{Nagy:2006kb}
Z.~Nagy and D.~E. Soper, {\it {A New parton shower algorithm: Shower evolution,
  matching at leading and next-to-leading order level}},
  \href{http://xxx.lanl.gov/abs/hep-ph/0601021}{{\tt hep-ph/0601021}}.

\bibitem{Dinsdale:2007mf}
M.~Dinsdale, M.~Ternick, and S.~Weinzierl, {\it {Parton showers from the dipole
  formalism}},  {\em Phys.Rev.} {\bf D76} (2007) 094003,
  [\href{http://xxx.lanl.gov/abs/0709.1026}{{\tt arXiv:0709.1026}}].

\bibitem{Schumann:2007mg}
S.~Schumann and F.~Krauss, {\it {A Parton shower algorithm based on
  Catani-Seymour dipole factorisation}},  {\em JHEP} {\bf 0803} (2008) 038,
  [\href{http://xxx.lanl.gov/abs/0709.1027}{{\tt arXiv:0709.1027}}].

\bibitem{Winter:2007ye}
J.-C. Winter and F.~Krauss, {\it {Initial-state showering based on colour
  dipoles connected to incoming parton lines}},  {\em JHEP} {\bf 0807} (2008)
  040, [\href{http://xxx.lanl.gov/abs/0712.3913}{{\tt arXiv:0712.3913}}].

\bibitem{Platzer:2011bc}
S.~Platzer and S.~Gieseke, {\it {Dipole Showers and Automated NLO Matching in
  Herwig++}},  {\em Eur.Phys.J.} {\bf C72} (2012) 2187,
  [\href{http://xxx.lanl.gov/abs/1109.6256}{{\tt arXiv:1109.6256}}].

\bibitem{Ritzmann:2012ca}
M.~Ritzmann, D.~Kosower, and P.~Skands, {\it {Antenna Showers with Hadronic
  Initial States}},  {\em Phys.Lett.} {\bf B718} (2013) 1345--1350,
  [\href{http://xxx.lanl.gov/abs/1210.6345}{{\tt arXiv:1210.6345}}].

\bibitem{Friberg:1996xc}
C.~Friberg, G.~Gustafson, and J.~Hakkinen, {\it {Color connections in $e^+e^-$
  annihilation}},  {\em Nucl.Phys.} {\bf B490} (1997) 289--305,
  [\href{http://xxx.lanl.gov/abs/hep-ph/9604347}{{\tt hep-ph/9604347}}].

\bibitem{Giele:2011cb}
W.~Giele, D.~Kosower, and P.~Skands, {\it {Higher-Order Corrections to Timelike
  Jets}},  {\em Phys.Rev.} {\bf D84} (2011) 054003,
  [\href{http://xxx.lanl.gov/abs/1102.2126}{{\tt arXiv:1102.2126}}].

\bibitem{Platzer:2012np}
S.~Platzer and M.~Sjodahl, {\it {Subleading $N_c$ improved Parton Showers}},
  {\em JHEP} {\bf 1207} (2012) 042,
  [\href{http://xxx.lanl.gov/abs/1201.0260}{{\tt arXiv:1201.0260}}].

\bibitem{Nagy:2012bt}
Z.~Nagy and D.~E. Soper, {\it {Parton shower evolution with subleading color}},
   {\em JHEP} {\bf 1206} (2012) 044,
  [\href{http://xxx.lanl.gov/abs/1202.4496}{{\tt arXiv:1202.4496}}].

\bibitem{Altarelli:1977zs}
G.~Altarelli and G.~Parisi, {\it {Asymptotic Freedom in Parton Language}},
  {\em Nucl.Phys.} {\bf B126} (1977) 298.

\bibitem{Odagiri:1998ep}
K.~Odagiri, {\it {Color connection structure of supersymmetric QCD ($2 \to 2$)
  processes}},  {\em JHEP} {\bf 9810} (1998) 006,
  [\href{http://xxx.lanl.gov/abs/hep-ph/9806531}{{\tt hep-ph/9806531}}].

\bibitem{Nason:2012pr}
P.~Nason and B.~Webber, {\it {Next-to-Leading-Order Event Generators}},  {\em
  Ann.Rev.Nucl.Part.Sci.} {\bf 62} (2012) 187--213,
  [\href{http://xxx.lanl.gov/abs/1202.1251}{{\tt arXiv:1202.1251}}].

\bibitem{Hoeche:2014lxa}
S.~Hoeche, F.~Krauss, and M.~Schonherr, {\it {Uncertainties in MEPS@NLO
  calculations of h+jets}},  \href{http://xxx.lanl.gov/abs/1401.7971}{{\tt
  arXiv:1401.7971}}.

\bibitem{Frederix:2012ps}
R.~Frederix and S.~Frixione, {\it {Merging meets matching in MC@NLO}},  {\em
  JHEP} {\bf 1212} (2012) 061, [\href{http://xxx.lanl.gov/abs/1209.6215}{{\tt
  arXiv:1209.6215}}].

\bibitem{Catani:2001cc}
S.~Catani, F.~Krauss, R.~Kuhn, and B.~Webber, {\it {QCD matrix elements +
  parton showers}},  {\em JHEP} {\bf 0111} (2001) 063,
  [\href{http://xxx.lanl.gov/abs/hep-ph/0109231}{{\tt hep-ph/0109231}}].

\bibitem{Krauss:2002up}
F.~Krauss, {\it {Matrix elements and parton showers in hadronic interactions}},
   {\em JHEP} {\bf 0208} (2002) 015,
  [\href{http://xxx.lanl.gov/abs/hep-ph/0205283}{{\tt hep-ph/0205283}}].

\bibitem{Mrenna:2003if}
S.~Mrenna and P.~Richardson, {\it {Matching matrix elements and parton showers
  with HERWIG and PYTHIA}},  {\em JHEP} {\bf 0405} (2004) 040,
  [\href{http://xxx.lanl.gov/abs/hep-ph/0312274}{{\tt hep-ph/0312274}}].

\bibitem{Lavesson:2005xu}
N.~Lavesson and L.~Lonnblad, {\it {$W+$jets matrix elements and the dipole
  cascade}},  {\em JHEP} {\bf 0507} (2005) 054,
  [\href{http://xxx.lanl.gov/abs/hep-ph/0503293}{{\tt hep-ph/0503293}}].

\bibitem{Hoeche:2009rj}
S.~Hoeche, F.~Krauss, S.~Schumann, and F.~Siegert, {\it {QCD matrix elements
  and truncated showers}},  {\em JHEP} {\bf 0905} (2009) 053,
  [\href{http://xxx.lanl.gov/abs/0903.1219}{{\tt arXiv:0903.1219}}].

\bibitem{Hamilton:2009ne}
K.~Hamilton, P.~Richardson, and J.~Tully, {\it {A Modified CKKW matrix element
  merging approach to angular-ordered parton showers}},  {\em JHEP} {\bf 0911}
  (2009) 038, [\href{http://xxx.lanl.gov/abs/0905.3072}{{\tt
  arXiv:0905.3072}}].

\bibitem{Lavesson:2008ah}
N.~Lavesson and L.~Lonnblad, {\it {Extending CKKW-merging to One-Loop Matrix
  Elements}},  {\em JHEP} {\bf 0812} (2008) 070,
  [\href{http://xxx.lanl.gov/abs/0811.2912}{{\tt arXiv:0811.2912}}].

\bibitem{Hamilton:2010wh}
K.~Hamilton and P.~Nason, {\it {Improving NLO-parton shower matched simulations
  with higher order matrix elements}},  {\em JHEP} {\bf 1006} (2010) 039,
  [\href{http://xxx.lanl.gov/abs/1004.1764}{{\tt arXiv:1004.1764}}].

\bibitem{Hoche:2010kg}
S.~Hoche, F.~Krauss, M.~Schonherr, and F.~Siegert, {\it {NLO matrix elements
  and truncated showers}},  {\em JHEP} {\bf 1108} (2011) 123,
  [\href{http://xxx.lanl.gov/abs/1009.1127}{{\tt arXiv:1009.1127}}].

\bibitem{Alioli:2011nr}
S.~Alioli, K.~Hamilton, and E.~Re, {\it {Practical improvements and merging of
  POWHEG simulations for vector boson production}},  {\em JHEP} {\bf 1109}
  (2011) 104, [\href{http://xxx.lanl.gov/abs/1108.0909}{{\tt
  arXiv:1108.0909}}].

\bibitem{Hoeche:2012yf}
S.~Hoeche, F.~Krauss, M.~Schonherr, and F.~Siegert, {\it {QCD matrix elements +
  parton showers: The NLO case}},  {\em JHEP} {\bf 1304} (2013) 027,
  [\href{http://xxx.lanl.gov/abs/1207.5030}{{\tt arXiv:1207.5030}}].

\bibitem{Platzer:2012bs}
S.~Plaetzer, {\it {Controlling inclusive cross sections in parton shower +
  matrix element merging}},  {\em JHEP} {\bf 1308} (2013) 114,
  [\href{http://xxx.lanl.gov/abs/1211.5467}{{\tt arXiv:1211.5467}}].

\bibitem{Alioli:2012fc}
S.~Alioli, C.~W. Bauer, C.~J. Berggren, A.~Hornig, F.~J. Tackmann, et~al., {\it
  {Combining Higher-Order Resummation with Multiple NLO Calculations and Parton
  Showers in GENEVA}},  {\em JHEP} {\bf 1309} (2013) 120,
  [\href{http://xxx.lanl.gov/abs/1211.7049}{{\tt arXiv:1211.7049}}].

\bibitem{Lonnblad:2012ix}
L.~Lonnblad and S.~Prestel, {\it {Merging Multi-leg NLO Matrix Elements with
  Parton Showers}},  {\em JHEP} {\bf 1303} (2013) 166,
  [\href{http://xxx.lanl.gov/abs/1211.7278}{{\tt arXiv:1211.7278}}].

\bibitem{Hamilton:2012rf}
K.~Hamilton, P.~Nason, C.~Oleari, and G.~Zanderighi, {\it {Merging H/W/Z + 0
  and 1 jet at NLO with no merging scale: a path to parton shower + NNLO
  matching}},  {\em JHEP} {\bf 1305} (2013) 082,
  [\href{http://xxx.lanl.gov/abs/1212.4504}{{\tt arXiv:1212.4504}}].

\bibitem{Alioli:2013hqa}
S.~Alioli, C.~W. Bauer, C.~Berggren, F.~J. Tackmann, J.~R. Walsh, et~al., {\it
  {Matching Fully Differential NNLO Calculations and Parton Showers}},
  \href{http://xxx.lanl.gov/abs/1311.0286}{{\tt arXiv:1311.0286}}.

\bibitem{Bozzi:2005wk}
G.~Bozzi, S.~Catani, D.~de~Florian, and M.~Grazzini, {\it {Transverse-momentum
  resummation and the spectrum of the Higgs boson at the LHC}},  {\em
  Nucl.Phys.} {\bf B737} (2006) 73--120,
  [\href{http://xxx.lanl.gov/abs/hep-ph/0508068}{{\tt hep-ph/0508068}}].

\bibitem{Stuart:1991xk}
R.~G. Stuart, {\it {Gauge invariance, analyticity and physical observables at
  the $Z^0$ resonance}},  {\em Phys.Lett.} {\bf B262} (1991) 113--119.

\bibitem{Aeppli:1993rs}
A.~Aeppli, G.~J. van Oldenborgh, and D.~Wyler, {\it {Unstable particles in one
  loop calculations}},  {\em Nucl.Phys.} {\bf B428} (1994) 126--146,
  [\href{http://xxx.lanl.gov/abs/hep-ph/9312212}{{\tt hep-ph/9312212}}].

\bibitem{Frixione:2007zp}
S.~Frixione, E.~Laenen, P.~Motylinski, and B.~R. Webber, {\it {Angular
  correlations of lepton pairs from vector boson and top quark decays in Monte
  Carlo simulations}},  {\em JHEP} {\bf 0704} (2007) 081,
  [\href{http://xxx.lanl.gov/abs/hep-ph/0702198}{{\tt hep-ph/0702198}}].

\bibitem{Richardson:2001df}
P.~Richardson, {\it {Spin correlations in Monte Carlo simulations}},  {\em
  JHEP} {\bf 0111} (2001) 029,
  [\href{http://xxx.lanl.gov/abs/hep-ph/0110108}{{\tt hep-ph/0110108}}].

\bibitem{Artoisenet:2012st}
P.~Artoisenet, R.~Frederix, O.~Mattelaer, and R.~Rietkerk, {\it {Automatic
  spin-entangled decays of heavy resonances in Monte Carlo simulations}},  {\em
  JHEP} {\bf 1303} (2013) 015, [\href{http://xxx.lanl.gov/abs/1212.3460}{{\tt
  arXiv:1212.3460}}].

\bibitem{Papanastasiou:2013dta}
A.~Papanastasiou, R.~Frederix, S.~Frixione, V.~Hirschi, and F.~Maltoni, {\it
  {Single-top $t$-channel production with off-shell and non-resonant effects}},
   {\em Phys.Lett.} {\bf B726} (2013) 223--227,
  [\href{http://xxx.lanl.gov/abs/1305.7088}{{\tt arXiv:1305.7088}}].

\bibitem{PGS}
J.~Conway, ``{Pretty Good Simulator}.''
  http://www.physics.ucdavis.edu/~conway/research/software/pgs/pgs4-general.ht%
m.

\bibitem{deFavereau:2013fsa}
J.~de~Favereau et~al., {\it {DELPHES 3, A modular framework for fast simulation
  of a generic collider experiment}},  {\em JHEP} {\bf 1402} (2014) 057,
  [\href{http://xxx.lanl.gov/abs/1307.6346}{{\tt arXiv:1307.6346}}].

\bibitem{Ovyn:2009tx}
S.~Ovyn, X.~Rouby, and V.~Lemaitre, {\it {DELPHES, a framework for fast
  simulation of a generic collider experiment}},
  \href{http://xxx.lanl.gov/abs/0903.2225}{{\tt arXiv:0903.2225}}.

\bibitem{LHEv30}
{\bf SM Working group} Collaboration, {\it Proceedings of the workshop physics
  at tev colliders, les houches, 2013, to appear}, .

\bibitem{Martin:2009iq}
A.~Martin, W.~Stirling, R.~Thorne, and G.~Watt, {\it {Parton distributions for
  the LHC}},  {\em Eur.Phys.J.} {\bf C63} (2009) 189--285,
  [\href{http://xxx.lanl.gov/abs/0901.0002}{{\tt arXiv:0901.0002}}].

\bibitem{Cacciari:2008gp}
M.~Cacciari, G.~P. Salam, and G.~Soyez, {\it {The anti-$k_t$ jet clustering
  algorithm}},  {\em JHEP} {\bf 0804} (2008) 063,
  [\href{http://xxx.lanl.gov/abs/0802.1189}{{\tt arXiv:0802.1189}}].

\bibitem{Frixione:1998jh}
S.~Frixione, {\it {Isolated photons in perturbative QCD}},  {\em Phys.Lett.}
  {\bf B429} (1998) 369--374,
  [\href{http://xxx.lanl.gov/abs/hep-ph/9801442}{{\tt hep-ph/9801442}}].

\bibitem{Frederix:2012dh}
R.~Frederix, E.~Re, and P.~Torrielli, {\it {Single-top t-channel
  hadroproduction in the four-flavour scheme with POWHEG and aMC@NLO}},  {\em
  JHEP} {\bf 1209} (2012) 130, [\href{http://xxx.lanl.gov/abs/1207.5391}{{\tt
  arXiv:1207.5391}}].

\bibitem{Campbell:2002tg}
J.~M. Campbell and R.~K. Ellis, {\it {Next-to-leading order corrections to
  $W^+$ 2 jet and $Z^+$ 2 jet production at hadron colliders}},  {\em
  Phys.Rev.} {\bf D65} (2002) 113007,
  [\href{http://xxx.lanl.gov/abs/hep-ph/0202176}{{\tt hep-ph/0202176}}].

\bibitem{Campbell:2003hd}
J.~M. Campbell, R.~K. Ellis, and D.~L. Rainwater, {\it {Next-to-leading order
  QCD predictions for $W$ + 2 jet and $Z$ + 2 jet production at the CERN LHC}},
   {\em Phys.Rev.} {\bf D68} (2003) 094021,
  [\href{http://xxx.lanl.gov/abs/hep-ph/0308195}{{\tt hep-ph/0308195}}].

\bibitem{Campbell:2013vha}
J.~M. Campbell, R.~K. Ellis, P.~Nason, and G.~Zanderighi, {\it {W and Z bosons
  in association with two jets using the POWHEG method}},  {\em JHEP} {\bf
  1308} (2013) 005, [\href{http://xxx.lanl.gov/abs/1303.5447}{{\tt
  arXiv:1303.5447}}].

\bibitem{Campbell:2008hh}
J.~M. Campbell, R.~K. Ellis, F.~Febres~Cordero, F.~Maltoni, L.~Reina, et~al.,
  {\it {Associated Production of a $W$ Boson and One $b$ Jet}},  {\em
  Phys.Rev.} {\bf D79} (2009) 034023,
  [\href{http://xxx.lanl.gov/abs/0809.3003}{{\tt arXiv:0809.3003}}].

\bibitem{Campbell:2003dd}
J.~M. Campbell, R.~K. Ellis, F.~Maltoni, and S.~Willenbrock, {\it {Associated
  production of a $Z$ Boson and a single heavy quark jet}},  {\em Phys.Rev.}
  {\bf D69} (2004) 074021, [\href{http://xxx.lanl.gov/abs/hep-ph/0312024}{{\tt
  hep-ph/0312024}}].

\bibitem{Campbell:2005zv}
J.~M. Campbell, R.~K. Ellis, F.~Maltoni, and S.~Willenbrock, {\it {Production
  of a $Z$ boson and two jets with one heavy-quark tag}},  {\em Phys.Rev.} {\bf
  D73} (2006) 054007, [\href{http://xxx.lanl.gov/abs/hep-ph/0510362}{{\tt
  hep-ph/0510362}}].

\bibitem{Campbell:2006cu}
J.~M. Campbell, R.~K. Ellis, F.~Maltoni, and S.~Willenbrock, {\it {Production
  of a $W$ boson and two jets with one $b$-quark tag}},  {\em Phys.Rev.} {\bf
  D75} (2007) 054015, [\href{http://xxx.lanl.gov/abs/hep-ph/0611348}{{\tt
  hep-ph/0611348}}].

\bibitem{Caola:2011pz}
J.~Campbell, F.~Caola, F.~Febres~Cordero, L.~Reina, and D.~Wackeroth, {\it {NLO
  QCD predictions for $W+1$ jet and $W+2$ jet production with at least one $b$
  jet at the 7 TeV LHC}},  {\em Phys.Rev.} {\bf D86} (2012) 034021,
  [\href{http://xxx.lanl.gov/abs/1107.3714}{{\tt arXiv:1107.3714}}].

\bibitem{Alioli:2008gx}
S.~Alioli, P.~Nason, C.~Oleari, and E.~Re, {\it {NLO vector-boson production
  matched with shower in POWHEG}},  {\em JHEP} {\bf 0807} (2008) 060,
  [\href{http://xxx.lanl.gov/abs/0805.4802}{{\tt arXiv:0805.4802}}].

\bibitem{Alioli:2010qp}
S.~Alioli, P.~Nason, C.~Oleari, and E.~Re, {\it {Vector boson plus one jet
  production in POWHEG}},  {\em JHEP} {\bf 1101} (2011) 095,
  [\href{http://xxx.lanl.gov/abs/1009.5594}{{\tt arXiv:1009.5594}}].

\bibitem{Re:2012zi}
E.~Re, {\it {NLO corrections merged with parton showers for Z+2 jets production
  using the POWHEG method}},  {\em JHEP} {\bf 1210} (2012) 031,
  [\href{http://xxx.lanl.gov/abs/1204.5433}{{\tt arXiv:1204.5433}}].

\bibitem{Ellis:2009zw}
R.~K. Ellis, K.~Melnikov, and G.~Zanderighi, {\it {Generalized unitarity at
  work: first NLO QCD results for hadronic $W^+$ 3jet production}},  {\em JHEP}
  {\bf 0904} (2009) 077, [\href{http://xxx.lanl.gov/abs/0901.4101}{{\tt
  arXiv:0901.4101}}].

\bibitem{Melnikov:2009wh}
K.~Melnikov and G.~Zanderighi, {\it {$W$+3 jet production at the LHC as a
  signal or background}},  {\em Phys.Rev.} {\bf D81} (2010) 074025,
  [\href{http://xxx.lanl.gov/abs/0910.3671}{{\tt arXiv:0910.3671}}].

\bibitem{Berger:2009zg}
C.~Berger, Z.~Bern, L.~J. Dixon, F.~Febres~Cordero, D.~Forde, et~al., {\it
  {Precise Predictions for $W$ + 3 Jet Production at Hadron Colliders}},  {\em
  Phys.Rev.Lett.} {\bf 102} (2009) 222001,
  [\href{http://xxx.lanl.gov/abs/0902.2760}{{\tt arXiv:0902.2760}}].

\bibitem{Berger:2009ep}
C.~Berger, Z.~Bern, L.~J. Dixon, F.~Febres~Cordero, D.~Forde, et~al., {\it
  {Next-to-Leading Order QCD Predictions for W+3-Jet Distributions at Hadron
  Colliders}},  {\em Phys.Rev.} {\bf D80} (2009) 074036,
  [\href{http://xxx.lanl.gov/abs/0907.1984}{{\tt arXiv:0907.1984}}].

\bibitem{Berger:2010vm}
C.~Berger, Z.~Bern, L.~J. Dixon, F.~Febres~Cordero, D.~Forde, et~al., {\it
  {Next-to-Leading Order QCD Predictions for $Z,\gamma^*+3$-Jet Distributions
  at the Tevatron}},  {\em Phys.Rev.} {\bf D82} (2010) 074002,
  [\href{http://xxx.lanl.gov/abs/1004.1659}{{\tt arXiv:1004.1659}}].

\bibitem{Berger:2010zx}
C.~Berger, Z.~Bern, L.~J. Dixon, F.~Febres~Cordero, D.~Forde, et~al., {\it
  {Precise Predictions for W + 4 Jet Production at the Large Hadron Collider}},
   {\em Phys.Rev.Lett.} {\bf 106} (2011) 092001,
  [\href{http://xxx.lanl.gov/abs/1009.2338}{{\tt arXiv:1009.2338}}].

\bibitem{Ita:2011wn}
H.~Ita, Z.~Bern, L.~Dixon, F.~Febres~Cordero, D.~Kosower, et~al., {\it {Precise
  Predictions for Z + 4 Jets at Hadron Colliders}},  {\em Phys.Rev.} {\bf D85}
  (2012) 031501, [\href{http://xxx.lanl.gov/abs/1108.2229}{{\tt
  arXiv:1108.2229}}].

\bibitem{Hoeche:2012ft}
S.~Hoeche, F.~Krauss, M.~Schonherr, and F.~Siegert, {\it {W+n-jet predictions
  at the Large Hadron Collider at next-to-leading order matched with a parton
  shower}},  {\em Phys.Rev.Lett.} {\bf 110} (2013) 052001,
  [\href{http://xxx.lanl.gov/abs/1201.5882}{{\tt arXiv:1201.5882}}].

\bibitem{Catani:2002ny}
S.~Catani, M.~Fontannaz, J.~Guillet, and E.~Pilon, {\it {Cross-section of
  isolated prompt photons in hadron hadron collisions}},  {\em JHEP} {\bf 0205}
  (2002) 028, [\href{http://xxx.lanl.gov/abs/hep-ph/0204023}{{\tt
  hep-ph/0204023}}].

\bibitem{Bern:2011pa}
Z.~Bern, G.~Diana, L.~Dixon, F.~Febres~Cordero, S.~Hoche, et~al., {\it {Driving
  Missing Data at Next-to-Leading Order}},  {\em Phys.Rev.} {\bf D84} (2011)
  114002, [\href{http://xxx.lanl.gov/abs/1106.1423}{{\tt arXiv:1106.1423}}].

\bibitem{Arnold:2011wj}
K.~Arnold, J.~Bellm, G.~Bozzi, M.~Brieg, F.~Campanario, et~al., {\it {VBFNLO: A
  Parton Level Monte Carlo for Processes with Electroweak Bosons -- Manual for
  Version 2.5.0}},  \href{http://xxx.lanl.gov/abs/1107.4038}{{\tt
  arXiv:1107.4038}}.

\bibitem{Jager:2012xk}
B.~Jager, S.~Schneider, and G.~Zanderighi, {\it {Next-to-leading order QCD
  corrections to electroweak $Zjj$ production in the POWHEG BOX}},  {\em JHEP}
  {\bf 1209} (2012) 083, [\href{http://xxx.lanl.gov/abs/1207.2626}{{\tt
  arXiv:1207.2626}}].

\bibitem{Mele:1990bq}
B.~Mele, P.~Nason, and G.~Ridolfi, {\it {QCD radiative corrections to Z boson
  pair production in hadronic collisions}},  {\em Nucl.Phys.} {\bf B357} (1991)
  409--438.

\bibitem{Ohnemus:1990za}
J.~Ohnemus and J.~Owens, {\it {An $O(\alpha_s)$ calculation of hadronic $ZZ$
  production}},  {\em Phys.Rev.} {\bf D43} (1991) 3626--3639.

\bibitem{Frixione:1992pj}
S.~Frixione, P.~Nason, and G.~Ridolfi, {\it {Strong corrections to $WZ$
  production at hadron colliders}},  {\em Nucl.Phys.} {\bf B383} (1992) 3--44.

\bibitem{Ohnemus:1991kk}
J.~Ohnemus, {\it {An $O(\alpha_s)$ calculation of hadronic $W^{-}W^{+}$
  production}},  {\em Phys.Rev.} {\bf D44} (1991) 1403--1414.

\bibitem{Ohnemus:1991gb}
J.~Ohnemus, {\it {An $O(\alpha_s)$ calculation of hadronic $W^\pm Z$
  production}},  {\em Phys.Rev.} {\bf D44} (1991) 3477--3489.

\bibitem{Frixione:1993yp}
S.~Frixione, {\it {A Next-to-leading order calculation of the cross-section for
  the production of $W^+ W^-$ pairs in hadronic collisions}},  {\em Nucl.Phys.}
  {\bf B410} (1993) 280--324.

\bibitem{Campbell:1999ah}
J.~M. Campbell and R.~K. Ellis, {\it {An update on vector boson pair production
  at hadron colliders}},  {\em Phys.Rev.} {\bf D60} (1999) 113006,
  [\href{http://xxx.lanl.gov/abs/hep-ph/9905386}{{\tt hep-ph/9905386}}].

\bibitem{Dixon:1999di}
L.~J. Dixon, Z.~Kunszt, and A.~Signer, {\it {Vector boson pair production in
  hadronic collisions at order $\alpha_s$ : Lepton correlations and anomalous
  couplings}},  {\em Phys.Rev.} {\bf D60} (1999) 114037,
  [\href{http://xxx.lanl.gov/abs/hep-ph/9907305}{{\tt hep-ph/9907305}}].

\bibitem{DeFlorian:2000sg}
D.~De~Florian and A.~Signer, {\it {$W \gamma$ and $Z \gamma$ production at
  hadron colliders}},  {\em Eur.Phys.J.} {\bf C16} (2000) 105--114,
  [\href{http://xxx.lanl.gov/abs/hep-ph/0002138}{{\tt hep-ph/0002138}}].

\bibitem{Campbell:2011bn}
J.~M. Campbell, R.~K. Ellis, and C.~Williams, {\it {Vector boson pair
  production at the LHC}},  {\em JHEP} {\bf 1107} (2011) 018,
  [\href{http://xxx.lanl.gov/abs/1105.0020}{{\tt arXiv:1105.0020}}].

\bibitem{Nason:2006hfa}
P.~Nason and G.~Ridolfi, {\it {A Positive-weight next-to-leading-order Monte
  Carlo for Z pair hadroproduction}},  {\em JHEP} {\bf 0608} (2006) 077,
  [\href{http://xxx.lanl.gov/abs/hep-ph/0606275}{{\tt hep-ph/0606275}}].

\bibitem{Melia:2011tj}
T.~Melia, P.~Nason, R.~Rontsch, and G.~Zanderighi, {\it {$W^+W^-$, $WZ$ and
  $ZZ$ production in the POWHEG BOX}},  {\em JHEP} {\bf 1111} (2011) 078,
  [\href{http://xxx.lanl.gov/abs/1107.5051}{{\tt arXiv:1107.5051}}].

\bibitem{Nason:2013ydw}
P.~Nason and G.~Zanderighi, {\it {$W^+ W^-$ , $W Z$ and $Z Z$ production in the
  POWHEG-BOX-V2}},  {\em Eur.Phys.J.} {\bf C74} (2014) 2702,
  [\href{http://xxx.lanl.gov/abs/1311.1365}{{\tt arXiv:1311.1365}}].

\bibitem{DelDuca:2003uz}
V.~Del~Duca, F.~Maltoni, Z.~Nagy, and Z.~Trocsanyi, {\it {QCD radiative
  corrections to prompt diphoton production in association with a jet at hadron
  colliders}},  {\em JHEP} {\bf 0304} (2003) 059,
  [\href{http://xxx.lanl.gov/abs/hep-ph/0303012}{{\tt hep-ph/0303012}}].

\bibitem{Gehrmann:2013bga}
T.~Gehrmann, N.~Greiner, and G.~Heinrich, {\it {Precise QCD predictions for the
  production of a photon pair in association with two jets}},  {\em
  Phys.Rev.Lett.} {\bf 111} (2013) 222002,
  [\href{http://xxx.lanl.gov/abs/1308.3660}{{\tt arXiv:1308.3660}}].

\bibitem{Badger:2013ava}
S.~Badger, A.~Guffanti, and V.~Yundin, {\it {Next-to-leading order QCD
  corrections to di-photon production in association with up to three jets at
  the Large Hadron Collider}},  {\em JHEP} {\bf 1403} (2014) 122,
  [\href{http://xxx.lanl.gov/abs/1312.5927}{{\tt arXiv:1312.5927}}].

\bibitem{Bern:2013bha}
Z.~Bern, L.~Dixon, F.~Febres~Cordero, S.~Hoeche, H.~Ita, et~al., {\it
  {Next-to-leading order diphoton+2-jet production at the LHC}},
  \href{http://xxx.lanl.gov/abs/1312.0592}{{\tt arXiv:1312.0592}}.

\bibitem{Bern:2014vza}
Z.~Bern, L.~Dixon, F.~Febres~Cordero, S.~Hoeche, H.~Ita, et~al., {\it
  {Next-to-Leading Order $\gamma \gamma + 2$-Jet Production at the LHC}},
  \href{http://xxx.lanl.gov/abs/1402.4127}{{\tt arXiv:1402.4127}}.

\bibitem{Greiner:2012im}
N.~Greiner, G.~Heinrich, P.~Mastrolia, G.~Ossola, T.~Reiter, et~al., {\it {NLO
  QCD corrections to the production of $W^+ W^-$ plus two jets at the LHC}},
  {\em Phys.Lett.} {\bf B713} (2012) 277--283,
  [\href{http://xxx.lanl.gov/abs/1202.6004}{{\tt arXiv:1202.6004}}].

\bibitem{Melia:2011dw}
T.~Melia, K.~Melnikov, R.~Rontsch, and G.~Zanderighi, {\it {NLO QCD corrections
  for $W^+W^-$ pair production in association with two jets at hadron
  colliders}},  {\em Phys.Rev.} {\bf D83} (2011) 114043,
  [\href{http://xxx.lanl.gov/abs/1104.2327}{{\tt arXiv:1104.2327}}].

\bibitem{Melia:2010bm}
T.~Melia, K.~Melnikov, R.~Rontsch, and G.~Zanderighi, {\it {Next-to-leading
  order QCD predictions for $W^+W^+jj$ production at the LHC}},  {\em JHEP}
  {\bf 1012} (2010) 053, [\href{http://xxx.lanl.gov/abs/1007.5313}{{\tt
  arXiv:1007.5313}}].

\bibitem{Campanario:2013gea}
F.~Campanario, M.~Kerner, L.~D. Ninh, and D.~Zeppenfeld, {\it {Next-to-leading
  order QCD corrections to $W^+W^+$ and $W^-W^-$ production in association with
  two jets}},  {\em Phys.Rev.} {\bf D89} (2014) 054009,
  [\href{http://xxx.lanl.gov/abs/1311.6738}{{\tt arXiv:1311.6738}}].

\bibitem{Campbell:2012ft}
J.~M. Campbell, H.~B. Hartanto, and C.~Williams, {\it {Next-to-leading order
  predictions for $Z \gamma$+jet and $Z \gamma \gamma$ final states at the
  LHC}},  {\em JHEP} {\bf 1211} (2012) 162,
  [\href{http://xxx.lanl.gov/abs/1208.0566}{{\tt arXiv:1208.0566}}].

\bibitem{Campanario:2014dpa}
F.~Campanario, M.~Kerner, L.~D. Ninh, and D.~Zeppenfeld, {\it {Next-to-leading
  order QCD corrections to $W \gamma$ production in association with two
  jets}},  \href{http://xxx.lanl.gov/abs/1402.0505}{{\tt arXiv:1402.0505}}.

\bibitem{Campanario:2013ysa}
F.~Campanario, M.~Kerner, L.~D. Ninh, and D.~Zeppenfeld, {\it {NLO QCD
  corrections to WZjj production at the LHC}},
  \href{http://xxx.lanl.gov/abs/1310.4369}{{\tt arXiv:1310.4369}}.

\bibitem{Campanario:2009um}
F.~Campanario, C.~Englert, M.~Spannowsky, and D.~Zeppenfeld, {\it {NLO-QCD
  corrections to $W \gamma j$ production}},  {\em Europhys.Lett.} {\bf 88}
  (2009) 11001, [\href{http://xxx.lanl.gov/abs/0908.1638}{{\tt
  arXiv:0908.1638}}].

\bibitem{Campanario:2010hv}
F.~Campanario, C.~Englert, and M.~Spannowsky, {\it {Precise predictions for
  (non-standard) $W \gamma$ + jet production}},  {\em Phys.Rev.} {\bf D83}
  (2011) 074009, [\href{http://xxx.lanl.gov/abs/1010.1291}{{\tt
  arXiv:1010.1291}}].

\bibitem{Campanario:2010hp}
F.~Campanario, C.~Englert, S.~Kallweit, M.~Spannowsky, and D.~Zeppenfeld, {\it
  {NLO QCD corrections to $WZ$+jet production with leptonic decays}},  {\em
  JHEP} {\bf 1007} (2010) 076, [\href{http://xxx.lanl.gov/abs/1006.0390}{{\tt
  arXiv:1006.0390}}].

\bibitem{Jager:2013mu}
B.~Jager and G.~Zanderighi, {\it {Electroweak $W^+W^-jj$ prodution at NLO in
  QCD matched with parton shower in the POWHEG-BOX}},  {\em JHEP} {\bf 1304}
  (2013) 024, [\href{http://xxx.lanl.gov/abs/1301.1695}{{\tt
  arXiv:1301.1695}}].

\bibitem{Schissler:2013nga}
F.~Schissler and D.~Zeppenfeld, {\it {Parton Shower Effects on $W$ and $Z$
  Production via Vector Boson Fusion at NLO QCD}},  {\em JHEP} {\bf 1304}
  (2013) 057, [\href{http://xxx.lanl.gov/abs/1302.2884}{{\tt
  arXiv:1302.2884}}].

\bibitem{Jager:2013iza}
B.~Jäger, A.~Karlberg, and G.~Zanderighi, {\it {Electroweak $ZZjj$ production
  in the Standard Model and beyond in the POWHEG-BOX V2}},  {\em JHEP} {\bf
  1403} (2014) 141, [\href{http://xxx.lanl.gov/abs/1312.3252}{{\tt
  arXiv:1312.3252}}].

\bibitem{Bozzi:2011en}
G.~Bozzi, F.~Campanario, M.~Rauch, and D.~Zeppenfeld, {\it {$Z \gamma\gamma$
  production with leptonic decays and triple photon production at
  next-to-leading order QCD}},  {\em Phys.Rev.} {\bf D84} (2011) 074028,
  [\href{http://xxx.lanl.gov/abs/1107.3149}{{\tt arXiv:1107.3149}}].

\bibitem{Bozzi:2011wwa}
G.~Bozzi, F.~Campanario, M.~Rauch, and D.~Zeppenfeld, {\it {$W^\pm\gamma
  \gamma$ production with leptonic decays at NLO QCD}},  {\em Phys.Rev.} {\bf
  D83} (2011) 114035, [\href{http://xxx.lanl.gov/abs/1103.4613}{{\tt
  arXiv:1103.4613}}].

\bibitem{Bozzi:2010sj}
G.~Bozzi, F.~Campanario, M.~Rauch, H.~Rzehak, and D.~Zeppenfeld, {\it {NLO QCD
  corrections to $W^\pm Z\gamma$ production with leptonic decays}},  {\em
  Phys.Lett.} {\bf B696} (2011) 380--385,
  [\href{http://xxx.lanl.gov/abs/1011.2206}{{\tt arXiv:1011.2206}}].

\bibitem{Bozzi:2009ig}
G.~Bozzi, F.~Campanario, V.~Hankele, and D.~Zeppenfeld, {\it {NLO QCD
  corrections to $W^+W^- \gamma$ and $Z Z \gamma$ production with leptonic
  decays}},  {\em Phys.Rev.} {\bf D81} (2010) 094030,
  [\href{http://xxx.lanl.gov/abs/0911.0438}{{\tt arXiv:0911.0438}}].

\bibitem{Binoth:2008kt}
T.~Binoth, G.~Ossola, C.~Papadopoulos, and R.~Pittau, {\it {NLO QCD corrections
  to tri-boson production}},  {\em JHEP} {\bf 0806} (2008) 082,
  [\href{http://xxx.lanl.gov/abs/0804.0350}{{\tt arXiv:0804.0350}}].

\bibitem{Campanario:2008yg}
F.~Campanario, V.~Hankele, C.~Oleari, S.~Prestel, and D.~Zeppenfeld, {\it {QCD
  corrections to charged triple vector boson production with leptonic decay}},
  {\em Phys.Rev.} {\bf D78} (2008) 094012,
  [\href{http://xxx.lanl.gov/abs/0809.0790}{{\tt arXiv:0809.0790}}].

\bibitem{Campbell:2014yka}
J.~M. Campbell and C.~Williams, {\it {Triphoton production at hadron
  colliders}},  \href{http://xxx.lanl.gov/abs/1403.2641}{{\tt
  arXiv:1403.2641}}.

\bibitem{Mandal:2014vpa}
M.~Mandal, P.~Mathews, V.~Ravindran, and S.~Seth, {\it {Three photon production
  to NLO+PS accuracy at the LHC}},
  \href{http://xxx.lanl.gov/abs/1403.2917}{{\tt arXiv:1403.2917}}.

\bibitem{Lazopoulos:2007ix}
A.~Lazopoulos, K.~Melnikov, and F.~Petriello, {\it {QCD corrections to
  tri-boson production}},  {\em Phys.Rev.} {\bf D76} (2007) 014001,
  [\href{http://xxx.lanl.gov/abs/hep-ph/0703273}{{\tt hep-ph/0703273}}].

\bibitem{Campanario:2011ud}
F.~Campanario, C.~Englert, M.~Rauch, and D.~Zeppenfeld, {\it {Precise
  predictions for $W \gamma \gamma$ +jet production at hadron colliders}},
  {\em Phys.Lett.} {\bf B704} (2011) 515--519,
  [\href{http://xxx.lanl.gov/abs/1106.4009}{{\tt arXiv:1106.4009}}].

\bibitem{Hoeche:2014rya}
S.~Hoeche, F.~Krauss, S.~Pozzorini, M.~Schoenherr, J.~Thompson, et~al., {\it
  {Triple vector boson production through Higgs-Strahlung with NLO multijet
  merging}},  \href{http://xxx.lanl.gov/abs/1403.7516}{{\tt arXiv:1403.7516}}.

\bibitem{Nason:1987xz}
P.~Nason, S.~Dawson, and R.~K. Ellis, {\it {The Total Cross-Section for the
  Production of Heavy Quarks in Hadronic Collisions}},  {\em Nucl.Phys.} {\bf
  B303} (1988) 607.

\bibitem{Beenakker:1988bq}
W.~Beenakker, H.~Kuijf, W.~van Neerven, and J.~Smith, {\it {QCD Corrections to
  Heavy Quark Production in $p \bar{p}$ Collisions}},  {\em Phys.Rev.} {\bf
  D40} (1989) 54--82.

\bibitem{Nason:1989zy}
P.~Nason, S.~Dawson, and R.~K. Ellis, {\it {The One Particle Inclusive
  Differential Cross-Section for Heavy Quark Production in Hadronic
  Collisions}},  {\em Nucl.Phys.} {\bf B327} (1989) 49--92.

\bibitem{Beenakker:1990maa}
W.~Beenakker, W.~van Neerven, R.~Meng, G.~Schuler, and J.~Smith, {\it {QCD
  corrections to heavy quark production in hadron hadron collisions}},  {\em
  Nucl.Phys.} {\bf B351} (1991) 507--560.

\bibitem{Mangano:1991jk}
M.~L. Mangano, P.~Nason, and G.~Ridolfi, {\it {Heavy quark correlations in
  hadron collisions at next-to-leading order}},  {\em Nucl.Phys.} {\bf B373}
  (1992) 295--345.

\bibitem{Frixione:2007nw}
S.~Frixione, P.~Nason, and G.~Ridolfi, {\it {A Positive-weight
  next-to-leading-order Monte Carlo for heavy flavour hadroproduction}},  {\em
  JHEP} {\bf 0709} (2007) 126, [\href{http://xxx.lanl.gov/abs/0707.3088}{{\tt
  arXiv:0707.3088}}].

\bibitem{Dittmaier:2007wz}
S.~Dittmaier, P.~Uwer, and S.~Weinzierl, {\it {NLO QCD corrections to $t
  \bar{t}$ + jet production at hadron colliders}},  {\em Phys.Rev.Lett.} {\bf
  98} (2007) 262002, [\href{http://xxx.lanl.gov/abs/hep-ph/0703120}{{\tt
  hep-ph/0703120}}].

\bibitem{Melnikov:2011qx}
K.~Melnikov, A.~Scharf, and M.~Schulze, {\it {Top quark pair production in
  association with a jet: QCD corrections and jet radiation in top quark
  decays}},  {\em Phys.Rev.} {\bf D85} (2012) 054002,
  [\href{http://xxx.lanl.gov/abs/1111.4991}{{\tt arXiv:1111.4991}}].

\bibitem{Alioli:2011as}
S.~Alioli, S.-O. Moch, and P.~Uwer, {\it {Hadronic top-quark pair-production
  with one jet and parton showering}},  {\em JHEP} {\bf 1201} (2012) 137,
  [\href{http://xxx.lanl.gov/abs/1110.5251}{{\tt arXiv:1110.5251}}].

\bibitem{Kardos:2011qa}
A.~Kardos, C.~Papadopoulos, and Z.~Trocsanyi, {\it {Top quark pair production
  in association with a jet with NLO parton showering}},  {\em Phys.Lett.} {\bf
  B705} (2011) 76--81, [\href{http://xxx.lanl.gov/abs/1101.2672}{{\tt
  arXiv:1101.2672}}].

\bibitem{Bevilacqua:2010ve}
G.~Bevilacqua, M.~Czakon, C.~Papadopoulos, and M.~Worek, {\it {Dominant QCD
  Backgrounds in Higgs Boson Analyses at the LHC: A Study of $pp \to t \bar{t}$
  + 2 jets at Next-To-Leading Order}},  {\em Phys.Rev.Lett.} {\bf 104} (2010)
  162002, [\href{http://xxx.lanl.gov/abs/1002.4009}{{\tt arXiv:1002.4009}}].

\bibitem{Schonherr:2013bpa}
M.~Schonherr, S.~Hoeche, J.~Huang, G.~Luisoni, and J.~Winter, {\it {NLO merging
  in $t\bar{t}$+jets}},  \href{http://xxx.lanl.gov/abs/1311.3621}{{\tt
  arXiv:1311.3621}}.

\bibitem{Hoeche:2014qda}
S.~Hoeche, F.~Krauss, P.~Maierhoefer, S.~Pozzorini, M.~Schonherr, et~al., {\it
  {Next-to-leading order QCD predictions for top-quark pair production with up
  to two jets merged with a parton shower}},
  \href{http://xxx.lanl.gov/abs/1402.6293}{{\tt arXiv:1402.6293}}.

\bibitem{Nagy:2003tz}
Z.~Nagy, {\it {Next-to-leading order calculation of three jet observables in
  hadron hadron collision}},  {\em Phys.Rev.} {\bf D68} (2003) 094002,
  [\href{http://xxx.lanl.gov/abs/hep-ph/0307268}{{\tt hep-ph/0307268}}].

\bibitem{Badger:2012pf}
S.~Badger, B.~Biedermann, P.~Uwer, and V.~Yundin, {\it {NLO QCD corrections to
  multi-jet production at the LHC with a centre-of-mass energy of $\sqrt{s}=8$
  TeV}},  {\em Phys.Lett.} {\bf B718} (2013) 965--978,
  [\href{http://xxx.lanl.gov/abs/1209.0098}{{\tt arXiv:1209.0098}}].

\bibitem{Badger:2013yda}
S.~Badger, B.~Biedermann, P.~Uwer, and V.~Yundin, {\it {Next-to-leading order
  QCD corrections to five jet production at the LHC}},  {\em Phys.Rev.} {\bf
  D89} (2014) 034019, [\href{http://xxx.lanl.gov/abs/1309.6585}{{\tt
  arXiv:1309.6585}}].

\bibitem{Alioli:2010xa}
S.~Alioli, K.~Hamilton, P.~Nason, C.~Oleari, and E.~Re, {\it {Jet pair
  production in POWHEG}},  {\em JHEP} {\bf 1104} (2011) 081,
  [\href{http://xxx.lanl.gov/abs/1012.3380}{{\tt arXiv:1012.3380}}].

\bibitem{Kardos:2014dua}
A.~Kardos, P.~Nason, and C.~Oleari, {\it {Three-jet production in POWHEG}},
  {\em JHEP} {\bf 1404} (2014) 043,
  [\href{http://xxx.lanl.gov/abs/1402.4001}{{\tt arXiv:1402.4001}}].

\bibitem{Greiner:2011mp}
N.~Greiner, A.~Guffanti, T.~Reiter, and J.~Reuter, {\it {NLO QCD corrections to
  the production of two bottom-antibottom pairs at the LHC}},  {\em
  Phys.Rev.Lett.} {\bf 107} (2011) 102002,
  [\href{http://xxx.lanl.gov/abs/1105.3624}{{\tt arXiv:1105.3624}}].

\bibitem{Bevilacqua:2013taa}
G.~Bevilacqua, M.~Czakon, M.~Kr{\"a}mer, M.~Kubocz, and M.~Worek, {\it
  {Quantifying quark mass effects at the LHC: A study of $pp \to
  b\bar{b}b\bar{b} + X$ at next-to-leading order}},  {\em JHEP} {\bf 1307}
  (2013) 095, [\href{http://xxx.lanl.gov/abs/1304.6860}{{\tt
  arXiv:1304.6860}}].

\bibitem{Bevilacqua:2009zn}
G.~Bevilacqua, M.~Czakon, C.~Papadopoulos, R.~Pittau, and M.~Worek, {\it
  {Assault on the NLO Wishlist: $pp \to t \bar{t} b\bar{b}$}},  {\em JHEP} {\bf
  0909} (2009) 109, [\href{http://xxx.lanl.gov/abs/0907.4723}{{\tt
  arXiv:0907.4723}}].

\bibitem{Bredenstein:2009aj}
A.~Bredenstein, A.~Denner, S.~Dittmaier, and S.~Pozzorini, {\it {NLO QCD
  corrections to $pp \to t\bar{t}b\bar{b} + X$ at the LHC}},  {\em
  Phys.Rev.Lett.} {\bf 103} (2009) 012002,
  [\href{http://xxx.lanl.gov/abs/0905.0110}{{\tt arXiv:0905.0110}}].

\bibitem{Bredenstein:2010rs}
A.~Bredenstein, A.~Denner, S.~Dittmaier, and S.~Pozzorini, {\it {NLO QCD
  Corrections to Top Anti-Top Bottom Anti-Bottom Production at the LHC: 2. full
  hadronic results}},  {\em JHEP} {\bf 1003} (2010) 021,
  [\href{http://xxx.lanl.gov/abs/1001.4006}{{\tt arXiv:1001.4006}}].

\bibitem{Kardos:2013vxa}
A.~Kardos and Z.~Trócsányi, {\it {Hadroproduction of t anti-t pair with a b
  anti-b pair using PowHel}},  {\em J.Phys.} {\bf G41} (2014) 075005,
  [\href{http://xxx.lanl.gov/abs/1303.6291}{{\tt arXiv:1303.6291}}].

\bibitem{Bevilacqua:2012em}
G.~Bevilacqua and M.~Worek, {\it {Constraining BSM Physics at the LHC: Four top
  final states with NLO accuracy in perturbative QCD}},  {\em JHEP} {\bf 1207}
  (2012) 111, [\href{http://xxx.lanl.gov/abs/1206.3064}{{\tt
  arXiv:1206.3064}}].

\bibitem{Ellis:1998fv}
R.~K. Ellis and S.~Veseli, {\it {Strong radiative corrections to $W b \bar{b}$
  production in $p \bar{p}$ collisions}},  {\em Phys.Rev.} {\bf D60} (1999)
  011501, [\href{http://xxx.lanl.gov/abs/hep-ph/9810489}{{\tt
  hep-ph/9810489}}].

\bibitem{Badger:2010mg}
S.~Badger, J.~M. Campbell, and R.~Ellis, {\it {QCD corrections to the hadronic
  production of a heavy quark pair and a W-boson including decay
  correlations}},  {\em JHEP} {\bf 1103} (2011) 027,
  [\href{http://xxx.lanl.gov/abs/1011.6647}{{\tt arXiv:1011.6647}}].

\bibitem{Frederix:2011qg}
R.~Frederix, S.~Frixione, V.~Hirschi, F.~Maltoni, R.~Pittau, et~al., {\it {$W$
  and $Z/\gamma^*$ boson production in association with a bottom-antibottom
  pair}},  {\em JHEP} {\bf 1109} (2011) 061,
  [\href{http://xxx.lanl.gov/abs/1106.6019}{{\tt arXiv:1106.6019}}].

\bibitem{Oleari:2011ey}
C.~Oleari and L.~Reina, {\it {$W^\pm b \bar{b}$ production in POWHEG}},  {\em
  JHEP} {\bf 1108} (2011) 061, [\href{http://xxx.lanl.gov/abs/1105.4488}{{\tt
  arXiv:1105.4488}}].

\bibitem{Campbell:2000bg}
J.~M. Campbell and R.~K. Ellis, {\it {Radiative corrections to $Z b \bar{b}$
  production}},  {\em Phys.Rev.} {\bf D62} (2000) 114012,
  [\href{http://xxx.lanl.gov/abs/hep-ph/0006304}{{\tt hep-ph/0006304}}].

\bibitem{Melnikov:2011ta}
K.~Melnikov, M.~Schulze, and A.~Scharf, {\it {QCD corrections to top quark pair
  production in association with a photon at hadron colliders}},  {\em
  Phys.Rev.} {\bf D83} (2011) 074013,
  [\href{http://xxx.lanl.gov/abs/1102.1967}{{\tt arXiv:1102.1967}}].

\bibitem{Lazopoulos:2008de}
A.~Lazopoulos, T.~McElmurry, K.~Melnikov, and F.~Petriello, {\it
  {Next-to-leading order QCD corrections to $t \bar{t} Z$ production at the
  LHC}},  {\em Phys.Lett.} {\bf B666} (2008) 62--65,
  [\href{http://xxx.lanl.gov/abs/0804.2220}{{\tt arXiv:0804.2220}}].

\bibitem{Garzelli:2011is}
M.~Garzelli, A.~Kardos, C.~Papadopoulos, and Z.~Trocsanyi, {\it {Z0 - boson
  production in association with a top anti-top pair at NLO accuracy with
  parton shower effects}},  {\em Phys.Rev.} {\bf D85} (2012) 074022,
  [\href{http://xxx.lanl.gov/abs/1111.1444}{{\tt arXiv:1111.1444}}].

\bibitem{Kardos:2011na}
A.~Kardos, Z.~Trocsanyi, and C.~Papadopoulos, {\it {Top quark pair production
  in association with a Z-boson at NLO accuracy}},  {\em Phys.Rev.} {\bf D85}
  (2012) 054015, [\href{http://xxx.lanl.gov/abs/1111.0610}{{\tt
  arXiv:1111.0610}}].

\bibitem{Garzelli:2012bn}
M.~Garzelli, A.~Kardos, C.~Papadopoulos, and Z.~Trocsanyi, {\it {$t \bar{t}
  W^\pm$ and $t \bar{t} Z$ Hadroproduction at NLO accuracy in QCD with Parton
  Shower and Hadronization effects}},  {\em JHEP} {\bf 1211} (2012) 056,
  [\href{http://xxx.lanl.gov/abs/1208.2665}{{\tt arXiv:1208.2665}}].

\bibitem{Campbell:2012dh}
J.~M. Campbell and R.~K. Ellis, {\it {$t \bar{t} W^\pm$ production and decay at
  NLO}},  {\em JHEP} {\bf 1207} (2012) 052,
  [\href{http://xxx.lanl.gov/abs/1204.5678}{{\tt arXiv:1204.5678}}].

\bibitem{Stelzer:1995mi}
T.~Stelzer and S.~Willenbrock, {\it {Single top quark production via $q \bar{q}
  \to t \bar{b}$}},  {\em Phys.Lett.} {\bf B357} (1995) 125--130,
  [\href{http://xxx.lanl.gov/abs/hep-ph/9505433}{{\tt hep-ph/9505433}}].

\bibitem{Stelzer:1997ns}
T.~Stelzer, Z.~Sullivan, and S.~Willenbrock, {\it {Single top quark production
  via $W$ - gluon fusion at next-to-leading order}},  {\em Phys.Rev.} {\bf D56}
  (1997) 5919--5927, [\href{http://xxx.lanl.gov/abs/hep-ph/9705398}{{\tt
  hep-ph/9705398}}].

\bibitem{Campbell:2004ch}
J.~M. Campbell, R.~K. Ellis, and F.~Tramontano, {\it {Single top production and
  decay at next-to-leading order}},  {\em Phys.Rev.} {\bf D70} (2004) 094012,
  [\href{http://xxx.lanl.gov/abs/hep-ph/0408158}{{\tt hep-ph/0408158}}].

\bibitem{Campbell:2009gj}
J.~M. Campbell, R.~Frederix, F.~Maltoni, and F.~Tramontano, {\it {NLO
  predictions for t-channel production of single top and fourth generation
  quarks at hadron colliders}},  {\em JHEP} {\bf 0910} (2009) 042,
  [\href{http://xxx.lanl.gov/abs/0907.3933}{{\tt arXiv:0907.3933}}].

\bibitem{Campbell:2009ss}
J.~M. Campbell, R.~Frederix, F.~Maltoni, and F.~Tramontano, {\it
  {Next-to-Leading-Order Predictions for t-Channel Single-Top Production at
  Hadron Colliders}},  {\em Phys.Rev.Lett.} {\bf 102} (2009) 182003,
  [\href{http://xxx.lanl.gov/abs/0903.0005}{{\tt arXiv:0903.0005}}].

\bibitem{Campbell:2005bb}
J.~M. Campbell and F.~Tramontano, {\it {Next-to-leading order corrections to Wt
  production and decay}},  {\em Nucl.Phys.} {\bf B726} (2005) 109--130,
  [\href{http://xxx.lanl.gov/abs/hep-ph/0506289}{{\tt hep-ph/0506289}}].

\bibitem{Frixione:2008yi}
S.~Frixione, E.~Laenen, P.~Motylinski, B.~R. Webber, and C.~D. White, {\it
  {Single-top hadroproduction in association with a $W$ boson}},  {\em JHEP}
  {\bf 0807} (2008) 029, [\href{http://xxx.lanl.gov/abs/0805.3067}{{\tt
  arXiv:0805.3067}}].

\bibitem{Alioli:2009je}
S.~Alioli, P.~Nason, C.~Oleari, and E.~Re, {\it {NLO single-top production
  matched with shower in POWHEG: s- and t-channel contributions}},  {\em JHEP}
  {\bf 0909} (2009) 111, [\href{http://xxx.lanl.gov/abs/0907.4076}{{\tt
  arXiv:0907.4076}}].

\bibitem{Re:2010bp}
E.~Re, {\it {Single-top Wt-channel production matched with parton showers using
  the POWHEG method}},  {\em Eur.Phys.J.} {\bf C71} (2011) 1547,
  [\href{http://xxx.lanl.gov/abs/1009.2450}{{\tt arXiv:1009.2450}}].

\bibitem{Campbell:2013yla}
J.~Campbell, R.~K. Ellis, and R.~Rontsch, {\it {Single top production in
  association with a Z boson at the LHC}},  {\em Phys.Rev.} {\bf D87} (2013)
  114006, [\href{http://xxx.lanl.gov/abs/1302.3856}{{\tt arXiv:1302.3856}}].

\bibitem{Dittmaier:2011ti}
{\bf LHC Higgs Cross Section Working Group} Collaboration, S.~Dittmaier et~al.,
  {\it {Handbook of LHC Higgs Cross Sections: 1. Inclusive Observables}},
  \href{http://xxx.lanl.gov/abs/1101.0593}{{\tt arXiv:1101.0593}}.

\bibitem{Dittmaier:2012vm}
S.~Dittmaier, S.~Dittmaier, C.~Mariotti, G.~Passarino, R.~Tanaka, et~al., {\it
  {Handbook of LHC Higgs Cross Sections: 2. Differential Distributions}},
  \href{http://xxx.lanl.gov/abs/1201.3084}{{\tt arXiv:1201.3084}}.

\bibitem{Heinemeyer:2013tqa}
{\bf LHC Higgs Cross Section Working Group} Collaboration, S.~Heinemeyer
  et~al., {\it {Handbook of LHC Higgs Cross Sections: 3. Higgs Properties}},
  \href{http://xxx.lanl.gov/abs/1307.1347}{{\tt arXiv:1307.1347}}.

\bibitem{Cullen:2013saa}
G.~Cullen, H.~van Deurzen, N.~Greiner, G.~Luisoni, P.~Mastrolia, et~al., {\it
  {Next-to-Leading-Order QCD Corrections to Higgs Boson Production Plus Three
  Jets in Gluon Fusion}},  {\em Phys.Rev.Lett.} {\bf 111} (2013), no.~13
  131801, [\href{http://xxx.lanl.gov/abs/1307.4737}{{\tt arXiv:1307.4737}}].

\bibitem{vanDeurzen:2013xla}
H.~van Deurzen, G.~Luisoni, P.~Mastrolia, E.~Mirabella, G.~Ossola, et~al., {\it
  {Next-to-Leading-Order QCD Corrections to Higgs Boson Production in
  Association with a Top Quark Pair and a Jet}},  {\em Phys.Rev.Lett.} {\bf
  111} (2013), no.~17 171801, [\href{http://xxx.lanl.gov/abs/1307.8437}{{\tt
  arXiv:1307.8437}}].

\bibitem{Baglio:2012np}
J.~Baglio, A.~Djouadi, R.~Gr{\"o}ber, M.~M{\"u}hlleitner, J.~Quevillon, et~al.,
  {\it {The measurement of the Higgs self-coupling at the LHC: theoretical
  status}},  {\em JHEP} {\bf 1304} (2013) 151,
  [\href{http://xxx.lanl.gov/abs/1212.5581}{{\tt arXiv:1212.5581}}].

\bibitem{Plehn:1996wb}
T.~Plehn, M.~Spira, and P.~Zerwas, {\it {Pair production of neutral Higgs
  particles in gluon-gluon collisions}},  {\em Nucl.Phys.} {\bf B479} (1996)
  46--64, [\href{http://xxx.lanl.gov/abs/hep-ph/9603205}{{\tt
  hep-ph/9603205}}].

\bibitem{Dawson:1998py}
S.~Dawson, S.~Dittmaier, and M.~Spira, {\it {Neutral Higgs boson pair
  production at hadron colliders: QCD corrections}},  {\em Phys.Rev.} {\bf D58}
  (1998) 115012, [\href{http://xxx.lanl.gov/abs/hep-ph/9805244}{{\tt
  hep-ph/9805244}}].

\bibitem{Signer:1996bf}
A.~Signer and L.~J. Dixon, {\it {Electron - positron annihilation into four
  jets at next-to-leading order in $\alpha_s$}},  {\em Phys.Rev.Lett.} {\bf 78}
  (1997) 811--814, [\href{http://xxx.lanl.gov/abs/hep-ph/9609460}{{\tt
  hep-ph/9609460}}].

\bibitem{Signer:1997dm}
A.~Signer, {\it {Next-to-leading order corrections to $e^+ e^- \to$ four
  jets}},  \href{http://xxx.lanl.gov/abs/hep-ph/9705218}{{\tt hep-ph/9705218}}.

\bibitem{Nagy:1997mf}
Z.~Nagy and Z.~Trocsanyi, {\it {Four jet production in $e^+e^-$ annihilation at
  next-to-leading order}},  {\em Nucl.Phys.Proc.Suppl.} {\bf 64} (1998) 63--67,
  [\href{http://xxx.lanl.gov/abs/hep-ph/9708344}{{\tt hep-ph/9708344}}].

\bibitem{Nagy:1997yn}
Z.~Nagy and Z.~Trocsanyi, {\it {Next-to-leading order calculation of four jet
  shape variables}},  {\em Phys.Rev.Lett.} {\bf 79} (1997) 3604--3607,
  [\href{http://xxx.lanl.gov/abs/hep-ph/9707309}{{\tt hep-ph/9707309}}].

\bibitem{Frederix:2010ne}
R.~Frederix, S.~Frixione, K.~Melnikov, and G.~Zanderighi, {\it {NLO QCD
  corrections to five-jet production at LEP and the extraction of
  $\alpha_s(M_Z)$}},  {\em JHEP} {\bf 1011} (2010) 050,
  [\href{http://xxx.lanl.gov/abs/1008.5313}{{\tt arXiv:1008.5313}}].

\bibitem{Bilenky:1994ad}
M.~S. Bilenky, G.~Rodrigo, and A.~Santamaria, {\it {Three jet production at LEP
  and the bottom quark mass}},  {\em Nucl.Phys.} {\bf B439} (1995) 505--535,
  [\href{http://xxx.lanl.gov/abs/hep-ph/9410258}{{\tt hep-ph/9410258}}].

\bibitem{Schmidt:1995mr}
C.~R. Schmidt, {\it {Top quark production and decay at next-to-leading order in
  $e^+e^-$ annihilation}},  {\em Phys.Rev.} {\bf D54} (1996) 3250--3265,
  [\href{http://xxx.lanl.gov/abs/hep-ph/9504434}{{\tt hep-ph/9504434}}].

\bibitem{Oleari:1997az}
C.~Oleari, {\it {Next-to-leading order corrections to the production of heavy
  flavor jets in $e^+e^-$ collisions}},
  \href{http://xxx.lanl.gov/abs/hep-ph/9802431}{{\tt hep-ph/9802431}}.

\bibitem{Nason:1997nw}
P.~Nason and C.~Oleari, {\it {Next-to-leading order corrections to the
  production of heavy flavor jets in $e^+e^-$ collisions}},  {\em Nucl.Phys.}
  {\bf B521} (1998) 237--273,
  [\href{http://xxx.lanl.gov/abs/hep-ph/9709360}{{\tt hep-ph/9709360}}].

\bibitem{Bernreuther:1997jn}
W.~Bernreuther, A.~Brandenburg, and P.~Uwer, {\it {Next-to-leading order QCD
  corrections to three jet cross-sections with massive quarks}},  {\em
  Phys.Rev.Lett.} {\bf 79} (1997) 189--192,
  [\href{http://xxx.lanl.gov/abs/hep-ph/9703305}{{\tt hep-ph/9703305}}].

\bibitem{Brandenburg:1997pu}
A.~Brandenburg and P.~Uwer, {\it {Next-to-leading order QCD corrections and
  massive quarks in $e^+ e^- \to$ three jets}},  {\em Nucl.Phys.} {\bf B515}
  (1998) 279--320, [\href{http://xxx.lanl.gov/abs/hep-ph/9708350}{{\tt
  hep-ph/9708350}}].

\bibitem{Brandenburg:1999gm}
A.~Brandenburg, {\it {The Reaction $e^+ e^- \to t \bar{t} g$ at next-to-leading
  order in $\alpha_s$}},  \href{http://xxx.lanl.gov/abs/hep-ph/9908383}{{\tt
  hep-ph/9908383}}.

\bibitem{Dittmaier:1998dz}
S.~Dittmaier, M.~Kramer, Y.~Liao, M.~Spira, and P.~Zerwas, {\it {Higgs
  radiation off top quarks in $e^+e^-$ collisions}},  {\em Phys.Lett.} {\bf
  B441} (1998) 383--388, [\href{http://xxx.lanl.gov/abs/hep-ph/9808433}{{\tt
  hep-ph/9808433}}].

\bibitem{Frixione:2013mta}
S.~Frixione, P.~Torrielli, and M.~Zaro, {\it {Higgs production through
  vector-boson fusion at the NLO matched with parton showers}},  {\em
  Phys.Lett.} {\bf B726} (2013) 273--282,
  [\href{http://xxx.lanl.gov/abs/1304.7927}{{\tt arXiv:1304.7927}}].

\bibitem{ATLASpttop}
{\bf ATLAS} Collaboration {\em ~} {\bf ATLAS-CONF-2013-099} ~.

\bibitem{Chatrchyan:2012saa}
{\bf CMS} Collaboration, S.~Chatrchyan et~al., {\it {Measurement of
  differential top-quark pair production cross sections in $pp$ colisions at
  $\sqrt{s}=7$ TeV}},  {\em Eur.Phys.J.} {\bf C73} (2013) 2339,
  [\href{http://xxx.lanl.gov/abs/1211.2220}{{\tt arXiv:1211.2220}}].

\bibitem{CMSpttop27}
{\bf CMS} Collaboration {\em ~} {\bf CMS-TOP-12-027} ~.

\bibitem{CMSpttop28}
{\bf CMS} Collaboration {\em ~} {\bf CMS-TOP-12-028} ~.

\bibitem{ATLAS:2012al}
{\bf ATLAS} Collaboration, G.~Aad et~al., {\it {Measurement of $t \bar{t}$
  production with a veto on additional central jet activity in pp collisions at
  $\sqrt{s}=7$ TeV using the ATLAS detector}},  {\em Eur.Phys.J.} {\bf C72}
  (2012) 2043, [\href{http://xxx.lanl.gov/abs/1203.5015}{{\tt
  arXiv:1203.5015}}].

\bibitem{CMSgf7tev}
{\bf CMS} Collaboration {\em ~} {\bf TOP-12-023} ~.

\bibitem{CMSgf8tev}
{\bf CMS} Collaboration {\em ~} {\bf TOP-12-041} ~.

\bibitem{Dittmaier:2001ay}
S.~Dittmaier and M.~Kramer, {\it {Electroweak radiative corrections to W boson
  production at hadron colliders}},  {\em Phys.Rev.} {\bf D65} (2002) 073007,
  [\href{http://xxx.lanl.gov/abs/hep-ph/0109062}{{\tt hep-ph/0109062}}].

\bibitem{Denner:1991kt}
A.~Denner, {\it {Techniques for calculation of electroweak radiative
  corrections at the one loop level and results for W physics at LEP-200}},
  {\em Fortsch.Phys.} {\bf 41} (1993) 307--420,
  [\href{http://xxx.lanl.gov/abs/0709.1075}{{\tt arXiv:0709.1075}}].

\bibitem{Beenakker:2002nc}
W.~Beenakker, S.~Dittmaier, M.~Kramer, B.~Plumper, M.~Spira, et~al., {\it {NLO
  QCD corrections to $t \bar{t} H$ production in hadron collisions}},  {\em
  Nucl.Phys.} {\bf B653} (2003) 151--203,
  [\href{http://xxx.lanl.gov/abs/hep-ph/0211352}{{\tt hep-ph/0211352}}].

\bibitem{Whalley:2005nh}
M.~Whalley, D.~Bourilkov, and R.~Group, {\it {The Les Houches accord PDFs
  (LHAPDF) and LHAGLUE}},  \href{http://xxx.lanl.gov/abs/hep-ph/0508110}{{\tt
  hep-ph/0508110}}.

\bibitem{Cacciari:2008zb}
M.~Cacciari, S.~Frixione, M.~L. Mangano, P.~Nason, and G.~Ridolfi, {\it
  {Updated predictions for the total production cross sections of top and of
  heavier quark pairs at the Tevatron and at the LHC}},  {\em JHEP} {\bf 0809}
  (2008) 127, [\href{http://xxx.lanl.gov/abs/0804.2800}{{\tt
  arXiv:0804.2800}}].

\bibitem{Ball:2010de}
R.~D. Ball, L.~Del~Debbio, S.~Forte, A.~Guffanti, J.~I. Latorre, et~al., {\it
  {A first unbiased global NLO determination of parton distributions and their
  uncertainties}},  {\em Nucl.Phys.} {\bf B838} (2010) 136--206,
  [\href{http://xxx.lanl.gov/abs/1002.4407}{{\tt arXiv:1002.4407}}].

\bibitem{Soper:2012pb}
D.~E. Soper and M.~Spannowsky, {\it {Finding top quarks with shower
  deconstruction}},  {\em Phys.Rev.} {\bf D87} (2013), no.~5 054012,
  [\href{http://xxx.lanl.gov/abs/1211.3140}{{\tt arXiv:1211.3140}}].

\bibitem{Soper:2014rya}
D.~E. Soper and M.~Spannowsky, {\it {Finding physics signals with event
  deconstruction}},  \href{http://xxx.lanl.gov/abs/1402.1189}{{\tt
  arXiv:1402.1189}}.

\bibitem{vanHameren:2010cp}
A.~van Hameren, {\it {OneLOop: For the evaluation of one-loop scalar
  functions}},  {\em Comput.Phys.Commun.} {\bf 182} (2011) 2427--2438,
  [\href{http://xxx.lanl.gov/abs/1007.4716}{{\tt arXiv:1007.4716}}].

\bibitem{Ellis:2007qk}
R.~K. Ellis and G.~Zanderighi, {\it {Scalar one-loop integrals for QCD}},  {\em
  JHEP} {\bf 0802} (2008) 002, [\href{http://xxx.lanl.gov/abs/0712.1851}{{\tt
  arXiv:0712.1851}}].

\bibitem{Denner:2005nn}
A.~Denner and S.~Dittmaier, {\it {Reduction schemes for one-loop tensor
  integrals}},  {\em Nucl. Phys.} {\bf B734} (2006) 62--115,
  [\href{http://xxx.lanl.gov/abs/hep-ph/0509141}{{\tt hep-ph/0509141}}].

\bibitem{Tarasov:1996br}
O.~Tarasov, {\it {Connection between Feynman integrals having different values
  of the space-time dimension}},  {\em Phys.Rev.} {\bf D54} (1996) 6479--6490,
  [\href{http://xxx.lanl.gov/abs/hep-th/9606018}{{\tt hep-th/9606018}}].

\bibitem{Duplancic:2003tv}
G.~Duplancic and B.~Nizic, {\it {Reduction method for dimensionally regulated
  one-loop N-point Feynman integrals}},  {\em Eur.Phys.J.} {\bf C35} (2004)
  105--118, [\href{http://xxx.lanl.gov/abs/hep-ph/0303184}{{\tt
  hep-ph/0303184}}].

\bibitem{Cacciari:2011ma}
M.~Cacciari, G.~P. Salam, and G.~Soyez, {\it {FastJet User Manual}},  {\em
  Eur.Phys.J.} {\bf C72} (2012) 1896,
  [\href{http://xxx.lanl.gov/abs/1111.6097}{{\tt arXiv:1111.6097}}].

\bibitem{Kleiss:1985gy}
R.~Kleiss, W.~J. Stirling, and S.~Ellis, {\it {A New Monte Carlo Treatment of
  Multiparticle Phase Space at High-energies}},  {\em Comput.Phys.Commun.} {\bf
  40} (1986) 359.

\bibitem{StdHEP}
L.~Garren and P.~Lebrun, {\it {StdHEP v5 manual}},
  \href{http://xxx.lanl.gov/abs/http://cepa.fnal.gov/psm/stdhep/}{{\tt
  http://cepa.fnal.gov/psm/stdhep/}}.

\bibitem{Lepage:1977sw}
G.~P. Lepage, {\it {A New Algorithm for Adaptive Multidimensional
  Integration}},  {\em J.Comput.Phys.} {\bf 27} (1978) 192.

\bibitem{Lepage:1980dq}
G.~P. Lepage, {\em {VEGAS: an adaptive multidimentional integration program}}.
\newblock CLNS-80/447, 1980.

\end{thebibliography}\endgroup

\end{document}